\documentclass[aps,prd,twocolumn,superscriptaddress,nofootinbib]{revtex4-2}
\usepackage[colorlinks=true,citecolor=red,urlcolor=blue]{hyperref}
\usepackage{amssymb}
\usepackage{amsmath,color}
\usepackage{graphicx}
\usepackage{bbm}
\usepackage{verbatim}
\usepackage[T1]{fontenc}
\usepackage[utf8]{inputenc}
\usepackage[normalem]{ulem}

\usepackage{url}
\usepackage{xcolor}
\usepackage{caption}
\usepackage{subcaption}

\usepackage{diagbox}
\usepackage{color}
\usepackage{multirow}
\usepackage{hhline}
\usepackage{tabularx}

\newcommand{\orcid}[1]{\href{https://orcid.org/#1}{\hspace{0.5mm}\raisebox{-0.5ex}{\includegraphics[height=2.0ex]{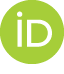}}}}

\begin{document}

%\preprint{}

\title{Parameter estimation of eccentric gravitational waves with a decihertz observatory and its cosmological implications}


\author{Tao Yang\orcid{0000-0002-2161-0495}}
\email[]{yangtao.lighink@gmail.com}

%\thanks{}
%\affiliation{}
\affiliation{Center for the Gravitational-Wave Universe, Astronomy Program Department of Physics and Astronomy, Seoul National University, 1 Gwanak-ro, Gwanak-gu, Seoul 08826, Korea}

\author{Rong-Gen Cai}
\email[]{cairg@itp.ac.cn}
\affiliation{CAS Key Laboratory of Theoretical Physics, Institute of Theoretical Physics, Chinese Academy of Sciences, Beijing 100190, China}
\affiliation{School of Physical Sciences, University of Chinese Academy of Sciences, No. 19A Yuquan Road, Beijing 100049, China}
\affiliation{School of Fundamental Physics and Mathematical Sciences, Hangzhou Institute for Advanced Study (HIAS), University of Chinese Academy of Sciences, Hangzhou 310024, China}

\author{Zhoujian Cao\orcid{0000-0002-1932-7295}}
\email[]{zjcao@bnu.edu.cn}
\affiliation{Department of Astronomy, Beijing Normal University, Beijing 100875, China}
\affiliation{School of Fundamental Physics and Mathematical Sciences, Hangzhou Institute for Advanced Study (HIAS), University of Chinese Academy of Sciences, Hangzhou 310024, China}

\author{Hyung Mok Lee\orcid{0000-0003-4412-7161}}
\email[]{hmlee@snu.ac.kr}
\affiliation{Center for the Gravitational-Wave Universe, Astronomy Program Department of Physics and Astronomy, Seoul National University, 1 Gwanak-ro, Gwanak-gu, Seoul 08826, Korea}

\date{\today}

\begin{abstract}
Eccentricity of compact binaries can improve the parameter estimation of gravitational waves (GWs), which is due to the fact that the multiple harmonics induced by eccentricity can provide more information and break the degeneracy between waveform parameters. In this paper, we first investigate the parameter estimation of eccentric GWs with decihertz observatory. We consider two scenarios for the configuration of DECIGO, i.e., the one cluster of DECIGO with its design sensitivity and B-DECIGO which also has one cluster but with inferior sensitivity as a comparison. We adopt the Fisher matrix to estimate the parameter errors.  By mocking up the typical binaries in GWTC-3, we find a nonvanishing eccentricity can significantly improve the estimation for almost all waveform parameters. In particular, the localization of typical binary black holes (BBH) can achieve $\mathcal{O}(10-10^{3.5})$ factors of improvement when the initial eccentricity $e_0=0.4$ at 0.1 Hz. The precise localization of binary neutron stars (BNS) and neutron star--black hole binaries (NSBH), together with the large improvement of localization of BBH from eccentricity in the mid-band, inspire us to construct the catalogs of golden dark sirens whose host galaxies can be uniquely identified. We find that with only one cluster of DECIGO running 1 year in its design sensitivity,  hundreds of golden dark BNS, NSBH, and tens of golden dark BBH can be observed. Eccentricity can greatly increase the population of golden dark BBH from $\sim 7~(e_0=0)$ to $\sim 65~(e_0=0.2)$. Such an increase of population of golden dark BBH events can improve the precision of Hubble constant measurement from  2.06\% to 0.68\%, matter density parameter from 64\% to 16\% in $\Lambda$CDM model. Through the phenomenological parameterization of GW propagation, the constraints of modified gravity can be improved from 6.2\% to 1.6\%. Our results show the remarkable significance of eccentricity for the detection and parameter estimation of GW events, allowing us to probe the Universe precisely.
\end{abstract}

\maketitle

\section{Introduction}

The gravitational waves (GWs) detected by LIGO-Virgo-KAGAR Collaboration have been widely applied to research on cosmology, astrophysics, and fundamental physics~\cite{LIGOScientific:2016lio,LIGOScientific:2016vpg,LIGOScientific:2017adf,LIGOScientific:2018cki,LIGOScientific:2018cki,LIGOScientific:2021aug,LIGOScientific:2021psn,LIGOScientific:2021sio}. After the third observing run (O3), the LIGO-Virgo-KAGAR collaboration released the third Gravitational-Wave Transient Catalog (GWTC-3), bringing the total number of GW events~\cite{LIGOScientific:2021djp} to 90. The events in GWTC-3 are dominated by the binary black holes (BBH), together with a few binary neutron stars (BNS) and neutron star–black hole binaries (NSBH). In particular, the observation of the GW170817 from a BNS merger~\cite{LIGOScientific:2017vwq} and its associated electromagnetic (EM) counterparts~\cite{LIGOScientific:2017ync,LIGOScientific:2017zic} announced the era of GW multi-messenger astronomy. GW170817 with its EM counterparts provided the first standard siren measurement of cosmic expansion history~\cite{LIGOScientific:2017adf}, which is independent from the traditional EM experiments such as the cosmic microwave background (CMB)~\cite{Planck:2018vyg}, baryon acoustic oscillations (BAO)~\cite{BOSS:2016wmc}, type Ia supernovae (SNe Ia)~\cite{Riess:2016jrr,Riess:2019cxk}, and strong gravitational lensing~\cite{Wong:2019kwg}. 

GW standard siren is supposed to be one of the most promising probes (though currently not precise enough) to arbitrate the Hubble tension~\cite{Verde:2019ivm,Chen:2017rfc,Feeney:2018mkj,Borhanian:2020vyr} which arises from the $4.4\sigma$ discrepancy of the Hubble constant measurements between Planck~\cite{Planck:2018vyg} and SH0ES project~\cite{Riess:2019cxk}. Besides, GW standard sirens can be widely utilized for the study of cosmology, astrophysics, and fundamental physics~\cite{Dalal:2006qt,Cutler:2009qv,Sathyaprakash:2009xt,Zhao:2010sz,Cai:2016sby,Yang:2017bkv,Cai:2017aea,Belgacem:2017ihm,Belgacem:2019zzu,Yang:2021qge}. These applications of GW standard sirens are guaranteed by the fact that the luminosity distance can be directly inferred from the amplitude and shape of the waveform. Compared to the traditional standard candles--SNe Ia, the physics of GW standard sirens is very clear and there is no need for the calibration of distance~\cite{Schutz:1986gp}. However, the usage of standard sirens is drastically limited by the measurement of redshift of the GW sources, due to the mass-redshift degeneracy. Several techniques are proposed to obtain the redshift information. For sources with confirmed EM counterparts, the host galaxy and hence its redshift can be determined directly~\cite{Holz:2005df,Dalal:2006qt,Nissanke:2009kt}. These GWs associated with EM counterparts are dubbed as ``bright sirens''. For sources without the detection of EM counterparts, i.e., the ``dark sirens'', alternative techniques are needed to infer the source redshift. Many methods have been proposed such as adopting the astrophysically-motivated source mass distribution~\cite{Taylor:2012db,Farr:2019twy,You:2020wju,Mastrogiovanni:2021wsd}, counting all the potential host galaxies in the localized region and obtaining the statistical redshift information from the galaxy catalogs~\cite{Schutz:1986gp,DelPozzo:2011vcw,Nair:2018ign,LIGOScientific:2018gmd,DES:2019ccw,Gray:2019ksv,DES:2020nay,Borhanian:2020vyr,Finke:2021aom}, cross-correlating between GWs and galaxies~\cite{Oguri:2016dgk,Mukherjee:2019wcg,Mukherjee:2020hyn,Bera:2020jhx,Mukherjee:2022afz}, and including the tidal effects of BNS mergers to break the mass-redshift degeneracy~\cite{Messenger:2011gi,Li:2013via,Messenger:2013fya,DelPozzo:2015bna,Wang:2020xwn,Chatterjee:2021xrm,Jin:2022qnj,Dhani:2022ulg}. Compared to the bright sirens, the constraints of cosmological parameters from dark sirens are much looser due to the undetermined redshift information. For using the galaxy catalogs, the large uncertainty of GW localization makes it very hard to pinpoint the true host galaxy and hence its redshift. In addition, the measurement of distance suffers from the degeneracy between distance and orbital inclination. While, for bright sirens, not only the host galaxy can be identified, but also the degeneracy can be broken with the help of the EM counterparts~\cite{Hotokezaka:2018dfi}. This makes the uncertainty of the Hubble constant from one dark siren much worse than that from one bright siren~\cite{LIGOScientific:2018gmd,DES:2019ccw,DES:2020nay}. However, the majority of dark sirens can compensate for this inferiority. For instance, the constraint of the Hubble constant from 46 selected dark sirens in GWTC-3 with GLADE+ K--band galaxy catalog information is comparable to that from bright siren GW170817~\cite{LIGOScientific:2021aug}. It also finds that the better sky localization of GW190814 makes this event more informative on the value of $H_0$ in comparison to the other GW events. Currently, due to the rarity of bright sirens (only one in total 90 events) and large sky errors of dark sirens (typically $10^{3}~\rm deg^2$), GW standard sirens of LIGO-Virgo-KAGRA are not precise enough to arbitrate the Hubble tension. Therefore, in order to improve the usage of GWs for the study of cosmic expansion history and other cosmological problems, one should resort to either more GW events with EM counterparts, namely bright sirens, or the precise localization of dark sirens. Considering the great challenge for the detection of EM counterparts and the very small fraction of bright sirens even with the future GW detector networks~\cite{Belgacem:2019tbw,Yang:2021qge}, the improvement of parameter estimation (in particular for the distance and localization) of dark sirens is very crucial.

In this paper, we would like to extend our research in~\cite{Yang:2022tig} which demonstrated that the eccentricity of long inspiralling compact binaries can significantly improve the distance inference and source localization of GWs. The non-negligible eccentricity of compact binaries that emit GWs is suggested in many investigations, which may contribute observational features in the sensitivity band of ground and space-based detectors~\cite{Antonini:2012ad,Samsing:2013kua,Thompson:2010dp,East:2012xq}. Different mechanisms of the dynamic formation of the compact binaries of black holes and neutron stars have been proposed to study their eccentricities~\cite{Rodriguez:2017pec,Samsing:2017xmd,Samsing:2017oij,Samsing:2018ykz,Wen:2002km,Pratten:2020fqn,OLeary:2008myb,Lee:2009ca,Lee:1994nq,Hong:2015aba}. 
%{\bf You may add some of our papers in the reference: Lee 1995, MNRAS, 272,605; Hong & Lee 2015, MNRAS, 448, 754). }
Orbital eccentricity is one of the most important features to distinguish between isolated and dynamical BBH formation scenarios~\cite{Nishizawa:2016jji,Nishizawa:2016eza,Breivik:2016ddj,Zevin:2021rtf}. Some studies indicate that a fraction of the binaries possesses eccentricities larger than 0.1 at 10 Hz~\cite{Wen:2002km,Silsbee:2016djf,Antonini:2017ash,Liu:2019gdc}. In addition to the implication of formation channels, the imprint of eccentricity in the waveform can also help improving the parameter estimation of GWs. The improvements of parameter estimation and source localization by eccentricity have been investigated in~\cite{Sun:2015bva,Ma:2017bux,Pan:2019anf} for the stellar-mass compact binaries with the ground-based detector networks and in~\cite{Mikoczi:2012qy} for the supermassive black hole binaries with space-borne LISA. For the stellar-mass binaries with ground-based detectors that are sensitive to high frequencies ($> 10$ Hz), the authors found the improvements in source localization increase with the eccentricity and mass of the binaries. For the case of the $100~M_{\odot}$ total mass BBH, the improvement factor is about 2 in general when eccentricity $e_0$ increases from 0.0 to 0.4. While for low-mass binaries, the improvement is negligible when total mass is smaller than $40~M_{\odot}$ and the localization is even worsened at some orientations when total mass is smaller than $5~M_{\odot}$~\cite{Pan:2019anf}. For the supermassive BBH observed by LISA ($10^{-4}-0.1$ Hz), the authors also found the source localization improves with increasing eccentricity and mass. In the case of ($\sim10^7~M_{\odot}$) supermassive BBH, the angular resolution is improved by $\sim 1$ order of magnitude for highly eccentric sources ($e=0.6$). Intriguingly, our recent research shows that in the mid-band ($0.1-10$ Hz), the source localization of the typical stellar-mass BBHs can achieve much more improvements from the nonvanishing eccentricities (as much as 1--3 orders of magnitude when $e_0=0.4$)~\cite{Yang:2022tig}. The distance inference can also be improved by more than 2 orders of magnitude in the near face-on orientations. These results suggest that eccentricity is of great significance for dark sirens as precise probes of cosmology. In this paper, we would like to present in some detail the parameter estimation of GWs emitted by the eccentric compact binaries in the mid-band. This is to complement our results in~\cite{Yang:2022tig}. Then the cosmological implications will be investigated, which is to extend our previous research.  

The motivation for conducting this research in the mid-band is as follows. From current LIGO-Virgo-KAGRA detections, there is no strong evidence of eccentric compact binaries in the high-frequency band (>10 Hz)~\cite{LIGOScientific:2019dag,Romero-Shaw:2019itr,Nitz:2019spj,Wu:2020zwr}. The only exception is GW190521 which has been reported to be consistent with an eccentric BBH by separate teams~\cite{Romero-Shaw:2020thy,Gayathri:2020coq}. In~\cite{Romero-Shaw:2020thy}, the authors found GW190521 prefers a signal with eccentricity $e>0.1$ at 10 Hz, to a non-precessing, quasi-circular signal. However, for the lack of available waveforms which include both eccentricity and orbital precession, they found there is a degeneracy between non-spinning, moderately eccentric waveform and quasi-circular, precessing waveform. While, in~\cite{Gayathri:2020coq} the authors used the numerical relativity simulations which incorporate both procession and eccentricity, and found that GW190521 is consistent with a highly eccentric ($e=0.69^{+0.17}_{-0.22}$) merger at 90\% confidence level.  
%Considering the heavy component mass of GW190521 ($m_1=85,~m2=66~M_{\odot}$) and the strong evidence of nonvanishing eccentricity (and also precession), it is supported (very likely) to be formed dynamically. 
The scarcity of eccentric binaries in the LIGO/Virgo band can be ascribed to the damping of eccentricity (at leading order $\sim f^{-19/18}$) in the inspiral period of binaries, even if they are born with high eccentricity. It means that the probability of observing eccentric binaries is much higher in the mid-band. 
Another great advantage in the mid-band is the long inspiral period (days to years) of the stellar-mass binaries. The motion of the space-borne detector can induce the modulation and Doppler effects in the phase of the waveform which yields important angular information. In addition, the effects of eccentricity can be accumulated over a long period and may lead to more improvements in parameter estimation. In the aspect of waveform modeling, the inpiral-only and frequency-domain waveform based on post-Newtonian approximation is accurate enough for the stellar-mass compact binaries in the mid-band. Finally, considering the nonsignificant improvement of localization by eccentricity in the LIGO/Virgo band~\cite{Pan:2019anf} (bear in mind that the typical localization of LIGO/Virgo binaries is around $10^2-10^3$ deg$^2$), the investigation of this issue in the mid-band is very essential. 

In the mid-frequency band (0.1--10 Hz), the space-borne laser interferometers like DECIGO~\cite{Kawamura:2006up,Kawamura:2020pcg} and BBO~\cite{Harry:2006fi}, and the atom interferometers like MAGIS~\cite{Graham:2017pmn} and AEDGE~\cite{AEDGE:2019nxb} have been proposed. In this frequency band, the stellar-mass binary usually has a long inspiral period and the motion of the detectors could provide a very precise localization for the sources~\cite{Cutler:2009qv,Graham:2017lmg,Yang:2021xox,Liu:2022rvk,Yang:2022iwn}. In this paper, we adopt DECIGO as our fiducial detector in the mid-band. DECIGO consists of 4 clusters and each cluster has three satellites. Its pathfinder B-DECIGO with only one cluster is planned to be launched earlier than DECIGO in the mid-2030s. 
We consider two scenarios of DECIGO for comparison. The first is the one cluster of DECIGO (hereafter we call it ``DECIGO-I'') with its design sensitivity $4\times 10^{-24}$ Hz$^{-1/2}$ at 1 Hz~\cite{Kawamura:2020pcg}. The second is the B-DECIGO whose sensitivity is around $6\times 10^{-23}$ Hz$^{-1/2}$ at 1 Hz. The main difference between DECIGO-I and B-DECIGO is the sensitivity, for which the former is 15 times better than the latter. One of the information we would like to provide in this research is how much we can benefit if B-DECIGO is upgraded to just one cluster of DECIGO. Our results can demonstrate the potential of DECIGO with only one cluster, which is more promising (e.g. for the launch time and cost) than the full configuration of DECIGO in the future. 

The organization of this paper is as follows. In section~\ref{sec:typical} we focus on the parameter estimation of the GWs with various eccentricities in the mid-band. The main parameters we would like to investigate are luminosity distances and sky locations, from which we will also derive the 3-D localization volume for the GW sources. The improvements of the estimation for these parameters from eccentricity will be presented. In addition, the constraints of other parameters such as the chirp mass, inclination angle, and eccentricity will also be discussed. This is to complete and elaborate the results in our previous work~\cite{Yang:2022tig}.  In section~\ref{sec:catalog}, we will simulate the catalogs of GWs with DECIGO-I and B-DECIGO based on current knowledge and on the assumption of various eccentricities. We focus on the golden dark sirens whose host galaxies can be uniquely identified. From the catalogs, we can assess the influence of eccentricity on the detection of golden dark sirens. In section~\ref{sec:cosmo}, we will discuss the implications of eccentricity for the applications of dark sirens in cosmology. Finally, we conclude our results in section~\ref{sec:conclusion} followed by some discussions.

\section{Parameter estimation of the typical binaries \label{sec:typical}}

\subsection{The eccentric waveform}
We adopt the non-spinning, inspiral-only EccentricFD waveform approximant available in {\sc LALSuite}~\cite{lalsuite} and use {\sc PyCBC}~\cite{alex_nitz_2022_6324278} to generate the eccentric waveform. EccentricFD corresponds to the enhanced post-circular (EPC) model~\cite{Huerta:2014eca}. To the zeroth order in the eccentricity, the model recovers the TaylorF2 post-Newtonian (PN) waveform at 3.5 PN order~\cite{Buonanno:2009zt}. To the zeroth PN order, the model recovers the PC expansion of~\cite{Yunes:2009yz}, including eccentricity corrections up to order $\mathcal{O}(e^8)$. In this eccentric waveform we have 11 parameters $P=\{\mathcal{M}_c,\eta,d_L,\iota,\theta,\phi,\psi,t_c,\phi_c,e_0,\beta\}$, with $\mathcal{M}_c$ the chirp mass, $\eta$ the symmetric mass ratio, $d_L$ the luminosity distance, ($\theta,~\phi$) the sky location of the source, $\psi$ the polarization angle, ($t_c,~\phi_c$) the time and phase at merger. $e_0$ and $\beta$ are the two additional parameters to the TaylorF2 waveform. The former is the initial eccentricity defined at the frequency $f_0$ and the latter is the azimuthal component of inclination angles (longitude of ascending nodes axis). 

The eccentric waveform consists of multiple harmonics induced by the eccentricity of the orbit~\cite{Huerta:2014eca},
\begin{equation}
\tilde{h}(f)=-\sqrt{\frac{5}{384}}\frac{\mathcal{M}_c^{5/6}}{\pi^{2/3}d_L}f^{-7/6}\sum_{\ell=1}^{10}\xi_{\ell}\left(\frac{\ell}{2}\right)^{2/3}e^{-i\Psi_{\ell}} \,.
\end{equation} 
The phase of each harmonic $\Psi_{\ell}$ is
\begin{equation}
\Psi_{\ell}=2\pi f t_c-\ell \phi_c+\left(\frac{\ell}{2}\right)^{8/3}\frac{3}{128\eta v_{\rm ecc}^5}\sum_{n=0}^{7}a_n v_{\rm ecc}^n \,.
\end{equation} 
When $e_0=0$ it recovers the circular TaylorF2 with only the quadrupole mode ($\ell=2$). $\xi_{\ell}$ are functions of $e_0$ and angular parameters $P_{\rm ang}=\{\iota,~\theta,~\phi,~\psi,~\beta\}$~\cite{Yunes:2009yz}.
$v_{\rm ecc}$ is the modified velocity (relative to the orbit velocity $v=(\pi M f)^{1/3}$ when $e=0$) which is a function of eccentricity, $v_{\rm ecc}(f;e_0)=g(f;e_0)(\pi M f)^{1/3}$. The function $g(f;e_0)$ is expanded to $e_0^8$ and its specific form can be found in Eq. (13) of~\cite{Huerta:2014eca}.  
$a_n$ is the corresponding coefficient of the 3.5 PN expansion~\cite{Buonanno:2009zt}. The waveform keeps up to 10 harmonics, which corresponds to a consistent expansion in the eccentricity to $\mathcal{O}(e^8)$ both in the amplitude and in the phase~\cite{Yunes:2009yz}. Eccentricity induces more harmonics to the waveform other than the dominant quadrupole mode. Multiple harmonics make the distance and angular parameters nontrivially coupled, enabling us to break the degeneracy between these parameters. 

\subsection{The antenna response and observation of the harmonics}

In the mid-band, the motion of the space-borne detector in the long inspiral period should be taken into account. The frequency of each harmonic at a specific time (or orbital frequency $F$) is $\ell F$. In other words, the frequency of each harmonic $f$ corresponds to a different detector time, thus a different antenna response. Instead of the orbital frequency, we use the frequency of quadrupole mode $2F$ as the baseline frequency to derive the relation between time and GW frequency $t(f)$, by numerically solving the phase evolution of the eccentric orbits~\cite{Yunes:2009yz}. The orbits evolve faster with larger eccentricities (at a given orbital frequency). Thus the inspiral period (within the detector band) is shorter for a binary with greater eccentricity.
We first derive the antenna response functions $F_{+,\times}(t(f))$ in terms of the quadrupole frequency.  Then for each harmonic, its corresponding antenna response functions should be $F_{+,\times}(t(2f/\ell))$. Here $f$ is the frequency of the harmonic $\ell$ and we transform it to the quadrupole frequency to derive the corresponding time and antenna response. Clearly, we can see the higher harmonics enter the detector band earlier, which could provide more angular information. 

We follow~\cite{Rubbo:2003ap} for the modeling of the space-borne detectors, with the arm length $L=1000~(100)$ km for DECIGO-I (B-DECIGO). We set the initial observation time to be 1 year. Then the starting frequency of the quadrupole mode $f_{\rm start}$ can be calculated by setting $t_c-t(f_{\rm start})=1$ year. The starting frequency of each harmonic is $\ell f_{\rm start}/2$. So we should truncate the contribution of each harmonic before  its starting frequency~\footnote{In the mid-band, we do not need to truncate the GWs after the inner-most stable circular orbit (ISCO) since the frequencies of the typical stellar-mass binaries at ISCO are much higher than the upper limit (10 Hz) of the detector band.},
\begin{equation}
\tilde{h}_{\rm 1~yr}(f)=\tilde{h}(f)\mathcal{H}(2f-\ell f_{\rm start}) \,,
\label{eq:h1yr}
\end{equation}
with the unit step function
\begin{equation}
\mathcal{H}(x)=
\begin{cases}
1 & {\rm if}~x\geq0 \,, \\
0 & {\rm otherwise} \,.
\end{cases}
\end{equation}
Now we use $\tilde{h}_{\rm 1~yr}(f)$ to present the GW signal during the 1-year observation. 

\subsection{Parameter estimation using the Fisher information matrix }

To estimate the uncertainty and covariance of the parameters in the waveform, we adopt the Fisher matrix technique for GWs~\cite{Cutler:1994ys},
\begin{equation}
\Gamma_{ij}=\left(\frac{\partial h}{\partial P_i},\frac{\partial h}{\partial P_j}\right)\,,
\label{eq:Gamma}
\end{equation}
with $P_i$ being one of the 11 waveform parameters. Note in the circular case we only have 9 parameters excluding $e_0$ and $\beta$~\footnote{$\beta$ is meaningless in the circular case. We can still use eccentric waveform and set $e_0$ as a free parameter to constrain it in the circular case. However, in this paper, we focus on the improvement of parameter estimation of the eccentric case relative to the circular case. It means that we know the eccentricity {\it a prior} (though we still set eccentricity as a free parameter to be constrained in the eccentric case). We have checked  whether including $e_0$  in the circular case has no obvious influence on our results.}.  
The inner product is defined as
\begin{equation}
(a,b)=4\int_{f_{\rm min}}^{f_{\rm max}}\frac{\tilde{a}^*(f)\tilde{b}(f)+\tilde{b}^*(f)\tilde{a}(f)}{2 S_n(f)}df\,.
\label{eq:innerp}
\end{equation}
For the sensitivity $S_n(f)$ of B-DECIGO and DECIGO-I, 
we use the fitting formula in~\cite{Yagi:2011wg} but rescale it according to the white paper of DECIGO~\cite{Kawamura:2020pcg}. We choose $f_{\rm min}=0.1$ Hz and $f_{\rm max}=10$ Hz, corresponding to the frequency band of DECIGO. 
Note the meaning of  $f_{\rm min}$ is different from the starting frequency $f_{\rm start}$. The latter is just to ensure that we truncate the contribution of each harmonic (set them to be 0) before the 1-year observation period through Eq.~(\ref{eq:h1yr}). While $f_{\rm min}=0.1$ Hz is for the frequency window of the detector. It means we only integrate the GW signal inside DECIGO frequency band (0.1 -- 10 Hz). For example, if we set $f_{\rm start}=0.08$ Hz (note it is the starting frequency for $\ell=2$ mode), it means for $\ell=1,~2,~3...$ modes, their starting frequencies are 0.04, 0.08, 0.12 Hz...(we truncate the part of each modes before their starting frequency). However, the detector's frequency window starts from 0.1 Hz. So when we calculate the SNR, we only integrate from 0.1 Hz in the inner product. Since we truncate $\ell=3$ modes at 0.12 Hz, it means the integral of $\ell=3$ mode between 0.1 and 0.12 Hz is 0. 

The covariance matrix of the parameters is $C_{ij}=(\Gamma^{-1})_{ij}$, from which the uncertainty of each parameter $\Delta P_i=\sqrt{C_{ii}}$. The error of the sky localization is~\cite{Cutler:1997ta}
\begin{equation}
\Delta \Omega=2\pi |\sin(\theta)|\sqrt{C_{\theta\theta}C_{\phi\phi}-C_{\theta\phi}^2}\,.
\end{equation}
We calculate the partial derivatives $\partial \tilde{h}/\partial P_i$ numerically by $[\tilde{h}(f,P_i+dP_i)-\tilde{h}(f,P_i)]/dP_i$, with $dP_i=10^{-n}$.  For each parameter, we need to optimize $n$ to make the derivative converge so that the Fisher matrix calculation is reliable. To check the robustness of our methodology, we first adopt EccentricFD waveform with $e_0=0$ and check its consistency with the analytical TaylorF2 waveform. We find that with a proper choice of $n$ for the numerical derivative of each parameter in Eq.~(\ref{eq:Gamma}), the Fisher matrix calculations from the waveform generated by {\sc PyCBC} and from the analytical TaylorF2 are very consistent with each other. This consistency check paves the way for the usage of EccentricFD waveform with a nonvanishing eccentricity.

\subsection{Mocking up typical binaries in GWTC-3}

We mock up five types of typical events from GWTC-3~\cite{LIGOScientific:2021djp}, i.e., a GW170817-like BNS with $(m_1,m_2)=(1.46,1.27)~M_{\odot}$, a GW200105-like neutron star–black hole binary (NSBH) with $(9.0,1.91)~M_{\odot}$, a GW191129-like light-mass BBH with $(10.7,6.7)~M_{\odot}$, a GW150914-like medium-mass BBH with $(35.6,30.6)~M_{\odot}$, and a GW190426-like heavy-mass BBH with $(106.9,76.6)~M_{\odot}$. Note the light, medium, and heavy masses are in the context of the mass range of the stellar-mass BBH in GWTC-3. The redshift and luminosity distance of each binary are also chosen to be consistent with the real event of GWTC-3, by assuming a $\Lambda$CDM cosmology with $H_0=67.9~\rm km~s^{-1}~Mpc^{-1}$ and $\Omega_{\rm m}=0.3065$~\cite{LIGOScientific:2021djp}. Then the chirp mass $\mathcal{M}_c$ and symmetric mass ratio $\eta$ can be derived from the component masses. Note the parameters in the waveform are defined in the detector frame. To count the influence of the source orientation, we sample 1000 random sets of the angular parameters $P_{\rm ang}$ from the uniform and isotropic distribution. Note regarding the symmetry, we only sample the inclination angle $\iota\in [0, \pi/2]$, with the distribution $\cos\iota\sim \mathcal{U}[0,1]$.  Since EPC waveform is tested valid with initial eccentricity up to 0.4~\cite{Huerta:2014eca}, we assign seven discrete initial eccentricities from 0 to 0.4 at $f_0=0.1$ Hz, i.e, $e_0=0$, 0.01, 0.05, 0.1, 0.2, 0.3, and 0.4. As a rough estimation, eccentricity evolves with $e\sim f^{-19/18}$.  So even the largest eccentricity $e_0=0.4$ at 0.1 Hz should decrease to $e_0\sim0.004$ at 10 Hz, which is undetectable with current LIGO/Virgo sensitivity~\cite{Lower:2018seu}. It means that our choices of initial eccentricities at 0.1 Hz will not conflict with current observations at LIGO/Virgo band. Without loss of generality, we fix the coalescence time and phase to be $t_c=\phi_c=0$. We then totally have $5\times 7\times 1000=35000$ cases.

Since we set a limit of the observation time $\sim 1$ year, we need to choose the proper starting frequency $f_{\rm start}$ for these typical binaries. By calculating $t_c-t(f)\sim 1$ year in the circular case, we set $f_{\rm start}$ to be 0.2, 0.1, 0.059, 0.026, and 0.0105 Hz for the typical BNS, NSBH, light BBH, medium BBH, and heavy BBH, respectively. 
Note in the same frequency range, the evolving time of the binary is shorter with a larger eccentricity. However, for a typical binary with each $e_0$, we set the same starting frequency. This means that in all cases the inspiral periods of the binaries can meet the limit of the observation time ($\leq 1$ year).

\begin{figure}
\includegraphics[width=0.45\textwidth]{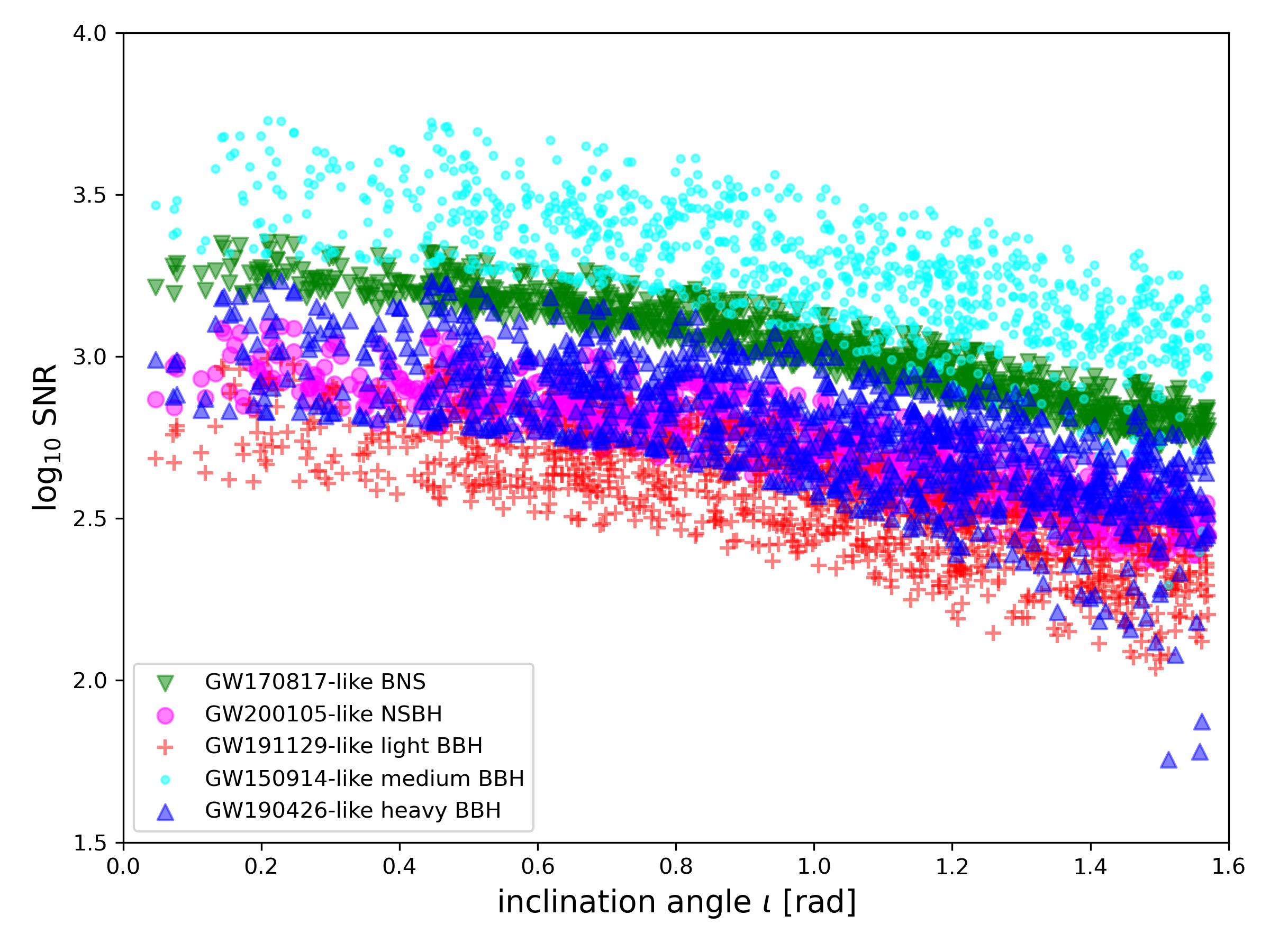} 
\caption{The SNR of the five typical binaries with DECIGO-I when $e_0=0$.}
\label{fig:SNR}
\end{figure}

The signal-to-noise ratio (SNR) of these typical binaries can be calculated by
\begin{equation}
\rho^2=(h_{\rm 1yr},h_{\rm 1yr}) \,.
\end{equation}
Figure~\ref{fig:SNR} shows the SNR of the five typical binaries in the circular case with DECIGO-I. The SNR decreases with increasing the inclination angle (remind that $\iota$ is defined in the range $\cos\iota\in \mathcal{U}[0,1]$). The corresponding SNR with B-DECIGO can be approximately estimated by rescaling the SNR with DECIGO-I by a factor of $1/15$. Note that the SNR of the heaviest BBH GW190426 is not the largest due to its large distance. For a typical binary with different eccentricities, we find the SNR does not change significantly.

\subsection{The inference of luminosity distance and source localization}

We calculate the Fisher matrix for the five typical binaries and collect the results of all the 35000 cases. We define the ratio
\begin{equation}
R_{\Delta P}=\frac{\Delta P|_{e_0={\rm nonzero}}}{\Delta P|_{e_0=0}} \,,
\end{equation} 
to indicate the improvement of the estimation of parameter $P$ induced by eccentricity relative to the quasi-circular case. $\Delta P$ is the error of the parameter $P$ derived from the Fisher matrix. If $R<1$, there is an improvement in the relevant parameter. A smaller $R$ indicates a  larger improvement. On the contrary, $R>1$ means the eccentricity will worsen the relevant parameter estimation. To assess the dependence of parameter estimation on the source's angular parameters, we will show the scatter plots of the error of the parameters against the inclination angle (orientation) $\iota$ of the sources. The reason for choosing $\iota$ as the representative of the angular parameters is that we find the results are most sensitive to the inclination angle. To see this, figure~\ref{fig:SNR} shows that the SNR decreases as $\iota$ increases. So, a larger $\iota$ would degrade the parameter estimation. In addition, the distance inference heavily depends on the inclination angle due to the degeneracy between these two parameters. To summarize the parameter estimation among the 1000 orientations, we define the minimum, mean, and maximum values of the quantity $x$ in the 1000 orientations as $\min(x)$, $\mathbb{E}(x)$, and $\max(x)$, respectively.

In this section, we choose GW170817-like BNS and GW150914-like medium BBH as the representatives and show their results. We only choose the cases with $e_0=0$, 0.1, and 0.4 to give a concise look of the figures. The complementary results of other typical binaries are shown in the appendix~\ref{app:A}. In this section, we mainly focus on the estimation of distance and localization since they are the key parameters for the identification of the host galaxies of GW dark sirens and their cosmological applications.

\begin{figure*}
\includegraphics[width=0.3\textwidth]{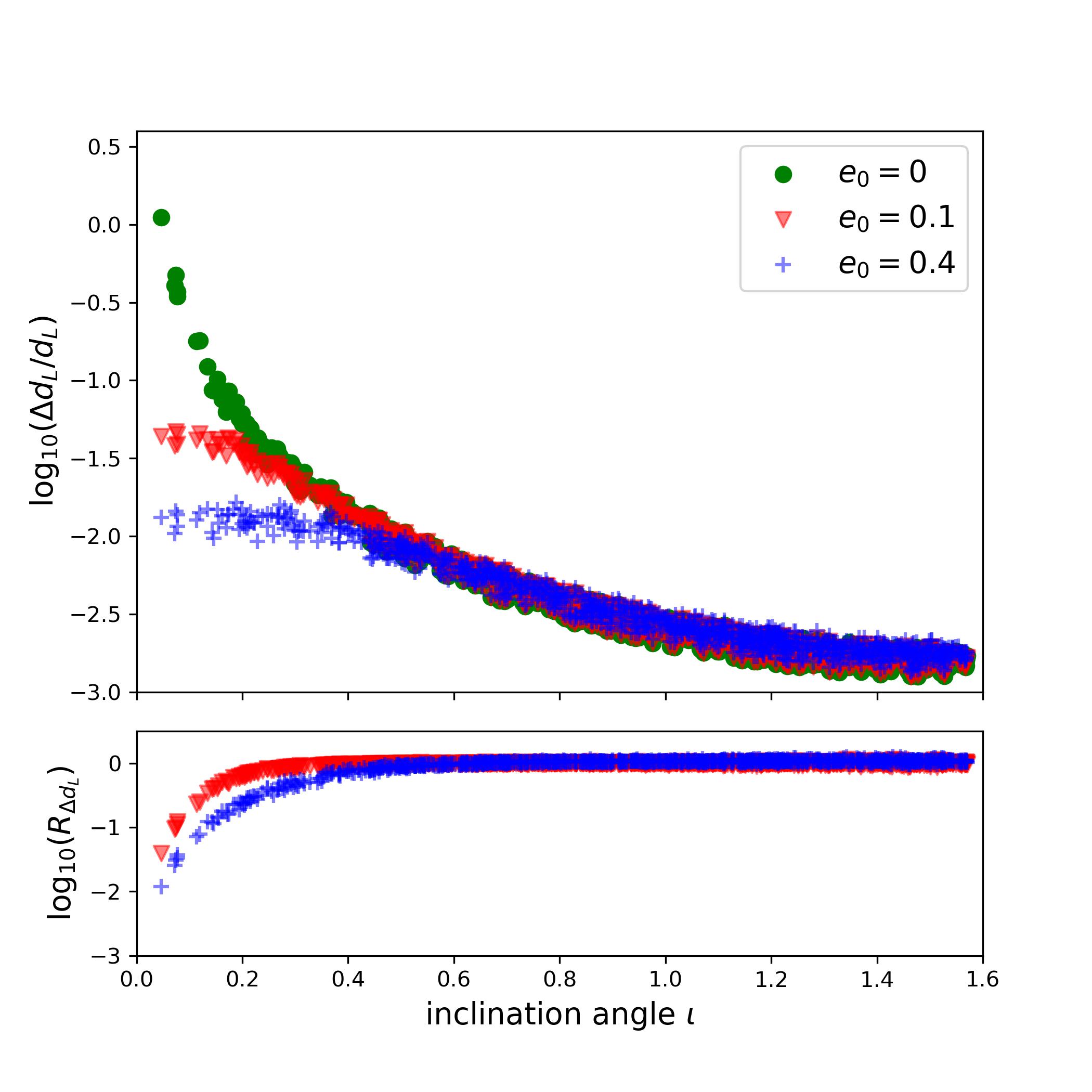}
\includegraphics[width=0.3\textwidth]{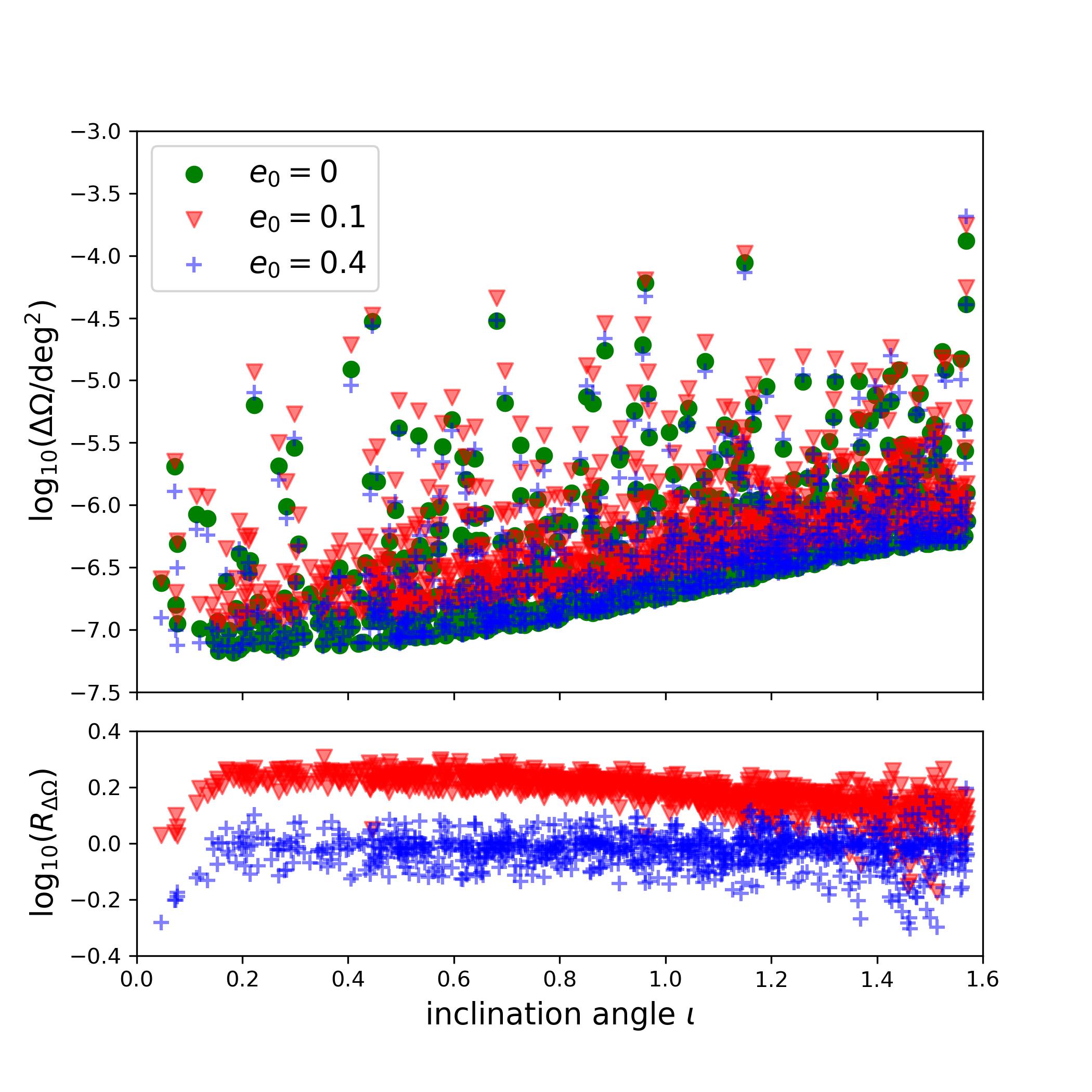}
\includegraphics[width=0.3\textwidth]{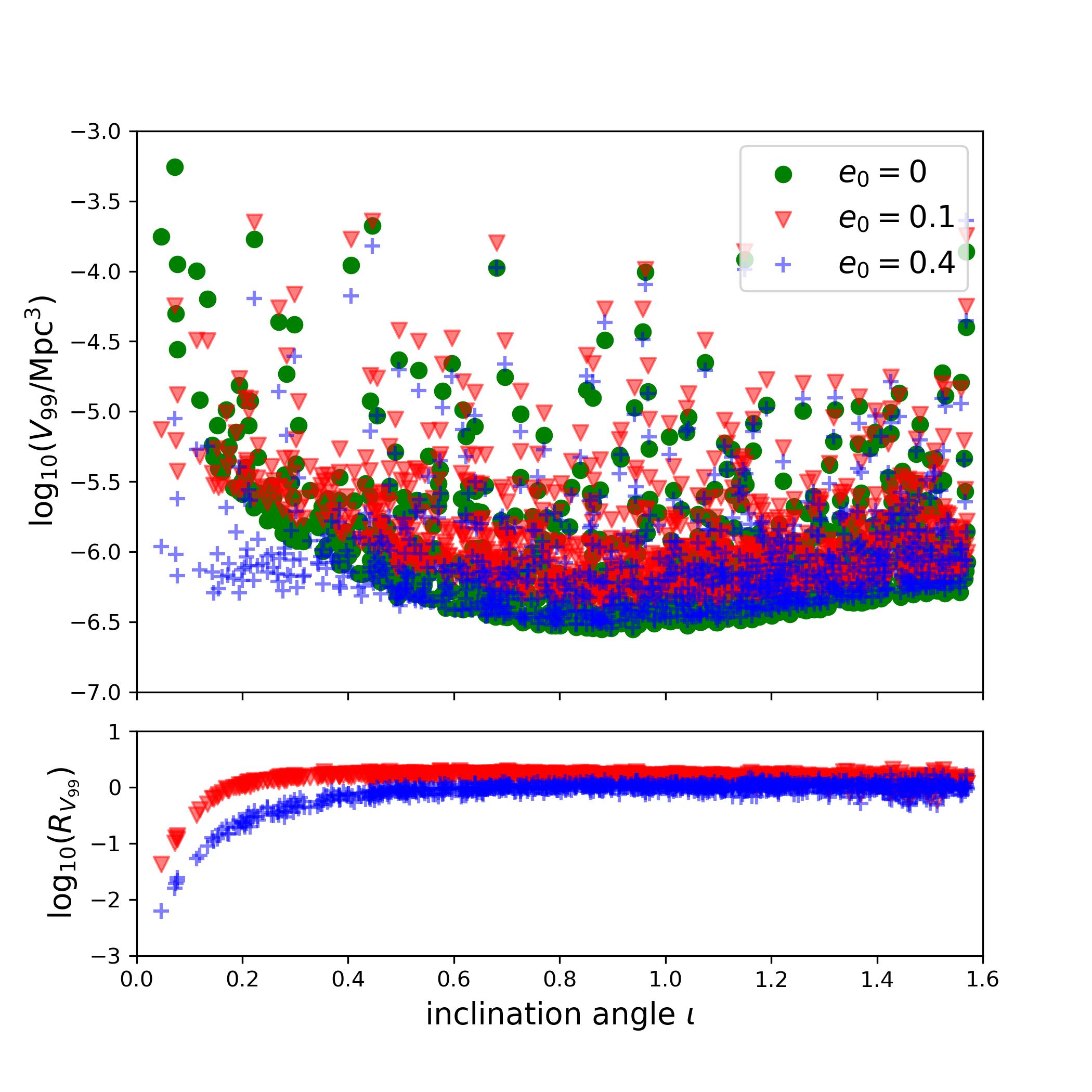}
\caption{The distance inference, sky localization, and 3-D localization volume of GW170817-like BNS with DECIGO-I. We also show the ratio $R$ of each parameter in the bottom panel.}
\label{fig:BNS_DI}
\end{figure*}

\begin{figure*}
\includegraphics[width=0.3\textwidth]{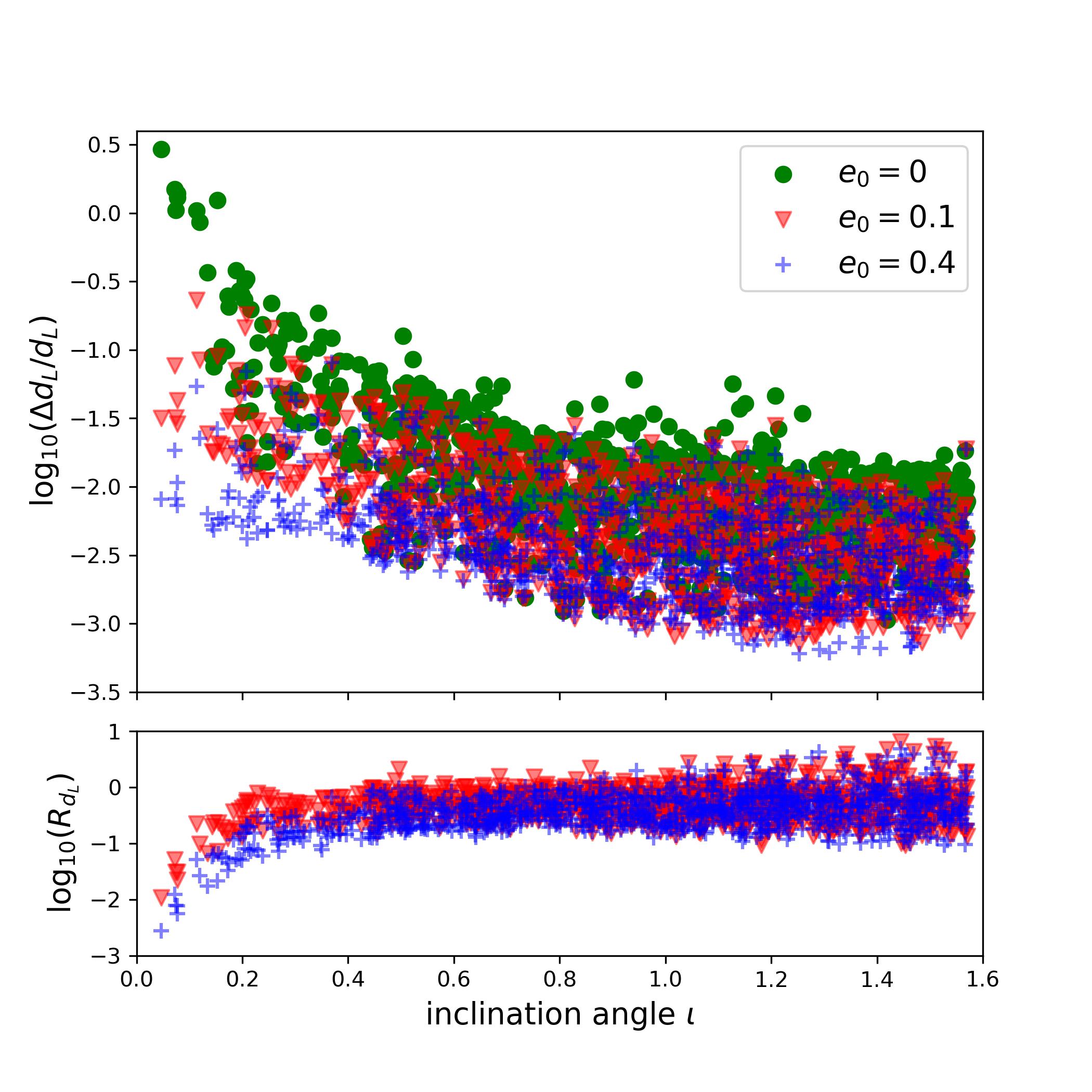}
\includegraphics[width=0.3\textwidth]{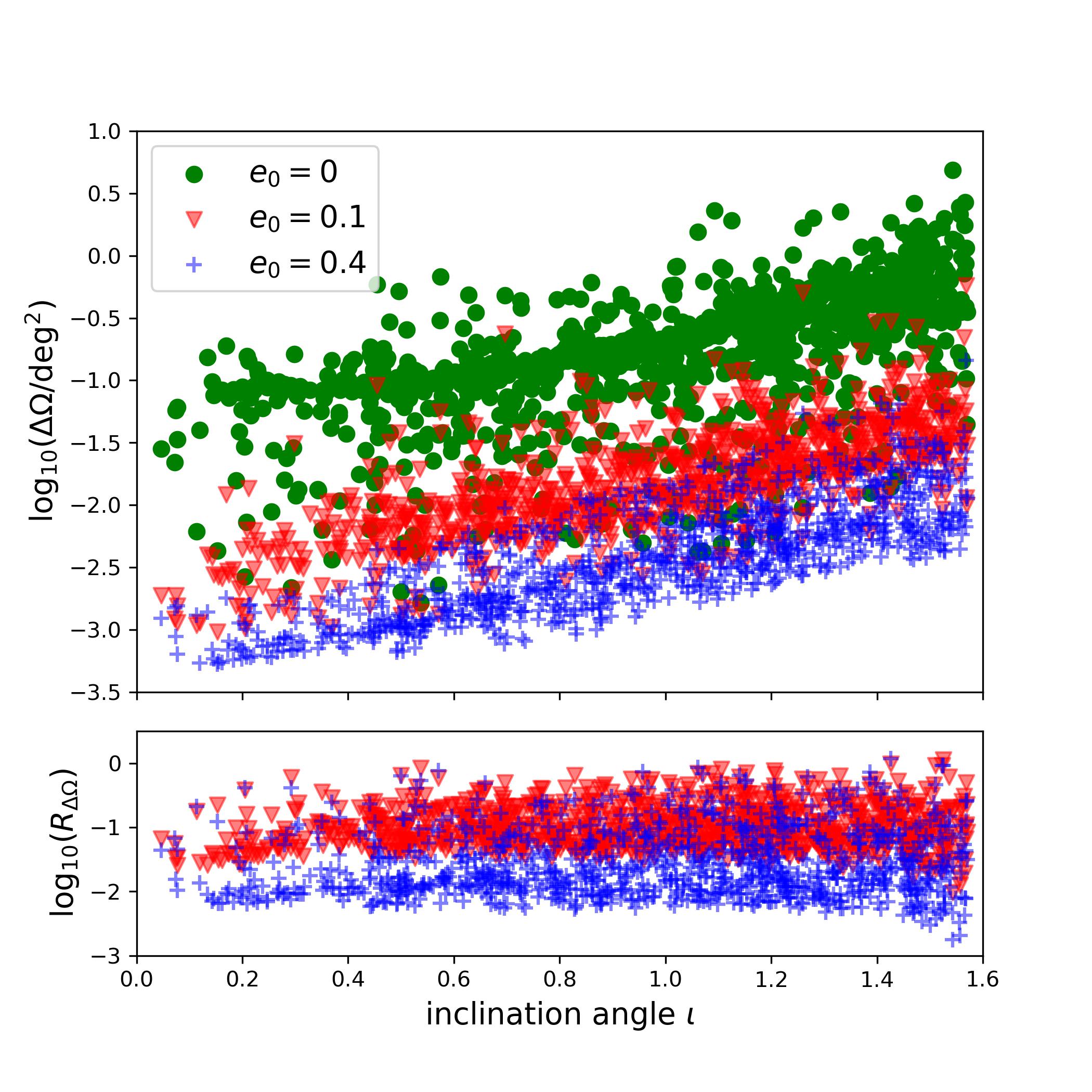}
\includegraphics[width=0.3\textwidth]{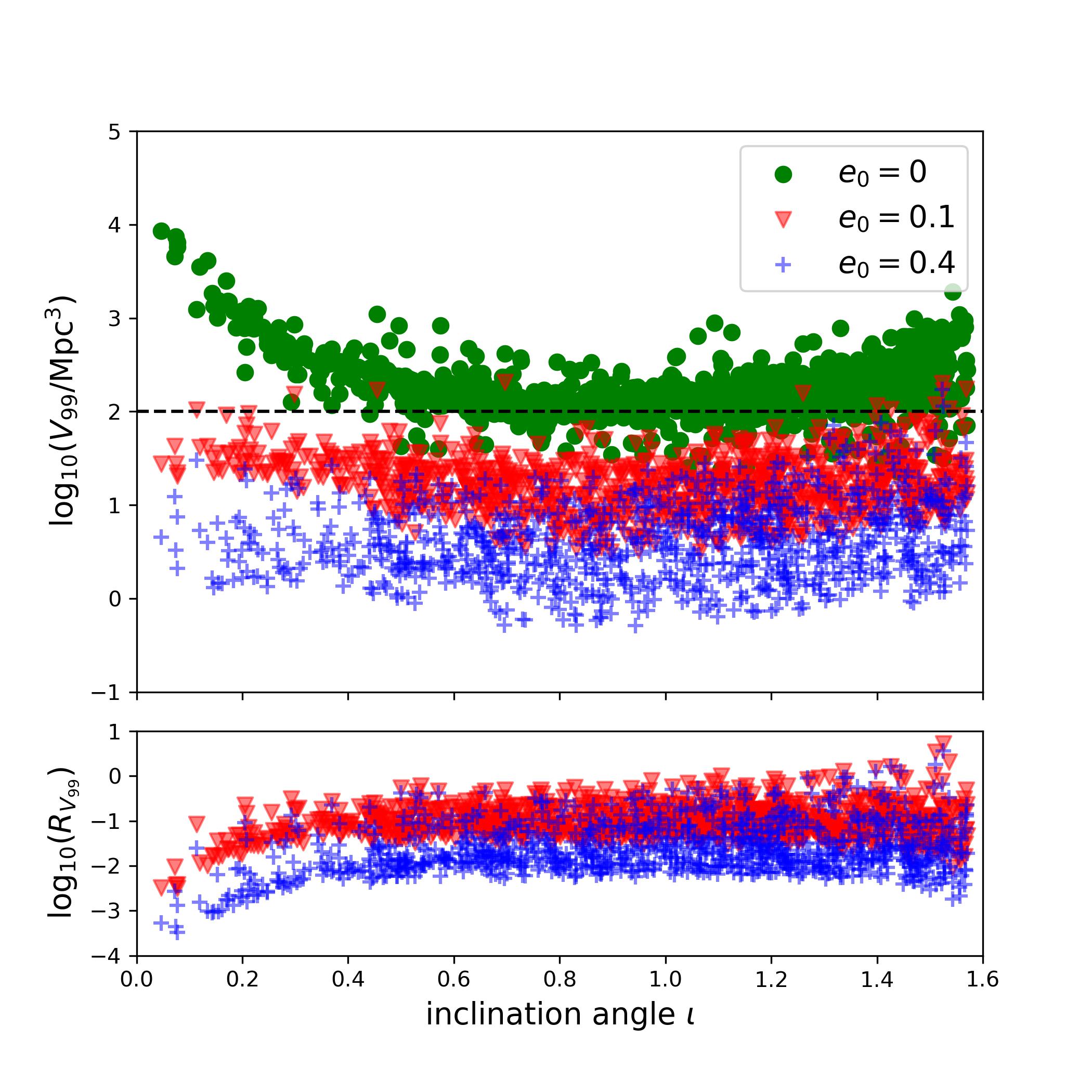}
\caption{The distance inference, sky localization, and 3-D localization volume of GW150914-like medium BBH with DECIGO-I. Note the dashed horizontal line in the right panel denotes the threshold volume below which there is only one potential host galaxy, by assuming a uniform number density of galaxy $n_g=0.01~\rm Mpc^{-3}$.}
\label{fig:BBHmedium_DI}
\end{figure*}

In figures~\ref{fig:BNS_DI} and~\ref{fig:BBHmedium_DI} we show the errors of luminosity distance $\Delta d_L/d_L$,  sky localization $\Delta \Omega$, and 3-dimensional (3-D) localization volume in 99\% confidence level (C.L.) $V_{99}$ against the orientation $\iota$ for GW170817-like BNS and GW150914-like medium BBH with DECIGO-I. We follow the method in~\cite{Yu:2020vyy} to transform $\Delta d_L$ and $\Delta \Omega$ (and the covariance between them) to the 99\% 3-D localization volume. We can clearly see the significant improvement of the distance inference by eccentricity in the near face-on (small $\iota$) orientations for both the BNS and medium BBH. As shown in the left panels of figures~\ref{fig:BNS_DI} and~\ref{fig:BBHmedium_DI}, the largest improvement is about 1.5 (2) orders of magnitude with $e_0=0.1~(0.4)$ for BNS and 2 (2.6) orders of magnitude with $e_0=0.1~(0.4)$ for medium BBH. The error of distance estimation is relatively smaller in larger $\iota$. For the BNS, the improvement of distance inference is almost negligible when $\iota$ is large. But for the medium BBH, there are still small improvements even in the near edge-on orientations. It suggests that the improvement is more distinct for a heavier binary with a larger eccentricity. 

For the sky localization, we can see, in the middle panel of figure~\ref{fig:BNS_DI}, the improvement is almost negligible for BNS. In some orientations, the localization even gets worse, especially for the $e_0=0.1$ case. The $e_0=0.4$ case has a better performance but the improvement is still not very prominent. However, though BNS benefits little from eccentricity for its localization, it can be precisely localized even in the circular case due to its long inspiral period in the detector band. 
While, for the medium BBH whose localization is much worse than that of BNS in the circular case (because of a much shorter inspiral in the detector band), a non-vanishing eccentricity can significantly improve the source localization in almost all orientations. The largest improvement is about 2 (3) orders of magnitude with $e_0=0.1$ (0.4). From the improvement of distance inference and source localization, we can expect the reduction of the 3-D localization volume which is shown in the right panels of figures~\ref{fig:BNS_DI} and~\ref{fig:BBHmedium_DI}. For BNS, the reduction of 3-D localization volume mainly comes from the improvement of distance inference in the near face-on orientations. While for the medium BBH, it benefits from the improvement of both distance inference and sky localization, which can reduce its 3-D localization volume more significantly. The largest improvement is about 2.5 (3.5) orders of magnitude with $e_0=0.1$ (0.4). If we assume the galaxy is uniformly distributed in the comoving volume and the number density $n_g= 0.01~\rm Mpc^{-3}$~\cite{Chen:2016tys}, then the threshold localization volume is $V_{\rm th}=100~\rm Mpc^3$. It means that the host galaxies of the GWs (dark sirens) whose localization volume $V_{99}\leq V_{\rm th}$ can be unambiguously identified. As shown in the right panel of figure~\ref{fig:BBHmedium_DI}, for the medium BBH with DECIGO-I, if $e_0=0$, the localization volume is larger than the threshold volume in most orientations. But a non-vanishing eccentricity can significantly improve its 3-D localization ($V_{99}< V_{\rm th}$ in almost all orientations) so that one can identify the host galaxy of the medium BBH. 

\begin{figure*}
\includegraphics[width=0.3\textwidth]{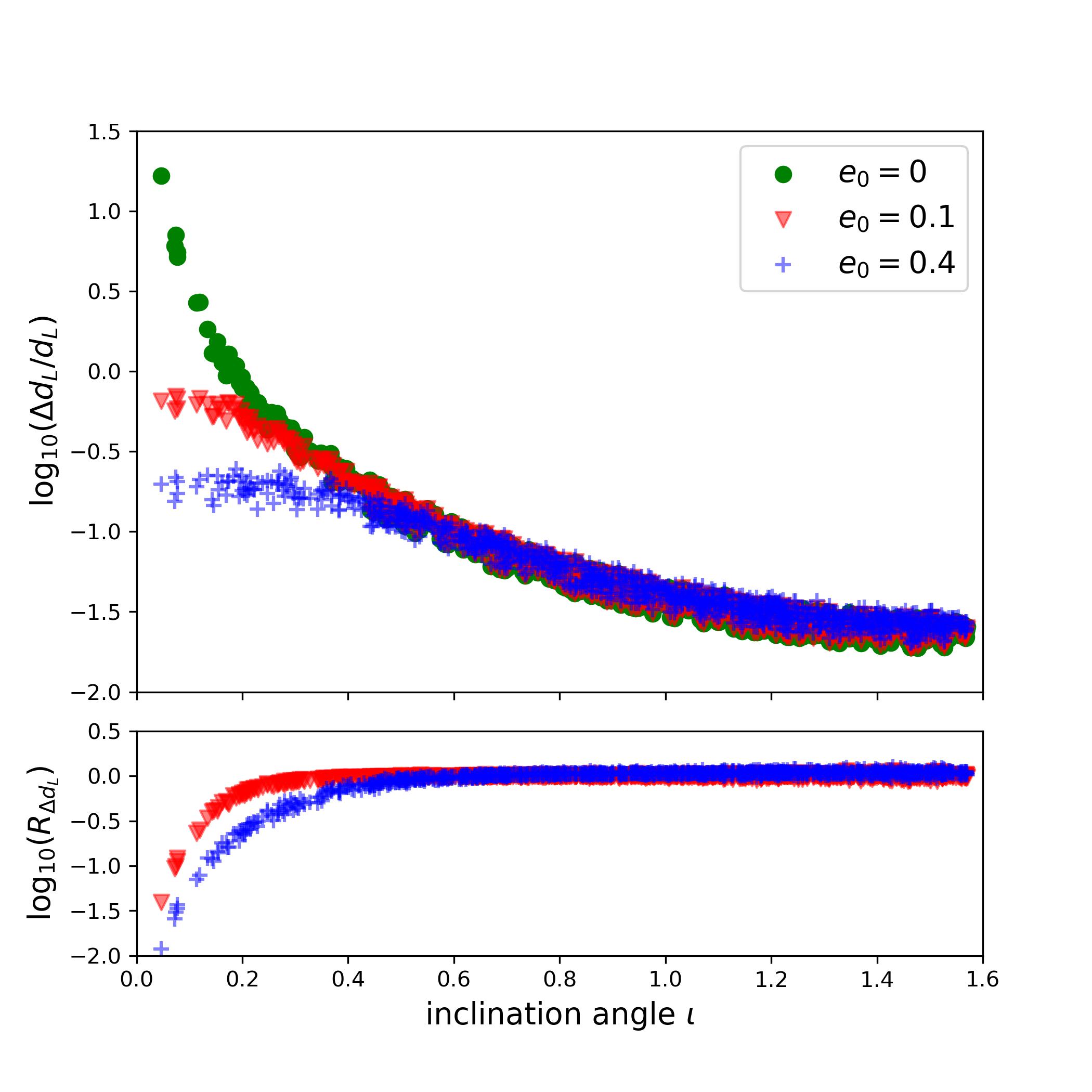}
\includegraphics[width=0.3\textwidth]{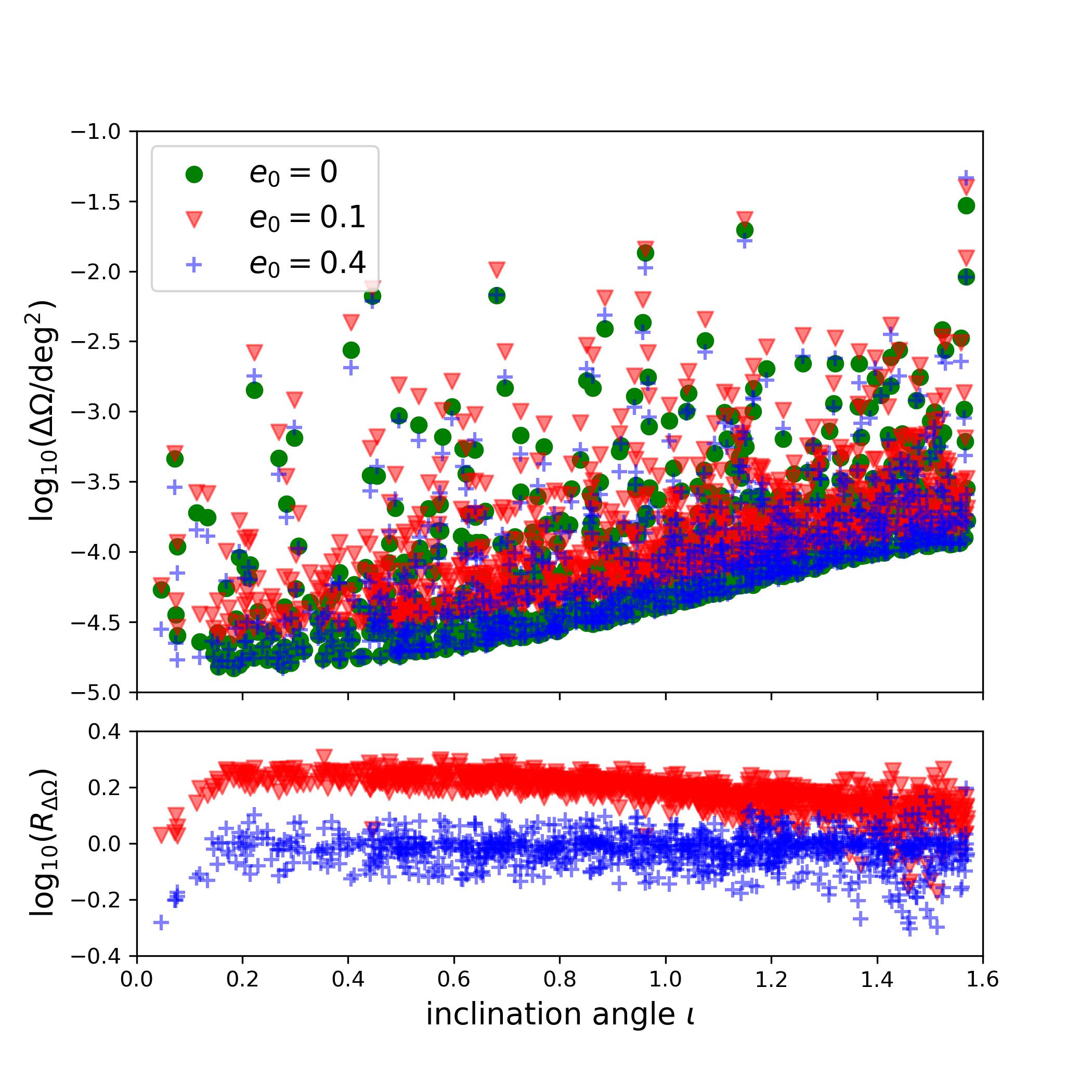}
\includegraphics[width=0.3\textwidth]{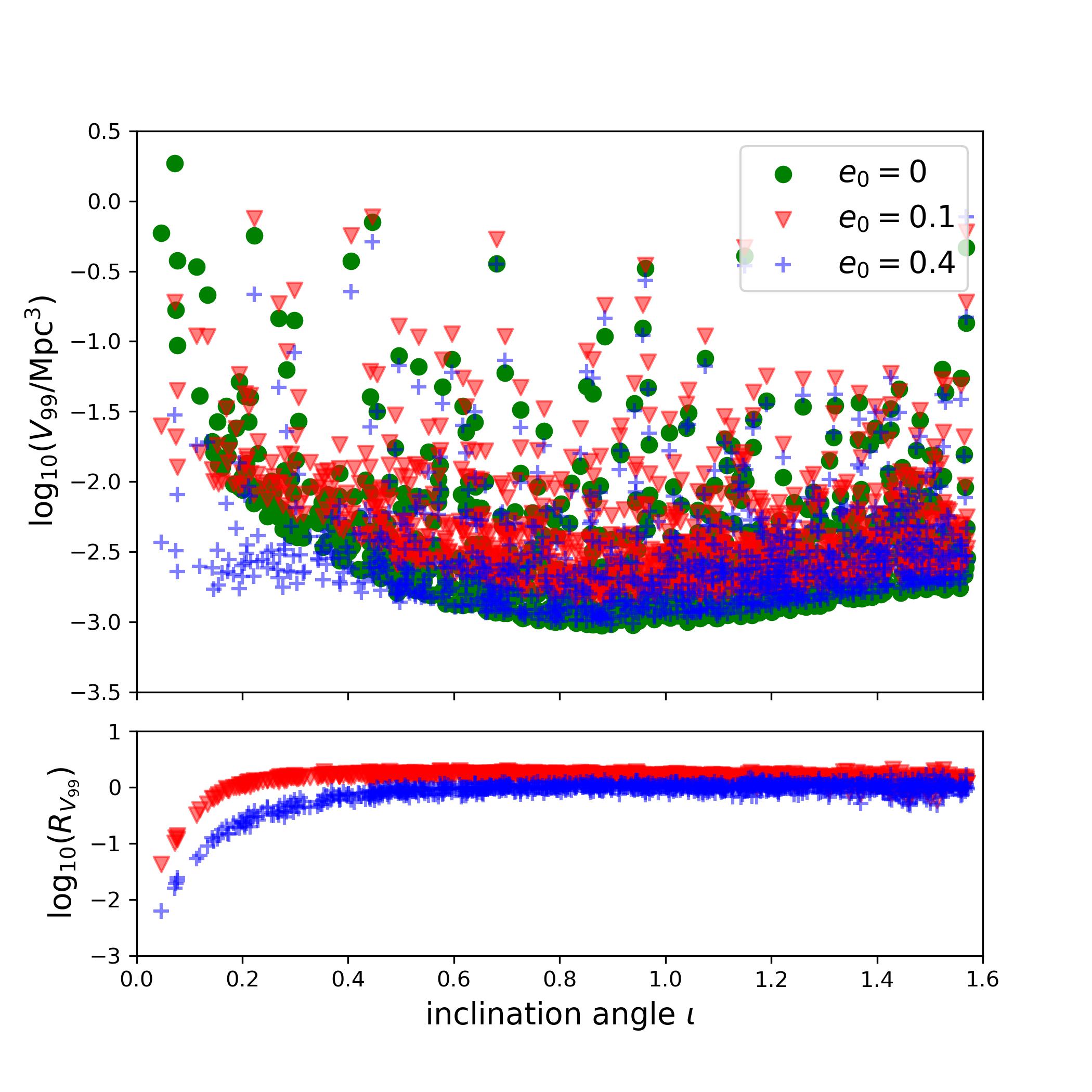}
\caption{The same as figure~\ref{fig:BNS_DI}, but with B-DECIGO.}
\label{fig:BNS_BD}
\end{figure*}

\begin{figure*}
\includegraphics[width=0.3\textwidth]{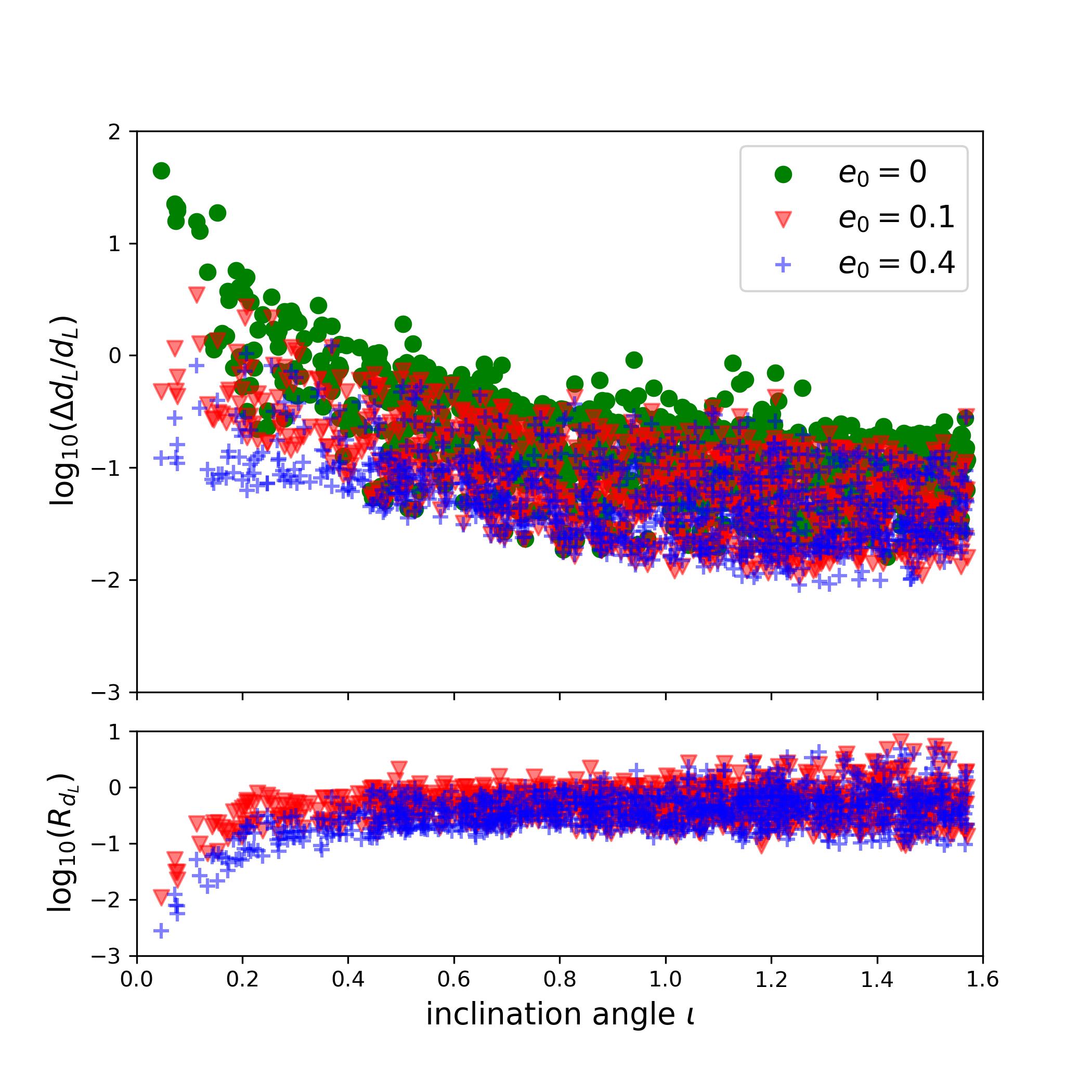}
\includegraphics[width=0.3\textwidth]{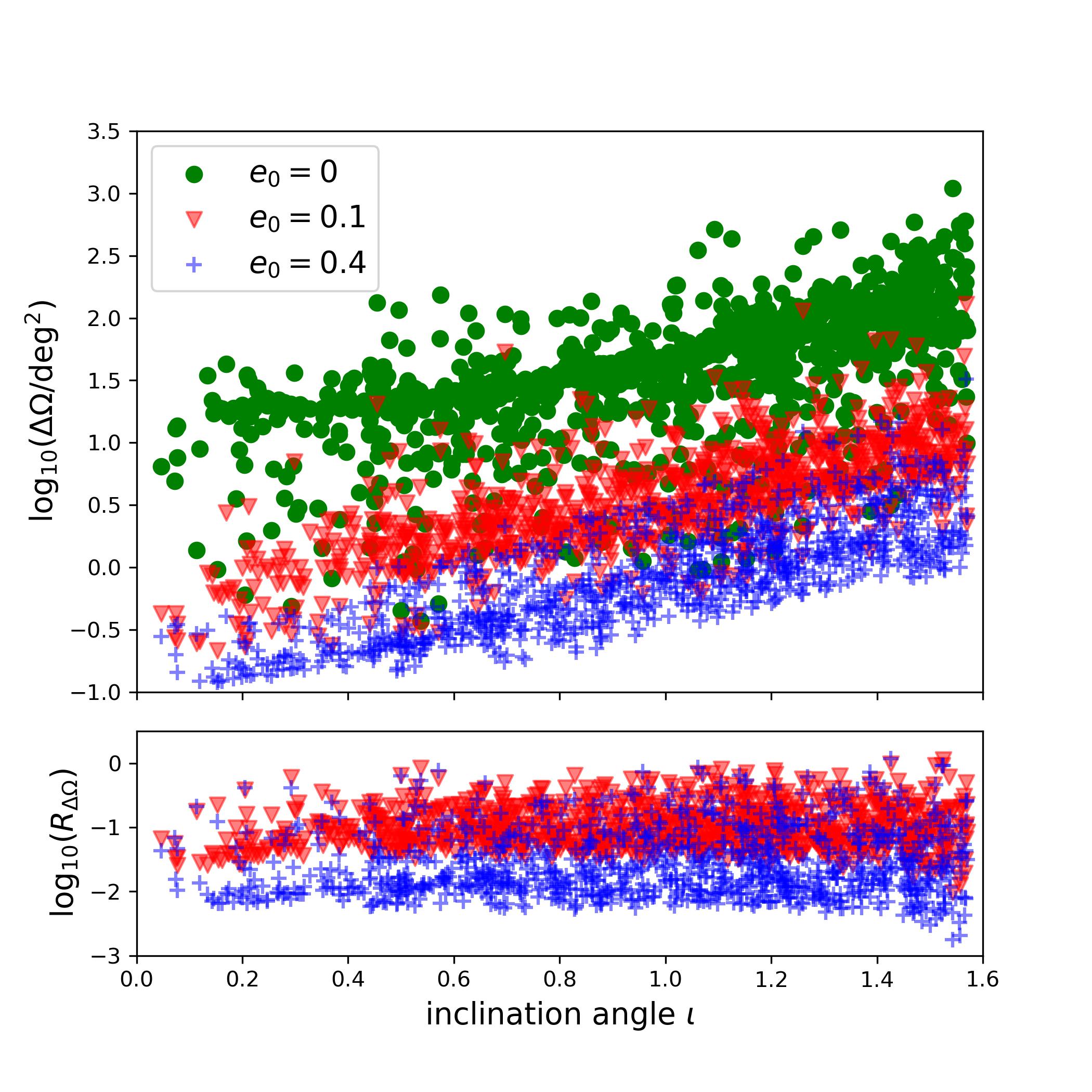}
\includegraphics[width=0.3\textwidth]{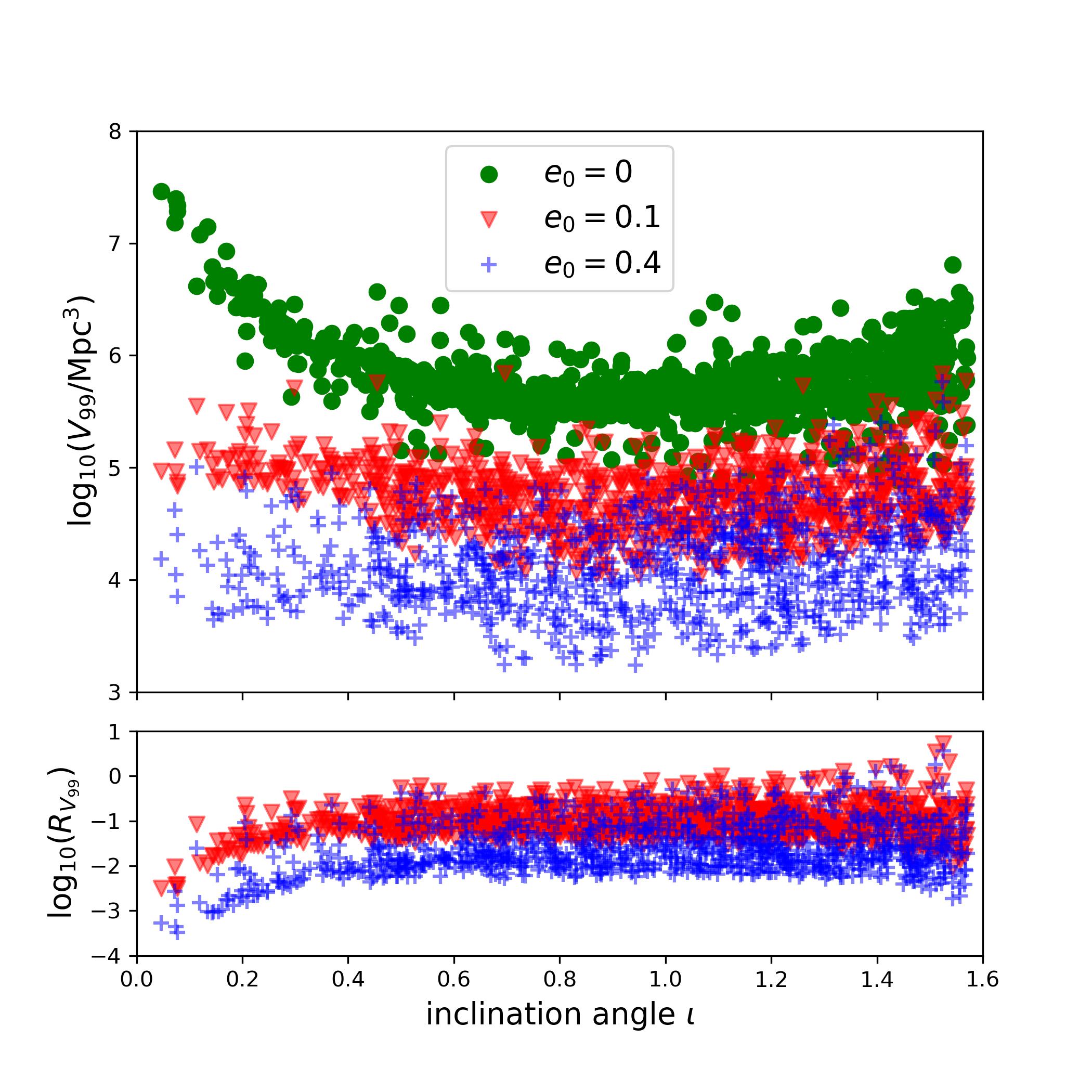}
\caption{The same as figure~\ref{fig:BBHmedium_DI}, but with B-DECIGO.}
\label{fig:BBHmedium_BD}
\end{figure*}

As comparisons, the similar results of GW170817-like BNS and GW150914-like medium BBH with B-DECIGO are shown in figures~\ref{fig:BNS_BD} and~\ref{fig:BBHmedium_BD}. Since the main difference between B-DECIGO and DECIGO-I only lies in their sensitivity (the latter is 15 times better than the former), we can see the parameter estimations based on these two detectors are very similar and only differ by a constant factor. So the ratio $R$ of each parameter for B-DECIGO is the same as that for DECIGO-I (see the bottom panel of the figures). It means that the improvement of parameter estimation from eccentricity does not depend on the sensitivity of the detector. However, due to better sensitivity, we can see the distance inference and localization of BNS and medium BBH with DECIGO-I are more precise than with B-DECIGO.

The results show different features and trends for distance inference and sky localization. For distance inference, the errors become smaller in the larger inclination angle. On the contrary, the source localization is better in the smaller inclination angle. In addition, eccentricity can only significantly reduce the error of distance in the near face-on orientations (when the error is the largest). While for localization, it can always be improved in almost all orientations. Finally, the localization of small-mass binaries like BNS can not be significantly improved (sometimes even worsen) by eccentricity. But for distance inference, all binaries can benefit from eccentricity, especially in the near face-on orientations. Such different features can be explained as follows. 
In the circular case, the large error of distance in the small inclination angle is due to the large degeneracy between distance and inclination angle when $\iota\sim 0$. The distance and inclination angle are simply coupled in the amplitude of the waveform as $h\sim \mathcal{A}_++\mathcal{A}_\times$, 
where $\mathcal{A}_+\sim \frac{1}{d_L} \frac{1+\cos^2(\iota)}{2}$ and $\mathcal{A}_\times\sim\frac{1}{d_L}\cos(\iota)$.  In order to identify the inclination of the binary system using the polarizations of the gravitational waves, we must distinguish the contributions of the plus ($+$) and cross ($\times$) polarizations. When the binary system is near face-on, the two amplitudes from plus and cross polarization have nearly identical contributions to the overall gravitational-wave amplitude. This is the main factor that leads to the strong degeneracy in the measurement of the distance and inclination when $\iota$ is small (see ~\cite{Usman:2018imj} for details). So, in the near face-on orientations, we can not measure the inclination angle precisely, as well as the distance. However, in the eccentric cases, the multiple harmonics make the distance and inclination angle ($\iota$ is written in the functions $\xi_{\ell}$) nontrivially coupled, allowing us to significantly break their degeneracy and hence improve the inference of both parameters. While, for a large inclination angle, the degeneracy between $d_L$ and $\iota$ is small. So the estimation of these two parameters is good enough and there is no further large room to improve it from eccentricity. As for the localization, there is no obvious degeneracy between the sky location parameters and inclination angle. From the Fisher matrix, the parameter estimation should be better for a higher SNR. From figure~\ref{fig:SNR}, we know the SNR is larger with a smaller $\iota$. So we can expect the localization (the estimation of ($\theta,~\phi$)) should be better when $\iota$ is smaller. The multiple harmonics induced by the eccentricity can add more information (e.g. the higher modes can enter the detector band much earlier) for the estimation of the sky location parameters, thus improving the localization regardless of the orientations. The 3-D localization ($V_{99}$) is just the combination of the distance inference ($\Delta d_L$) and sky localization ($\Delta \Omega$). Finally, the fact that eccentricity has little effect on the localization of small-mass binaries has also been reported in~\cite{Mikoczi:2012qy,Pan:2019anf}. We are not going to give an in-depth discussion here since it is beyond the scope of this paper.

\begin{figure}
\includegraphics[width=0.45\textwidth]{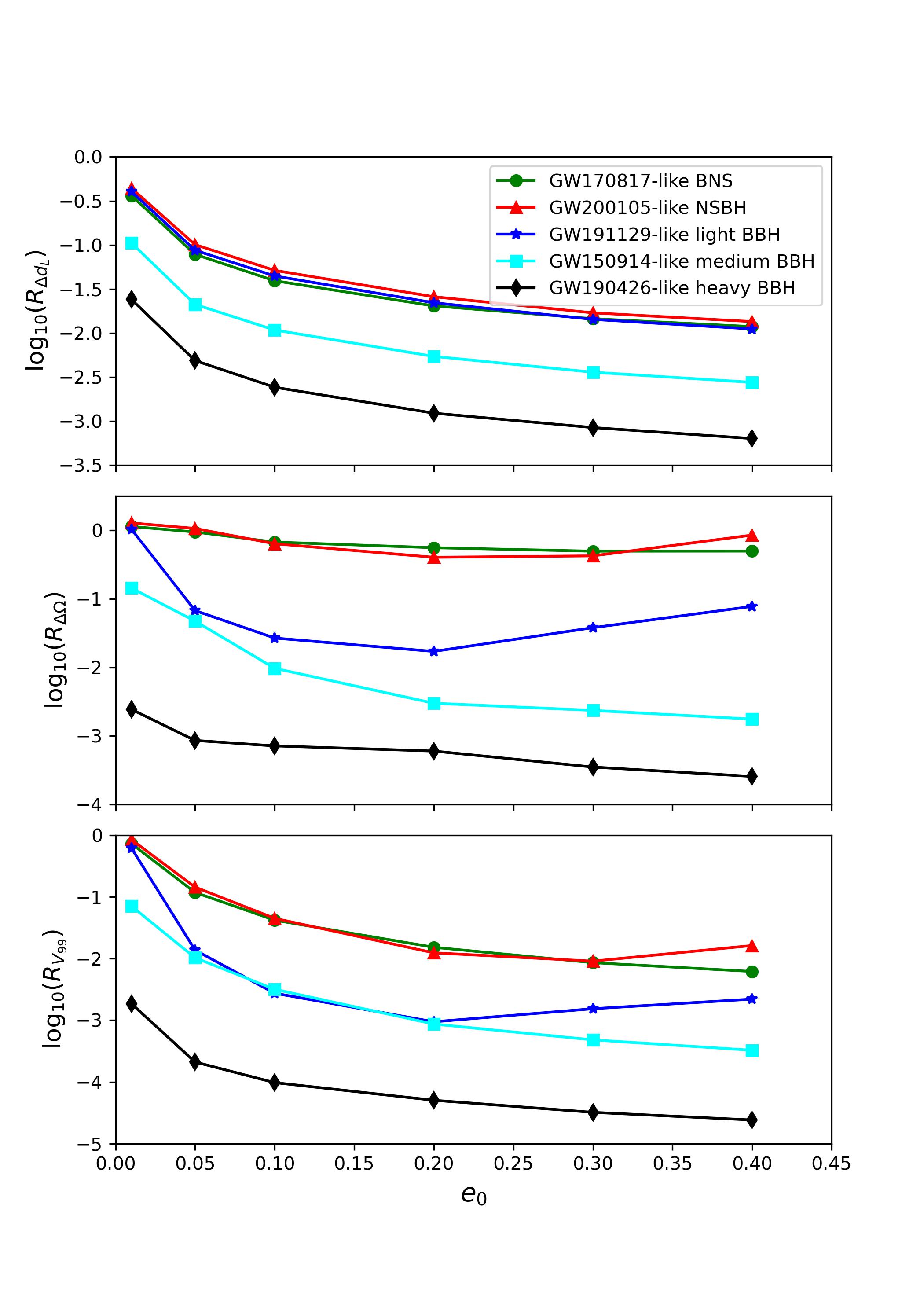}
\caption{The largest improvement ($\min(R)$ among the 1000 orientations) of distance inference, localization, and 3-D localization volume for each binary with eccentricities from 0.01 to 0.4. The detector is either DECIGO-I or B-DECIGO.}
\label{fig:Rwe}
\end{figure}

To illustrate the improvement of distance inference and localization for these typical binaries with variable eccentricities, we show the largest improvement (the $\min(R)$ among the 1000 orientations) for each case in figure~\ref{fig:Rwe}. As we mentioned above, the values of $R$ with DECIGO-I and B-DECIGO should be the same and we only show one of them. Generally, a heavier binary with larger eccentricity can achieve more improvement for the distance inference and source localization. Note this can not be explained by the SNR of each binary. As suggested in figure~\ref{fig:SNR}, the heavier binary does not have to have a larger SNR. A simple explanation is that a larger eccentricity can make the non-quadrupole modes ($\ell\neq 2$) more prominent so that it can break the degeneracy of the parameters and provide more information for the parameter estimation.  While, a heavier mass means a larger orbital velocity which can make the effects of eccentricity more distinct~\footnote{For instance, this can be suggested in the relation $v_{\rm ecc}(f;e_0)=g(f;e_0)(\pi M f)^{1/3}$. Note the fact that a higher-mass binary can achieve more improvement of localization from eccentricity has been indicated in previous work such as~\cite{Mikoczi:2012qy,Sun:2015bva,Ma:2017bux,Pan:2019anf}, but with different explanations.}. With $e_0=0.4$, the distance inference of these typical binaries (from BNS to heavy BBH) in the near face-on orientations can achieve 1.5--3 orders of magnitude improvements~\footnote{Note the largest improvement of distance inference in the near face-on orientations is sensitive to the sampling of the small $\iota$. The numbers here are different from that of~\cite{Yang:2022tig} because we use a different random seed for the sampling of $P_{\rm ang}$.}. For the typical BNS and NSBH, the improvement of localization is negligible. But for the typical BBH, the localization can achieve 1--3.5 orders of magnitude improvement. By combining the improvement of distance inference and sky localization, the 3-D localization of these typical binaries (from BNS to heavy BBH) can achieve 2--4.5 orders of magnitude improvement.

We should note that there are some anomalies in figure~\ref{fig:Rwe}. For instance, the distance inference of BNS is a little better than that of NSBH. We also find that eccentricity has almost negligible effects on the localization of BNS and NSBH. It may even slightly worsen the localization in some cases. In addition, the localization of NSBH and light BBH achieve the largest improvement with $e_0=0.2$, but a larger eccentricity will reduce the improvement.
The reason for these nontrivial features in distance inference and source localization is as follows. On the one hand, eccentricity adds more harmonics to the waveform hence breaking the parameter degeneracy and improving the parameter estimation. The higher harmonics which enter the detector band earlier can also provide more angular information. On the other hand, in a specific frequency band with a higher eccentricity, the binary evolves faster thus the inspiral time (observation time) is shorter. This can downgrade the parameter estimation, especially for localization. We should also bear in mind that, in the eccentric case, there are two additional free parameters that make the parameter estimation be harder than in the circular case. In addition, with different starting frequencies, the frequency ranges of the multiple harmonics of these typical binaries covered in the detector band (0.1--10 Hz) are also different. These factors compete with each other and make the parameter inference differ from case to case.

In this section, we only choose the typical BNS and medium BBH with $e_0=0$, 0.1, and 0.4 as the representatives and show their distance inference and source localization as well as the corresponding improvement from eccentricity. The results for other typical binaries with different eccentricities can be inferred from the trends suggested in figure~\ref{fig:Rwe}. The full results such as the distance inference and localization of other typical binaries and the tables of the statistical results are summarized in the appendix~\ref{app:A}. 

\subsection{The estimation of other parameters}

Apart from the distance and localization, the estimations of other parameters are also important. As we mentioned above, the determination of inclination angle is crucial for the distance inference hence the cosmological applications such as the measurement of the Hubble constant~\cite{LIGOScientific:2017adf,Hotokezaka:2018dfi}. In addition, it is also helpful for the modeling of associated EM counterpart like the narrow-beamed short gamma-ray burst~\cite{LIGOScientific:2017zic}. We have already shown that eccentricity can significantly improve distance inference in near face-on orientations. One can also expect that eccentricity has the same effect on the estimation of the inclination angle $\iota$. Besides, the chirp mass $\mathcal{M}_c$, symmetric mass ratio $\eta$, the time to coalescence $t_c$, and the eccentricity itself $e_0$ are also the key parameters for studying the properties of the binary and its astrophysical implications. Since these parameters are not the primary parameters we focus on in this paper, we only present some typical results for them. We choose BNS and medium BBH with $e_0=0$, 0.1, and 0.4 as the representatives and only show the results based on DECIGO-I (the parameter estimation with B-DECIGO can be inferred by rescaling the results with DECIGO-I based on the sensitivity of these two detectors).

\begin{figure}
\includegraphics[width=0.45\textwidth]{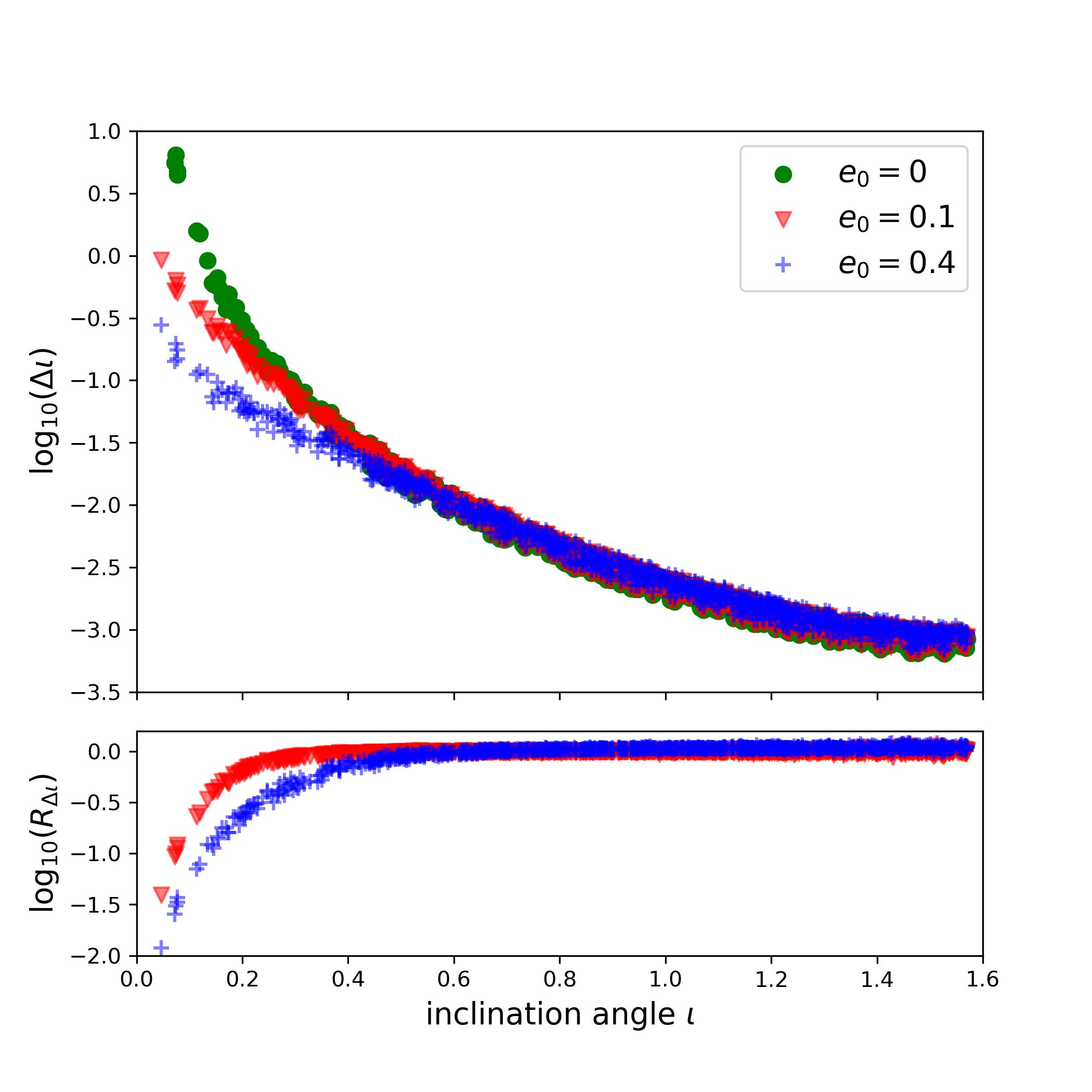}
\caption{The error of inclination angle $\iota$ of GW170817-like BNS with DECIGO-I. As in the case of distance inference, the large errors in the near face-on orientations are due to the huge degeneracy between distance and inclination angle. A nonvanishing eccentricity can alleviate this degeneracy hence improve the constraints.}
\label{fig:diota_BNS170817_DI}
\end{figure}

Figure~\ref{fig:diota_BNS170817_DI} shows the error of inclination angle $\iota$ of GW170817-like BNS with DECIGO-I. As in the case of distance inference, we can see the significant improvement of the measurement of $\iota$ from eccentricity in the near face-on orientations. With $e_0=0.4$, the improvement can be as large as 2 orders of magnitude. The same feature in the inference of distance and inclination angle supports our argument above -- eccentricity can break the large degeneracy between distance and inclination angle in the near face-on orientations and thus improve their measurements at the same time. In the medium BBH case, the improvement of the inclination angle is relatively more significant than in the BNS case, which can be expected from the results of the distance inference.

\begin{figure*}
\includegraphics[width=0.45\textwidth]{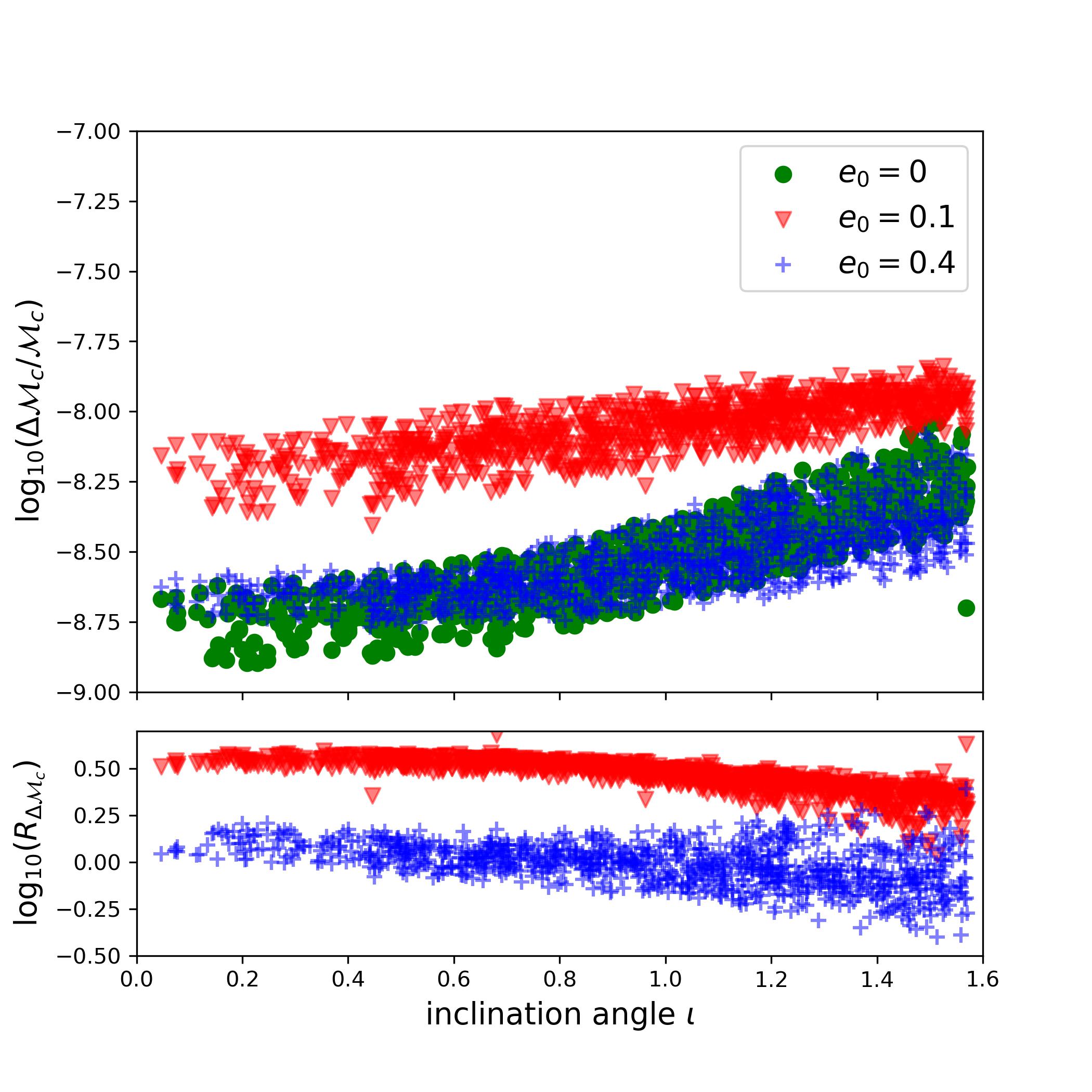}
\includegraphics[width=0.45\textwidth]{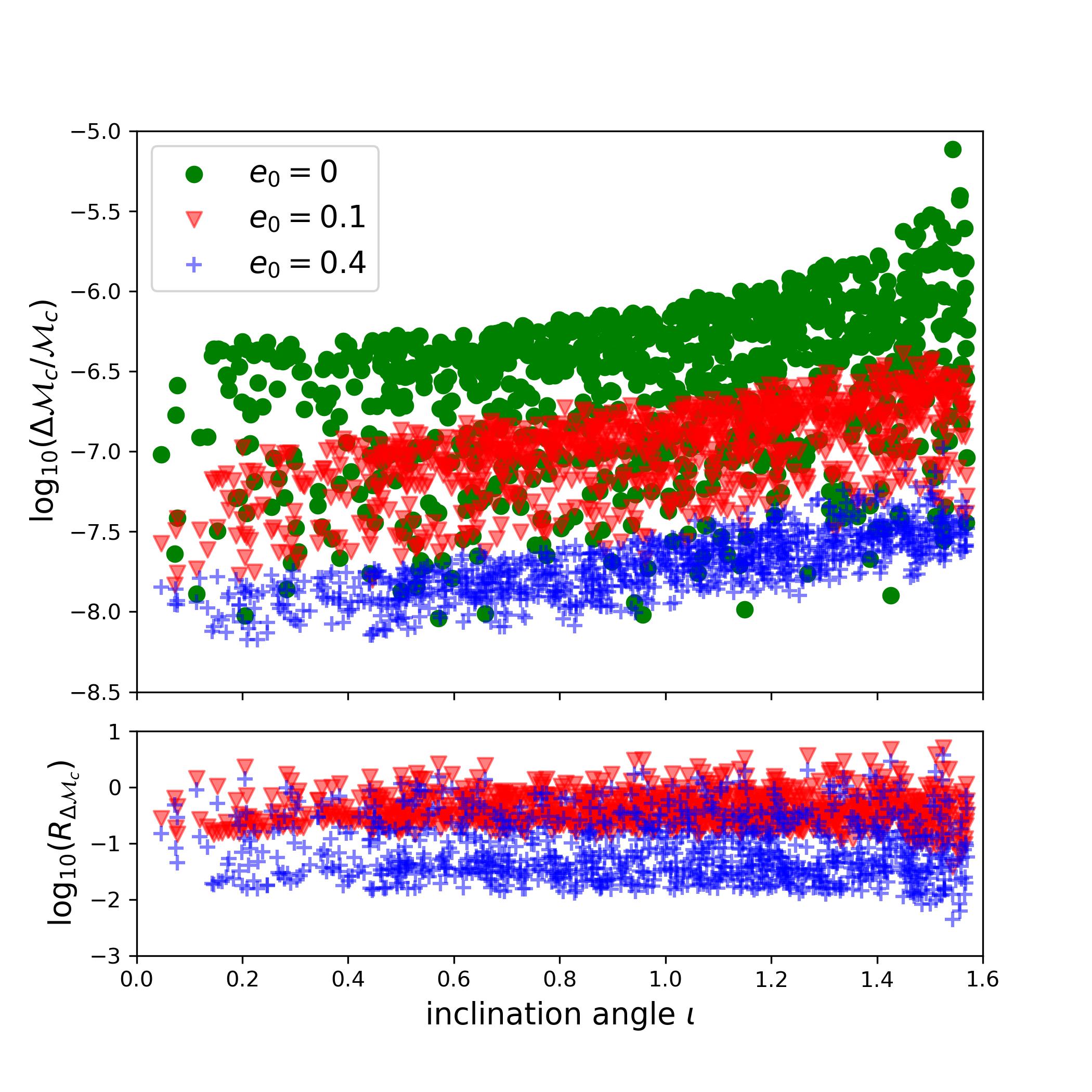}
\caption{The estimation of chirp mass for GW170817-like BNS (left) and GW150914-like medium BBH (right) with DECIGO-I.}
\label{fig:dMc}
\end{figure*}

The errors of chirp mass for GW170817-like BNS and GW150914-like medium BBH are shown in figure~\ref{fig:dMc}. In the decihertz band, the chirp mass can be tightly constrained with $\Delta \mathcal{M}_c/\mathcal{M}_c\sim 10^{-6}-10^{-8}$. Similar to the case of distance inference, eccentricity has little effect on the estimation of chirp mass for small-mass binary like BNS. With $e_0=0.1$, it even slightly worsens the chirp mass estimation. The improvement is still negligible in the case $e_0=0.4$. However, for the larger-mass binary like the medium BBH, eccentricity can significantly improve the chirp mass estimation. As shown in the right panel of figure~\ref{fig:dMc}, the improvement is 1-2 orders of magnitude with $e_0=0.4$. Note the feature that eccentricity has little effect on small-mass binary and greater effects on larger-mass binary is also indicated in~\cite{Mikoczi:2012qy} for the massive BBH in the LISA band. For the estimation of symmetric mass ratio $\eta$, we find a similar result as for the chirp mass.  For the medium BBH, $\Delta \eta \sim10^{-5}-10^{-6}$, but the improvement of $\eta$ from eccentricity is much smaller than that of chirp mass. 

\begin{figure*}
\includegraphics[width=0.45\textwidth]{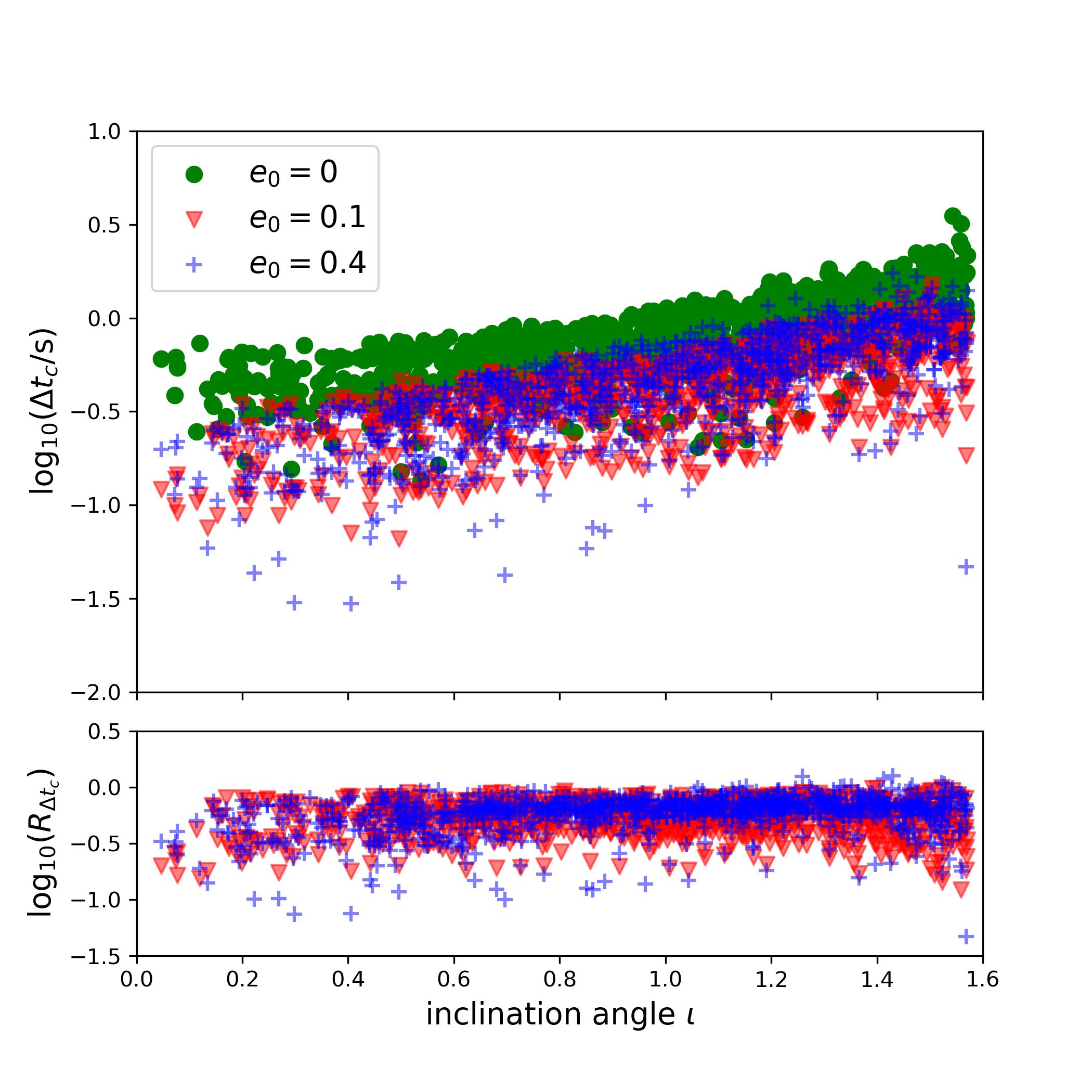}
\includegraphics[width=0.45\textwidth]{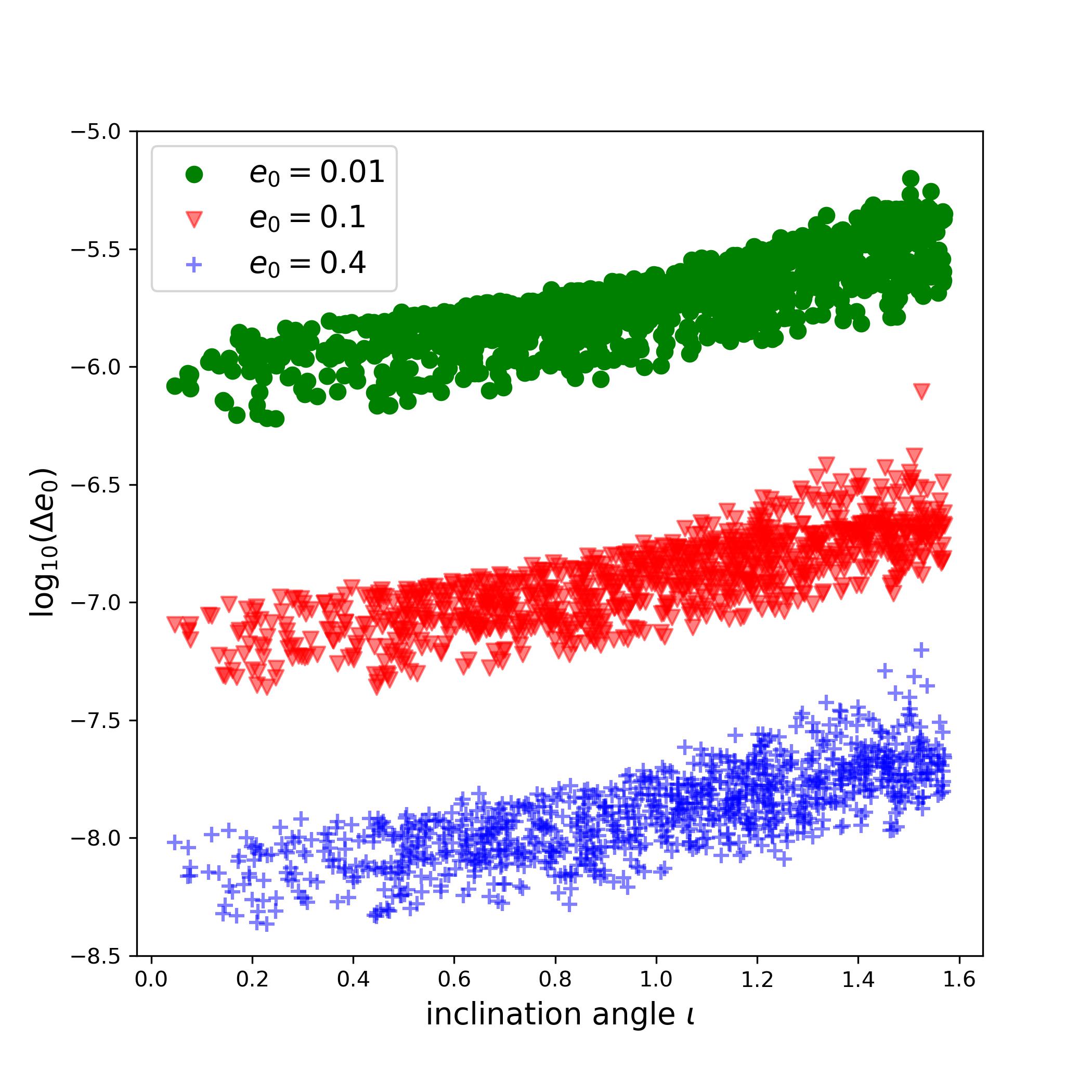}
\caption{The estimation of coalescence time $t_c$ (left) and initial eccentricity $e_0$ (right) for GW150914-like medium BBH with DECIGO-I. Note for the estimation of eccentricity, we choose $e_0=0.01$ as a small-eccentricity case since $e_0$ is not a free parameter in the circular case.}
\label{fig:dtcde0}
\end{figure*}

Figure~\ref{fig:dtcde0} shows an example of the estimation of coalescence time and initial eccentricity. We choose the medium BBH case with DECIGO-I as a representative. We can see an obvious improvement in the estimation of coalescence time from eccentricity. With $e_0=0.4$, the improvement can be as large as a factor of 10. We also find that in the BNS case, the improvement is relatively smaller but the estimation is more precise ($\min(\Delta t_c) \sim 10^{-3}$ s) due to the longer inspiral period. The estimation of eccentricity is shown in the right panel of figure~\ref{fig:dtcde0}. Since in the circular case ($e_0=0$), eccentricity is not a free parameter to be constrained, we choose $e_0=0.01$ to represent the small-eccentricity case. We can see the initial eccentricity can be very tightly constrained with $\Delta e_0\sim 10^{-6}-10^{-8}$ for $e_0=0.01$ to 0.4. A larger eccentricity can significantly improve the estimation of itself. We also find a similar result in the BNS case.

In this section, we take a close look at the GW parameter estimation for the typical binaries in the circular and eccentric cases. We find that eccentricity can improve the estimation of almost all the waveform parameters. Generally, the improvement is more significant for a heavier binary with a larger eccentricity.
One notable point is that except for the distances and inclination angles, the parameter estimation is better in the smaller $\iota$ cases. As we have explained, the anomalous trend of the distance error against the inclination angle is due to large $d_L-\iota$ degeneracy when $\iota$ is small. As for other parameters, there is no obvious degeneracy with the inclination angle. A larger SNR in the smaller $\iota$ cases can provide a better parameter estimation. Though we only choose BNS and medium BBH with $e_0=0$, 0.1, and 0.4 as representatives, the results of other binaries and with various eccentricities can be inferred from the improvements suggested in figure~\ref{fig:Rwe}. In this paper, distance and localization are the key parameters we would like to investigate since they are crucial for dark sirens as accurate and precise probes of the Universe.

\section{Construction of the golden dark siren catalogs \label{sec:catalog}}

In the mid-band, the precise localization of BNS and NSBH, and the significant improvement of localization from eccentricity for BBH inspire us to investigate the possibility of unambiguously identifying their host galaxies. As shown in the right panel of figure~\ref{fig:BBHmedium_DI}, with the help of eccentricity, the GW150914-like BBH can be localized in a 3-D volume which is smaller than the threshold volume so that its host galaxy can be uniquely identified. For the lower-mass binaries like BNS and NSBH, though eccentricity has almost negligible effects on their localization, they have been precisely localized even without eccentricity (see figure~\ref{fig:BNS_DI}). So, in the mid-band with a non-vanishing eccentricity, it is possible for all three types of binaries (BNS, NSBH, and BBH) to be precisely localized so that their host galaxies can be uniquely identified.
In this section, we would like to mock up the GW catalogs with decihertz observatory like DECIGO. Based on our current knowledge, we forecast the population of GWs on the assumption of various eccentricities. We assume that the dark sirens like BNS, NSBH, and BBH, are not accompanied by the EM counterparts or the EM counterparts can not be detected. We focus on the well-localized dark sirens whose host galaxy can be uniquely identified (hereafter the golden dark sirens). The redshift of these golden dark sirens (from their host galaxies) can be unambiguously measured from the present galaxy catalog or follow-up spectroscopic observations.  This makes golden dark sirens as good quality as bright sirens in terms of the redshift inference, thus can be used as precise probes in cosmology. We still assume two scenarios of DECIGO. We would like to compare these two configurations, to see the improvement of the population of GW detections (and also the golden dark sirens) if B-DECIGO is upgraded to DECIGO-I. 

\subsection{Population of simulated GW detections}

We follow~\cite{Yang:2021xox,Yang:2022iwn} to simulate the GW detections and construct the catalogs of BNS, NSBH, and BBH. The merge rate per comoving volume at a specific redshift $R_m(z_m)$ is related to the formation rate of massive binaries and the time delay distribution $P(t_d,\tau)=\frac{1}{\tau}\exp(-t_d/\tau)$ with an e-fold time of $\tau=100$ Myr~\cite{Vitale:2018yhm},
\begin{equation}
R_m(z_m)=\int_{z_m}^{\infty}dz_f\frac{dt_f}{dz_f}R_f(z_f)P(t_d) \,.
\label{eq:Rm}
\end{equation}
Here $t_m$ (corresponding to redshift $z_m$) and $t_f$ are the look-back times when the systems merged and formed,  $t_d=t_f-t_m$ is the time delay between the formation and merger, and $R_f$ is the formation rate of massive binaries. We assume the formation of compact binaries tracks the star formation rate. So $R_f$ is proportional to the Madau-Dickinson (MD) star formation rate~\cite{Madau:2014bja},
\begin{equation}
\psi_{\rm MD}=\psi_0\frac{(1+z)^{\alpha}}{1+[(1+z)/C]^{\beta}} \,,
\label{eq:psiMD}
\end{equation}
with parameters $\alpha=2.7$, $\beta=5.6$ and $C=2.9$. The normalization factor $\psi_0$ is determined by the local merger rates. We adopt the local merger rates of BNS, NSBH, and BBH inferred from GWTC-3, with $\mathcal{R}_{\rm BNS}=105.5^{+190.2}_{-83.9}~\rm Gpc^{-3}~\rm yr^{-1}$, $\mathcal{R}_{\rm NSBH}=45^{+75}_{-33}~\rm Gpc^{-3}~\rm yr^{-1}$, and $\mathcal{R}_{\rm BBH}=23.9^{+14.3}_{-8.6}~\rm Gpc^{-3}~\rm yr^{-1}$~\cite{LIGOScientific:2021psn}. Note the merger rate of NSBH is based on the assumption that the observed NSBH GW200105 and GW200115 are representatives of the population of NSBH. Then we convert the merger rate per comoving volume in the source frame to merger rate density per unit redshift in the observer frame through
\begin{equation}
R_z(z)=\frac{R_m(z)}{1+z}\frac{dV(z)}{dz} \,,
\label{eq:Rz}
\end{equation}
where $dV/dz$ is the comoving volume element. 

Having the merger rates $R_z(z)$ for BNS, NSBH, and BBH, we can sample the redshift distribution of them respectively. In this paper, we just adopt the median $\mathcal{R}$ for the construction of the catalogs. We have 11 parameters in the waveform (for vanishing eccentricity there are 9 except $e_0$ and $\beta$). The luminosity distance $d_L$ is calculated from the sampled redshift by assuming a fiducial cosmological model $\Lambda$CDM with $H_0=67.72~\rm km~s^{-1}~Mpc^{-1}$ and $\Omega_m=0.3104$, corresponding to the mean values obtained from the latest \textit{Planck} experiment~\cite{Planck:2018vyg}. The sky localization ($\theta$, $\phi$), inclination angle $\iota$, polarization $\psi$, and $\beta$ are drawn from the uniform and isotropic distribution. Without loss of generality, we set the time and phase at coalescence to be $t_c=\phi_c=0$. As for the chirp mass and symmetric mass ratio, we consider  different strategies for these three binary types. In the BNS case, we assume a uniform distribution of mass in [1, 2.5] $M_{\odot}$, which is consistent with the assumption for the prediction of the BNS merger rate in GWTC-3~\cite{LIGOScientific:2021psn}. In the NSBH case, since the merger rate is inferred by assuming the observed NSBH GW200105 and GW200115 are representatives of the population of NSBH, we just randomly choose the component mass to be one of these two events. As for the BBH case, we adopt the same strategy in~\cite{Yang:2021xox} but with the BBH population given by GWTC-3. We draw the distribution of component masses of simulated BBH based on the histogram of mass distribution from the real BBH detections in GWTC-3~\footnote{We first infer the histograms of primary mass $m_1$ and mass ratio $q$ from GWTC-3. The distribution of $m_1$ and $q$ for the simulated BBH are sampled accordingly. Then the second mass is just $m_2=m_1q$. We should make sure that $m_2\ge3~M_{\odot}$.}. The primary mass and mass ratio peak around $30-40~M_{\odot}$ and 0.7, respectively.

We set the initial observation time of B-DECIGO and DECIGO-I to be 1 year. The population of each type binary within 1 year in every redshift bin is sampled from $R_z(z)$. The starting frequency of each type binary is chosen to be the same as in section~\ref{sec:typical}. For BBH, since the peak primary mass is around $30-40~M_{\odot}$, we choose their starting frequencies to be that of the typical medium BBH. We consider five discrete eccentricities $e_0=0$, 0.1, 0.2, 0.3, and 0.4 and in each case, we assume all binaries have the same eccentricity. This is however not a realistic assumption and we will give a discussion on this later.  For each assumption of eccentricity, we select the binaries with SNR>8 as the GW detections. We can assess the influence of eccentricity on the population of GW catalogs with DECIGO in 1-year observation time. 

\begin{figure*}
\includegraphics[width=\textwidth]{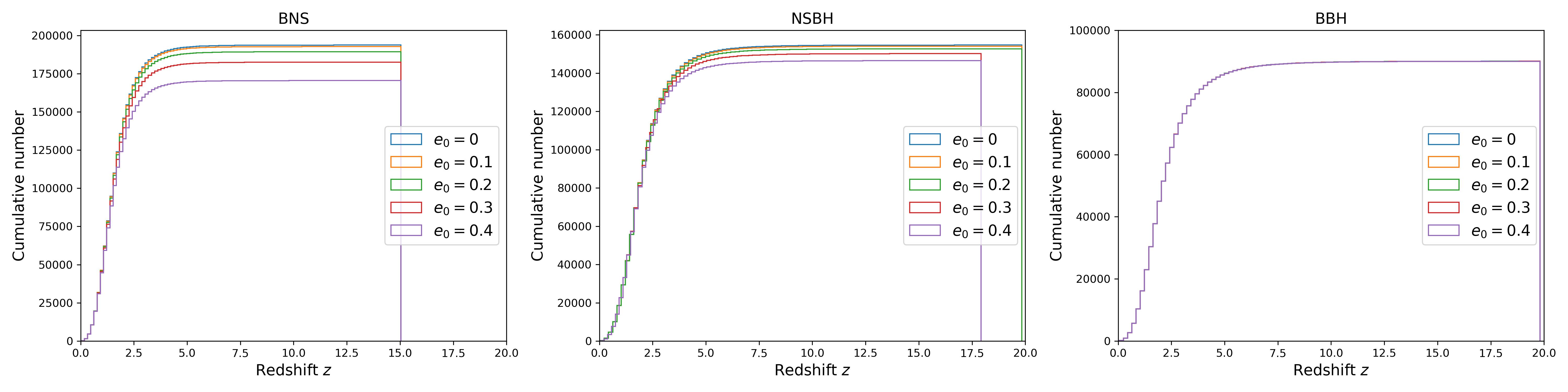}
\caption{The cumulative histograms of the simulated detections on the assumption of various eccentricities for BNS (left), NSBH (middle), and BBH (right) by 1-year observation of DECIGO-I. Note in the BBH case the differences among different  eccentricities are too small and hence invisible.}
\label{fig:hist_DI}
\end{figure*}

Figure~\ref{fig:hist_DI} shows the cumulative histograms of the simulated BNS, NSBH, and BBH detections by 1-year observation of DECIGO-I and on the assumption of various eccentricities. The numbers of the detections are accumulated quickly in the redshift [0, 5] and there are sparse events up to redshift 15, 18, and 20 for BNS, NSBH, and BBH, respectively. Note we set a cut-off at $z=20$ for all binaries. The total number of events is around $1.8\times 10^5$ for BNS, $1.5\times 10^5$ for NSBH, and $9\times 10^4$ for BBH. The smaller number of BBH is due to its lower predicted merger rate. For BNS and NSBH, a non-vanishing eccentricity can slightly decrease the number of detections. While for BBH, the decrease is almost negligible.

\begin{figure*}
\includegraphics[width=\textwidth]{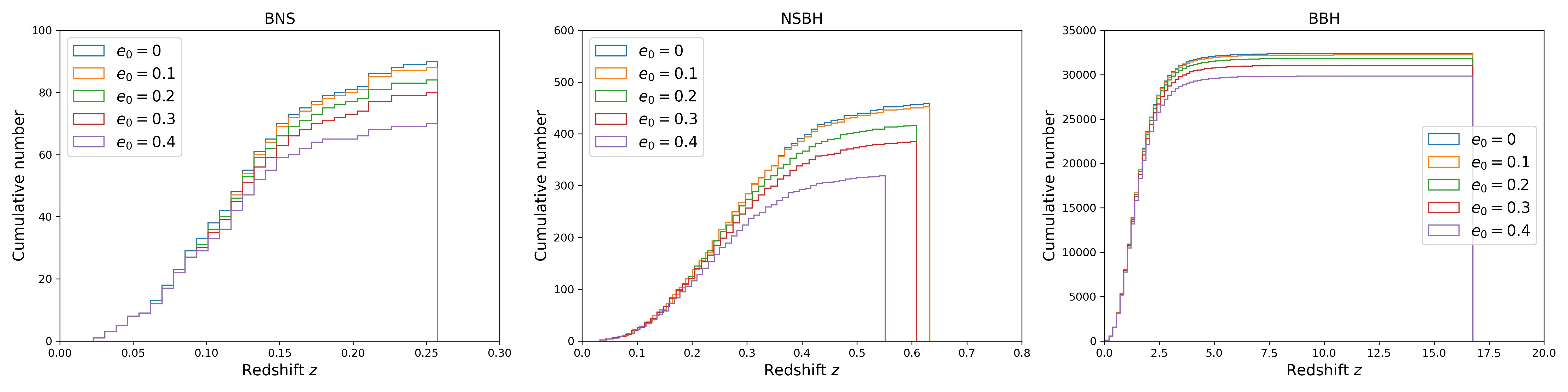}
\caption{The same as figure~\ref{fig:hist_DI}, but with B-DECIGO. }
\label{fig:hist_BD}
\end{figure*}

As a comparison, the cumulative histograms of BNS, NSBH, and BBH by 1-year observation of B-DECIGO are shown in figure~\ref{fig:hist_BD}. Since the sensitivity of B-DECIGO is inferior to that of DECIGO-I, we can see the numbers of detections are drastically decreased.
For BNS and NSBH, the largest redshift B-DECIGO can reach is around 0.25 and 0.6, respectively. The total number of detections is around one hundred and a few hundreds for BNS and NSBH. While, the horizon of BBH is much larger and can reach redshift more than 15. Though the merger rate of BBH is lower than that of BNS and NSBH, the total number of BBH detections can achieve around $3\times 10^4$. For all these three types of binaries, we can see the number of detections will be slightly decreased due to eccentricity. This can be explained by the fact that the eccentricity can shorten the inspiral period hence the observation time of the binary, causing a slight reduction of the SNR. Unlike BNS and BBH, the largest redshift of NSBH is also reduced by eccentricity (this also happens in the DECIGO-I case). This is due to the fixed component mass (either of GW200105 or GW200115) we choose when we construct the catalog of NSBH. We can not sample a larger mass to compensate for the decrease of the SNR by eccentricity. Thus the threshold redshift of NSBH in the circular case is not reachable in the eccentric case.

\subsection{Golden dark sirens}

As we have mentioned, in the mid-band, eccentricity can make the localization of BNS, NSBH, and BBH so precise that their host galaxies can be uniquely identified.  Now, we can demonstrate this argument and estimate how many such golden events can be observed by DECIGO in 1-year observing run from the catalogs in figures~\ref{fig:hist_DI} and~\ref{fig:hist_BD}.
Since we only focus on the well-localized events, we limit the catalogs to small redshifts $z<0.5$. For BNS, NSBH, and BBH with $z<0.5$, we transform the errors of distance and sky localization to the 3-D localization volumes. To estimate the numbers of potential host galaxies in the localization volume, we assume the galaxy is uniformly distributed in the comoving volume and the number density $n_g= 0.01~\rm Mpc^{-3}$. This number is derived by taking the Schechter function parameters in B-band $\phi_*=1.6\times 10^{-2} h^3 {\rm Mpc^{-3}}, \alpha=-1.07, L_*=1.2\times 10^{10} h^{-2} L_{B,\odot}$ and $h=0.7$, integrating down to 0.12 $L_*$ and comprising 86\% of the total luminosity~\cite{Chen:2016tys}. Then the threshold localization volume is $V_{\rm th}=100~\rm Mpc^3$. We choose the ones with 3-D localization volumes smaller than the threshold $V_{\rm th}$ as the golden dark sirens.

\begin{figure*}
\includegraphics[width=\textwidth]{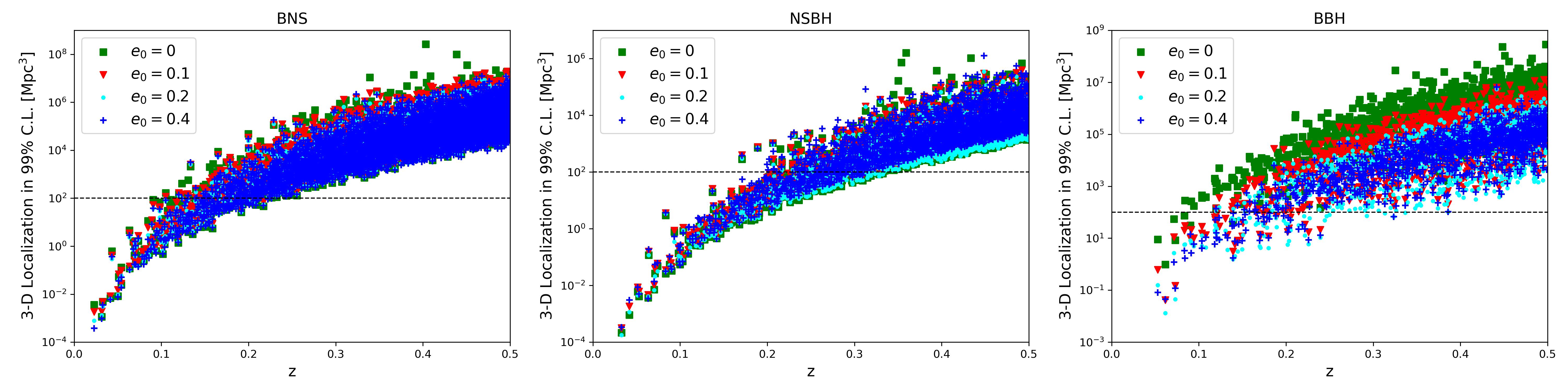}
\caption{The 3-D localization volume of BNS, and NSBH, and BBH detected by DECIGO-I within 1-year operation time. We set a cut-off at $z=0.5$.  The dashed line is the threshold volume $V_{\rm th}=100~\rm Mpc^3$. }
\label{fig:V_loc_DI}
\end{figure*}

In figure~\ref{fig:V_loc_DI}, we show the 3-D localization volume of BNS, NSBH, and BBH detected by DECIGO-I within 1-year operation time. For the readability of the plots we only show the cases with $e_0=0$, 0.1, 0.2, and 0.4, which can present the main feature and tendency of the results (see figure~\ref{fig:Rwe}). For BNS and NSBH, there is no obvious improvement of the 3-D localization volume from eccentricity because the improvement only comes from the distance inference in the near face-on orientations. Though the improvement is negligible, BNS and NSBH are precisely localized and there are hundreds of golden events regardless of the eccentricity. The largest redshifts of golden dark BNS and NSBH are 0.25 and 0.35, respectively. While, in the BBH case, we can also detect a few golden dark sirens in the circular case below redshift 0.1. However, a nonvanishing eccentricity can significantly improve the localization so that much more golden dark BBH can be detected. The largest redshift of golden dark BBH is around 0.4 when $e_0=0.4$. We move the results for the distance inference and sky localization of BNS, NSBH, and BBH to appendix~\ref{app:B}. 

\begin{figure*}
\includegraphics[width=\textwidth]{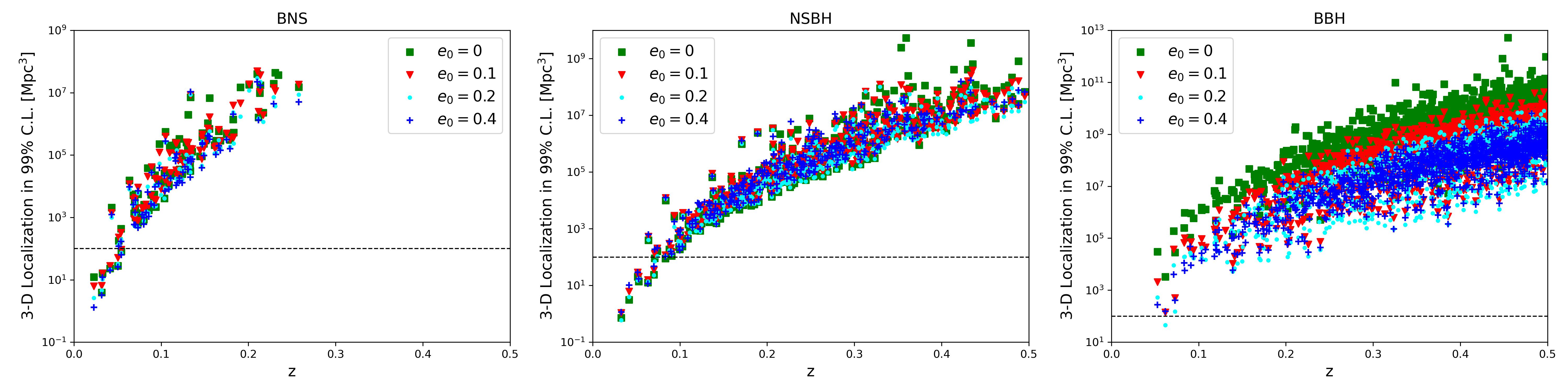}
\caption{The same as figure~\ref{fig:V_loc_DI}, but with B-DECIGO.}
\label{fig:V_loc_BD}
\end{figure*}

Figure~\ref{fig:V_loc_BD} shows the results of 3-D localization volume with B-DECIGO. As expected, in the B-DECIGO case, the numbers of golden dark sirens are much smaller.  We can only detect a few golden BNS and NSBH up to redshift 0.05 and 0.1, respectively. For BBH, the golden event is unlikely in the circular case but a nonvanishing $e_0=0.2$ can make it possible to detect the golden dark BBH.

In table~\ref{tab:golden}, we summarize the number of golden dark sirens detected by DECIGO-I and B-DECIGO within 1-year observation time. As expected from the results in section~\ref{sec:typical}, eccentricity leads much more improvement for the localization of BBH than that of BNS and NSBH. With only one cluster of DECIGO running for 1 year in its design sensitivity, the observations of hundreds of golden BNS, NSBH, and tens of golden BBH are very promising. As a pathfinder in the near future, B-DECIGO can also observe a few golden BNS and NSBH. The detection of golden BBH is possible if it is eccentric. We can see the number of golden NSBH is relatively larger than that of BNS and BBH. This is due to the longer inspiral time of NSBHs  (compared to BBH) and heavier mass (compared to BNS). The former provides good sky localization and the latter ensures well-inferred distance. Our results also suggest that, for NSBH and BBH, we can observe a little more golden events with $e_0=0.2$ than other eccentricities. This can be explained by the trends of the improvement of 3-D localization volume for NSBH and light BBH in figure~\ref{fig:Rwe}.

\begin{table*}
\centering
\begin{tabular}{c|cccc|cccc|cccc} 
\hline\hline
         & \multicolumn{4}{c|}{Golden BNS}             & \multicolumn{4}{c|}{Golden NSBH}            & \multicolumn{4}{c}{Golden BBH}               \\ 
\cline{2-13}
         & $e_0=0$ & $e_0=0.1$ & $e_0=0.2$ & $e_0=0.4$ & $e_0=0$ & $e_0=0.1$ & $e_0=0.2$ & $e_0=0.4$ & $e_0=0$ & $e_0=0.1$ & $e_0=0.2$ & $e_0=0.4$  \\ 
\hline
DECIGO-I & 181     & 190       & 225       & 230       & 367     & 328       & 396       & 310       & 7       & 38        & 65        & 55         \\
\hline
B-DECIGO & 6       & 6         & 6         & 6         & 8       & 6         & 7         & 6         & 0       & 0         & 1         & 0          \\ 
\hline
\end{tabular}
\caption{The number of golden dark sirens for 1-year observation of DECIGO-I and B-DECIGO on the assumption of various eccentricities.}
\label{tab:golden}
\end{table*}

Here we would like to give a discussion on the assumptions of the eccentricity we made in the construction of the GW catalogs. In each case of $e_0$, we assume that the initial eccentricity is uniform for all the binaries. This is good for the comparison between different scenarios of initial eccentricity but not a realistic assumption.  In principle, we should also sample the value of $e_0$ according to a given distribution, along with other waveform parameters when constructing the GW catalogs. However, the formation channel of compact binaries and the distribution of eccentricity at a specific frequency (in our case $f_0=0.1$ Hz) are still under debate and not clear~\cite{Wen:2002km,Kowalska:2010qg,Takatsy:2018euo}. Simulations suggest that eccentric mergers from strong gravitational encounters account for 10\% of the underlying population of BBH mergers in globular clusters, and about half of these having eccentricities larger than 0.1 at 10 Hz~\cite{Samsing:2013kua,Samsing:2017xmd,Samsing:2017rat} (and references therein). While, back to decihertz, the population of eccentric binaries and their eccentricities should be much larger. For the isolated formation scenario, even binaries born with high eccentricity could be fully circularized when entering the LIGO/Virgo band, by efficiently damping orbital eccentricity through angular momentum loss from GW emission. However, the eccentricity of the field binary at mid-band is still uncertain. Since eccentricity has small effects on the population of GW detections (see figures~\ref{fig:hist_BD} and \ref{fig:hist_DI}), the total number of events in the GW catalogs will not be largely altered. As for the number of golden dark sirens in a realistic situation (considering the distribution of eccentricity), we can estimate it by weighting the numbers of the circular and eccentric cases in table~\ref{tab:golden}, roughly as $N_{\rm gold}=\sum P_{e_0}N_{e_0}$. Here, $N_{e_0}$ is the number of golden dark sirens by assuming the same eccentricity $e_0$ for all binaries (the numbers in table~\ref{tab:golden}). $P_{e_0}$ is the actual fraction of the events which hold eccentricity $e_0$.  Since we focus on the comparison of the results between different eccentricity scenarios, we prefer to present the results based on different $e_0$ separately. 

When constructing the GW catalogs and estimating the numbers of golden dark sirens, we only adopt the median value of the predicted merger rates and neglect their uncertainties.  The merger rates of BNS and NSBH are much more uncertain than that of BBH since we have lots of BBH detections and only a few BNS and NSBH detections in current LIGO-Virgo-KAGRA catalogs. This means the simulated BBH catalogs are more realistic than that of BNS and NSBH. We will give a discussion for this in section~\ref{sec:conclusion}. Also, considering the fact that eccentricity can significantly improve the localization for BBH only, we prefer concentrating on BBH golden dark sirens to show the implications of eccentric dark sirens on cosmology in section~\ref{sec:cosmo}.

\section{Cosmological implications \label{sec:cosmo}}

To make dark sirens accurate and precise cosmological probes, distance inference and source localization are two crucial factors. The GWs of long inspiraling compact binaries observed by the space-borne detector in the mid-band can provide much tighter constraints in these two aspects, thus the more precise measurement of the cosmological parameters, compared to the LIGO-Virgo band (see some examples~\cite{Yang:2021xox,Yang:2022iwn,Liu:2022rvk} and also in the millihertz band~\cite{Wang:2020dkc,Zhu:2021aat,Zhu:2021bpp}). Without EM counterparts, dark sirens in the mid-band still face the issues like large degeneracy between luminosity distance and inclination angle in the near face-on orientations, as well as the uncertain localization for the larger-mass BBH. In section~\ref{sec:typical} we demonstrate that the eccentricity, which is more likely to be non-vanishing in the mid-band than in the LIGO-Virgo band, can greatly alleviate these issues. The localization is more precise in the near face-on orientations where the eccentricity happens to significantly improve the distance inference. The precise localization for these compact binaries (BNS, NSBH, and BBH) makes it possible that we can identify their host galaxies without the help of the EM counterparts. Bear in mind that EM counterparts can help to identify the host galaxy~\cite{LIGOScientific:2017vwq,LIGOScientific:2017ync}, and also improve the constraint of inclination angle (and hence the distance inference)~\cite{Hotokezaka:2018dfi}. It means that eccentricity in the mid-band can make dark sirens equivalent to the bright sirens in the context of the distance and redshift measurement, or greatly reduce their uncertainties. These features can significantly improve the constraints of cosmological parameters/models from dark sirens. Also, the improvement in the estimation of other parameters like inclination angle and coalescence time can help for the follow-up search of EM counterparts and observation of GWs at high frequency.

In section~\ref{sec:catalog}, we construct the catalogs of GWs based on DECIGO-I and B-DECIGO. We demonstrate that a certain amount of golden dark BNS, NSBH, and BBH can be observed in the mid-band.  In particular,  a nonvanishing eccentricity can significantly increase the number of golden dark BBH. In this section, we would like to focus on these golden dark sirens and assess the improvement for the constraints of cosmological parameters/models from eccentricity, especially for BBH dark sirens. Since the sensitivity of B-DECIGO is not good enough, it can only observe $\mathcal{O}(1)$ golden dark sirens and the numbers are not greatly changed in the eccentric cases. Therefore, we only use the golden dark sirens with DECIGO-I to exemplify the improvement of cosmological constraints from eccentricity.

\subsection{The Hubble diagram of golden dark sirens}

To construct the Hubble diagram of these golden dark sirens, we assume the fiducial cosmological model to be $\Lambda$CDM with $H_0=67.72~\rm km~s^{-1}~Mpc^{-1}$ and $\Omega_{\rm m}=0.3104$, corresponding to the mean values obtained from the latest \textit{Planck} TT,TE,EE+lowE+lensing+BAO+Pantheon data combination~\cite{Planck:2018vyg}. We also fix the present CMB temperature $T_{\rm CMB}=2.7255~\rm K$, the sum of neutrino masses $\Sigma_{\nu}m_{\nu}=0.06~\rm eV$, and the effective extra relativistic degrees of freedom $N_{\rm eff}=3.046$, as in the {\it Planck} baseline analysis.  We include the errors from weak lensing and the peculiar velocity of the source galaxy. For weak lensing, we adopt the analytically fitting formula~\cite{Hirata:2010ba,Tamanini:2016zlh}
\begin{equation}
\left(\frac{\Delta d_L(z)}{d_L(z)}\right)_{\rm lens}=0.066\left(\frac{1-(1+z)^{-0.25}}{0.25}\right)^{1.8} \,.
\end{equation}
We consider a delensing factor. Following~\cite{Speri:2020hwc} we adopt a phenomenological formula
\begin{equation}
F_{\rm delens}(z)=1-\frac{0.3}{\pi/2}\arctan(z/z_*) \,,
\end{equation}
where $z_*=0.073$. Then the final error from weak lensing is
\begin{equation}
\left(\frac{\Delta d_L(z)}{d_L(z)}\right)_{\rm delens}=F_{\rm delens}(z)\left(\frac{\Delta d_L(z)}{d_L(z)}\right)_{\rm lens} \,.
\end{equation}
For the peculiar velocity uncertainty, we use the fitting formula~\cite{Kocsis:2005vv},
\begin{equation}
\left(\frac{\Delta d_L(z)}{d_L(z)}\right)_{\rm pec}=\left[1+\frac{c(1+z)^2}{H(z)d_L(z)}\right]\frac{\sqrt{\langle v^2\rangle}}{c} \,,
\end{equation}
here we set the peculiar velocity value to be 500 km/s, in agreement with average values observed in galaxy catalogs. For the redshift measurement, we assume the host galaxies of all the golden dark sirens can be unambiguously identified and their redshifts are measured spectroscopically.

\begin{figure*}
\includegraphics[width=0.45\textwidth]{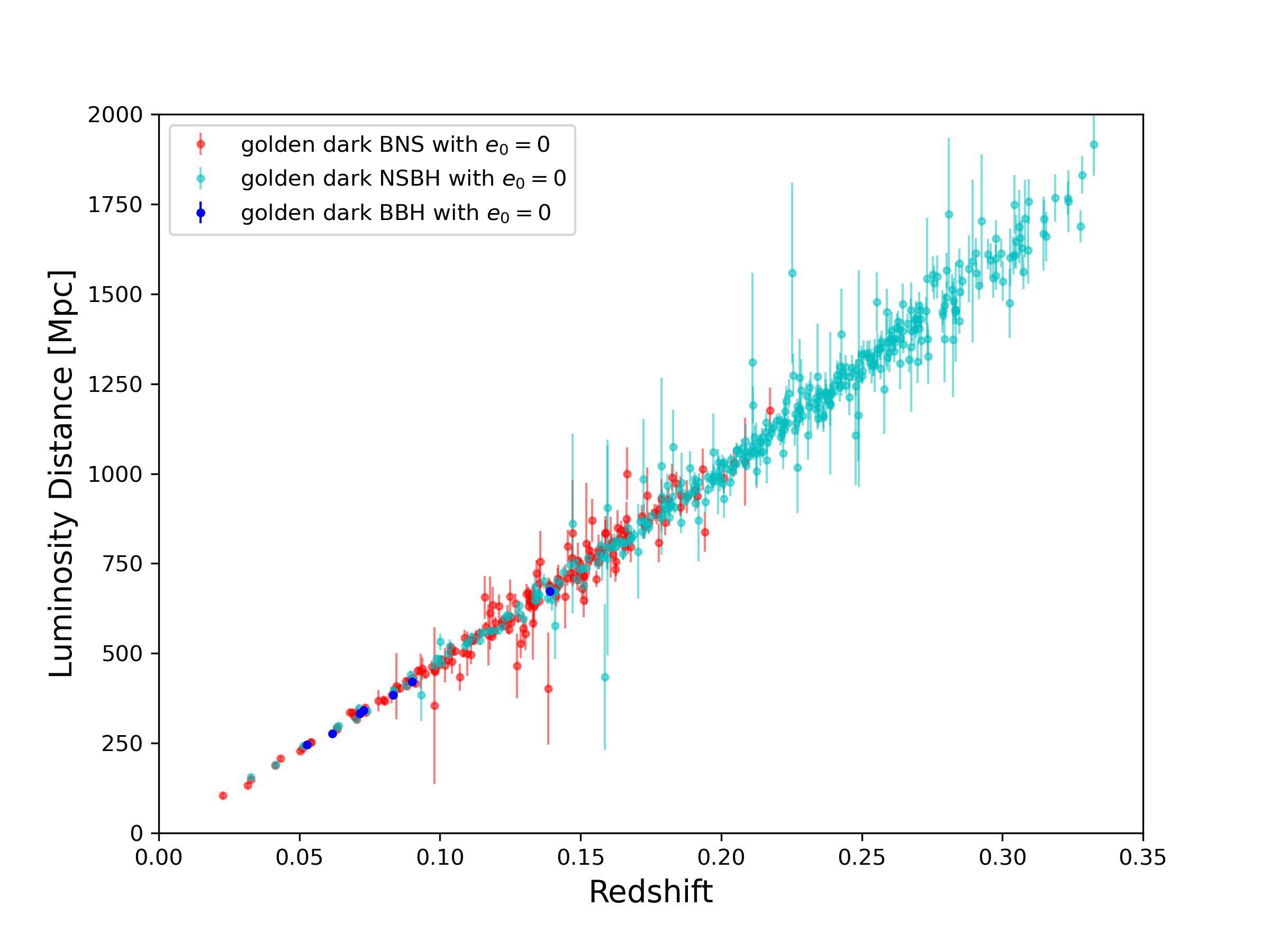}
\includegraphics[width=0.45\textwidth]{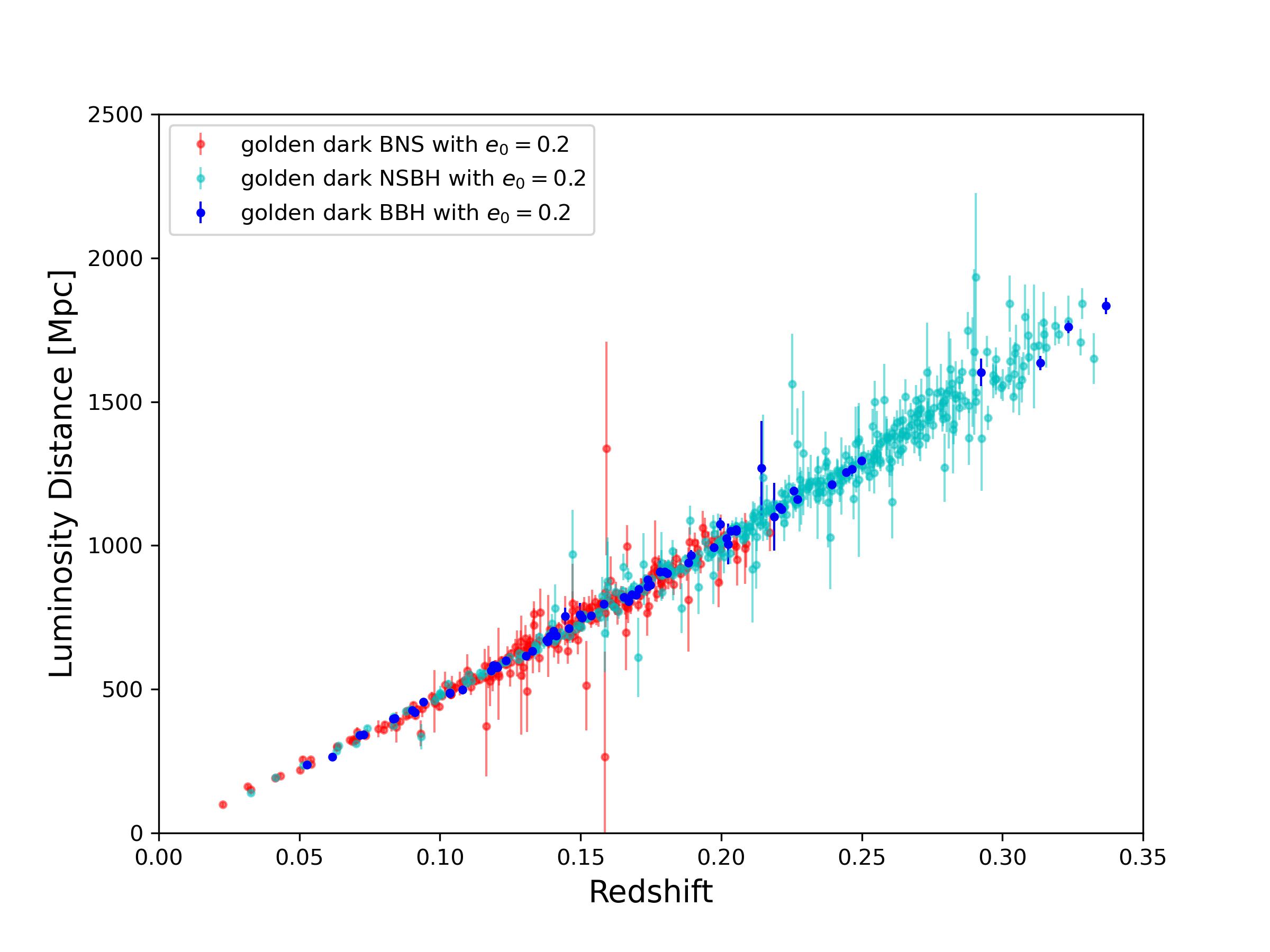}
\caption{The Hubble diagram of the golden dark sirens observed by DECIGO-I in 1 year with $e_0=0$ (left) and $e_0=0.2$ (right). Note in the $e_0=0$ case,  we discard three BNS events whose distance errors are larger than 50\%.}
\label{fig:HD}
\end{figure*}

Figure~\ref{fig:HD} shows one realization of the Hubble diagram of the golden dark sirens observed by DECIGO-I in 1 year with $e_0=0$ and $e_0=0.2$. We have discarded three BNS events whose distance errors $\Delta d_L/d_L$ are larger than 50\% in the $e_0=0$ case since they are not useful in constraining cosmological parameters. While in $e_0=0.2$ case, all the events have distance errors smaller than $50\%$. In this section, we choose $e_0=0.2$ to represent the eccentric case. We can clearly see the golden BBHs have much smaller distance errors and they can reach higher redshifts (in the $e_0=0.2$ case), which means BBH should be more efficient than BNS and NSBH in measuring the cosmological parameters. 
We have shown that BBH dark sirens benefit most from eccentricity. While BNS and NSBH dark sirens only get non-substantial help from eccentricity (e.g. for the distance inference in the near face-on orientations). Bear in mind that the possible associate EM counterparts can also help BNS and NSBH identify their host galaxies. Therefore, the improvement of cosmological constraints from eccentric dark sirens is mainly manifest in the case of BBH dark sirens. In addition, the predicted errors for the merger rates of BNS and NSBH are very uncertain, which makes the numbers of golden BNS and NSBH less affirmative than that of BBH (see discussion in section~\ref{sec:conclusion}). Therefore, in this section, we stick to the BBH dark sirens in measuring the cosmological parameters and move the results of BNS and NSBH to appendix~\ref{app:C}. 
Since the redshift of golden dark sirens mainly reside in the low redshift region where the dynamics of dark energy is not very sensitive, we only focus on the following two cosmological models.  

\subsection{Constraints on $\Lambda$CDM model from BBH golden dark sirens}

In the baseline $\Lambda$CDM model with dark energy as a cosmological constant, there are two free parameters, namely, the Hubble constant $H_0$ and the matter density parameter $\Omega_{\rm m}$. 
To infer the posteriors of the cosmological parameters, we run Markov-Chain Monte-Carlo (MCMC) using the package {\sc Cobaya}~\cite{Torrado:2020dgo,2019ascl.soft10019T}. The marginalized statistics of the parameters and the plots are produced by the Python package {\sc GetDist}~\cite{Lewis:2019xzd}.
When we sample the distance measurement in the Hubble diagram, the scatter of the random sampling should be taken into account. For each type of golden dark siren, we randomly sample 10 sets of Hubble diagrams (repeat the sampling 10 times). We run MCMC for each data set and choose the median result as the representative. 

\begin{figure}
\includegraphics[width=0.45\textwidth]{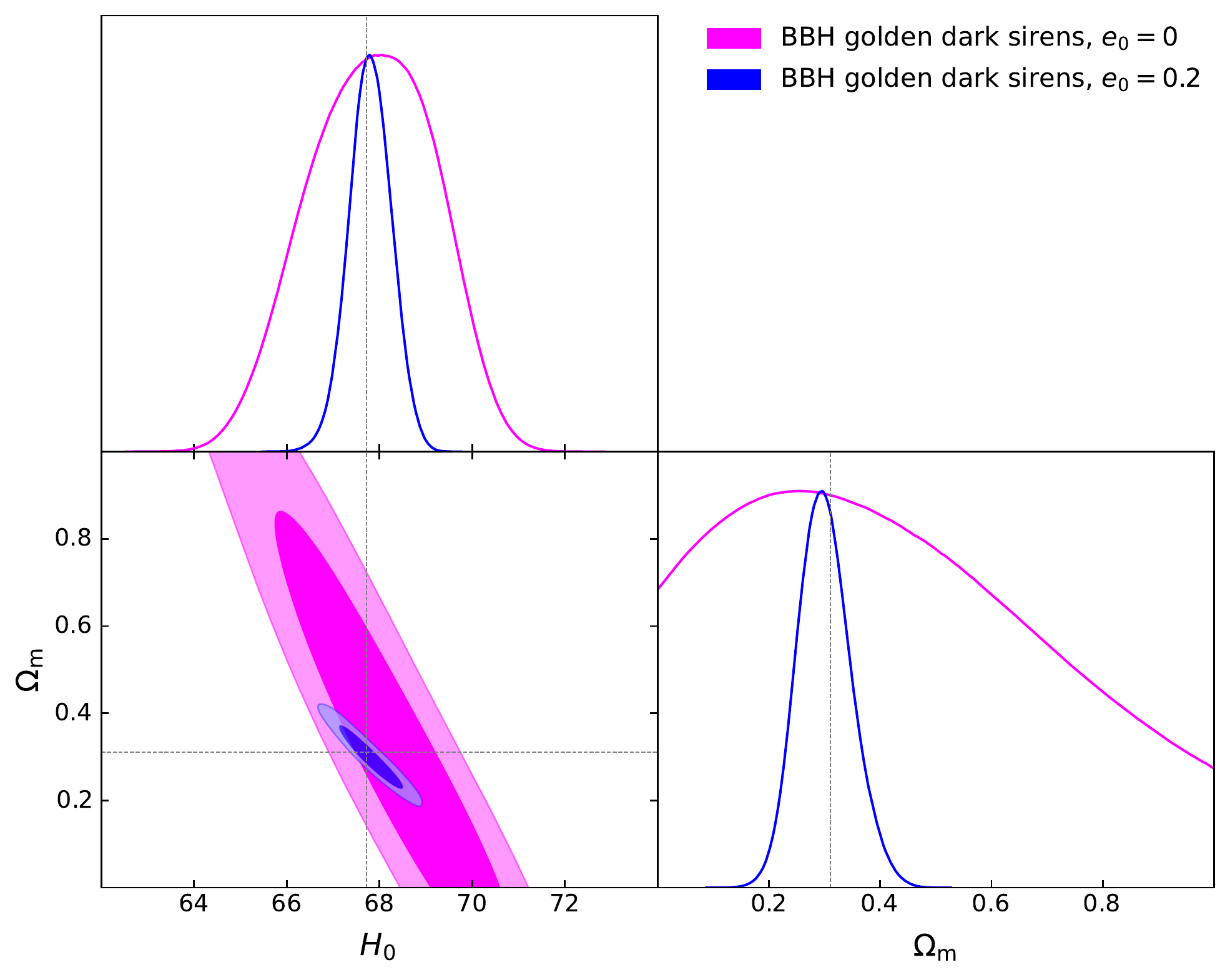}
\caption{The constraints of Hubble constant $H_0$ and matter density parameter $\Omega_{\rm m}$ with $\Lambda$CDM model from 1-year observation of the BBH golden dark sirens by DECIGO-I. Contours contain 68 \% and 95 \% of the probability. The dashed lines denote the fiducial value of the parameters.}
\label{fig:LCDM_BBH}
\end{figure}

Figure~\ref{fig:LCDM_BBH} shows the constraints of the Hubble constant $H_0$ and matter density parameter $\Omega_{\rm m}$ by the golden dark BBH from 1-year observation of DECIGO-I. In the circular case, 7 golden dark BBH can constrain $H_0=67.80^{+1.50}_{-1.30}~\rm km~s^{-1}~Mpc^{-1}$. While in the eccentric case with $e_0=0.2$, 65 golden dark BBH can constrain $H_0=67.80\pm 0.46~\rm km~s^{-1}~Mpc^{-1}$. The precision of the Hubble constant measurement is improved from 2.06\% to 0.68\%. For the matter density parameter, $\Omega_{\rm m}=0.43^{+0.14}_{-0.41}$ with $e_0=0$ and $\Omega_{\rm m}=0.30^{+0.043}_{-0.051}$ with $e_0=0.2$. In the eccentric case, the golden dark BBH can reach higher redshifts, where $\Omega_{\rm m}$ is more sensitive. Therefore, eccentricity can greatly improve the precision of $\Omega_{\rm m}$ from 64\% to 16\%.

\subsection{Constraints on modified gravity from BBH golden dark sirens}

The general test of general relativity (GR) by the propagation of GWs across cosmological distances has been investigated in recent research~\cite{Belgacem:2017ihm,Belgacem:2018lbp,Nishizawa:2017nef,Arai:2017hxj,Nishizawa:2019rra,LISACosmologyWorkingGroup:2019mwx,Belgacem:2019tbw,Belgacem:2019zzu,Mukherjee:2019wcg,DAgostino:2019hvh,Bonilla:2019mbm,Mukherjee:2020mha,Kalomenopoulos:2020klp,Mastrogiovanni:2020mvm,Mastrogiovanni:2020gua,Finke:2021aom}.  In a generic modified gravity model the linearized evolution equation for GWs traveling on an Friedmann–Robertson–Walker (FRW) background in four-dimensional space-time is~\cite{LISACosmologyWorkingGroup:2019mwx}
\begin{equation}
\tilde{h}_A^{\prime\prime}+2[1-\delta(\eta)]\mathcal{H}\tilde{h}_A^{\prime}+[c_T^2(\eta)k^2+m_T^2(\eta)]\tilde{h}_A=\Pi_A \,,
\label{eq:GWpropa}
\end{equation}
where $\tilde{h}_A$ are the Fourier modes of the GW amplitude. $\mathcal{H}=a^{\prime}/a$ is the Hubble parameter in conformal time, the primes indicate derivatives with respect to conformal time $\eta$. $A=+,\times$ label the two polarizations of GWs. $\Pi_A$ is the source term, related to the anisotropic stress tensor. The function $\delta (\eta)$ modifies the frication term in the propagation equation. $c_T$ corresponds to the speed of gravitational waves.  In theories of modified gravity (MG), the tensor mode can be massive, with $m_T$ being its mass. In GR, we have $\delta=0$, $c_T=c$,  and $m_T=0$. The observation of GW170817/GRB170817A puts a very tight constraint on the speed of the gravitational wave, $(c_T-c)/c<\mathcal{O}(10^{-15})$~\cite{LIGOScientific:2017zic}. 

In this paper, we only retain the deviations from GR induced by the friction term and set $c_T=c$ and $m_T=0$~\cite{Belgacem:2018lbp,Belgacem:2019tbw}. Then one can show the inferred ``GW luminosity distance'' in modified gravity theories is different from the traditional ``electromagnetic luminosity distance''~\cite{Belgacem:2017ihm,Belgacem:2018lbp},
\begin{equation}
d_L^{\rm gw}(z)=d_L^{\rm em}(z)\exp\left\{-\int_0^z\frac{dz^{\prime}}{1+z^{\prime}}\delta(z^{\prime})\right\} \,.
\end{equation}
To constrain the modified gravity theory (or to test GR), we need to constrain the $\delta$ function. In this section, we follow~\cite{Belgacem:2018lbp,Belgacem:2019tbw} and adopt the 2-parameter phenomenological parameterization for the deviation of GR,
\begin{equation}
\Xi(z)\equiv\frac{d_L^{\rm gw}(z)}{d_L^{\rm em}(z)}=\Xi_0+\frac{1-\Xi_0}{(1+z)^n} \,.
\label{eq:Xi}
\end{equation}
From GW and EM measurements one can therefore access the quantity $\delta(z)$, or equivalently $\Xi(z)$, a smoking gun of modified gravity. $\Xi_0=1$ recovers GR. This parametrization is very general and expected to fit the predictions from a large class of MG models~\cite{LISACosmologyWorkingGroup:2019mwx}. We set $n=2.5$ since it plays in general a lesser role in the parametrization~\cite{Belgacem:2018lbp}.

To constrain $\Xi_0$, we combine GWs from BBH golden dark sirens and EM data from CMB, BAO, and SNe Ia. We use CMB data from the latest {\it Planck} release~\cite{Planck:2018vyg}, that is, \textit{Planck} TT,TE,EE+lowE+lensing (hereafter {\it Planck}).  For BAO, we adopt the isotropic constraints provided by 6dFGS at $z_{\rm eff}=0.106$~\cite{Beutler:2011hx}, SDSS-MGS DR7 at $z_{\rm eff}=0.15$~\cite{Ross:2014qpa}, and ``consensus'' BAOs in three redshift slices with effective redshifts $z_{\rm eff}$ = 0.38, 0.51, and 0.61~\cite{BOSS:2016apd,Vargas-Magana:2016imr,BOSS:2016hvq}. We use the Pantheon data~\cite{Pan-STARRS1:2017jku} as the latest compilation of SNe Ia. We also set the equation of state of dark energy $w_{\rm DE}$ as a free parameter. We incorporate the MG parameter $\Xi_0$ into  {\sc Cobaya} and run MCMC to infer the posteriors.

\begin{figure}
\includegraphics[width=0.45\textwidth]{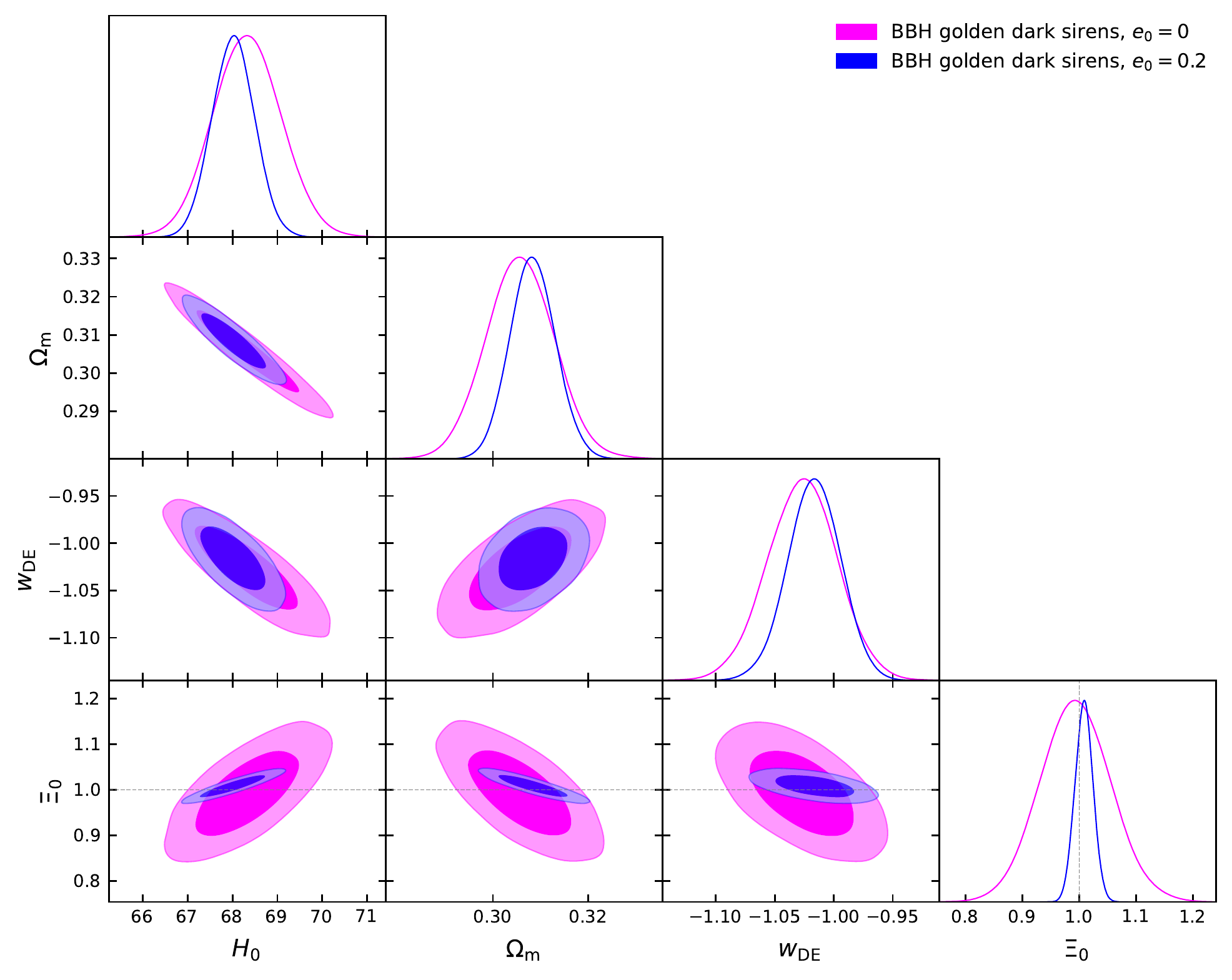}
\caption{Constraints on parameters of a phenomenological parameterization of modified GW propagation from 1-year observation of the BBH golden dark sirens by DECIGO-I combined with the EM experiments {\it Planck}, BAO, and Pantheon. Contours contain 68 \% and 95 \% of the probability. The dashed lines denote the fiducial value of $\Xi_0$.}
\label{fig:MG_BBH}
\end{figure}

Figure~\ref{fig:MG_BBH} shows the posteriors of the Hubble constant $H_0$, matter density parameter $\Omega_{\rm m}$, equation of state of dark energy $w_{\rm DE}$ and the MG parameter $\Xi_0$ inferred from the BBH golden dark sirens by 1-year observation of DECIGO-I, together with the EM anchors from {\it Planck}, BAO, and Pantheon data sets. With $e_0=0$, the golden dark BBH can constrain $\Xi_0=0.994\pm 0.062$. A nonvanishing eccentricity $e_0=0.2$ can improve the constraint to $\Xi_0=1.008\pm 0.016$. The precision of constraining MG effects through the phenomenological parameterization of GW propagation is improved from 6.2\% to 1.6\%. With more BBH golden dark sirens in the eccentric case, the constraints of other parameters are also slightly tighter. Since the MG parameter is more sensitive to the GW data, it can achieve more improvement than other parameters. 

In this section, we discuss the implications of eccentricity for dark sirens on cosmology. We estimate the number of golden dark sirens on the assumption of various eccentricities. We focus on the BBH golden dark sirens and investigate their constraints on cosmological parameters in the circular and eccentric cases. The golden dark BBH observed in a larger number and at higher redshifts with eccentricity can greatly improve the constraints of cosmological parameters and MG effects. Note in this section we only investigate the cases of golden dark BBH whose host galaxies can be uniquely identified. There are numerous dark BBH with multiple host galaxies and they can reach much higher redshifts. A statistical method should be adopted for the applications of these dark sirens on cosmology~\cite{LIGOScientific:2018gmd,DES:2019ccw,LIGOScientific:2021aug,Finke:2021aom}. These BBH dark sirens with significantly improved localization from eccentricity can also achieve large improvements in constraining cosmological parameters, especially for dynamics of dark energy and MG at high redshift. We leave this for future research.

\section{Conclusions and discussions \label{sec:conclusion}}

\subsection{A summary of the results}

In the first part of this paper, we investigate the parameter estimation of GWs emitted by eccentric compact binaries with decihertz observatory. The multiple harmonics induced by eccentricity can provide more information and break the degeneracy between parameters. We find eccentricity of compact binaries in the mid-band can greatly improve the parameter estimation of GWs. In particular, with $e_0=0.4$ at $f_0=0.1$ Hz, the typical binaries (BNS, NSBH, and BBH) can achieve 1.5-3 orders of magnitude improvement for the distance inference in the near face-on orientations and 1-3.5 orders of magnitude improvement for the localization. We find that other parameters like inclination angle, chirp mass, coalescence time, and eccentricity can also be estimated more precisely with a non-vanishing eccentricity. In this paper, we use DECIGO as an example detector in the mid-band. We consider two scenarios for the configuration of DECIGO. The first is the one cluster of DECIGO with its designed sensitivity, which we call DECIGO-I. The second is B-DECIGO which also has one cluster but with inferior sensitivity. The physical findings, i.e., the improvement of parameter estimation of GWs from eccentricity, do not rely on a specific detector.

In the second part, we construct the GW catalogs observed by DECIGO in 1 year. On the assumption of various eccentricities, we estimate the population of golden dark sirens whose localization is precise enough so that their host galaxies can be uniquely identified. We find that, with only one cluster of DECIGO running 1 year in its designed sensitivity, the observations of hundreds of golden BNS, NSBH, and several tens of golden BBH are very promising. A non-vanishing eccentricity can significantly enlarge the number of golden dark BBH. This can greatly improve the constraints of cosmological parameters and modified gravity from BBH dark sirens. For instance, the number of BBH golden dark sirens is 65 in $e_0=0.2$ case, compared to 7 in $e_0=0$ case. As a result, in the $\Lambda$CDM model, the constraints of Hubble constant and matter density parameter from  BBH golden dark sirens can be improved from 2.06\% to 0.68\% and from 64\% to 16\%, respectively. Through the phenomenological parameterization of GW propagation, the constraint of modified gravity can be improved from 6.2\% to 1.6\%. While, due to the worse sensitivity, the results based on B-DECIGO are much inferior to that of DECIGO-I. The results in this paper can serve as a forecast for one cluster of DECIGO as well as for the early launched pathfinder B-DECIGO. These configurations are much more promising than the full configuration (four clusters) of DECIGO in the future.

Our result shows the remarkable significance of eccentricity for GWs on detection, data analysis, and physics including cosmology, astrophysics, and fundamental physics. The eccentric frequency-domain waveform adopted in this work is limited up to $e_0=0.4$. A ready-to-use more accurate and widely applicable eccentric waveform deserves extensive investigations by the community~\cite{Cao:2017ndf,Liu:2019jpg,Liu:2021pkr,Ramos-Buades:2021adz,Hinder:2017sxy,Moore:2019xkm,Joshi:2022ocr}, especially for GW detections in the mid-frequency band.

\subsection{The uncertainty of the golden dark sirens}

When constructing the GW catalogs and estimating the number of golden dark sirens, some assumptions have been made. We neglect the uncertainty of merger rates of $\mathcal{R}_{\rm BNS}=105.5^{+190.2}_{-83.9}~\rm Gpc^{-3}~\rm yr^{-1}$, $\mathcal{R}_{\rm NSBH}=45^{+75}_{-33}~\rm Gpc^{-3}~\rm yr^{-1}$, and $\mathcal{R}_{\rm BBH}=23.9^{+14.3}_{-8.6}~\rm Gpc^{-3}~\rm yr^{-1}$ and only adopt the median value. Due to the fewer events of BNS and NSBH in GWTC-3, their predicted event rates are more uncertain than that of BBH. In addition, the predicted merge rate of NSBH is based on the assumption that the observed NSBH GW200105 and GW200115 are representatives of the population of NSBH, which is not very realistic. We can roughly estimate the uncertainty of the numbers of golden dark sirens in table~\ref{tab:golden} using $\Delta N/N\sim\Delta \mathcal{R}/\mathcal{R}$~\footnote{Also note that the numbers of GW events in the catalogs are based on one realization of the sampling. However, the errors due to random sampling (Poisson) should be much smaller than the uncertainty of merger rates.}. The large uncertainty for the merger rates of BNS and NSBH makes the catalogs of golden dark BNS and NSBH very uncertain, as well as their applications on cosmology. The future observing run of LIGO-Virgo-KAGAR can provide a tighter prediction for the merger rates. Furthermore, from the observation perspective, it takes months to years to monitor the BNS and NSBH events in the mid-band, which is quite challenging in observing hundreds of events due to the time limits and overlap between signals. In contrast, the inspiral period of BBH is very short (hours to days), which makes it promising to search for tens of golden dark BBH in 1-year observation time. Considering the factors above, the results of BBH in our paper are more realistic and reliable than that of BNS and NSBH. 

As we have mentioned, we estimate the number of golden dark sirens by assuming the same value of eccentricity for all the binaries. The realistic number should be a weighted summation of the numbers in circular and eccentric cases, according to the distribution of eccentricity for each type of binary in the mid-band. Thus the results of the cosmological constraints should also be averaged accordingly. 
One can roughly estimate that $\Delta P_{\rm cosmo}\sim 1/\sqrt{N_{\rm gold}}$. For instance, even if only 10\% binaries have a nonvanishing eccentricity in the mid-band, e.g. $e_0=0.2$, the number of golden BBH can achieve $\sim 13$ with DECIGO-I. Then the Hubble constant in the $\Lambda$CDM model can be constrained at 1.5\% precision level. While, if eccentric BBH can take up 30\% of the total BBH in the mid-band, $H_0$ can be constrained at sub-percent level. 
Another factor in the uncertainty of golden dark sirens is that we assume the galaxies are uniformly distributed in the Universe. However, the clustering and grouping feature of galaxies can make it much easier to infer the redshift information of GWs from the cluster and group of its host galaxy instead of the host galaxy itself~\cite{Yu:2020vyy}. This means that our estimation for the number of golden dark sirens can be very conservative. 

\subsection{Issues in host identification}

For the host identification of golden GWs (hence the inference of their redshifts), we assume that the true host galaxy can be accurately identified if it solely resides in the localized volume. However, in the real situation, one needs to match ($d_{L\rm gw}, \theta_{\rm gw}, \phi_{\rm gw}$) from GWs to ($z_{\rm gal}, \theta_{\rm gal}, \phi_{\rm gal}$) from the galaxies. In this case, the cosmology (e.g. the Hubble constant) should be marginalized to infer the posterior of redshift $z_{\rm gw}$. Usually, we can neglect the uncertainty of $z_{\rm gal}$, $\theta_{\rm gal}$, and $\phi_{\rm gal}$ from either targeted observation campaign or galaxy catalogs (note the peculiar motion can add an uncertainty in $z_{\rm gal}$). By assuming a prior of $H_0$ (either from Planck or SH0ES, or a given prior we believe based on current knowledge), if ($d_{L\rm gw}, \theta_{\rm gw}, \phi_{\rm gw}$) are precisely measured (in our case the golden events), the posterior of $z_{\rm gw}$ should be very consistent with $z_{\rm gal}$ of the true galaxy which solely resides in the localized volume. In addition, most of the golden events in our paper are in the low redshift region $z\ll 1$. Therefore, when converting $d_{L\rm gw}$ to $z_{\rm gw}$ by $z_{\rm gw}\sim H_0 d_{L\rm gw}$, the uncertainty of cosmology $\Delta H_0$ will lead a smaller $\Delta z_{\rm gw}$ compared to large redshift/distance. If the galaxy is uniformly distributed in 3-D volume, in the line of sight, $z_{\rm gal}$ is more sparsely distributed in low redshift than in high redshift. This means that for the golden events at low redshift, it will be easier to match $z_{\rm gw}$ to the true $z_{\rm gal}$. Furthermore, as we have mentioned, the clustering of galaxies can make it easier to infer the redshifts of GWs, which will somehow downplay the effect of cosmology.

In this paper, we focus on the 3-D localization of GWs and would like to estimate the number of GW events for which there is only one galaxy inside their localized volumes. For these golden events, we assume that their true host galaxies and redshifts can be accurately inferred (note we include the uncertainty from peculiar motion when constructing the Hubble diagram). The specific analysis in the host identification (match between $z_{\rm gw}$ and $z_{\rm gal}$, peculiar motion, etc.) is beyond the scope of this paper. We will address these issues in future research.

\subsection{A caveat in the Fisher matrix}

Throughout this paper, we use the Fisher matrix to do the parameter estimation of GW waveform. Fisher matrix is a widely-used data analysis technique for the estimation of the waveform parameters and should be consistent with the more robust method like Markov chain Monte Carlo (MCMC) when the SNR of the signal is high enough (the strength of the signal or the sensitivity of the detector is good enough). Fisher matrix inherently assumes the errors (and the correlation) are Gaussian. Its nature is to assess the sensitivity of the waveform to the change of the parameters, from which one can get a bound of the error for each parameter. The estimation is more accurate (consistent with the full MCMC) for a larger SNR, in which case the parameters are tightly constrained. In this paper, the main purpose is to propose the idea that the eccentricity of the compact binary can significantly improve the parameter estimation of the GW waveform in the mid-band. Therefore we just stick to the theoretical (or physical) improvement of the parameter estimation of GW waveform from circular to eccentric cases, for which the Fisher matrix is adequate. In terms of the theoretical improvement, we say the improvement of the parameter estimation that the eccentric waveform ``should'' achieve (compared to circular waveform) in the ideal situation (Gaussian errors, adequate SNR). What we are interested in and would like to present in this paper is the physical improvement of parameter estimation from eccentricity in the waveform (by providing more information and breaking the degeneracy between parameters). This physical implication should not be limited by the condition of the detector (poor sensitivity and SNR, etc.). In this paper, we only adopt DECIGO as an example detector to demonstrate this improvement. We should bear in mind that if we conduct our data analysis with a more realistic technique such as MCMC which involves the prior information and the non-Gaussian posterior, especially when the SNR is low and the parameter is poorly constrained, the improvement could be inferior to the theoretical improvement derived from the Fisher matrix. However, the theoretical improvement we present in this paper can tell us that the constraints of the waveform parameters could be ``easier'' by that level if the binary is eccentric, no matter what techniques and detectors we adopt.

In section~\ref{sec:typical}, based on the example detectors, we compare the eccentric waveform with the circular one to show the improvement of the parameter estimation from eccentricity. This improvement is theoretical in the context of the Fisher matrix. Since the localization-relevant parameters are precisely constrained with the space-borne detector in the mid-band, the theoretical improvement of the localization derived from the Fisher matrix should be consistent with the full MCMC analysis. For the distance inference with large inclination angles, the error is very small and there is also no serious issue. While in the near face-on orientations, the distance error is quite large and there could be a discrepancy between the Fisher matrix and full MCMC which includes the prior information~\footnote{Note if the error of the parameter is even larger than the prior bound, the validity of the Fisher matrix which requires the Gaussian posteriors (even if we add a Gaussian prior of the bound to the Fisher matrix) should be in doubt.}. 
However, in the case of DECIGO-I which has a higher sensitivity (hence a larger SNR and relatively smaller error), this discrepancy is quite small. Again, for the distance inference, we just report the theoretical improvement in the near face-on orientations. The issues we discuss here should not influence the highlight of our paper -- the significant improvement of localization from eccentricity. 

In the second part of this paper, we focus on the construction of golden dark sirens (section~\ref{sec:catalog}) and the application of them on cosmology (section~\ref{sec:cosmo}). Since the distance and localization of golden dark sirens are most tightly constrained, they will not be influenced by the issues we discussed above.
Due to the lack of well-developed MCMC techniques for the mid-band GW parameter estimation in the literature, we just present the results solely based on the Fisher matrix. The more realistic MCMC approach of GW parameter estimation in the mid-band for the eccentric waveforms is to be developed and we leave this for future research.

\appendix

\section{The distance inference and source localization of the typical binaries and the statistical results \label{app:A}}

\begin{figure*}
\includegraphics[width=0.3\textwidth]{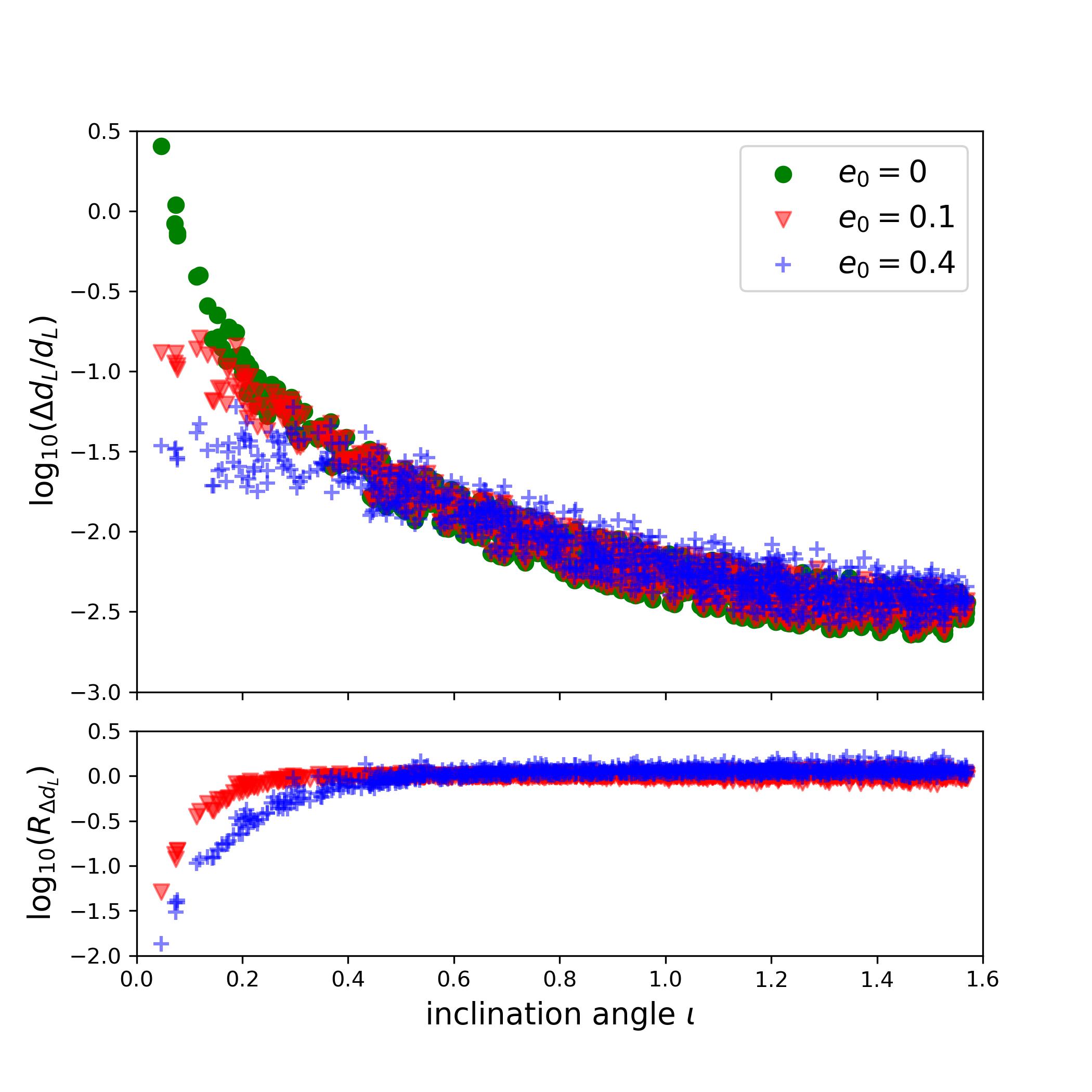}
\includegraphics[width=0.3\textwidth]{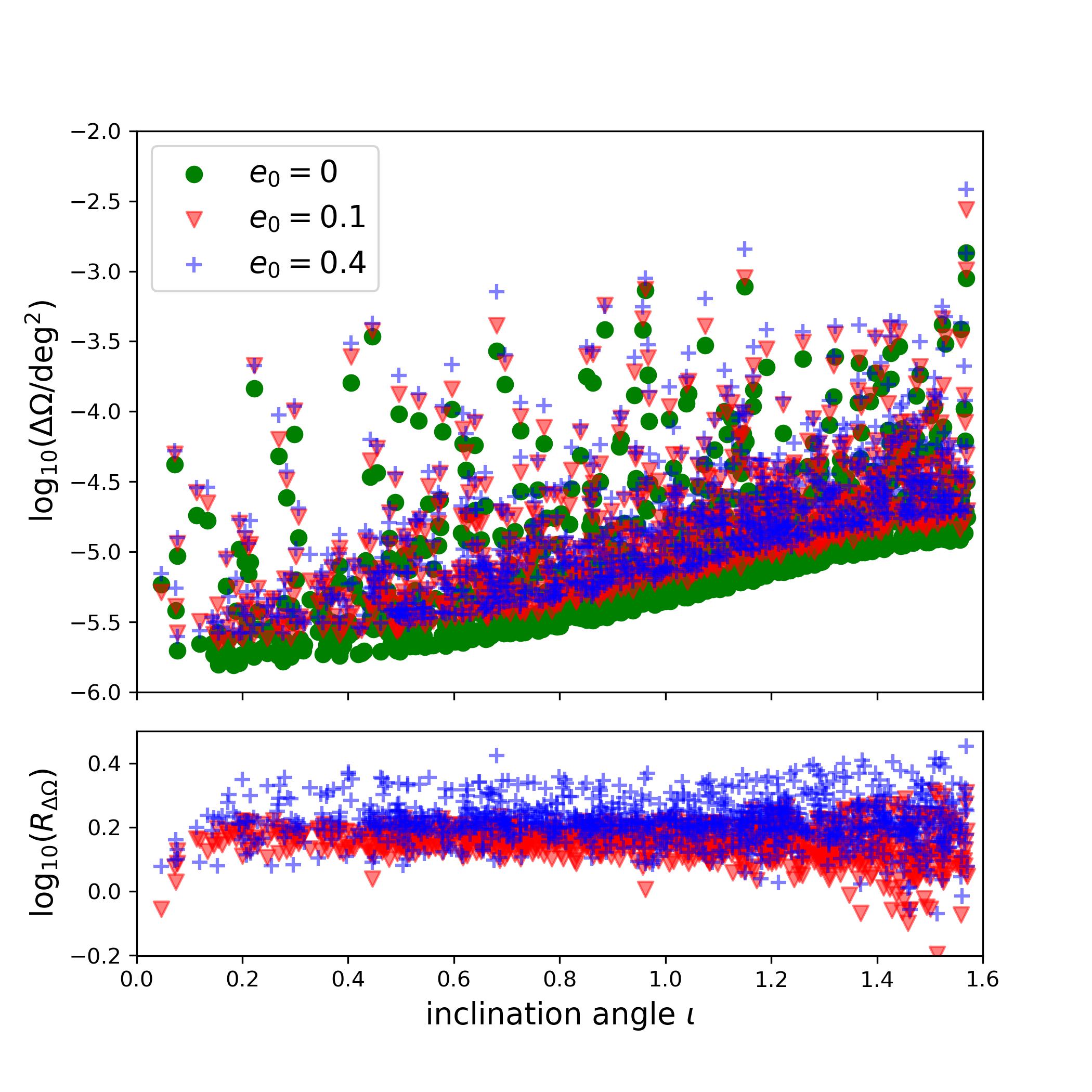}
\includegraphics[width=0.3\textwidth]{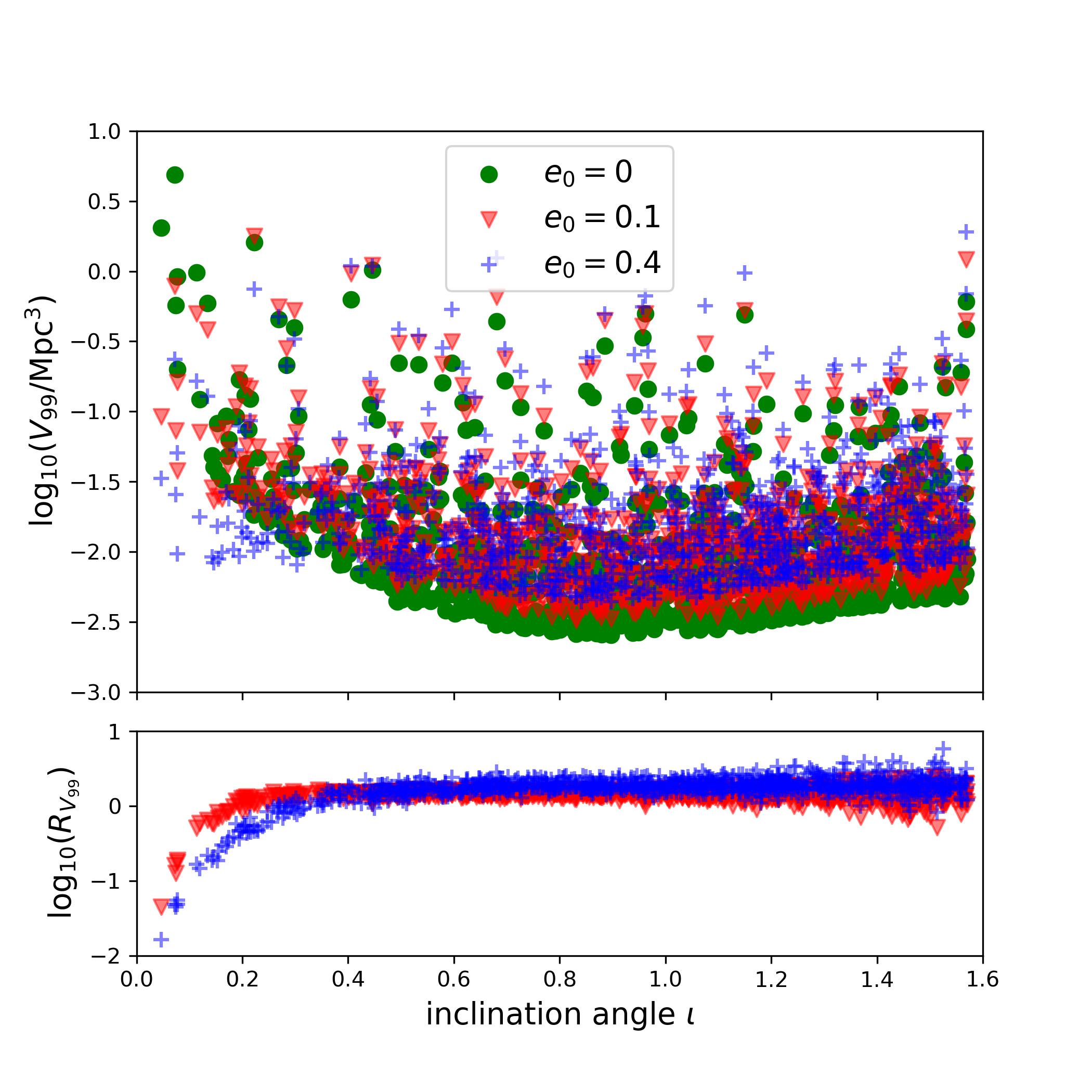}
\caption{The distance inference, sky localization, and 3-D localization volume of GW200105-like NSBH with DECIGO-I.}
\label{fig:NSBH_DI}
\end{figure*}

\begin{figure*}
\includegraphics[width=0.3\textwidth]{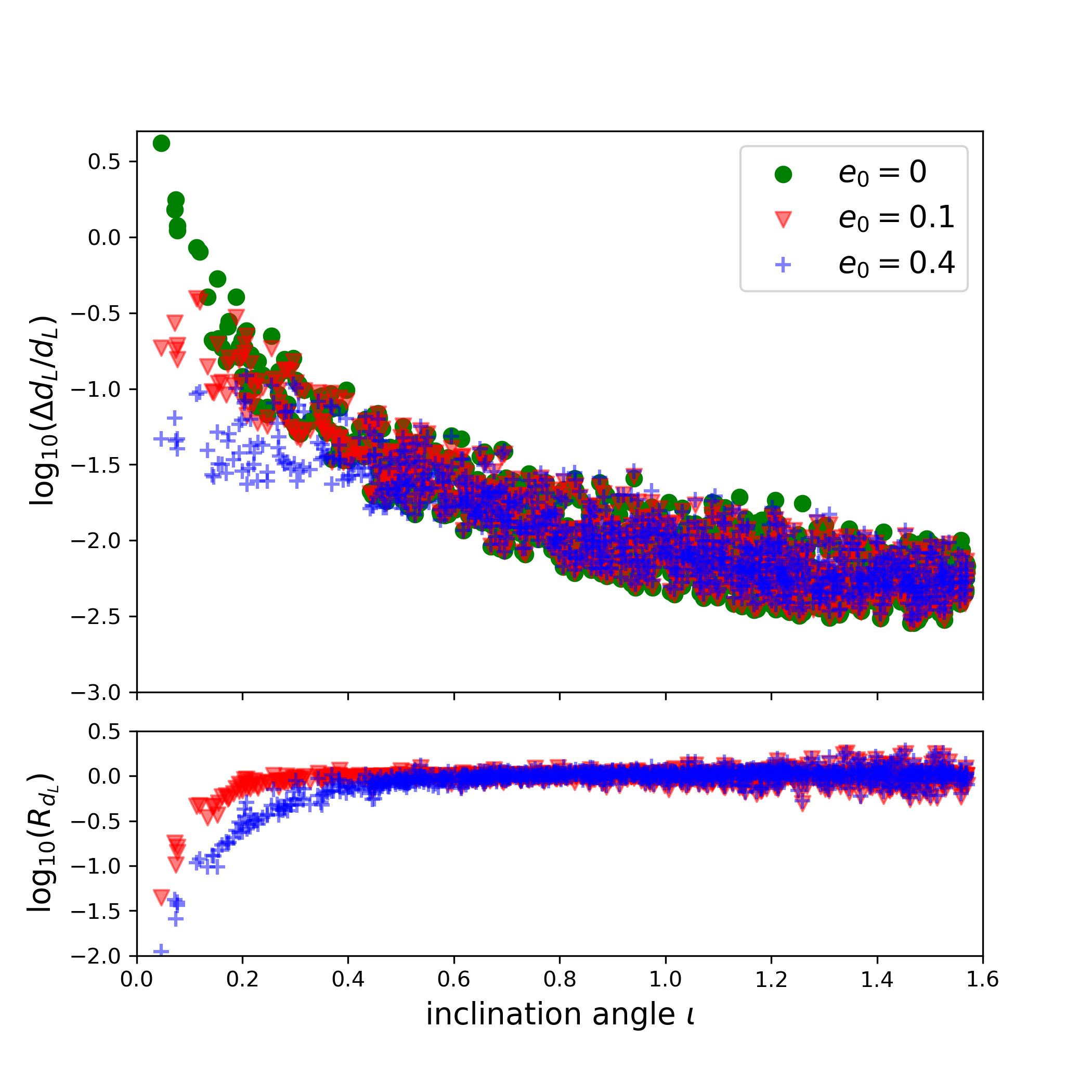}
\includegraphics[width=0.3\textwidth]{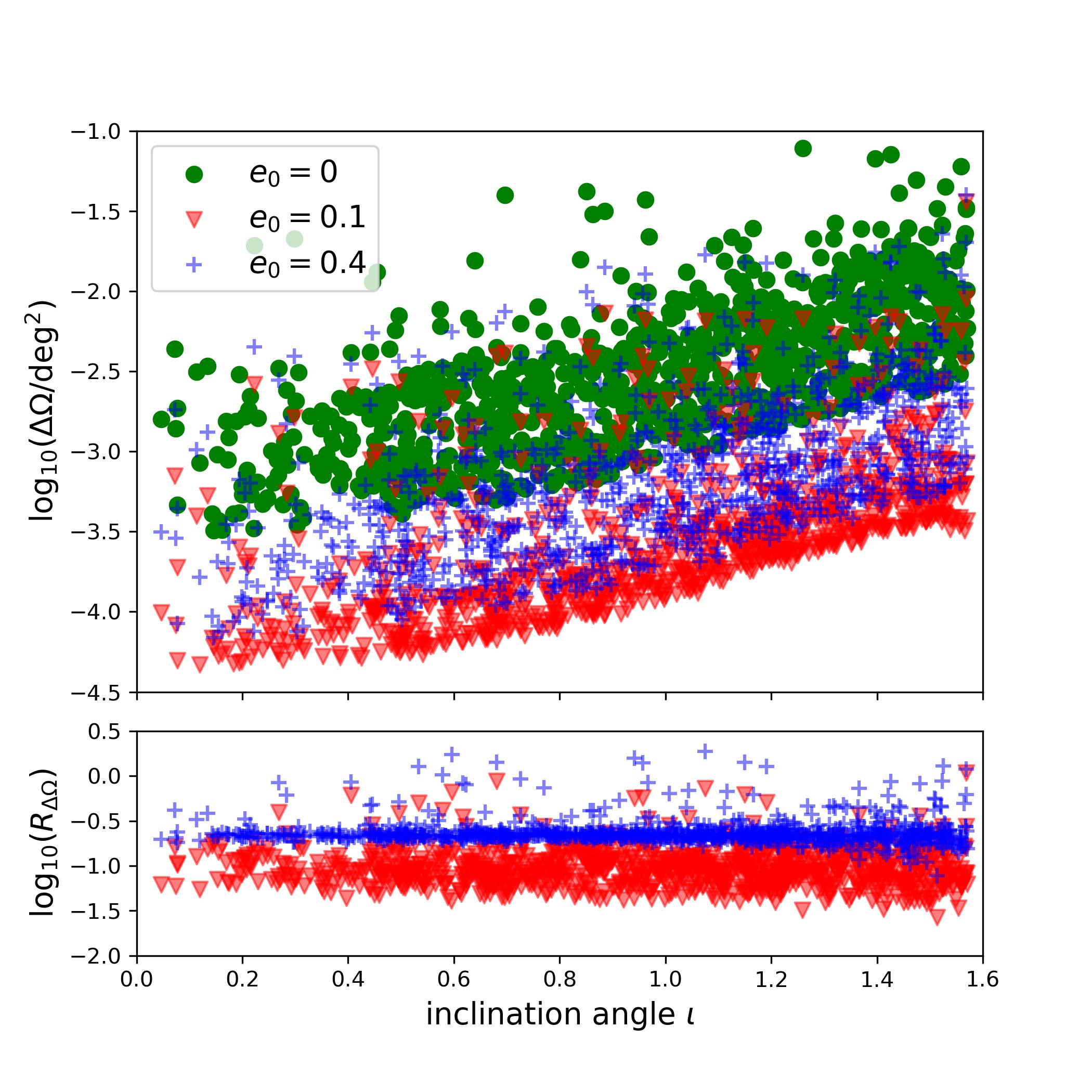}
\includegraphics[width=0.3\textwidth]{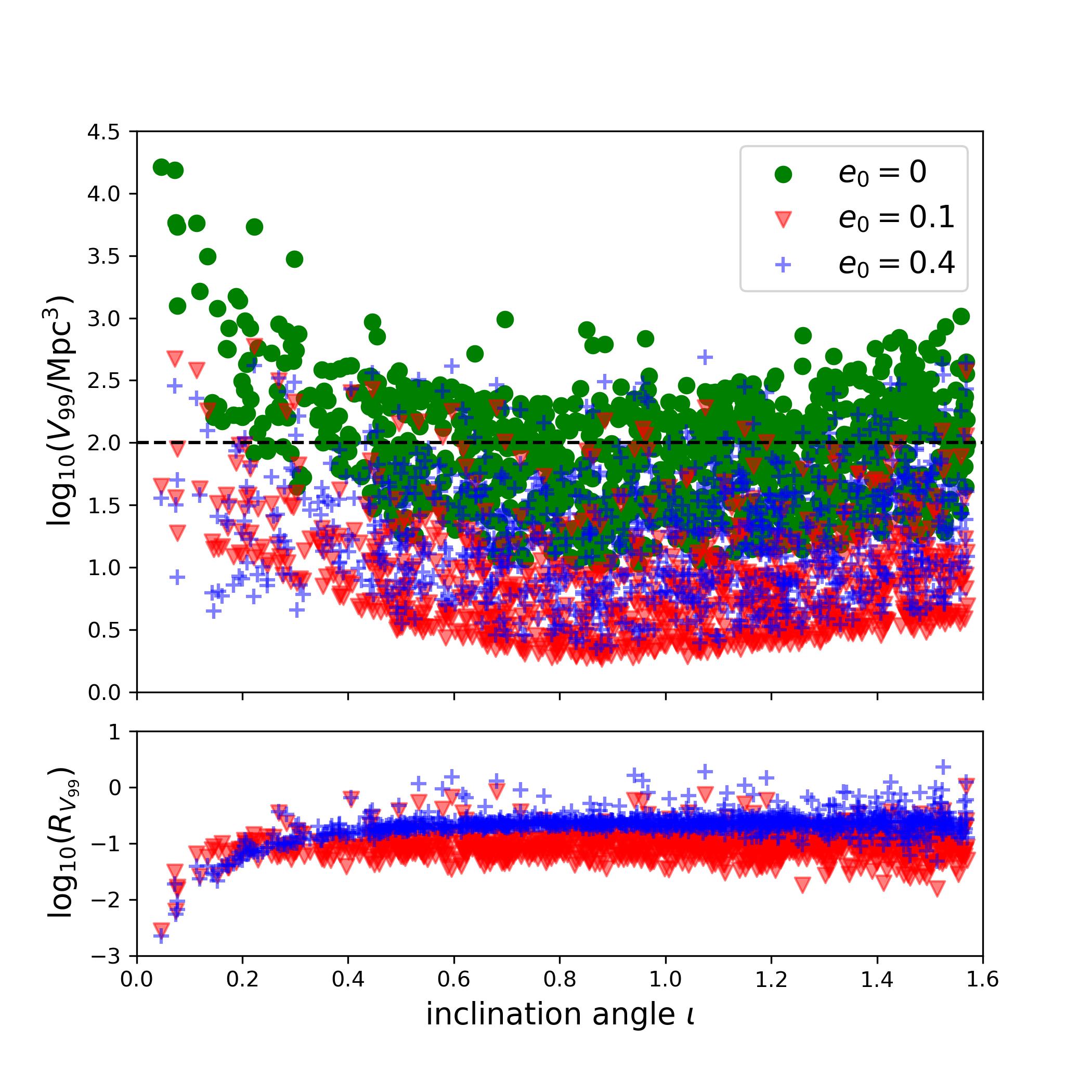}
\caption{The distance inference, sky localization, and 3-D localization volume of GW191129-like light BBH with DECIGO-I.}
\label{fig:BBHlight_DI}
\end{figure*}

\begin{figure*}
\includegraphics[width=0.3\textwidth]{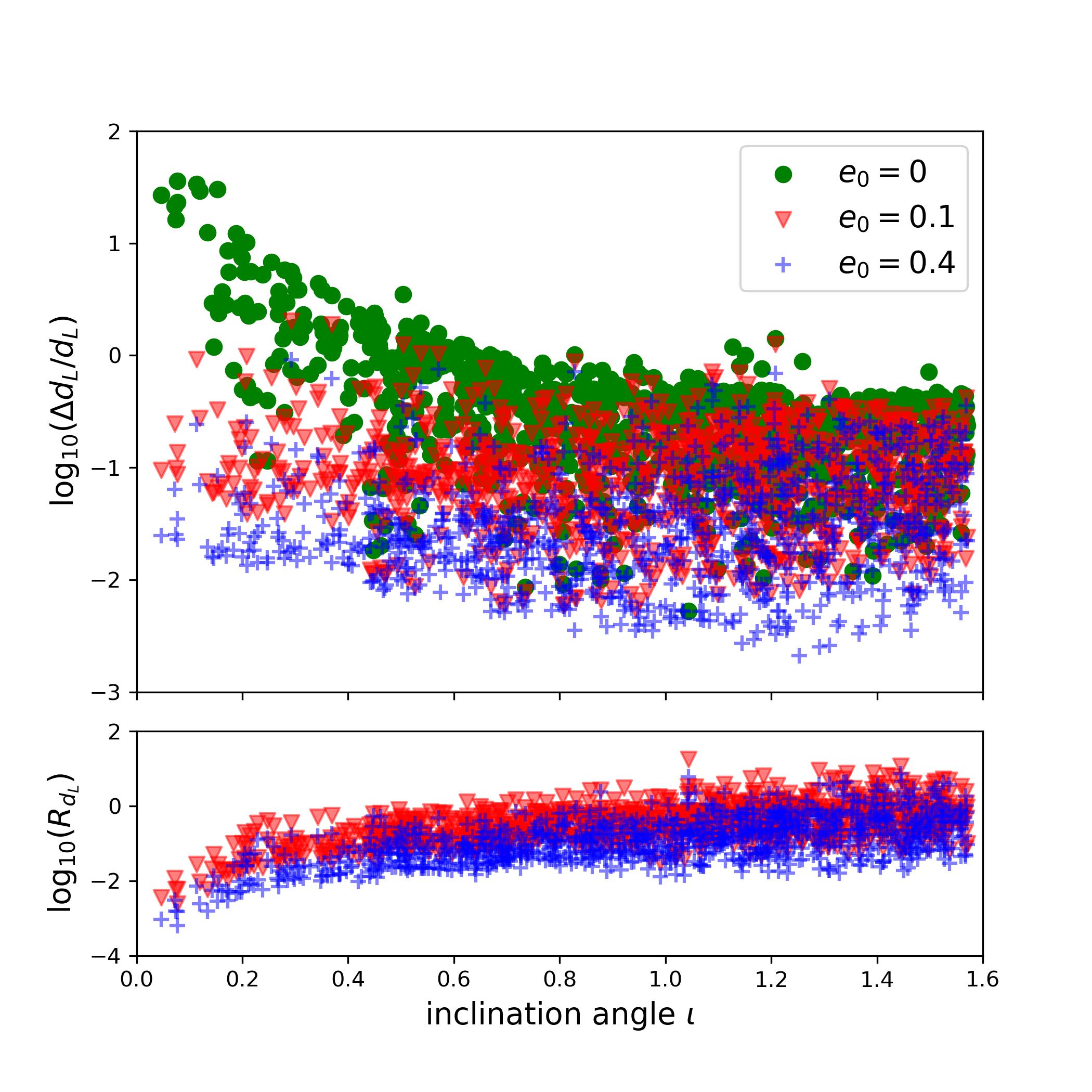}
\includegraphics[width=0.3\textwidth]{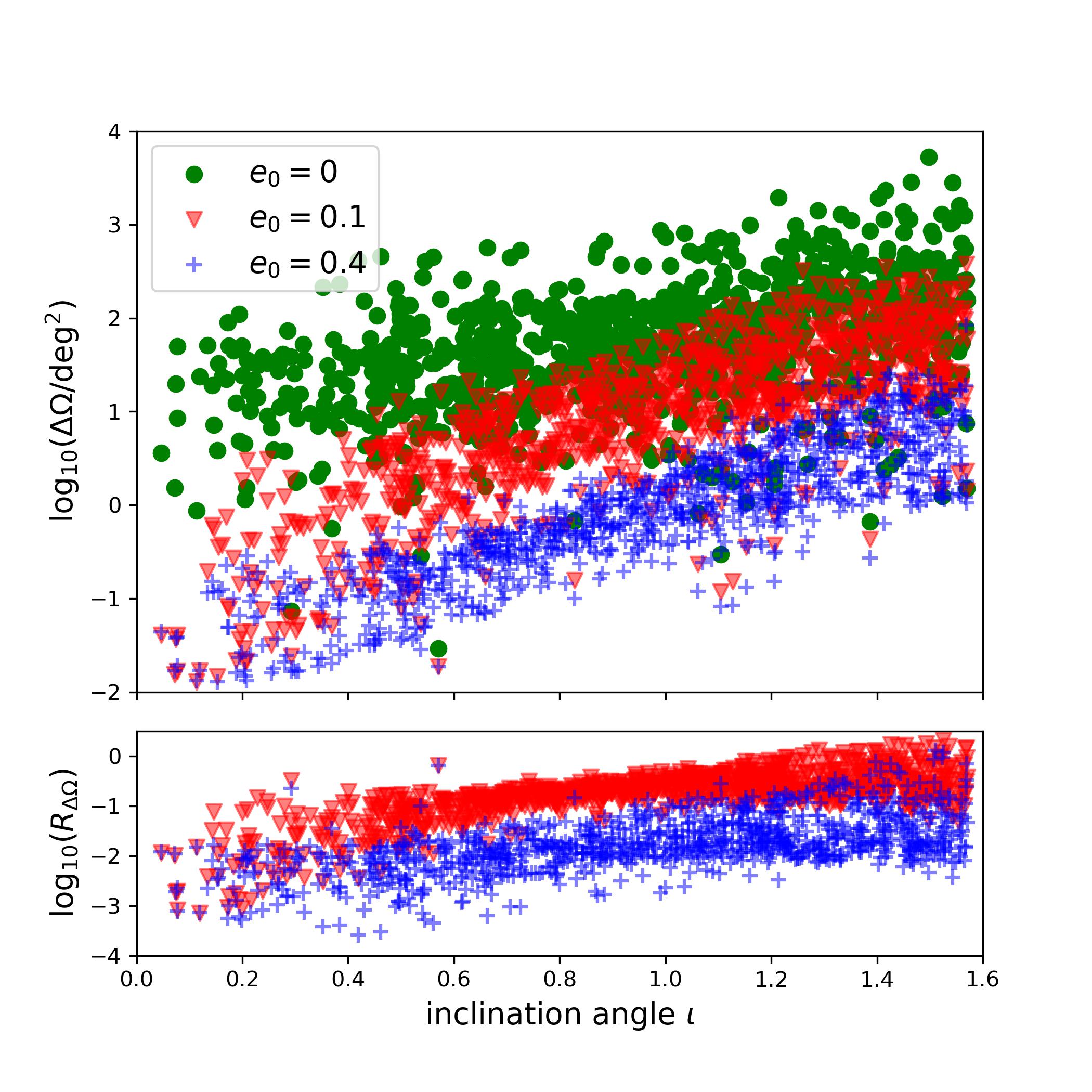}
\includegraphics[width=0.3\textwidth]{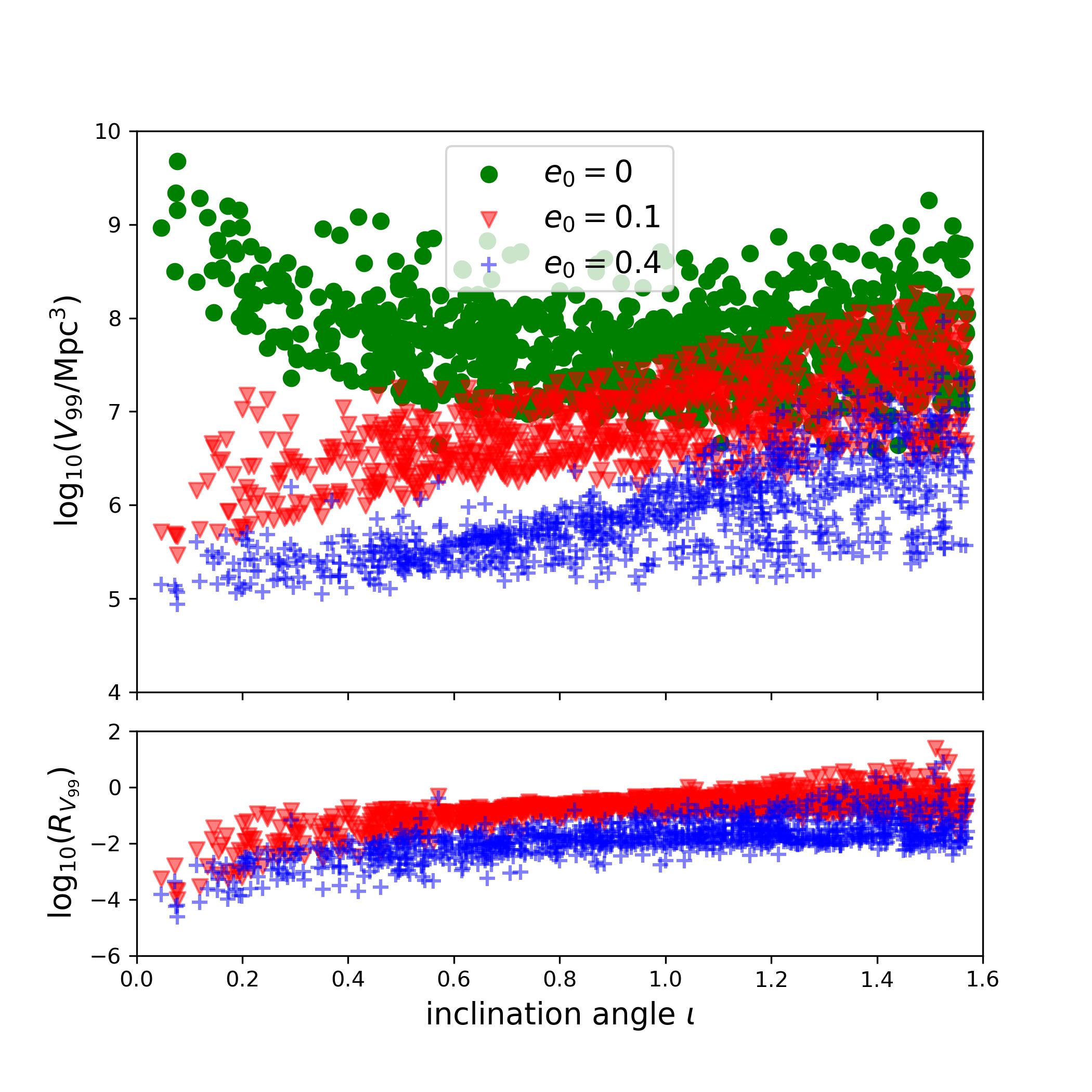}
\caption{The distance inference, sky localization, and 3-D localization volume of GW190426-like heavy BBH with DECIGO-I.}
\label{fig:BBHheavy_DI}
\end{figure*}

In figures~\ref{fig:NSBH_DI} -- \ref{fig:BBHheavy_DI}, we show the results of distance inference and source localization of other typical binaries that have not been presented in section~\ref{sec:typical}. Since the parameter estimations of B-DECIGO and DECIGO-I only differ by a factor of sensitivity ratio, we only show the results based on DECIGO-I.  There are several points which should be addressed. For NSBH, either $e_0=0.1$ or $e_0=0.4$ slightly worsens the source localization. The latter performs even worse than the former. We have explained in section~\ref{sec:typical} that many factors can contribute to the parameter estimation of the eccentric waveform. For the heavy BBH, we find the improvement of distance inference is still obvious in the large inclination angle. This suggests that eccentricity can improve the distance inference overall other than only breaking the $d_L-\iota$ degeneracy in the near face-on orientations. Finally, we can see the localization of NSBH and light BBH are worse with $e_0=0.4$ than that  with $e_0=0.1$. This is also suggested in figure~\ref{fig:Rwe} and we have given a brief explanation there.

We summarize the statistical results for the source localization of each typical binary in table~\ref{tab:stat}. Since in most cases the distance inference is only improved in the near face-on orientations, the statistical information is less meaningful and we are not going to list them in the table. We select the cases with $e_0=0$, 0.1, 0.2, and 0.4 as these four values can present the main feature and tendency of the results (see figure~\ref{fig:Rwe}). Again, we only give the results based on DECIGO-I, the results for B-DECIGO can be easily derived from the ratio of the sensitivity. A remarkable point in the table is that for the typical BBH, eccentricity can significantly improve their localizations. The typical light and medium BBH whose 3-D localization volumes are beyond the threshold volume in the circular case can be improved and well-localized inside the $V_{\rm th}$ so that their host galaxies can be uniquely identified in almost all orientations.

\begin{table*}
\centering
\begin{tabular}{cc|c|c|c|c|c|c} 
\hline\hline
                                          &           & $\min(\Delta\Omega)~\rm deg^2$ & $\mathbb{E}(\Delta\Omega)~\rm deg^2$ & $\max(\Delta\Omega)~\rm deg^2$ & $\min(V_{99})~\rm Mpc^3$ & $\mathbb{E}(V_{99})~\rm Mpc^3$ & $\max(V_{99})~\rm Mpc^3$  \\ 
\hline
\multirow{4}{*}{GW170817-like BNS}        & $e_0=0$   & $6.57\times 10^{-8}$           & $1.20\times 10^{-6}$                 & $1.32\times 10^{-4}$           & $2.79\times 10^{-7}$     & $3.71\times 10^{-6}$           & $5.53\times 10^{-4}$      \\
                                          & $e_0=0.1$ & $1.11\times 10^{-7}$           & $1.63\times 10^{-6}$                 & $1.76\times 10^{-4}$           & $4.32\times 10^{-7}$     & $3.73\times 10^{-6}$           & $2.28\times 10^{-4}$      \\
                                          & $e_0=0.2$ & $7.99\times 10^{-8}$           & $1.28\times 10^{-6}$                 & $1.62\times 10^{-4}$           & $3.16\times 10^{-7}$     & $2.61\times 10^{-6}$           & $1.96\times 10^{-4}$      \\
                                          & $e_0=0.4$ & $6.54\times 10^{-8}$           & $1.18\times 10^{-6}$                 & $2.06\times 10^{-4}$           & $2.84\times 10^{-7}$     & $2.23\time 10^{-6}$            & $2.30\time 10^{-4}$       \\ 
\hline
\multirow{4}{*}{GW200105-like NSBH}       & $e_0=0$   & $1.56\times 10^{-6}$           & $2.39\times 10^{-5}$                 & $1.35\times 10^{-3}$           & $2.56\times 10^{-3}$     & $3.29\times 10^{-2}$           & $4.88$                    \\
                                          & $e_0=0.1$ & $2.27\times 10^{-6}$           & $3.28\times 10^{-5}$                 & $2.75\times 10^{-3}$           & $3.25\times 10^{-3}$     & $3.21\times 10^{-2}$           & $1.78$                    \\
                                          & $e_0=0.2$ & $1.49\times 10^{-6}$           & $2.14\times 10^{-5}$                 & $2.42\times 10^{-3}$           & $2.33\times 10^{-3}$     & $1.92\times 10^{-2}$           & $1.10$                    \\
                                          & $e_0=0.4$ & $2.48\times 10^{-6}$           & $4.13\times 10^{-5}$                 & $3.85\times 10^{-3}$           & $4.42\times 10^{-3}$     & $3.74\times 10^{-2}$           & $1.90$                    \\ 
\hline
\multirow{4}{*}{GW191129-like light BBH}  & $e_0=0$   & $3.22\times 10^{-4}$           & $5.01\times 10^{-3}$                 & $7.81\times 10^{-2}$           & $10.5$                   & $1.89 \times 10^2$              & $1.63\times 10^4$         \\
                                          & $e_0=0.1$ & $4.68\times 10^{-5}$           & $5.61\times 10^{-4}$                 & $3.63\times 10^{-2}$           & $1.84$                   & $15.5$                         & $5.92\times 10^2$         \\
                                          & $e_0=0.2$ & $2.73\times 10^{-5}$           & $3.73\times 10^{-4}$                 & $2.70\times 10^{-2}$           & $1.02$                   & $9.50$                         & $3.01\times 10^2$         \\
                                          & $e_0=0.4$ & $6.91\times 10^{-5}$           & $1.41\times 10^{-3}$                 & $3.99\times 10^{-2}$           & $2.23$                   & $34.4$                         & $4.84\times 10^2$         \\ 
\hline
\multirow{4}{*}{GW150914-like medium BBH} & $e_0=0$   & $1.65\times 10^{-3}$           & $0.273$                              & $4.88$                         & $22.6$                   & $2.60\times 10^2$              & $8.54\times 10^3$         \\
                                          & $e_0=0.1$ & $9.58\times 10^{-4}$           & $2.45\times 10^{-2}$                 & $0.575$                        & $3.15$                   & $21.4$                         & $2.04\times 10^2$         \\
                                          & $e_0=0.2$ & $6.36\times 10^{-4}$           & $9.24\times 10^{-3}$                 & $0.180$                        & $0.946$                  & $8.97$                         & $1.62\times 10^2$         \\
                                          & $e_0=0.4$ & $5.31\times 10^{-4}$           & $6.39\times 10^{-3}$                 & $0.144$                        & $0.515$                  & $6.60$                         & $1.73\times 10^2$         \\ 
\hline
\multirow{4}{*}{GW190426-like heavy BBH}  & $e_0=0$   & $2.90\times 10^{-2}$           & $1.34\times 10^2$                    & $5.25\times 10^3$              & $3.98\times 10^6$        & $1.11\times 10^8$              & $4.78\times 10^9$         \\
                                          & $e_0=0.1$ & $1.29\times 10^{-2}$           & $27.6$                               & $3.78\times 10^2$              & $2.94\times 10^5$        & $1.56\times 10^7$              & $1.85\times 10^8$         \\
                                          & $e_0=0.2$ & $1.20\times 10^{-2}$           & $5.82$                               & $1.77\times 10^2$              & $1.49\times 10^5$        & $3.83\times 10^6$              & $9.81\times 10^7$         \\
                                          & $e_0=0.4$ & $1.29\times 10^{-2}$           & $2.49$                               & $84.6$                         & $8.67\times 10^4$        & $1.80\times 10^6$              & $9.24\times 10^7$         \\
\hline\hline
\end{tabular}
\caption{The statistical results of the sky localization $\Delta\Omega$ and 3-D localization volume in the 99\% confidence level $V_{99}$ for each typical binary with various eccentricities by 1-year observation of DECIGO-I.}
\label{tab:stat}
\end{table*}

\section{The distance inference and source localization of the simulated GWs \label{app:B}}
\begin{figure*}
\includegraphics[width=\textwidth]{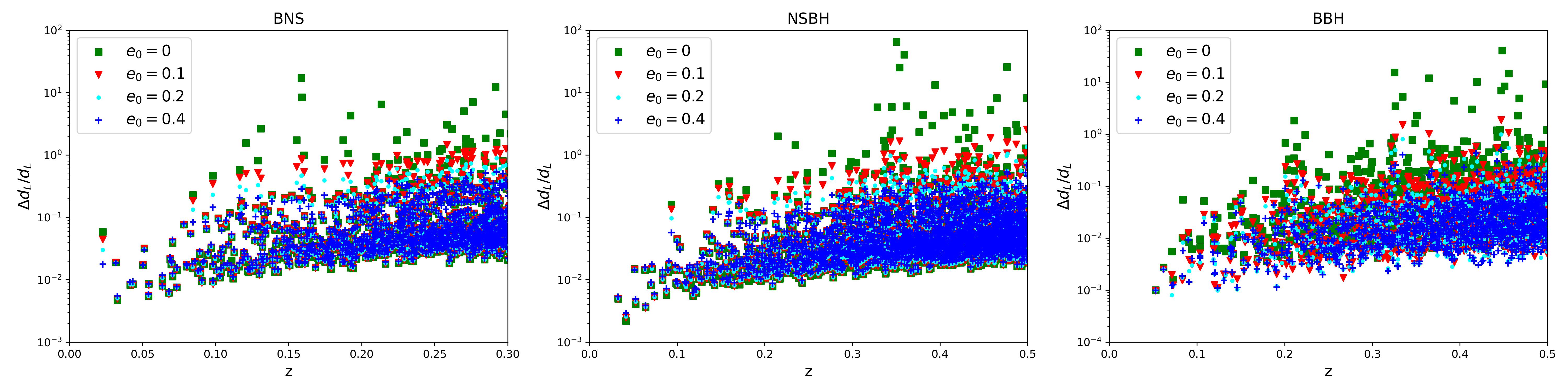}
\caption{The luminosity distance errors of BNS, NSBH, and BBH detected by DECIGO-I within 1 year.}
\label{fig:err_dL_DI}
\end{figure*}

\begin{figure*}
\includegraphics[width=\textwidth]{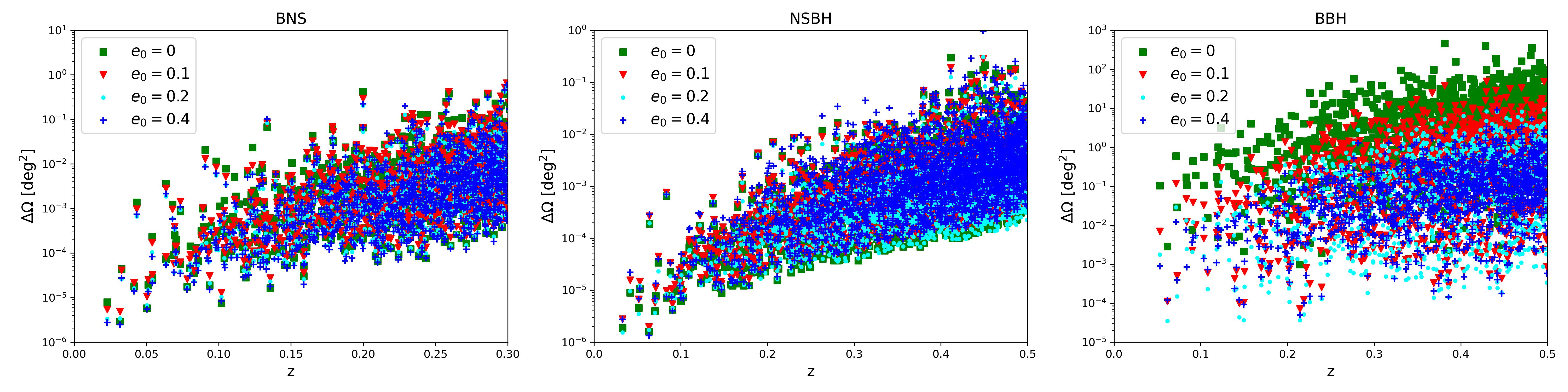}	
\caption{The localization of BNS, NSBH, and BBH detected by DECIGO-I within 1 year.}
\label{fig:err_Omega_DI}
\end{figure*}

In figures~\ref{fig:err_dL_DI} and~\ref{fig:err_Omega_DI} we show the distance errors and localization of BNS, NSBH, and BBH observed by DECIGO-I within 1-year observation time. With a nonvanishing eccentricity, the uncontrollably large errors of distance (corresponding to the near face-on orientations) in the circular case can be significantly suppressed. In the eccentric cases, the distance errors in the whole redshift range ($z\sim 0-5$) are rational. For the three types of binaries, the distance can be constrained at percent and sub-percent level, with BBH's distance measurement more precise. As for the localization, we can clearly see the significant improvement by eccentricity in the whole redshift range for BBH. While for BNS and NSBH, the effects of eccentricity are negligible. However, the best localization of BNS and NSBH can achieve $\sim10^{-5}-10^{-6}~\rm deg^2$ even in the circular case. With a nonvanishing eccentricity, the best localization of BBH can be improved from $\sim10^{-2}~\rm deg^2$ to $\sim10^{-4}~\rm deg^2$.

\section{The cosmological constraints from golden dark BNS and NSBH \label{app:C}}

Though eccentricity has little effects on the localization of BNS and NSBH, they have already been precisely localized even in the circular case. From table~\ref{tab:golden} we can see there are hundreds of golden dark BNS and NSBH regardless of the value of eccentricity. Here we would like to present the cosmological constraints from these golden dark BNS and NSBH. We still only show the results with DECIGO-I and choose $e_0=0$ and 0.2 as the representatives.  

\begin{figure*}
\includegraphics[width=0.45\textwidth]{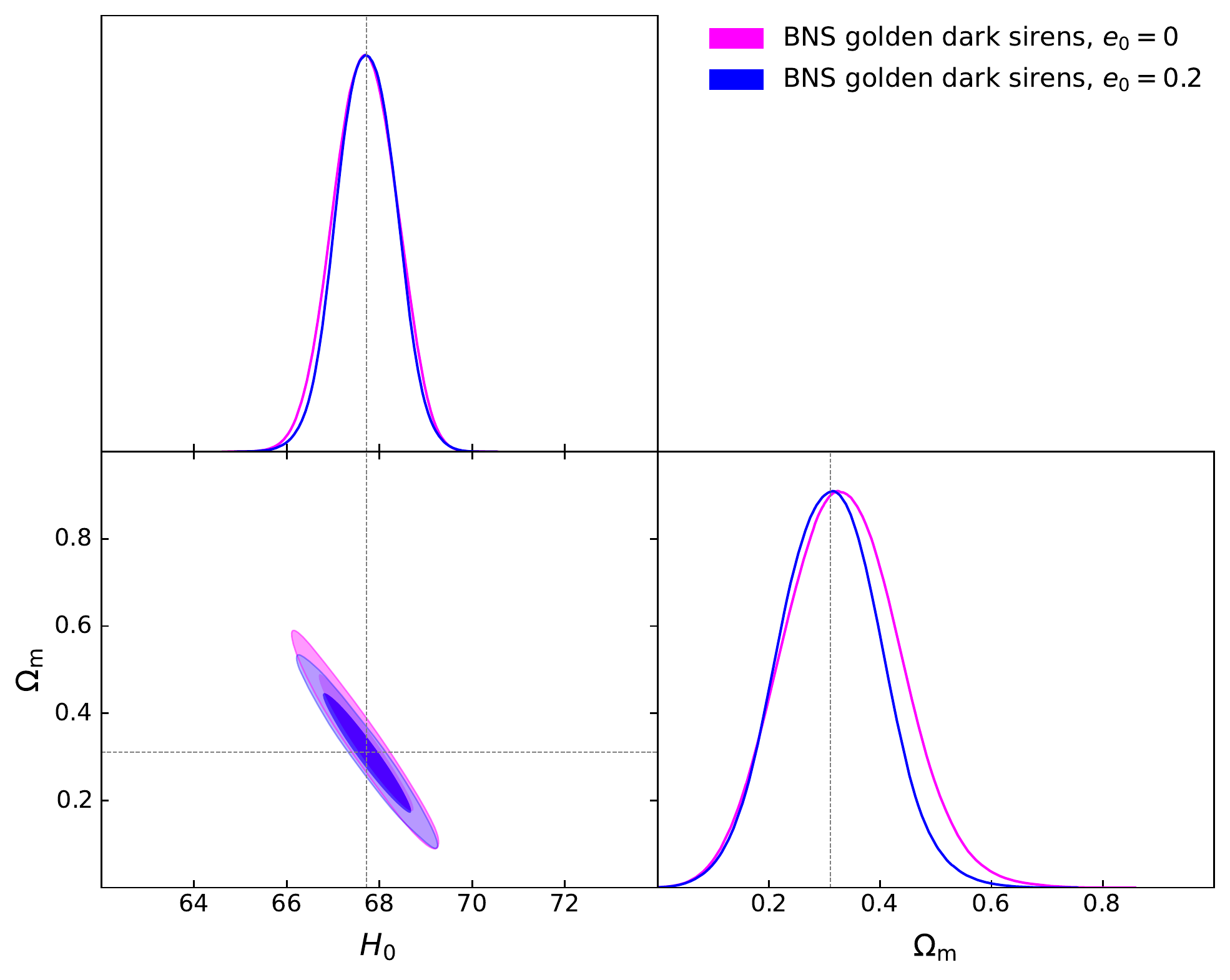}
\includegraphics[width=0.45\textwidth]{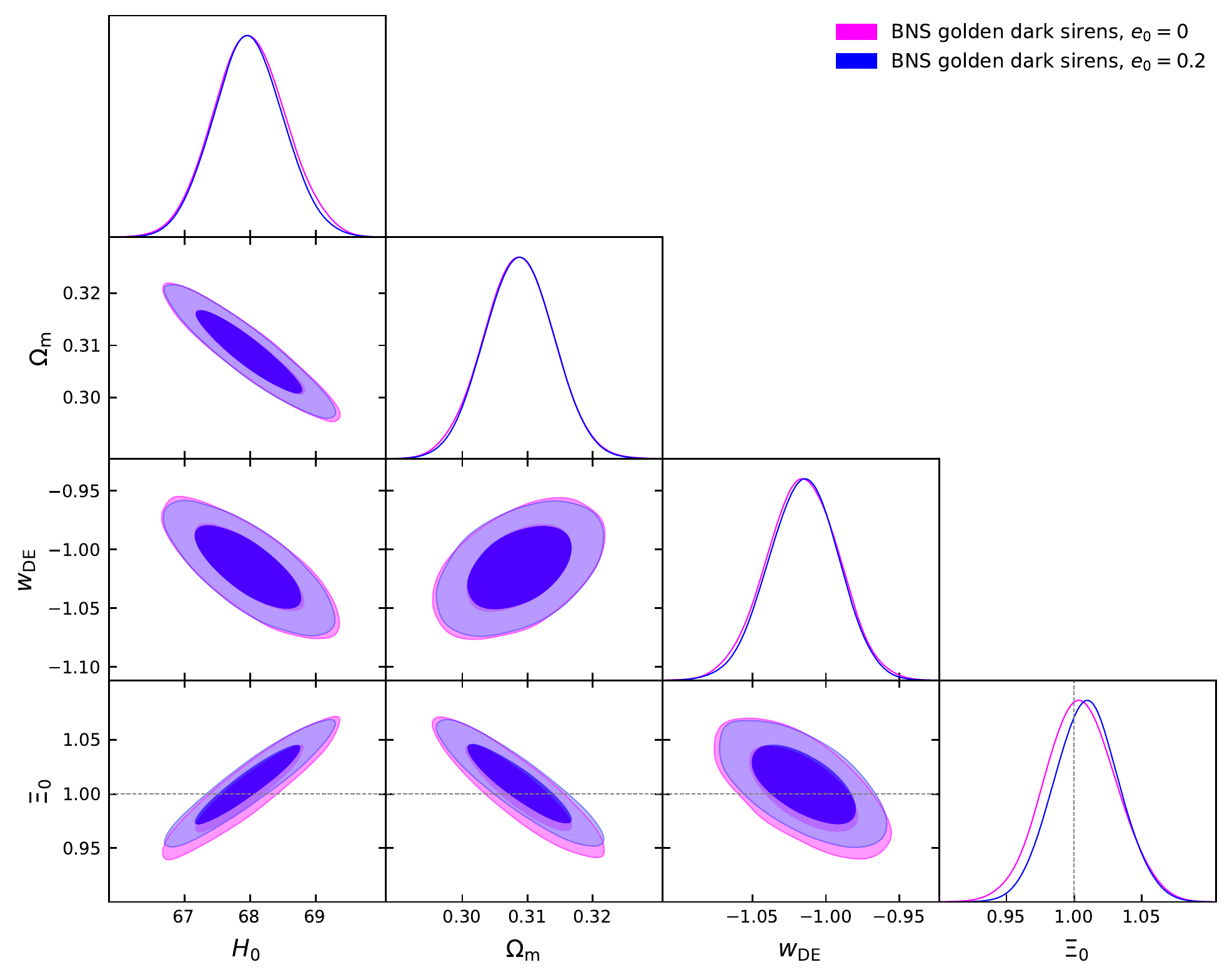}
\caption{The constraints of $\Lambda$CDM model (left panel) and MG (right panel) from 1-year observation of golden dark BNS by DECIGO-I.}
\label{fig:cosmo_BNS}
\end{figure*}

Figure~\ref{fig:cosmo_BNS} shows the cosmological constraints from golden dark BNS with 1-year observation of DECIGO-I. We can see the improvement from eccentricity is only subtle. For $\Lambda$CDM model, $H_0=67.69\pm 0.66~(67.72\pm 0.61)~\rm km~s^{-1}~Mpc^{-1}$ and $\Omega_{\rm m}=0.33\pm 0.10~(0.311\pm 0.089)$  in the $e_0=0~(0.2)$ case. In MG model, $\Xi_0=1.005\pm 0.026$ with $e_0=0$ and $\Xi_0=1.009\pm 0.024$ with $e_0=0.2$.

\begin{figure*}
\includegraphics[width=0.45\textwidth]{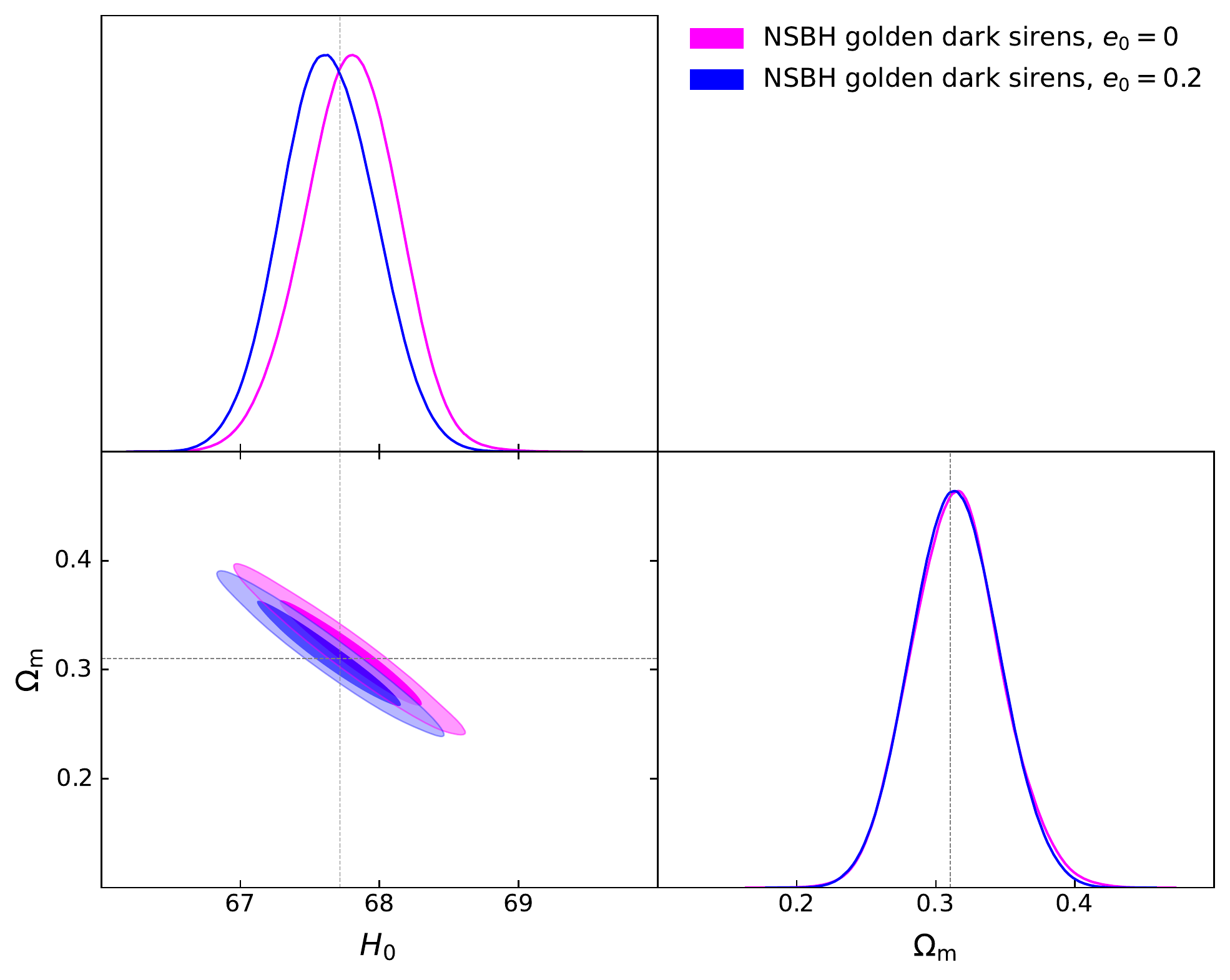}
\includegraphics[width=0.45\textwidth]{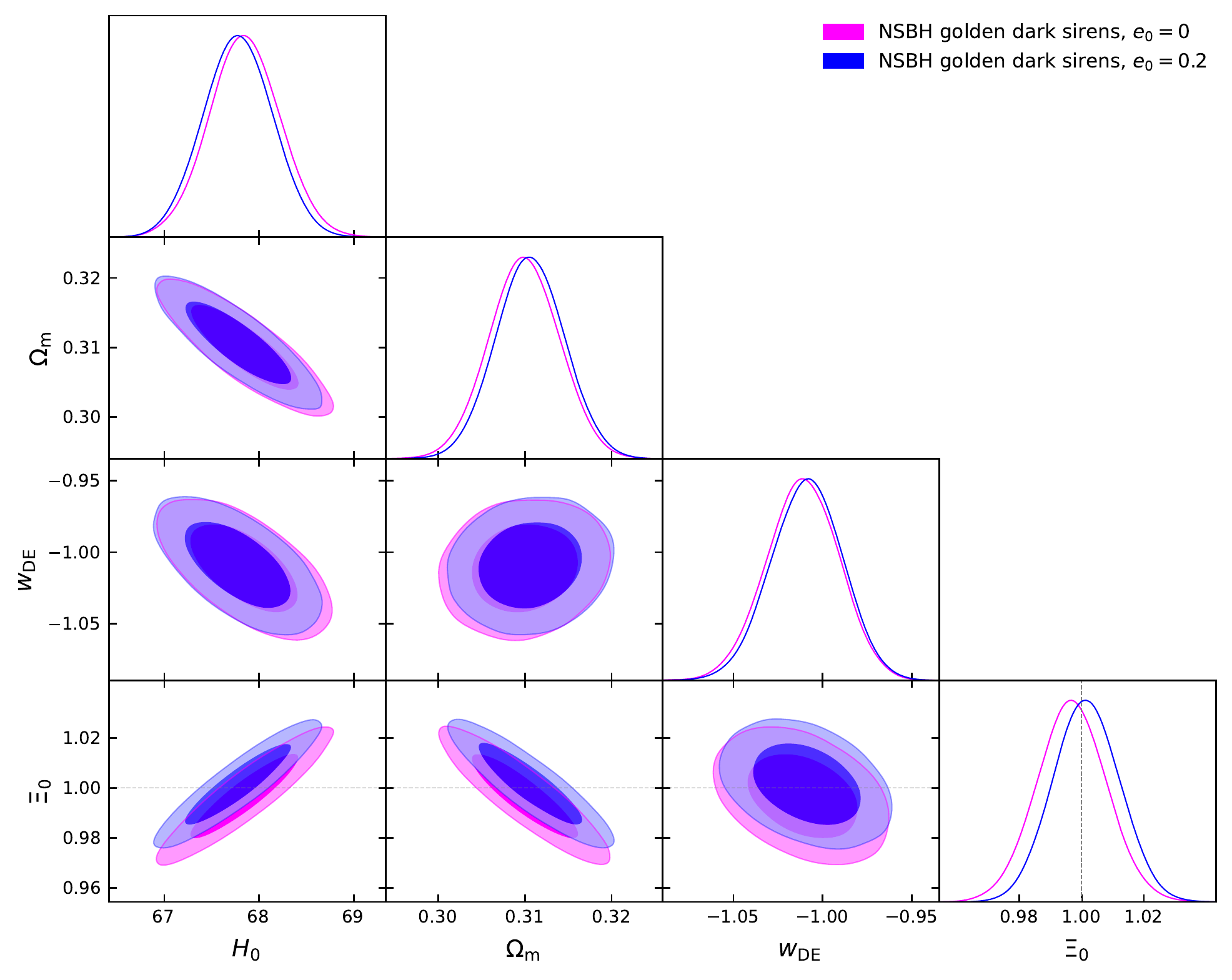}
\caption{The constraints of $\Lambda$CDM model (left panel) and MG (right panel) from 1-year observation of golden dark NSBH by DECIGO-I.}
\label{fig:cosmo_NSBH}
\end{figure*}

Figure~\ref{fig:cosmo_NSBH} shows the cosmological constraints from golden dark NSBH with 1-year observation of DECIGO-I. As expected, the improvement from eccentricity is still negligible. For $\Lambda$CDM model, $H_0=67.80\pm 0.34~(67.63\pm 0.33)~\rm km~s^{-1}~Mpc^{-1}$ and $\Omega_{\rm m}=0.315\pm 0.032~(0.314\pm 0.031)$  in the $e_0=0~(0.2)$ case. In MG model, $\Xi_0=0.997\pm 0.011$ with $e_0=0$ and $\Xi_0=1.002\pm 0.011$ with $e_0=0.2$.

As we have argued, since the merger rates of BNS and NSBH are very uncertain, the results of the population of golden dark BNS and NSBH, and their applications on cosmology are not ensured. A nonvanishing eccentricity can not offer much help for BNS and NSBH on these two aspects. However, the precise localization of BNS and NSBH in the mid-band can help for the search of their EM counterparts. We can use the golden dark BNS and NSBH together with their possible EM counterparts to cross-check the host galaxies of the golden dark sirens. This can also provide useful information to BBH golden dark sirens.


\begin{acknowledgments}
This work is supported by the National Research Foundation of Korea NRF-2021R1A2C2012473.
R.G.C is supported by the National Natural Science Foundation of China Grants No. 11821505, No. 11991052, No. 11947302, No. 12235019 and by the National Key Research and Development Program of China Grant No. 2020YFC2201502. 
Z.C is supported in part by the National Key Research and Development Program of China Grant
No. 2021YFC2203001, in part by the NSFC (No.~11920101003 and No.~12021003), in part by ``the Interdiscipline Research Funds of Beijing Normal University'' and in part by CAS Project for Young Scientists in Basic Research YSBR-006.
\end{acknowledgments}
\bibliography{ref}

\begin{thebibliography}{152}%
\makeatletter
\providecommand \@ifxundefined [1]{%
 \@ifx{#1\undefined}
}%
\providecommand \@ifnum [1]{%
 \ifnum #1\expandafter \@firstoftwo
 \else \expandafter \@secondoftwo
 \fi
}%
\providecommand \@ifx [1]{%
 \ifx #1\expandafter \@firstoftwo
 \else \expandafter \@secondoftwo
 \fi
}%
\providecommand \natexlab [1]{#1}%
\providecommand \enquote  [1]{``#1''}%
\providecommand \bibnamefont  [1]{#1}%
\providecommand \bibfnamefont [1]{#1}%
\providecommand \citenamefont [1]{#1}%
\providecommand \href@noop [0]{\@secondoftwo}%
\providecommand \href [0]{\begingroup \@sanitize@url \@href}%
\providecommand \@href[1]{\@@startlink{#1}\@@href}%
\providecommand \@@href[1]{\endgroup#1\@@endlink}%
\providecommand \@sanitize@url [0]{\catcode `\\12\catcode `\$12\catcode
  `\&12\catcode `\#12\catcode `\^12\catcode `\_12\catcode `\%12\relax}%
\providecommand \@@startlink[1]{}%
\providecommand \@@endlink[0]{}%
\providecommand \url  [0]{\begingroup\@sanitize@url \@url }%
\providecommand \@url [1]{\endgroup\@href {#1}{\urlprefix }}%
\providecommand \urlprefix  [0]{URL }%
\providecommand \Eprint [0]{\href }%
\providecommand \doibase [0]{https://doi.org/}%
\providecommand \selectlanguage [0]{\@gobble}%
\providecommand \bibinfo  [0]{\@secondoftwo}%
\providecommand \bibfield  [0]{\@secondoftwo}%
\providecommand \translation [1]{[#1]}%
\providecommand \BibitemOpen [0]{}%
\providecommand \bibitemStop [0]{}%
\providecommand \bibitemNoStop [0]{.\EOS\space}%
\providecommand \EOS [0]{\spacefactor3000\relax}%
\providecommand \BibitemShut  [1]{\csname bibitem#1\endcsname}%
\let\auto@bib@innerbib\@empty
%</preamble>
\bibitem [{\citenamefont {Abbott}\ \emph
  {et~al.}(2016{\natexlab{a}})\citenamefont {Abbott} \emph
  {et~al.}}]{LIGOScientific:2016lio}%
  \BibitemOpen
  \bibfield  {author} {\bibinfo {author} {\bibfnamefont {B.~P.}\ \bibnamefont
  {Abbott}} \emph {et~al.} (\bibinfo {collaboration} {LIGO Scientific,
  Virgo}),\ }\bibfield  {title} {\bibinfo {title} {{Tests of general relativity
  with GW150914}},\ }\href {https://doi.org/10.1103/PhysRevLett.116.221101}
  {\bibfield  {journal} {\bibinfo  {journal} {Phys. Rev. Lett.}\ }\textbf
  {\bibinfo {volume} {116}},\ \bibinfo {pages} {221101} (\bibinfo {year}
  {2016}{\natexlab{a}})},\ \bibinfo {note} {[Erratum: Phys.Rev.Lett. 121,
  129902 (2018)]},\ \Eprint {https://arxiv.org/abs/1602.03841}
  {arXiv:1602.03841 [gr-qc]} \BibitemShut {NoStop}%
\bibitem [{\citenamefont {Abbott}\ \emph
  {et~al.}(2016{\natexlab{b}})\citenamefont {Abbott} \emph
  {et~al.}}]{LIGOScientific:2016vpg}%
  \BibitemOpen
  \bibfield  {author} {\bibinfo {author} {\bibfnamefont {B.~P.}\ \bibnamefont
  {Abbott}} \emph {et~al.} (\bibinfo {collaboration} {LIGO Scientific,
  Virgo}),\ }\bibfield  {title} {\bibinfo {title} {{Astrophysical Implications
  of the Binary Black-Hole Merger GW150914}},\ }\href
  {https://doi.org/10.3847/2041-8205/818/2/L22} {\bibfield  {journal} {\bibinfo
   {journal} {Astrophys. J. Lett.}\ }\textbf {\bibinfo {volume} {818}},\
  \bibinfo {pages} {L22} (\bibinfo {year} {2016}{\natexlab{b}})},\ \Eprint
  {https://arxiv.org/abs/1602.03846} {arXiv:1602.03846 [astro-ph.HE]}
  \BibitemShut {NoStop}%
\bibitem [{\citenamefont {Abbott}\ \emph
  {et~al.}(2017{\natexlab{a}})\citenamefont {Abbott} \emph
  {et~al.}}]{LIGOScientific:2017adf}%
  \BibitemOpen
  \bibfield  {author} {\bibinfo {author} {\bibfnamefont {B.~P.}\ \bibnamefont
  {Abbott}} \emph {et~al.} (\bibinfo {collaboration} {LIGO Scientific, Virgo,
  1M2H, Dark Energy Camera GW-E, DES, DLT40, Las Cumbres Observatory, VINROUGE,
  MASTER}),\ }\bibfield  {title} {\bibinfo {title} {{A gravitational-wave
  standard siren measurement of the Hubble constant}},\ }\href
  {https://doi.org/10.1038/nature24471} {\bibfield  {journal} {\bibinfo
  {journal} {Nature}\ }\textbf {\bibinfo {volume} {551}},\ \bibinfo {pages}
  {85} (\bibinfo {year} {2017}{\natexlab{a}})},\ \Eprint
  {https://arxiv.org/abs/1710.05835} {arXiv:1710.05835 [astro-ph.CO]}
  \BibitemShut {NoStop}%
\bibitem [{\citenamefont {Abbott}\ \emph {et~al.}(2018)\citenamefont {Abbott}
  \emph {et~al.}}]{LIGOScientific:2018cki}%
  \BibitemOpen
  \bibfield  {author} {\bibinfo {author} {\bibfnamefont {B.~P.}\ \bibnamefont
  {Abbott}} \emph {et~al.} (\bibinfo {collaboration} {LIGO Scientific,
  Virgo}),\ }\bibfield  {title} {\bibinfo {title} {{GW170817: Measurements of
  neutron star radii and equation of state}},\ }\href
  {https://doi.org/10.1103/PhysRevLett.121.161101} {\bibfield  {journal}
  {\bibinfo  {journal} {Phys. Rev. Lett.}\ }\textbf {\bibinfo {volume} {121}},\
  \bibinfo {pages} {161101} (\bibinfo {year} {2018})},\ \Eprint
  {https://arxiv.org/abs/1805.11581} {arXiv:1805.11581 [gr-qc]} \BibitemShut
  {NoStop}%
\bibitem [{\citenamefont {Abbott}\ \emph
  {et~al.}(2021{\natexlab{a}})\citenamefont {Abbott} \emph
  {et~al.}}]{LIGOScientific:2021aug}%
  \BibitemOpen
  \bibfield  {author} {\bibinfo {author} {\bibfnamefont {R.}~\bibnamefont
  {Abbott}} \emph {et~al.} (\bibinfo {collaboration} {LIGO Scientific, VIRGO,
  KAGRA}),\ }\bibfield  {title} {\bibinfo {title} {{Constraints on the cosmic
  expansion history from GWTC-3}},\ }\href@noop {} {\  (\bibinfo {year}
  {2021}{\natexlab{a}})},\ \Eprint {https://arxiv.org/abs/2111.03604}
  {arXiv:2111.03604 [astro-ph.CO]} \BibitemShut {NoStop}%
\bibitem [{\citenamefont {Abbott}\ \emph
  {et~al.}(2021{\natexlab{b}})\citenamefont {Abbott} \emph
  {et~al.}}]{LIGOScientific:2021psn}%
  \BibitemOpen
  \bibfield  {author} {\bibinfo {author} {\bibfnamefont {R.}~\bibnamefont
  {Abbott}} \emph {et~al.} (\bibinfo {collaboration} {LIGO Scientific, VIRGO,
  KAGRA}),\ }\bibfield  {title} {\bibinfo {title} {{The population of merging
  compact binaries inferred using gravitational waves through GWTC-3}},\
  }\href@noop {} {\  (\bibinfo {year} {2021}{\natexlab{b}})},\ \Eprint
  {https://arxiv.org/abs/2111.03634} {arXiv:2111.03634 [astro-ph.HE]}
  \BibitemShut {NoStop}%
\bibitem [{\citenamefont {Abbott}\ \emph
  {et~al.}(2021{\natexlab{c}})\citenamefont {Abbott} \emph
  {et~al.}}]{LIGOScientific:2021sio}%
  \BibitemOpen
  \bibfield  {author} {\bibinfo {author} {\bibfnamefont {R.}~\bibnamefont
  {Abbott}} \emph {et~al.} (\bibinfo {collaboration} {LIGO Scientific, VIRGO,
  KAGRA}),\ }\bibfield  {title} {\bibinfo {title} {{Tests of General Relativity
  with GWTC-3}},\ }\href@noop {} {\  (\bibinfo {year} {2021}{\natexlab{c}})},\
  \Eprint {https://arxiv.org/abs/2112.06861} {arXiv:2112.06861 [gr-qc]}
  \BibitemShut {NoStop}%
\bibitem [{\citenamefont {Abbott}\ \emph
  {et~al.}(2021{\natexlab{d}})\citenamefont {Abbott} \emph
  {et~al.}}]{LIGOScientific:2021djp}%
  \BibitemOpen
  \bibfield  {author} {\bibinfo {author} {\bibfnamefont {R.}~\bibnamefont
  {Abbott}} \emph {et~al.} (\bibinfo {collaboration} {LIGO Scientific, VIRGO,
  KAGRA}),\ }\bibfield  {title} {\bibinfo {title} {{GWTC-3: Compact Binary
  Coalescences Observed by LIGO and Virgo During the Second Part of the Third
  Observing Run}},\ }\href@noop {} {\  (\bibinfo {year}
  {2021}{\natexlab{d}})},\ \Eprint {https://arxiv.org/abs/2111.03606}
  {arXiv:2111.03606 [gr-qc]} \BibitemShut {NoStop}%
\bibitem [{\citenamefont {Abbott}\ \emph
  {et~al.}(2017{\natexlab{b}})\citenamefont {Abbott} \emph
  {et~al.}}]{LIGOScientific:2017vwq}%
  \BibitemOpen
  \bibfield  {author} {\bibinfo {author} {\bibfnamefont {B.~P.}\ \bibnamefont
  {Abbott}} \emph {et~al.} (\bibinfo {collaboration} {LIGO Scientific,
  Virgo}),\ }\bibfield  {title} {\bibinfo {title} {{GW170817: Observation of
  Gravitational Waves from a Binary Neutron Star Inspiral}},\ }\href
  {https://doi.org/10.1103/PhysRevLett.119.161101} {\bibfield  {journal}
  {\bibinfo  {journal} {Phys. Rev. Lett.}\ }\textbf {\bibinfo {volume} {119}},\
  \bibinfo {pages} {161101} (\bibinfo {year} {2017}{\natexlab{b}})},\ \Eprint
  {https://arxiv.org/abs/1710.05832} {arXiv:1710.05832 [gr-qc]} \BibitemShut
  {NoStop}%
\bibitem [{\citenamefont {Abbott}\ \emph
  {et~al.}(2017{\natexlab{c}})\citenamefont {Abbott} \emph
  {et~al.}}]{LIGOScientific:2017ync}%
  \BibitemOpen
  \bibfield  {author} {\bibinfo {author} {\bibfnamefont {B.~P.}\ \bibnamefont
  {Abbott}} \emph {et~al.} (\bibinfo {collaboration} {LIGO Scientific, Virgo,
  Fermi GBM, INTEGRAL, IceCube, AstroSat Cadmium Zinc Telluride Imager Team,
  IPN, Insight-Hxmt, ANTARES, Swift, AGILE Team, 1M2H Team, Dark Energy Camera
  GW-EM, DES, DLT40, GRAWITA, Fermi-LAT, ATCA, ASKAP, Las Cumbres Observatory
  Group, OzGrav, DWF (Deeper Wider Faster Program), AST3, CAASTRO, VINROUGE,
  MASTER, J-GEM, GROWTH, JAGWAR, CaltechNRAO, TTU-NRAO, NuSTAR, Pan-STARRS,
  MAXI Team, TZAC Consortium, KU, Nordic Optical Telescope, ePESSTO, GROND,
  Texas Tech University, SALT Group, TOROS, BOOTES, MWA, CALET, IKI-GW
  Follow-up, H.E.S.S., LOFAR, LWA, HAWC, Pierre Auger, ALMA, Euro VLBI Team, Pi
  of Sky, Chandra Team at McGill University, DFN, ATLAS Telescopes, High Time
  Resolution Universe Survey, RIMAS, RATIR, SKA South Africa/MeerKAT}),\
  }\bibfield  {title} {\bibinfo {title} {{Multi-messenger Observations of a
  Binary Neutron Star Merger}},\ }\href
  {https://doi.org/10.3847/2041-8213/aa91c9} {\bibfield  {journal} {\bibinfo
  {journal} {Astrophys. J. Lett.}\ }\textbf {\bibinfo {volume} {848}},\
  \bibinfo {pages} {L12} (\bibinfo {year} {2017}{\natexlab{c}})},\ \Eprint
  {https://arxiv.org/abs/1710.05833} {arXiv:1710.05833 [astro-ph.HE]}
  \BibitemShut {NoStop}%
\bibitem [{\citenamefont {Abbott}\ \emph
  {et~al.}(2017{\natexlab{d}})\citenamefont {Abbott} \emph
  {et~al.}}]{LIGOScientific:2017zic}%
  \BibitemOpen
  \bibfield  {author} {\bibinfo {author} {\bibfnamefont {B.~P.}\ \bibnamefont
  {Abbott}} \emph {et~al.} (\bibinfo {collaboration} {LIGO Scientific, Virgo,
  Fermi-GBM, INTEGRAL}),\ }\bibfield  {title} {\bibinfo {title} {{Gravitational
  Waves and Gamma-rays from a Binary Neutron Star Merger: GW170817 and GRB
  170817A}},\ }\href {https://doi.org/10.3847/2041-8213/aa920c} {\bibfield
  {journal} {\bibinfo  {journal} {Astrophys. J. Lett.}\ }\textbf {\bibinfo
  {volume} {848}},\ \bibinfo {pages} {L13} (\bibinfo {year}
  {2017}{\natexlab{d}})},\ \Eprint {https://arxiv.org/abs/1710.05834}
  {arXiv:1710.05834 [astro-ph.HE]} \BibitemShut {NoStop}%
\bibitem [{\citenamefont {Aghanim}\ \emph {et~al.}(2020)\citenamefont {Aghanim}
  \emph {et~al.}}]{Planck:2018vyg}%
  \BibitemOpen
  \bibfield  {author} {\bibinfo {author} {\bibfnamefont {N.}~\bibnamefont
  {Aghanim}} \emph {et~al.} (\bibinfo {collaboration} {Planck}),\ }\bibfield
  {title} {\bibinfo {title} {{Planck 2018 results. VI. Cosmological
  parameters}},\ }\href {https://doi.org/10.1051/0004-6361/201833910}
  {\bibfield  {journal} {\bibinfo  {journal} {Astron. Astrophys.}\ }\textbf
  {\bibinfo {volume} {641}},\ \bibinfo {pages} {A6} (\bibinfo {year} {2020})},\
  \bibinfo {note} {[Erratum: Astron.Astrophys. 652, C4 (2021)]},\ \Eprint
  {https://arxiv.org/abs/1807.06209} {arXiv:1807.06209 [astro-ph.CO]}
  \BibitemShut {NoStop}%
\bibitem [{\citenamefont {Alam}\ \emph {et~al.}(2017)\citenamefont {Alam} \emph
  {et~al.}}]{BOSS:2016wmc}%
  \BibitemOpen
  \bibfield  {author} {\bibinfo {author} {\bibfnamefont {S.}~\bibnamefont
  {Alam}} \emph {et~al.} (\bibinfo {collaboration} {BOSS}),\ }\bibfield
  {title} {\bibinfo {title} {{The clustering of galaxies in the completed
  SDSS-III Baryon Oscillation Spectroscopic Survey: cosmological analysis of
  the DR12 galaxy sample}},\ }\href {https://doi.org/10.1093/mnras/stx721}
  {\bibfield  {journal} {\bibinfo  {journal} {Mon. Not. Roy. Astron. Soc.}\
  }\textbf {\bibinfo {volume} {470}},\ \bibinfo {pages} {2617} (\bibinfo {year}
  {2017})},\ \Eprint {https://arxiv.org/abs/1607.03155} {arXiv:1607.03155
  [astro-ph.CO]} \BibitemShut {NoStop}%
\bibitem [{\citenamefont {Riess}\ \emph {et~al.}(2016)\citenamefont {Riess}
  \emph {et~al.}}]{Riess:2016jrr}%
  \BibitemOpen
  \bibfield  {author} {\bibinfo {author} {\bibfnamefont {A.~G.}\ \bibnamefont
  {Riess}} \emph {et~al.},\ }\bibfield  {title} {\bibinfo {title} {{A 2.4\%
  Determination of the Local Value of the Hubble Constant}},\ }\href
  {https://doi.org/10.3847/0004-637X/826/1/56} {\bibfield  {journal} {\bibinfo
  {journal} {Astrophys. J.}\ }\textbf {\bibinfo {volume} {826}},\ \bibinfo
  {pages} {56} (\bibinfo {year} {2016})},\ \Eprint
  {https://arxiv.org/abs/1604.01424} {arXiv:1604.01424 [astro-ph.CO]}
  \BibitemShut {NoStop}%
\bibitem [{\citenamefont {Riess}\ \emph {et~al.}(2019)\citenamefont {Riess},
  \citenamefont {Casertano}, \citenamefont {Yuan}, \citenamefont {Macri},\ and\
  \citenamefont {Scolnic}}]{Riess:2019cxk}%
  \BibitemOpen
  \bibfield  {author} {\bibinfo {author} {\bibfnamefont {A.~G.}\ \bibnamefont
  {Riess}}, \bibinfo {author} {\bibfnamefont {S.}~\bibnamefont {Casertano}},
  \bibinfo {author} {\bibfnamefont {W.}~\bibnamefont {Yuan}}, \bibinfo {author}
  {\bibfnamefont {L.~M.}\ \bibnamefont {Macri}},\ and\ \bibinfo {author}
  {\bibfnamefont {D.}~\bibnamefont {Scolnic}},\ }\bibfield  {title} {\bibinfo
  {title} {{Large Magellanic Cloud Cepheid Standards Provide a 1\% Foundation
  for the Determination of the Hubble Constant and Stronger Evidence for
  Physics beyond $\Lambda$CDM}},\ }\href
  {https://doi.org/10.3847/1538-4357/ab1422} {\bibfield  {journal} {\bibinfo
  {journal} {Astrophys. J.}\ }\textbf {\bibinfo {volume} {876}},\ \bibinfo
  {pages} {85} (\bibinfo {year} {2019})},\ \Eprint
  {https://arxiv.org/abs/1903.07603} {arXiv:1903.07603 [astro-ph.CO]}
  \BibitemShut {NoStop}%
\bibitem [{\citenamefont {Wong}\ \emph {et~al.}(2020)\citenamefont {Wong} \emph
  {et~al.}}]{Wong:2019kwg}%
  \BibitemOpen
  \bibfield  {author} {\bibinfo {author} {\bibfnamefont {K.~C.}\ \bibnamefont
  {Wong}} \emph {et~al.},\ }\bibfield  {title} {\bibinfo {title} {{H0LiCOW
  \textendash{} XIII. A 2.4 per cent measurement of H0 from lensed quasars:
  5.3\ensuremath{\sigma} tension between early- and late-Universe probes}},\
  }\href {https://doi.org/10.1093/mnras/stz3094} {\bibfield  {journal}
  {\bibinfo  {journal} {Mon. Not. Roy. Astron. Soc.}\ }\textbf {\bibinfo
  {volume} {498}},\ \bibinfo {pages} {1420} (\bibinfo {year} {2020})},\ \Eprint
  {https://arxiv.org/abs/1907.04869} {arXiv:1907.04869 [astro-ph.CO]}
  \BibitemShut {NoStop}%
\bibitem [{\citenamefont {Verde}\ \emph {et~al.}(2019)\citenamefont {Verde},
  \citenamefont {Treu},\ and\ \citenamefont {Riess}}]{Verde:2019ivm}%
  \BibitemOpen
  \bibfield  {author} {\bibinfo {author} {\bibfnamefont {L.}~\bibnamefont
  {Verde}}, \bibinfo {author} {\bibfnamefont {T.}~\bibnamefont {Treu}},\ and\
  \bibinfo {author} {\bibfnamefont {A.~G.}\ \bibnamefont {Riess}},\ }\bibfield
  {title} {\bibinfo {title} {{Tensions between the Early and the Late
  Universe}},\ }\href {https://doi.org/10.1038/s41550-019-0902-0} {\bibfield
  {journal} {\bibinfo  {journal} {Nature Astron.}\ }\textbf {\bibinfo {volume}
  {3}},\ \bibinfo {pages} {891} (\bibinfo {year} {2019})},\ \Eprint
  {https://arxiv.org/abs/1907.10625} {arXiv:1907.10625 [astro-ph.CO]}
  \BibitemShut {NoStop}%
\bibitem [{\citenamefont {Chen}\ \emph {et~al.}(2018)\citenamefont {Chen},
  \citenamefont {Fishbach},\ and\ \citenamefont {Holz}}]{Chen:2017rfc}%
  \BibitemOpen
  \bibfield  {author} {\bibinfo {author} {\bibfnamefont {H.-Y.}\ \bibnamefont
  {Chen}}, \bibinfo {author} {\bibfnamefont {M.}~\bibnamefont {Fishbach}},\
  and\ \bibinfo {author} {\bibfnamefont {D.~E.}\ \bibnamefont {Holz}},\
  }\bibfield  {title} {\bibinfo {title} {{A two per cent Hubble constant
  measurement from standard sirens within five years}},\ }\href
  {https://doi.org/10.1038/s41586-018-0606-0} {\bibfield  {journal} {\bibinfo
  {journal} {Nature}\ }\textbf {\bibinfo {volume} {562}},\ \bibinfo {pages}
  {545} (\bibinfo {year} {2018})},\ \Eprint {https://arxiv.org/abs/1712.06531}
  {arXiv:1712.06531 [astro-ph.CO]} \BibitemShut {NoStop}%
\bibitem [{\citenamefont {Feeney}\ \emph {et~al.}(2019)\citenamefont {Feeney},
  \citenamefont {Peiris}, \citenamefont {Williamson}, \citenamefont {Nissanke},
  \citenamefont {Mortlock}, \citenamefont {Alsing},\ and\ \citenamefont
  {Scolnic}}]{Feeney:2018mkj}%
  \BibitemOpen
  \bibfield  {author} {\bibinfo {author} {\bibfnamefont {S.~M.}\ \bibnamefont
  {Feeney}}, \bibinfo {author} {\bibfnamefont {H.~V.}\ \bibnamefont {Peiris}},
  \bibinfo {author} {\bibfnamefont {A.~R.}\ \bibnamefont {Williamson}},
  \bibinfo {author} {\bibfnamefont {S.~M.}\ \bibnamefont {Nissanke}}, \bibinfo
  {author} {\bibfnamefont {D.~J.}\ \bibnamefont {Mortlock}}, \bibinfo {author}
  {\bibfnamefont {J.}~\bibnamefont {Alsing}},\ and\ \bibinfo {author}
  {\bibfnamefont {D.}~\bibnamefont {Scolnic}},\ }\bibfield  {title} {\bibinfo
  {title} {{Prospects for resolving the Hubble constant tension with standard
  sirens}},\ }\href {https://doi.org/10.1103/PhysRevLett.122.061105} {\bibfield
   {journal} {\bibinfo  {journal} {Phys. Rev. Lett.}\ }\textbf {\bibinfo
  {volume} {122}},\ \bibinfo {pages} {061105} (\bibinfo {year} {2019})},\
  \Eprint {https://arxiv.org/abs/1802.03404} {arXiv:1802.03404 [astro-ph.CO]}
  \BibitemShut {NoStop}%
\bibitem [{\citenamefont {Borhanian}\ \emph {et~al.}(2020)\citenamefont
  {Borhanian}, \citenamefont {Dhani}, \citenamefont {Gupta}, \citenamefont
  {Arun},\ and\ \citenamefont {Sathyaprakash}}]{Borhanian:2020vyr}%
  \BibitemOpen
  \bibfield  {author} {\bibinfo {author} {\bibfnamefont {S.}~\bibnamefont
  {Borhanian}}, \bibinfo {author} {\bibfnamefont {A.}~\bibnamefont {Dhani}},
  \bibinfo {author} {\bibfnamefont {A.}~\bibnamefont {Gupta}}, \bibinfo
  {author} {\bibfnamefont {K.~G.}\ \bibnamefont {Arun}},\ and\ \bibinfo
  {author} {\bibfnamefont {B.~S.}\ \bibnamefont {Sathyaprakash}},\ }\bibfield
  {title} {\bibinfo {title} {{Dark Sirens to Resolve the
  Hubble\textendash{}Lema\^\i{}tre Tension}},\ }\href
  {https://doi.org/10.3847/2041-8213/abcaf5} {\bibfield  {journal} {\bibinfo
  {journal} {Astrophys. J. Lett.}\ }\textbf {\bibinfo {volume} {905}},\
  \bibinfo {pages} {L28} (\bibinfo {year} {2020})},\ \Eprint
  {https://arxiv.org/abs/2007.02883} {arXiv:2007.02883 [astro-ph.CO]}
  \BibitemShut {NoStop}%
\bibitem [{\citenamefont {Dalal}\ \emph {et~al.}(2006)\citenamefont {Dalal},
  \citenamefont {Holz}, \citenamefont {Hughes},\ and\ \citenamefont
  {Jain}}]{Dalal:2006qt}%
  \BibitemOpen
  \bibfield  {author} {\bibinfo {author} {\bibfnamefont {N.}~\bibnamefont
  {Dalal}}, \bibinfo {author} {\bibfnamefont {D.~E.}\ \bibnamefont {Holz}},
  \bibinfo {author} {\bibfnamefont {S.~A.}\ \bibnamefont {Hughes}},\ and\
  \bibinfo {author} {\bibfnamefont {B.}~\bibnamefont {Jain}},\ }\bibfield
  {title} {\bibinfo {title} {{Short grb and binary black hole standard sirens
  as a probe of dark energy}},\ }\href
  {https://doi.org/10.1103/PhysRevD.74.063006} {\bibfield  {journal} {\bibinfo
  {journal} {Phys. Rev. D}\ }\textbf {\bibinfo {volume} {74}},\ \bibinfo
  {pages} {063006} (\bibinfo {year} {2006})},\ \Eprint
  {https://arxiv.org/abs/astro-ph/0601275} {arXiv:astro-ph/0601275}
  \BibitemShut {NoStop}%
\bibitem [{\citenamefont {Cutler}\ and\ \citenamefont
  {Holz}(2009)}]{Cutler:2009qv}%
  \BibitemOpen
  \bibfield  {author} {\bibinfo {author} {\bibfnamefont {C.}~\bibnamefont
  {Cutler}}\ and\ \bibinfo {author} {\bibfnamefont {D.~E.}\ \bibnamefont
  {Holz}},\ }\bibfield  {title} {\bibinfo {title} {{Ultra-high precision
  cosmology from gravitational waves}},\ }\href
  {https://doi.org/10.1103/PhysRevD.80.104009} {\bibfield  {journal} {\bibinfo
  {journal} {Phys. Rev. D}\ }\textbf {\bibinfo {volume} {80}},\ \bibinfo
  {pages} {104009} (\bibinfo {year} {2009})},\ \Eprint
  {https://arxiv.org/abs/0906.3752} {arXiv:0906.3752 [astro-ph.CO]}
  \BibitemShut {NoStop}%
\bibitem [{\citenamefont {Sathyaprakash}\ \emph {et~al.}(2010)\citenamefont
  {Sathyaprakash}, \citenamefont {Schutz},\ and\ \citenamefont {Van
  Den~Broeck}}]{Sathyaprakash:2009xt}%
  \BibitemOpen
  \bibfield  {author} {\bibinfo {author} {\bibfnamefont {B.~S.}\ \bibnamefont
  {Sathyaprakash}}, \bibinfo {author} {\bibfnamefont {B.~F.}\ \bibnamefont
  {Schutz}},\ and\ \bibinfo {author} {\bibfnamefont {C.}~\bibnamefont {Van
  Den~Broeck}},\ }\bibfield  {title} {\bibinfo {title} {{Cosmography with the
  Einstein Telescope}},\ }\href
  {https://doi.org/10.1088/0264-9381/27/21/215006} {\bibfield  {journal}
  {\bibinfo  {journal} {Class. Quant. Grav.}\ }\textbf {\bibinfo {volume}
  {27}},\ \bibinfo {pages} {215006} (\bibinfo {year} {2010})},\ \Eprint
  {https://arxiv.org/abs/0906.4151} {arXiv:0906.4151 [astro-ph.CO]}
  \BibitemShut {NoStop}%
\bibitem [{\citenamefont {Zhao}\ \emph {et~al.}(2011)\citenamefont {Zhao},
  \citenamefont {Van Den~Broeck}, \citenamefont {Baskaran},\ and\ \citenamefont
  {Li}}]{Zhao:2010sz}%
  \BibitemOpen
  \bibfield  {author} {\bibinfo {author} {\bibfnamefont {W.}~\bibnamefont
  {Zhao}}, \bibinfo {author} {\bibfnamefont {C.}~\bibnamefont {Van
  Den~Broeck}}, \bibinfo {author} {\bibfnamefont {D.}~\bibnamefont
  {Baskaran}},\ and\ \bibinfo {author} {\bibfnamefont {T.~G.~F.}\ \bibnamefont
  {Li}},\ }\bibfield  {title} {\bibinfo {title} {{Determination of Dark Energy
  by the Einstein Telescope: Comparing with CMB, BAO and SNIa Observations}},\
  }\href {https://doi.org/10.1103/PhysRevD.83.023005} {\bibfield  {journal}
  {\bibinfo  {journal} {Phys. Rev. D}\ }\textbf {\bibinfo {volume} {83}},\
  \bibinfo {pages} {023005} (\bibinfo {year} {2011})},\ \Eprint
  {https://arxiv.org/abs/1009.0206} {arXiv:1009.0206 [astro-ph.CO]}
  \BibitemShut {NoStop}%
\bibitem [{\citenamefont {Cai}\ and\ \citenamefont {Yang}(2017)}]{Cai:2016sby}%
  \BibitemOpen
  \bibfield  {author} {\bibinfo {author} {\bibfnamefont {R.-G.}\ \bibnamefont
  {Cai}}\ and\ \bibinfo {author} {\bibfnamefont {T.}~\bibnamefont {Yang}},\
  }\bibfield  {title} {\bibinfo {title} {{Estimating cosmological parameters by
  the simulated data of gravitational waves from the Einstein Telescope}},\
  }\href {https://doi.org/10.1103/PhysRevD.95.044024} {\bibfield  {journal}
  {\bibinfo  {journal} {Phys. Rev. D}\ }\textbf {\bibinfo {volume} {95}},\
  \bibinfo {pages} {044024} (\bibinfo {year} {2017})},\ \Eprint
  {https://arxiv.org/abs/1608.08008} {arXiv:1608.08008 [astro-ph.CO]}
  \BibitemShut {NoStop}%
\bibitem [{\citenamefont {Yang}\ \emph {et~al.}(2019)\citenamefont {Yang},
  \citenamefont {Holanda},\ and\ \citenamefont {Hu}}]{Yang:2017bkv}%
  \BibitemOpen
  \bibfield  {author} {\bibinfo {author} {\bibfnamefont {T.}~\bibnamefont
  {Yang}}, \bibinfo {author} {\bibfnamefont {R.~F.~L.}\ \bibnamefont
  {Holanda}},\ and\ \bibinfo {author} {\bibfnamefont {B.}~\bibnamefont {Hu}},\
  }\bibfield  {title} {\bibinfo {title} {{Constraints on the cosmic distance
  duality relation with simulated data of gravitational waves from the Einstein
  Telescope}},\ }\href {https://doi.org/10.1016/j.astropartphys.2019.01.005}
  {\bibfield  {journal} {\bibinfo  {journal} {Astropart. Phys.}\ }\textbf
  {\bibinfo {volume} {108}},\ \bibinfo {pages} {57} (\bibinfo {year} {2019})},\
  \Eprint {https://arxiv.org/abs/1710.10929} {arXiv:1710.10929 [astro-ph.CO]}
  \BibitemShut {NoStop}%
\bibitem [{\citenamefont {Cai}\ \emph {et~al.}(2018)\citenamefont {Cai},
  \citenamefont {Liu}, \citenamefont {Liu}, \citenamefont {Wang},\ and\
  \citenamefont {Yang}}]{Cai:2017aea}%
  \BibitemOpen
  \bibfield  {author} {\bibinfo {author} {\bibfnamefont {R.-G.}\ \bibnamefont
  {Cai}}, \bibinfo {author} {\bibfnamefont {T.-B.}\ \bibnamefont {Liu}},
  \bibinfo {author} {\bibfnamefont {X.-W.}\ \bibnamefont {Liu}}, \bibinfo
  {author} {\bibfnamefont {S.-J.}\ \bibnamefont {Wang}},\ and\ \bibinfo
  {author} {\bibfnamefont {T.}~\bibnamefont {Yang}},\ }\bibfield  {title}
  {\bibinfo {title} {{Probing cosmic anisotropy with gravitational waves as
  standard sirens}},\ }\href {https://doi.org/10.1103/PhysRevD.97.103005}
  {\bibfield  {journal} {\bibinfo  {journal} {Phys. Rev. D}\ }\textbf {\bibinfo
  {volume} {97}},\ \bibinfo {pages} {103005} (\bibinfo {year} {2018})},\
  \Eprint {https://arxiv.org/abs/1712.00952} {arXiv:1712.00952 [astro-ph.CO]}
  \BibitemShut {NoStop}%
\bibitem [{\citenamefont {Belgacem}\ \emph
  {et~al.}(2018{\natexlab{a}})\citenamefont {Belgacem}, \citenamefont {Dirian},
  \citenamefont {Foffa},\ and\ \citenamefont {Maggiore}}]{Belgacem:2017ihm}%
  \BibitemOpen
  \bibfield  {author} {\bibinfo {author} {\bibfnamefont {E.}~\bibnamefont
  {Belgacem}}, \bibinfo {author} {\bibfnamefont {Y.}~\bibnamefont {Dirian}},
  \bibinfo {author} {\bibfnamefont {S.}~\bibnamefont {Foffa}},\ and\ \bibinfo
  {author} {\bibfnamefont {M.}~\bibnamefont {Maggiore}},\ }\bibfield  {title}
  {\bibinfo {title} {{Gravitational-wave luminosity distance in modified
  gravity theories}},\ }\href {https://doi.org/10.1103/PhysRevD.97.104066}
  {\bibfield  {journal} {\bibinfo  {journal} {Phys. Rev. D}\ }\textbf {\bibinfo
  {volume} {97}},\ \bibinfo {pages} {104066} (\bibinfo {year}
  {2018}{\natexlab{a}})},\ \Eprint {https://arxiv.org/abs/1712.08108}
  {arXiv:1712.08108 [astro-ph.CO]} \BibitemShut {NoStop}%
\bibitem [{\citenamefont {Belgacem}\ \emph {et~al.}(2020)\citenamefont
  {Belgacem}, \citenamefont {Foffa}, \citenamefont {Maggiore},\ and\
  \citenamefont {Yang}}]{Belgacem:2019zzu}%
  \BibitemOpen
  \bibfield  {author} {\bibinfo {author} {\bibfnamefont {E.}~\bibnamefont
  {Belgacem}}, \bibinfo {author} {\bibfnamefont {S.}~\bibnamefont {Foffa}},
  \bibinfo {author} {\bibfnamefont {M.}~\bibnamefont {Maggiore}},\ and\
  \bibinfo {author} {\bibfnamefont {T.}~\bibnamefont {Yang}},\ }\bibfield
  {title} {\bibinfo {title} {{Gaussian processes reconstruction of modified
  gravitational wave propagation}},\ }\href
  {https://doi.org/10.1103/PhysRevD.101.063505} {\bibfield  {journal} {\bibinfo
   {journal} {Phys. Rev. D}\ }\textbf {\bibinfo {volume} {101}},\ \bibinfo
  {pages} {063505} (\bibinfo {year} {2020})},\ \Eprint
  {https://arxiv.org/abs/1911.11497} {arXiv:1911.11497 [astro-ph.CO]}
  \BibitemShut {NoStop}%
\bibitem [{\citenamefont {Yang}(2021)}]{Yang:2021qge}%
  \BibitemOpen
  \bibfield  {author} {\bibinfo {author} {\bibfnamefont {T.}~\bibnamefont
  {Yang}},\ }\bibfield  {title} {\bibinfo {title} {{Gravitational-Wave Detector
  Networks: Standard Sirens on Cosmology and Modified Gravity Theory}},\ }\href
  {https://doi.org/10.1088/1475-7516/2021/05/044} {\bibfield  {journal}
  {\bibinfo  {journal} {JCAP}\ }\textbf {\bibinfo {volume} {05}},\ \bibinfo
  {pages} {044}},\ \Eprint {https://arxiv.org/abs/2103.01923} {arXiv:2103.01923
  [astro-ph.CO]} \BibitemShut {NoStop}%
\bibitem [{\citenamefont {Schutz}(1986)}]{Schutz:1986gp}%
  \BibitemOpen
  \bibfield  {author} {\bibinfo {author} {\bibfnamefont {B.~F.}\ \bibnamefont
  {Schutz}},\ }\bibfield  {title} {\bibinfo {title} {{Determining the Hubble
  Constant from Gravitational Wave Observations}},\ }\href
  {https://doi.org/10.1038/323310a0} {\bibfield  {journal} {\bibinfo  {journal}
  {Nature}\ }\textbf {\bibinfo {volume} {323}},\ \bibinfo {pages} {310}
  (\bibinfo {year} {1986})}\BibitemShut {NoStop}%
\bibitem [{\citenamefont {Holz}\ and\ \citenamefont
  {Hughes}(2005)}]{Holz:2005df}%
  \BibitemOpen
  \bibfield  {author} {\bibinfo {author} {\bibfnamefont {D.~E.}\ \bibnamefont
  {Holz}}\ and\ \bibinfo {author} {\bibfnamefont {S.~A.}\ \bibnamefont
  {Hughes}},\ }\bibfield  {title} {\bibinfo {title} {{Using gravitational-wave
  standard sirens}},\ }\href {https://doi.org/10.1086/431341} {\bibfield
  {journal} {\bibinfo  {journal} {Astrophys. J.}\ }\textbf {\bibinfo {volume}
  {629}},\ \bibinfo {pages} {15} (\bibinfo {year} {2005})},\ \Eprint
  {https://arxiv.org/abs/astro-ph/0504616} {arXiv:astro-ph/0504616}
  \BibitemShut {NoStop}%
\bibitem [{\citenamefont {Nissanke}\ \emph {et~al.}(2010)\citenamefont
  {Nissanke}, \citenamefont {Holz}, \citenamefont {Hughes}, \citenamefont
  {Dalal},\ and\ \citenamefont {Sievers}}]{Nissanke:2009kt}%
  \BibitemOpen
  \bibfield  {author} {\bibinfo {author} {\bibfnamefont {S.}~\bibnamefont
  {Nissanke}}, \bibinfo {author} {\bibfnamefont {D.~E.}\ \bibnamefont {Holz}},
  \bibinfo {author} {\bibfnamefont {S.~A.}\ \bibnamefont {Hughes}}, \bibinfo
  {author} {\bibfnamefont {N.}~\bibnamefont {Dalal}},\ and\ \bibinfo {author}
  {\bibfnamefont {J.~L.}\ \bibnamefont {Sievers}},\ }\bibfield  {title}
  {\bibinfo {title} {{Exploring short gamma-ray bursts as gravitational-wave
  standard sirens}},\ }\href {https://doi.org/10.1088/0004-637X/725/1/496}
  {\bibfield  {journal} {\bibinfo  {journal} {Astrophys. J.}\ }\textbf
  {\bibinfo {volume} {725}},\ \bibinfo {pages} {496} (\bibinfo {year}
  {2010})},\ \Eprint {https://arxiv.org/abs/0904.1017} {arXiv:0904.1017
  [astro-ph.CO]} \BibitemShut {NoStop}%
\bibitem [{\citenamefont {Taylor}\ and\ \citenamefont
  {Gair}(2012)}]{Taylor:2012db}%
  \BibitemOpen
  \bibfield  {author} {\bibinfo {author} {\bibfnamefont {S.~R.}\ \bibnamefont
  {Taylor}}\ and\ \bibinfo {author} {\bibfnamefont {J.~R.}\ \bibnamefont
  {Gair}},\ }\bibfield  {title} {\bibinfo {title} {{Cosmology with the lights
  off: standard sirens in the Einstein Telescope era}},\ }\href
  {https://doi.org/10.1103/PhysRevD.86.023502} {\bibfield  {journal} {\bibinfo
  {journal} {Phys. Rev. D}\ }\textbf {\bibinfo {volume} {86}},\ \bibinfo
  {pages} {023502} (\bibinfo {year} {2012})},\ \Eprint
  {https://arxiv.org/abs/1204.6739} {arXiv:1204.6739 [astro-ph.CO]}
  \BibitemShut {NoStop}%
\bibitem [{\citenamefont {Farr}\ \emph {et~al.}(2019)\citenamefont {Farr},
  \citenamefont {Fishbach}, \citenamefont {Ye},\ and\ \citenamefont
  {Holz}}]{Farr:2019twy}%
  \BibitemOpen
  \bibfield  {author} {\bibinfo {author} {\bibfnamefont {W.~M.}\ \bibnamefont
  {Farr}}, \bibinfo {author} {\bibfnamefont {M.}~\bibnamefont {Fishbach}},
  \bibinfo {author} {\bibfnamefont {J.}~\bibnamefont {Ye}},\ and\ \bibinfo
  {author} {\bibfnamefont {D.}~\bibnamefont {Holz}},\ }\bibfield  {title}
  {\bibinfo {title} {{A Future Percent-Level Measurement of the Hubble
  Expansion at Redshift 0.8 With Advanced LIGO}},\ }\href
  {https://doi.org/10.3847/2041-8213/ab4284} {\bibfield  {journal} {\bibinfo
  {journal} {Astrophys. J. Lett.}\ }\textbf {\bibinfo {volume} {883}},\
  \bibinfo {pages} {L42} (\bibinfo {year} {2019})},\ \Eprint
  {https://arxiv.org/abs/1908.09084} {arXiv:1908.09084 [astro-ph.CO]}
  \BibitemShut {NoStop}%
\bibitem [{\citenamefont {You}\ \emph {et~al.}(2021)\citenamefont {You},
  \citenamefont {Zhu}, \citenamefont {Ashton}, \citenamefont {Thrane},\ and\
  \citenamefont {Zhu}}]{You:2020wju}%
  \BibitemOpen
  \bibfield  {author} {\bibinfo {author} {\bibfnamefont {Z.-Q.}\ \bibnamefont
  {You}}, \bibinfo {author} {\bibfnamefont {X.-J.}\ \bibnamefont {Zhu}},
  \bibinfo {author} {\bibfnamefont {G.}~\bibnamefont {Ashton}}, \bibinfo
  {author} {\bibfnamefont {E.}~\bibnamefont {Thrane}},\ and\ \bibinfo {author}
  {\bibfnamefont {Z.-H.}\ \bibnamefont {Zhu}},\ }\bibfield  {title} {\bibinfo
  {title} {{Standard-siren cosmology using gravitational waves from binary
  black holes}},\ }\href {https://doi.org/10.3847/1538-4357/abd4d4} {\bibfield
  {journal} {\bibinfo  {journal} {Astrophys. J.}\ }\textbf {\bibinfo {volume}
  {908}},\ \bibinfo {pages} {215} (\bibinfo {year} {2021})},\ \Eprint
  {https://arxiv.org/abs/2004.00036} {arXiv:2004.00036 [astro-ph.CO]}
  \BibitemShut {NoStop}%
\bibitem [{\citenamefont {Mastrogiovanni}\ \emph
  {et~al.}(2021{\natexlab{a}})\citenamefont {Mastrogiovanni}, \citenamefont
  {Leyde}, \citenamefont {Karathanasis}, \citenamefont {Chassande-Mottin},
  \citenamefont {Steer}, \citenamefont {Gair}, \citenamefont {Ghosh},
  \citenamefont {Gray}, \citenamefont {Mukherjee},\ and\ \citenamefont
  {Rinaldi}}]{Mastrogiovanni:2021wsd}%
  \BibitemOpen
  \bibfield  {author} {\bibinfo {author} {\bibfnamefont {S.}~\bibnamefont
  {Mastrogiovanni}}, \bibinfo {author} {\bibfnamefont {K.}~\bibnamefont
  {Leyde}}, \bibinfo {author} {\bibfnamefont {C.}~\bibnamefont {Karathanasis}},
  \bibinfo {author} {\bibfnamefont {E.}~\bibnamefont {Chassande-Mottin}},
  \bibinfo {author} {\bibfnamefont {D.~A.}\ \bibnamefont {Steer}}, \bibinfo
  {author} {\bibfnamefont {J.}~\bibnamefont {Gair}}, \bibinfo {author}
  {\bibfnamefont {A.}~\bibnamefont {Ghosh}}, \bibinfo {author} {\bibfnamefont
  {R.}~\bibnamefont {Gray}}, \bibinfo {author} {\bibfnamefont {S.}~\bibnamefont
  {Mukherjee}},\ and\ \bibinfo {author} {\bibfnamefont {S.}~\bibnamefont
  {Rinaldi}},\ }\bibfield  {title} {\bibinfo {title} {{On the importance of
  source population models for gravitational-wave cosmology}},\ }\href
  {https://doi.org/10.1103/PhysRevD.104.062009} {\bibfield  {journal} {\bibinfo
   {journal} {Phys. Rev. D}\ }\textbf {\bibinfo {volume} {104}},\ \bibinfo
  {pages} {062009} (\bibinfo {year} {2021}{\natexlab{a}})},\ \Eprint
  {https://arxiv.org/abs/2103.14663} {arXiv:2103.14663 [gr-qc]} \BibitemShut
  {NoStop}%
\bibitem [{\citenamefont {Del~Pozzo}(2012)}]{DelPozzo:2011vcw}%
  \BibitemOpen
  \bibfield  {author} {\bibinfo {author} {\bibfnamefont {W.}~\bibnamefont
  {Del~Pozzo}},\ }\bibfield  {title} {\bibinfo {title} {{Inference of the
  cosmological parameters from gravitational waves: application to second
  generation interferometers}},\ }\href
  {https://doi.org/10.1103/PhysRevD.86.043011} {\bibfield  {journal} {\bibinfo
  {journal} {Phys. Rev. D}\ }\textbf {\bibinfo {volume} {86}},\ \bibinfo
  {pages} {043011} (\bibinfo {year} {2012})},\ \Eprint
  {https://arxiv.org/abs/1108.1317} {arXiv:1108.1317 [astro-ph.CO]}
  \BibitemShut {NoStop}%
\bibitem [{\citenamefont {Nair}\ \emph {et~al.}(2018)\citenamefont {Nair},
  \citenamefont {Bose},\ and\ \citenamefont {Saini}}]{Nair:2018ign}%
  \BibitemOpen
  \bibfield  {author} {\bibinfo {author} {\bibfnamefont {R.}~\bibnamefont
  {Nair}}, \bibinfo {author} {\bibfnamefont {S.}~\bibnamefont {Bose}},\ and\
  \bibinfo {author} {\bibfnamefont {T.~D.}\ \bibnamefont {Saini}},\ }\bibfield
  {title} {\bibinfo {title} {{Measuring the Hubble constant: Gravitational wave
  observations meet galaxy clustering}},\ }\href
  {https://doi.org/10.1103/PhysRevD.98.023502} {\bibfield  {journal} {\bibinfo
  {journal} {Phys. Rev. D}\ }\textbf {\bibinfo {volume} {98}},\ \bibinfo
  {pages} {023502} (\bibinfo {year} {2018})},\ \Eprint
  {https://arxiv.org/abs/1804.06085} {arXiv:1804.06085 [astro-ph.CO]}
  \BibitemShut {NoStop}%
\bibitem [{\citenamefont {Fishbach}\ \emph {et~al.}(2019)\citenamefont
  {Fishbach} \emph {et~al.}}]{LIGOScientific:2018gmd}%
  \BibitemOpen
  \bibfield  {author} {\bibinfo {author} {\bibfnamefont {M.}~\bibnamefont
  {Fishbach}} \emph {et~al.} (\bibinfo {collaboration} {LIGO Scientific,
  Virgo}),\ }\bibfield  {title} {\bibinfo {title} {{A Standard Siren
  Measurement of the Hubble Constant from GW170817 without the Electromagnetic
  Counterpart}},\ }\href {https://doi.org/10.3847/2041-8213/aaf96e} {\bibfield
  {journal} {\bibinfo  {journal} {Astrophys. J. Lett.}\ }\textbf {\bibinfo
  {volume} {871}},\ \bibinfo {pages} {L13} (\bibinfo {year} {2019})},\ \Eprint
  {https://arxiv.org/abs/1807.05667} {arXiv:1807.05667 [astro-ph.CO]}
  \BibitemShut {NoStop}%
\bibitem [{\citenamefont {Soares-Santos}\ \emph {et~al.}(2019)\citenamefont
  {Soares-Santos} \emph {et~al.}}]{DES:2019ccw}%
  \BibitemOpen
  \bibfield  {author} {\bibinfo {author} {\bibfnamefont {M.}~\bibnamefont
  {Soares-Santos}} \emph {et~al.} (\bibinfo {collaboration} {DES, LIGO
  Scientific, Virgo}),\ }\bibfield  {title} {\bibinfo {title} {{First
  Measurement of the Hubble Constant from a Dark Standard Siren using the Dark
  Energy Survey Galaxies and the LIGO/Virgo Binary\textendash{}Black-hole
  Merger GW170814}},\ }\href {https://doi.org/10.3847/2041-8213/ab14f1}
  {\bibfield  {journal} {\bibinfo  {journal} {Astrophys. J. Lett.}\ }\textbf
  {\bibinfo {volume} {876}},\ \bibinfo {pages} {L7} (\bibinfo {year} {2019})},\
  \Eprint {https://arxiv.org/abs/1901.01540} {arXiv:1901.01540 [astro-ph.CO]}
  \BibitemShut {NoStop}%
\bibitem [{\citenamefont {Gray}\ \emph {et~al.}(2020)\citenamefont {Gray} \emph
  {et~al.}}]{Gray:2019ksv}%
  \BibitemOpen
  \bibfield  {author} {\bibinfo {author} {\bibfnamefont {R.}~\bibnamefont
  {Gray}} \emph {et~al.},\ }\bibfield  {title} {\bibinfo {title} {{Cosmological
  inference using gravitational wave standard sirens: A mock data analysis}},\
  }\href {https://doi.org/10.1103/PhysRevD.101.122001} {\bibfield  {journal}
  {\bibinfo  {journal} {Phys. Rev. D}\ }\textbf {\bibinfo {volume} {101}},\
  \bibinfo {pages} {122001} (\bibinfo {year} {2020})},\ \Eprint
  {https://arxiv.org/abs/1908.06050} {arXiv:1908.06050 [gr-qc]} \BibitemShut
  {NoStop}%
\bibitem [{\citenamefont {Palmese}\ \emph {et~al.}(2020)\citenamefont {Palmese}
  \emph {et~al.}}]{DES:2020nay}%
  \BibitemOpen
  \bibfield  {author} {\bibinfo {author} {\bibfnamefont {A.}~\bibnamefont
  {Palmese}} \emph {et~al.} (\bibinfo {collaboration} {DES}),\ }\bibfield
  {title} {\bibinfo {title} {{A statistical standard siren measurement of the
  Hubble constant from the LIGO/Virgo gravitational wave compact object merger
  GW190814 and Dark Energy Survey galaxies}},\ }\href
  {https://doi.org/10.3847/2041-8213/abaeff} {\bibfield  {journal} {\bibinfo
  {journal} {Astrophys. J. Lett.}\ }\textbf {\bibinfo {volume} {900}},\
  \bibinfo {pages} {L33} (\bibinfo {year} {2020})},\ \Eprint
  {https://arxiv.org/abs/2006.14961} {arXiv:2006.14961 [astro-ph.CO]}
  \BibitemShut {NoStop}%
\bibitem [{\citenamefont {Finke}\ \emph {et~al.}(2021)\citenamefont {Finke},
  \citenamefont {Foffa}, \citenamefont {Iacovelli}, \citenamefont {Maggiore},\
  and\ \citenamefont {Mancarella}}]{Finke:2021aom}%
  \BibitemOpen
  \bibfield  {author} {\bibinfo {author} {\bibfnamefont {A.}~\bibnamefont
  {Finke}}, \bibinfo {author} {\bibfnamefont {S.}~\bibnamefont {Foffa}},
  \bibinfo {author} {\bibfnamefont {F.}~\bibnamefont {Iacovelli}}, \bibinfo
  {author} {\bibfnamefont {M.}~\bibnamefont {Maggiore}},\ and\ \bibinfo
  {author} {\bibfnamefont {M.}~\bibnamefont {Mancarella}},\ }\bibfield  {title}
  {\bibinfo {title} {{Cosmology with LIGO/Virgo dark sirens: Hubble parameter
  and modified gravitational wave propagation}},\ }\href
  {https://doi.org/10.1088/1475-7516/2021/08/026} {\bibfield  {journal}
  {\bibinfo  {journal} {JCAP}\ }\textbf {\bibinfo {volume} {08}},\ \bibinfo
  {pages} {026}},\ \Eprint {https://arxiv.org/abs/2101.12660} {arXiv:2101.12660
  [astro-ph.CO]} \BibitemShut {NoStop}%
\bibitem [{\citenamefont {Oguri}(2016)}]{Oguri:2016dgk}%
  \BibitemOpen
  \bibfield  {author} {\bibinfo {author} {\bibfnamefont {M.}~\bibnamefont
  {Oguri}},\ }\bibfield  {title} {\bibinfo {title} {{Measuring the
  distance-redshift relation with the cross-correlation of gravitational wave
  standard sirens and galaxies}},\ }\href
  {https://doi.org/10.1103/PhysRevD.93.083511} {\bibfield  {journal} {\bibinfo
  {journal} {Phys. Rev. D}\ }\textbf {\bibinfo {volume} {93}},\ \bibinfo
  {pages} {083511} (\bibinfo {year} {2016})},\ \Eprint
  {https://arxiv.org/abs/1603.02356} {arXiv:1603.02356 [astro-ph.CO]}
  \BibitemShut {NoStop}%
\bibitem [{\citenamefont {Mukherjee}\ \emph {et~al.}(2020)\citenamefont
  {Mukherjee}, \citenamefont {Wandelt},\ and\ \citenamefont
  {Silk}}]{Mukherjee:2019wcg}%
  \BibitemOpen
  \bibfield  {author} {\bibinfo {author} {\bibfnamefont {S.}~\bibnamefont
  {Mukherjee}}, \bibinfo {author} {\bibfnamefont {B.~D.}\ \bibnamefont
  {Wandelt}},\ and\ \bibinfo {author} {\bibfnamefont {J.}~\bibnamefont
  {Silk}},\ }\bibfield  {title} {\bibinfo {title} {{Probing the theory of
  gravity with gravitational lensing of gravitational waves and galaxy
  surveys}},\ }\href {https://doi.org/10.1093/mnras/staa827} {\bibfield
  {journal} {\bibinfo  {journal} {Mon. Not. Roy. Astron. Soc.}\ }\textbf
  {\bibinfo {volume} {494}},\ \bibinfo {pages} {1956} (\bibinfo {year}
  {2020})},\ \Eprint {https://arxiv.org/abs/1908.08951} {arXiv:1908.08951
  [astro-ph.CO]} \BibitemShut {NoStop}%
\bibitem [{\citenamefont {Mukherjee}\ \emph
  {et~al.}(2021{\natexlab{a}})\citenamefont {Mukherjee}, \citenamefont
  {Wandelt}, \citenamefont {Nissanke},\ and\ \citenamefont
  {Silvestri}}]{Mukherjee:2020hyn}%
  \BibitemOpen
  \bibfield  {author} {\bibinfo {author} {\bibfnamefont {S.}~\bibnamefont
  {Mukherjee}}, \bibinfo {author} {\bibfnamefont {B.~D.}\ \bibnamefont
  {Wandelt}}, \bibinfo {author} {\bibfnamefont {S.~M.}\ \bibnamefont
  {Nissanke}},\ and\ \bibinfo {author} {\bibfnamefont {A.}~\bibnamefont
  {Silvestri}},\ }\bibfield  {title} {\bibinfo {title} {{Accurate precision
  Cosmology with redshift unknown gravitational wave sources}},\ }\href
  {https://doi.org/10.1103/PhysRevD.103.043520} {\bibfield  {journal} {\bibinfo
   {journal} {Phys. Rev. D}\ }\textbf {\bibinfo {volume} {103}},\ \bibinfo
  {pages} {043520} (\bibinfo {year} {2021}{\natexlab{a}})},\ \Eprint
  {https://arxiv.org/abs/2007.02943} {arXiv:2007.02943 [astro-ph.CO]}
  \BibitemShut {NoStop}%
\bibitem [{\citenamefont {Bera}\ \emph {et~al.}(2020)\citenamefont {Bera},
  \citenamefont {Rana}, \citenamefont {More},\ and\ \citenamefont
  {Bose}}]{Bera:2020jhx}%
  \BibitemOpen
  \bibfield  {author} {\bibinfo {author} {\bibfnamefont {S.}~\bibnamefont
  {Bera}}, \bibinfo {author} {\bibfnamefont {D.}~\bibnamefont {Rana}}, \bibinfo
  {author} {\bibfnamefont {S.}~\bibnamefont {More}},\ and\ \bibinfo {author}
  {\bibfnamefont {S.}~\bibnamefont {Bose}},\ }\bibfield  {title} {\bibinfo
  {title} {{Incompleteness Matters Not: Inference of $H_0$ from Binary Black
  Hole\textendash{}Galaxy Cross-correlations}},\ }\href
  {https://doi.org/10.3847/1538-4357/abb4e0} {\bibfield  {journal} {\bibinfo
  {journal} {Astrophys. J.}\ }\textbf {\bibinfo {volume} {902}},\ \bibinfo
  {pages} {79} (\bibinfo {year} {2020})},\ \Eprint
  {https://arxiv.org/abs/2007.04271} {arXiv:2007.04271 [astro-ph.CO]}
  \BibitemShut {NoStop}%
\bibitem [{\citenamefont {Mukherjee}\ \emph {et~al.}(2022)\citenamefont
  {Mukherjee}, \citenamefont {Krolewski}, \citenamefont {Wandelt},\ and\
  \citenamefont {Silk}}]{Mukherjee:2022afz}%
  \BibitemOpen
  \bibfield  {author} {\bibinfo {author} {\bibfnamefont {S.}~\bibnamefont
  {Mukherjee}}, \bibinfo {author} {\bibfnamefont {A.}~\bibnamefont
  {Krolewski}}, \bibinfo {author} {\bibfnamefont {B.~D.}\ \bibnamefont
  {Wandelt}},\ and\ \bibinfo {author} {\bibfnamefont {J.}~\bibnamefont
  {Silk}},\ }\bibfield  {title} {\bibinfo {title} {{Cross-correlating dark
  sirens and galaxies: measurement of $H_0$ from GWTC-3 of LIGO-Virgo-KAGRA}},\
  }\href@noop {} {\  (\bibinfo {year} {2022})},\ \Eprint
  {https://arxiv.org/abs/2203.03643} {arXiv:2203.03643 [astro-ph.CO]}
  \BibitemShut {NoStop}%
\bibitem [{\citenamefont {Messenger}\ and\ \citenamefont
  {Read}(2012)}]{Messenger:2011gi}%
  \BibitemOpen
  \bibfield  {author} {\bibinfo {author} {\bibfnamefont {C.}~\bibnamefont
  {Messenger}}\ and\ \bibinfo {author} {\bibfnamefont {J.}~\bibnamefont
  {Read}},\ }\bibfield  {title} {\bibinfo {title} {{Measuring a cosmological
  distance-redshift relationship using only gravitational wave observations of
  binary neutron star coalescences}},\ }\href
  {https://doi.org/10.1103/PhysRevLett.108.091101} {\bibfield  {journal}
  {\bibinfo  {journal} {Phys. Rev. Lett.}\ }\textbf {\bibinfo {volume} {108}},\
  \bibinfo {pages} {091101} (\bibinfo {year} {2012})},\ \Eprint
  {https://arxiv.org/abs/1107.5725} {arXiv:1107.5725 [gr-qc]} \BibitemShut
  {NoStop}%
\bibitem [{\citenamefont {Li}\ \emph {et~al.}(2015)\citenamefont {Li},
  \citenamefont {Del~Pozzo},\ and\ \citenamefont {Messenger}}]{Li:2013via}%
  \BibitemOpen
  \bibfield  {author} {\bibinfo {author} {\bibfnamefont {T.~G.~F.}\
  \bibnamefont {Li}}, \bibinfo {author} {\bibfnamefont {W.}~\bibnamefont
  {Del~Pozzo}},\ and\ \bibinfo {author} {\bibfnamefont {C.}~\bibnamefont
  {Messenger}},\ }\bibfield  {title} {\bibinfo {title} {{Measuring the redshift
  of standard sirens using the neutron star deformability}},\ }in\ \href
  {https://doi.org/10.1142/9789814623995_0346} {\emph {\bibinfo {booktitle}
  {{13th Marcel Grossmann Meeting on Recent Developments in Theoretical and
  Experimental General Relativity, Astrophysics, and Relativistic Field
  Theories}}}}\ (\bibinfo {year} {2015})\ pp.\ \bibinfo {pages} {2019--2021},\
  \Eprint {https://arxiv.org/abs/1303.0855} {arXiv:1303.0855 [gr-qc]}
  \BibitemShut {NoStop}%
\bibitem [{\citenamefont {Messenger}\ \emph {et~al.}(2014)\citenamefont
  {Messenger}, \citenamefont {Takami}, \citenamefont {Gossan}, \citenamefont
  {Rezzolla},\ and\ \citenamefont {Sathyaprakash}}]{Messenger:2013fya}%
  \BibitemOpen
  \bibfield  {author} {\bibinfo {author} {\bibfnamefont {C.}~\bibnamefont
  {Messenger}}, \bibinfo {author} {\bibfnamefont {K.}~\bibnamefont {Takami}},
  \bibinfo {author} {\bibfnamefont {S.}~\bibnamefont {Gossan}}, \bibinfo
  {author} {\bibfnamefont {L.}~\bibnamefont {Rezzolla}},\ and\ \bibinfo
  {author} {\bibfnamefont {B.~S.}\ \bibnamefont {Sathyaprakash}},\ }\bibfield
  {title} {\bibinfo {title} {{Source Redshifts from Gravitational-Wave
  Observations of Binary Neutron Star Mergers}},\ }\href
  {https://doi.org/10.1103/PhysRevX.4.041004} {\bibfield  {journal} {\bibinfo
  {journal} {Phys. Rev. X}\ }\textbf {\bibinfo {volume} {4}},\ \bibinfo {pages}
  {041004} (\bibinfo {year} {2014})},\ \Eprint
  {https://arxiv.org/abs/1312.1862} {arXiv:1312.1862 [gr-qc]} \BibitemShut
  {NoStop}%
\bibitem [{\citenamefont {Del~Pozzo}\ \emph {et~al.}(2017)\citenamefont
  {Del~Pozzo}, \citenamefont {Li},\ and\ \citenamefont
  {Messenger}}]{DelPozzo:2015bna}%
  \BibitemOpen
  \bibfield  {author} {\bibinfo {author} {\bibfnamefont {W.}~\bibnamefont
  {Del~Pozzo}}, \bibinfo {author} {\bibfnamefont {T.~G.~F.}\ \bibnamefont
  {Li}},\ and\ \bibinfo {author} {\bibfnamefont {C.}~\bibnamefont
  {Messenger}},\ }\bibfield  {title} {\bibinfo {title} {{Cosmological inference
  using only gravitational wave observations of binary neutron stars}},\ }\href
  {https://doi.org/10.1103/PhysRevD.95.043502} {\bibfield  {journal} {\bibinfo
  {journal} {Phys. Rev. D}\ }\textbf {\bibinfo {volume} {95}},\ \bibinfo
  {pages} {043502} (\bibinfo {year} {2017})},\ \Eprint
  {https://arxiv.org/abs/1506.06590} {arXiv:1506.06590 [gr-qc]} \BibitemShut
  {NoStop}%
\bibitem [{\citenamefont {Wang}\ \emph {et~al.}(2020)\citenamefont {Wang},
  \citenamefont {Zhu}, \citenamefont {Li},\ and\ \citenamefont
  {Zhao}}]{Wang:2020xwn}%
  \BibitemOpen
  \bibfield  {author} {\bibinfo {author} {\bibfnamefont {B.}~\bibnamefont
  {Wang}}, \bibinfo {author} {\bibfnamefont {Z.}~\bibnamefont {Zhu}}, \bibinfo
  {author} {\bibfnamefont {A.}~\bibnamefont {Li}},\ and\ \bibinfo {author}
  {\bibfnamefont {W.}~\bibnamefont {Zhao}},\ }\bibfield  {title} {\bibinfo
  {title} {{Comprehensive analysis of the tidal effect in gravitational waves
  and implication for cosmology}},\ }\href
  {https://doi.org/10.3847/1538-4365/aba2f3} {\bibfield  {journal} {\bibinfo
  {journal} {Astrophys. J. Suppl.}\ }\textbf {\bibinfo {volume} {250}},\
  \bibinfo {pages} {6} (\bibinfo {year} {2020})},\ \Eprint
  {https://arxiv.org/abs/2005.12875} {arXiv:2005.12875 [gr-qc]} \BibitemShut
  {NoStop}%
\bibitem [{\citenamefont {Chatterjee}\ \emph {et~al.}(2021)\citenamefont
  {Chatterjee}, \citenamefont {Hegade K~R}, \citenamefont {Holder},
  \citenamefont {Holz}, \citenamefont {Perkins}, \citenamefont {Yagi},\ and\
  \citenamefont {Yunes}}]{Chatterjee:2021xrm}%
  \BibitemOpen
  \bibfield  {author} {\bibinfo {author} {\bibfnamefont {D.}~\bibnamefont
  {Chatterjee}}, \bibinfo {author} {\bibfnamefont {A.}~\bibnamefont {Hegade
  K~R}}, \bibinfo {author} {\bibfnamefont {G.}~\bibnamefont {Holder}}, \bibinfo
  {author} {\bibfnamefont {D.~E.}\ \bibnamefont {Holz}}, \bibinfo {author}
  {\bibfnamefont {S.}~\bibnamefont {Perkins}}, \bibinfo {author} {\bibfnamefont
  {K.}~\bibnamefont {Yagi}},\ and\ \bibinfo {author} {\bibfnamefont
  {N.}~\bibnamefont {Yunes}},\ }\bibfield  {title} {\bibinfo {title}
  {{Cosmology with Love: Measuring the Hubble constant using neutron star
  universal relations}},\ }\href {https://doi.org/10.1103/PhysRevD.104.083528}
  {\bibfield  {journal} {\bibinfo  {journal} {Phys. Rev. D}\ }\textbf {\bibinfo
  {volume} {104}},\ \bibinfo {pages} {083528} (\bibinfo {year} {2021})},\
  \Eprint {https://arxiv.org/abs/2106.06589} {arXiv:2106.06589 [gr-qc]}
  \BibitemShut {NoStop}%
\bibitem [{\citenamefont {Jin}\ \emph {et~al.}(2022)\citenamefont {Jin},
  \citenamefont {Li}, \citenamefont {Zhang},\ and\ \citenamefont
  {Zhang}}]{Jin:2022qnj}%
  \BibitemOpen
  \bibfield  {author} {\bibinfo {author} {\bibfnamefont {S.-J.}\ \bibnamefont
  {Jin}}, \bibinfo {author} {\bibfnamefont {T.-N.}\ \bibnamefont {Li}},
  \bibinfo {author} {\bibfnamefont {J.-F.}\ \bibnamefont {Zhang}},\ and\
  \bibinfo {author} {\bibfnamefont {X.}~\bibnamefont {Zhang}},\ }\bibfield
  {title} {\bibinfo {title} {{Precisely measuring the Hubble constant and dark
  energy using only gravitational-wave dark sirens}},\ }\href@noop {} {\
  (\bibinfo {year} {2022})},\ \Eprint {https://arxiv.org/abs/2202.11882}
  {arXiv:2202.11882 [gr-qc]} \BibitemShut {NoStop}%
\bibitem [{\citenamefont {Dhani}\ \emph {et~al.}(2022)\citenamefont {Dhani},
  \citenamefont {Borhanian}, \citenamefont {Gupta},\ and\ \citenamefont
  {Sathyaprakash}}]{Dhani:2022ulg}%
  \BibitemOpen
  \bibfield  {author} {\bibinfo {author} {\bibfnamefont {A.}~\bibnamefont
  {Dhani}}, \bibinfo {author} {\bibfnamefont {S.}~\bibnamefont {Borhanian}},
  \bibinfo {author} {\bibfnamefont {A.}~\bibnamefont {Gupta}},\ and\ \bibinfo
  {author} {\bibfnamefont {B.}~\bibnamefont {Sathyaprakash}},\ }\bibfield
  {title} {\bibinfo {title} {{Cosmography with bright and Love sirens}},\
  }\href@noop {} {\  (\bibinfo {year} {2022})},\ \Eprint
  {https://arxiv.org/abs/2212.13183} {arXiv:2212.13183 [gr-qc]} \BibitemShut
  {NoStop}%
\bibitem [{\citenamefont {Hotokezaka}\ \emph {et~al.}(2019)\citenamefont
  {Hotokezaka}, \citenamefont {Nakar}, \citenamefont {Gottlieb}, \citenamefont
  {Nissanke}, \citenamefont {Masuda}, \citenamefont {Hallinan}, \citenamefont
  {Mooley},\ and\ \citenamefont {Deller}}]{Hotokezaka:2018dfi}%
  \BibitemOpen
  \bibfield  {author} {\bibinfo {author} {\bibfnamefont {K.}~\bibnamefont
  {Hotokezaka}}, \bibinfo {author} {\bibfnamefont {E.}~\bibnamefont {Nakar}},
  \bibinfo {author} {\bibfnamefont {O.}~\bibnamefont {Gottlieb}}, \bibinfo
  {author} {\bibfnamefont {S.}~\bibnamefont {Nissanke}}, \bibinfo {author}
  {\bibfnamefont {K.}~\bibnamefont {Masuda}}, \bibinfo {author} {\bibfnamefont
  {G.}~\bibnamefont {Hallinan}}, \bibinfo {author} {\bibfnamefont {K.~P.}\
  \bibnamefont {Mooley}},\ and\ \bibinfo {author} {\bibfnamefont {A.~T.}\
  \bibnamefont {Deller}},\ }\bibfield  {title} {\bibinfo {title} {{A Hubble
  constant measurement from superluminal motion of the jet in GW170817}},\
  }\href {https://doi.org/10.1038/s41550-019-0820-1} {\bibfield  {journal}
  {\bibinfo  {journal} {Nature Astron.}\ }\textbf {\bibinfo {volume} {3}},\
  \bibinfo {pages} {940} (\bibinfo {year} {2019})},\ \Eprint
  {https://arxiv.org/abs/1806.10596} {arXiv:1806.10596 [astro-ph.CO]}
  \BibitemShut {NoStop}%
\bibitem [{\citenamefont {Belgacem}\ \emph
  {et~al.}(2019{\natexlab{a}})\citenamefont {Belgacem}, \citenamefont {Dirian},
  \citenamefont {Foffa}, \citenamefont {Howell}, \citenamefont {Maggiore},\
  and\ \citenamefont {Regimbau}}]{Belgacem:2019tbw}%
  \BibitemOpen
  \bibfield  {author} {\bibinfo {author} {\bibfnamefont {E.}~\bibnamefont
  {Belgacem}}, \bibinfo {author} {\bibfnamefont {Y.}~\bibnamefont {Dirian}},
  \bibinfo {author} {\bibfnamefont {S.}~\bibnamefont {Foffa}}, \bibinfo
  {author} {\bibfnamefont {E.~J.}\ \bibnamefont {Howell}}, \bibinfo {author}
  {\bibfnamefont {M.}~\bibnamefont {Maggiore}},\ and\ \bibinfo {author}
  {\bibfnamefont {T.}~\bibnamefont {Regimbau}},\ }\bibfield  {title} {\bibinfo
  {title} {{Cosmology and dark energy from joint gravitational wave-GRB
  observations}},\ }\href {https://doi.org/10.1088/1475-7516/2019/08/015}
  {\bibfield  {journal} {\bibinfo  {journal} {JCAP}\ }\textbf {\bibinfo
  {volume} {08}},\ \bibinfo {pages} {015}},\ \Eprint
  {https://arxiv.org/abs/1907.01487} {arXiv:1907.01487 [astro-ph.CO]}
  \BibitemShut {NoStop}%
\bibitem [{\citenamefont {Yang}\ \emph
  {et~al.}(2022{\natexlab{a}})\citenamefont {Yang}, \citenamefont {Cai},
  \citenamefont {Cao},\ and\ \citenamefont {Lee}}]{Yang:2022tig}%
  \BibitemOpen
  \bibfield  {author} {\bibinfo {author} {\bibfnamefont {T.}~\bibnamefont
  {Yang}}, \bibinfo {author} {\bibfnamefont {R.-G.}\ \bibnamefont {Cai}},
  \bibinfo {author} {\bibfnamefont {Z.}~\bibnamefont {Cao}},\ and\ \bibinfo
  {author} {\bibfnamefont {H.~M.}\ \bibnamefont {Lee}},\ }\bibfield  {title}
  {\bibinfo {title} {{Eccentricity of Long Inspiraling Compact Binaries Sheds
  Light on Dark Sirens}},\ }\href
  {https://doi.org/10.1103/PhysRevLett.129.191102} {\bibfield  {journal}
  {\bibinfo  {journal} {Phys. Rev. Lett.}\ }\textbf {\bibinfo {volume} {129}},\
  \bibinfo {pages} {191102} (\bibinfo {year} {2022}{\natexlab{a}})},\ \Eprint
  {https://arxiv.org/abs/2202.08608} {arXiv:2202.08608 [gr-qc]} \BibitemShut
  {NoStop}%
\bibitem [{\citenamefont {Antonini}\ and\ \citenamefont
  {Perets}(2012)}]{Antonini:2012ad}%
  \BibitemOpen
  \bibfield  {author} {\bibinfo {author} {\bibfnamefont {F.}~\bibnamefont
  {Antonini}}\ and\ \bibinfo {author} {\bibfnamefont {H.~B.}\ \bibnamefont
  {Perets}},\ }\bibfield  {title} {\bibinfo {title} {{Secular evolution of
  compact binaries near massive black holes: Gravitational wave sources and
  other exotica}},\ }\href {https://doi.org/10.1088/0004-637X/757/1/27}
  {\bibfield  {journal} {\bibinfo  {journal} {Astrophys. J.}\ }\textbf
  {\bibinfo {volume} {757}},\ \bibinfo {pages} {27} (\bibinfo {year} {2012})},\
  \Eprint {https://arxiv.org/abs/1203.2938} {arXiv:1203.2938 [astro-ph.GA]}
  \BibitemShut {NoStop}%
\bibitem [{\citenamefont {Samsing}\ \emph {et~al.}(2014)\citenamefont
  {Samsing}, \citenamefont {MacLeod},\ and\ \citenamefont
  {Ramirez-Ruiz}}]{Samsing:2013kua}%
  \BibitemOpen
  \bibfield  {author} {\bibinfo {author} {\bibfnamefont {J.}~\bibnamefont
  {Samsing}}, \bibinfo {author} {\bibfnamefont {M.}~\bibnamefont {MacLeod}},\
  and\ \bibinfo {author} {\bibfnamefont {E.}~\bibnamefont {Ramirez-Ruiz}},\
  }\bibfield  {title} {\bibinfo {title} {{The Formation of Eccentric Compact
  Binary Inspirals and the Role of Gravitational Wave Emission in Binary-Single
  Stellar Encounters}},\ }\href {https://doi.org/10.1088/0004-637X/784/1/71}
  {\bibfield  {journal} {\bibinfo  {journal} {Astrophys. J.}\ }\textbf
  {\bibinfo {volume} {784}},\ \bibinfo {pages} {71} (\bibinfo {year} {2014})},\
  \Eprint {https://arxiv.org/abs/1308.2964} {arXiv:1308.2964 [astro-ph.HE]}
  \BibitemShut {NoStop}%
\bibitem [{\citenamefont {Thompson}(2011)}]{Thompson:2010dp}%
  \BibitemOpen
  \bibfield  {author} {\bibinfo {author} {\bibfnamefont {T.~A.}\ \bibnamefont
  {Thompson}},\ }\bibfield  {title} {\bibinfo {title} {{Accelerating Compact
  Object Mergers in Triple Systems with the Kozai Resonance: A Mechanism for
  'Prompt'' Type Ia Supernovae, Gamma-Ray Bursts, and Other Exotica}},\ }\href
  {https://doi.org/10.1088/0004-637X/741/2/82} {\bibfield  {journal} {\bibinfo
  {journal} {Astrophys. J.}\ }\textbf {\bibinfo {volume} {741}},\ \bibinfo
  {pages} {82} (\bibinfo {year} {2011})},\ \Eprint
  {https://arxiv.org/abs/1011.4322} {arXiv:1011.4322 [astro-ph.HE]}
  \BibitemShut {NoStop}%
\bibitem [{\citenamefont {East}\ \emph {et~al.}(2013)\citenamefont {East},
  \citenamefont {McWilliams}, \citenamefont {Levin},\ and\ \citenamefont
  {Pretorius}}]{East:2012xq}%
  \BibitemOpen
  \bibfield  {author} {\bibinfo {author} {\bibfnamefont {W.~E.}\ \bibnamefont
  {East}}, \bibinfo {author} {\bibfnamefont {S.~T.}\ \bibnamefont
  {McWilliams}}, \bibinfo {author} {\bibfnamefont {J.}~\bibnamefont {Levin}},\
  and\ \bibinfo {author} {\bibfnamefont {F.}~\bibnamefont {Pretorius}},\
  }\bibfield  {title} {\bibinfo {title} {{Observing complete gravitational wave
  signals from dynamical capture binaries}},\ }\href
  {https://doi.org/10.1103/PhysRevD.87.043004} {\bibfield  {journal} {\bibinfo
  {journal} {Phys. Rev. D}\ }\textbf {\bibinfo {volume} {87}},\ \bibinfo
  {pages} {043004} (\bibinfo {year} {2013})},\ \Eprint
  {https://arxiv.org/abs/1212.0837} {arXiv:1212.0837 [gr-qc]} \BibitemShut
  {NoStop}%
\bibitem [{\citenamefont {Rodriguez}\ \emph {et~al.}(2018)\citenamefont
  {Rodriguez}, \citenamefont {Amaro-Seoane}, \citenamefont {Chatterjee},\ and\
  \citenamefont {Rasio}}]{Rodriguez:2017pec}%
  \BibitemOpen
  \bibfield  {author} {\bibinfo {author} {\bibfnamefont {C.~L.}\ \bibnamefont
  {Rodriguez}}, \bibinfo {author} {\bibfnamefont {P.}~\bibnamefont
  {Amaro-Seoane}}, \bibinfo {author} {\bibfnamefont {S.}~\bibnamefont
  {Chatterjee}},\ and\ \bibinfo {author} {\bibfnamefont {F.~A.}\ \bibnamefont
  {Rasio}},\ }\bibfield  {title} {\bibinfo {title} {{Post-Newtonian Dynamics in
  Dense Star Clusters: Highly-Eccentric, Highly-Spinning, and Repeated Binary
  Black Hole Mergers}},\ }\href
  {https://doi.org/10.1103/PhysRevLett.120.151101} {\bibfield  {journal}
  {\bibinfo  {journal} {Phys. Rev. Lett.}\ }\textbf {\bibinfo {volume} {120}},\
  \bibinfo {pages} {151101} (\bibinfo {year} {2018})},\ \Eprint
  {https://arxiv.org/abs/1712.04937} {arXiv:1712.04937 [astro-ph.HE]}
  \BibitemShut {NoStop}%
\bibitem [{\citenamefont {Samsing}(2018)}]{Samsing:2017xmd}%
  \BibitemOpen
  \bibfield  {author} {\bibinfo {author} {\bibfnamefont {J.}~\bibnamefont
  {Samsing}},\ }\bibfield  {title} {\bibinfo {title} {{Eccentric Black Hole
  Mergers Forming in Globular Clusters}},\ }\href
  {https://doi.org/10.1103/PhysRevD.97.103014} {\bibfield  {journal} {\bibinfo
  {journal} {Phys. Rev. D}\ }\textbf {\bibinfo {volume} {97}},\ \bibinfo
  {pages} {103014} (\bibinfo {year} {2018})},\ \Eprint
  {https://arxiv.org/abs/1711.07452} {arXiv:1711.07452 [astro-ph.HE]}
  \BibitemShut {NoStop}%
\bibitem [{\citenamefont {Samsing}\ \emph
  {et~al.}(2018{\natexlab{a}})\citenamefont {Samsing}, \citenamefont {Askar},\
  and\ \citenamefont {Giersz}}]{Samsing:2017oij}%
  \BibitemOpen
  \bibfield  {author} {\bibinfo {author} {\bibfnamefont {J.}~\bibnamefont
  {Samsing}}, \bibinfo {author} {\bibfnamefont {A.}~\bibnamefont {Askar}},\
  and\ \bibinfo {author} {\bibfnamefont {M.}~\bibnamefont {Giersz}},\
  }\bibfield  {title} {\bibinfo {title} {{MOCCA-SURVEY Database. I. Eccentric
  Black Hole Mergers during Binary\textendash{}Single Interactions in Globular
  Clusters}},\ }\href {https://doi.org/10.3847/1538-4357/aaab52} {\bibfield
  {journal} {\bibinfo  {journal} {Astrophys. J.}\ }\textbf {\bibinfo {volume}
  {855}},\ \bibinfo {pages} {124} (\bibinfo {year} {2018}{\natexlab{a}})},\
  \Eprint {https://arxiv.org/abs/1712.06186} {arXiv:1712.06186 [astro-ph.HE]}
  \BibitemShut {NoStop}%
\bibitem [{\citenamefont {Samsing}\ \emph
  {et~al.}(2018{\natexlab{b}})\citenamefont {Samsing}, \citenamefont
  {D'Orazio}, \citenamefont {Askar},\ and\ \citenamefont
  {Giersz}}]{Samsing:2018ykz}%
  \BibitemOpen
  \bibfield  {author} {\bibinfo {author} {\bibfnamefont {J.}~\bibnamefont
  {Samsing}}, \bibinfo {author} {\bibfnamefont {D.~J.}\ \bibnamefont
  {D'Orazio}}, \bibinfo {author} {\bibfnamefont {A.}~\bibnamefont {Askar}},\
  and\ \bibinfo {author} {\bibfnamefont {M.}~\bibnamefont {Giersz}},\
  }\bibfield  {title} {\bibinfo {title} {{Black Hole Mergers from Globular
  Clusters Observable by LISA and LIGO: Results from post-Newtonian
  Binary-Single Scatterings}},\ }\href@noop {} {\  (\bibinfo {year}
  {2018}{\natexlab{b}})},\ \Eprint {https://arxiv.org/abs/1802.08654}
  {arXiv:1802.08654 [astro-ph.HE]} \BibitemShut {NoStop}%
\bibitem [{\citenamefont {Wen}(2003)}]{Wen:2002km}%
  \BibitemOpen
  \bibfield  {author} {\bibinfo {author} {\bibfnamefont {L.}~\bibnamefont
  {Wen}},\ }\bibfield  {title} {\bibinfo {title} {{On the eccentricity
  distribution of coalescing black hole binaries driven by the Kozai mechanism
  in globular clusters}},\ }\href {https://doi.org/10.1086/378794} {\bibfield
  {journal} {\bibinfo  {journal} {Astrophys. J.}\ }\textbf {\bibinfo {volume}
  {598}},\ \bibinfo {pages} {419} (\bibinfo {year} {2003})},\ \Eprint
  {https://arxiv.org/abs/astro-ph/0211492} {arXiv:astro-ph/0211492}
  \BibitemShut {NoStop}%
\bibitem [{\citenamefont {Pratten}\ \emph {et~al.}(2020)\citenamefont
  {Pratten}, \citenamefont {Husa}, \citenamefont {Garcia-Quiros}, \citenamefont
  {Colleoni}, \citenamefont {Ramos-Buades}, \citenamefont {Estelles},\ and\
  \citenamefont {Jaume}}]{Pratten:2020fqn}%
  \BibitemOpen
  \bibfield  {author} {\bibinfo {author} {\bibfnamefont {G.}~\bibnamefont
  {Pratten}}, \bibinfo {author} {\bibfnamefont {S.}~\bibnamefont {Husa}},
  \bibinfo {author} {\bibfnamefont {C.}~\bibnamefont {Garcia-Quiros}}, \bibinfo
  {author} {\bibfnamefont {M.}~\bibnamefont {Colleoni}}, \bibinfo {author}
  {\bibfnamefont {A.}~\bibnamefont {Ramos-Buades}}, \bibinfo {author}
  {\bibfnamefont {H.}~\bibnamefont {Estelles}},\ and\ \bibinfo {author}
  {\bibfnamefont {R.}~\bibnamefont {Jaume}},\ }\bibfield  {title} {\bibinfo
  {title} {{Setting the cornerstone for a family of models for gravitational
  waves from compact binaries: The dominant harmonic for nonprecessing
  quasicircular black holes}},\ }\href
  {https://doi.org/10.1103/PhysRevD.102.064001} {\bibfield  {journal} {\bibinfo
   {journal} {Phys. Rev. D}\ }\textbf {\bibinfo {volume} {102}},\ \bibinfo
  {pages} {064001} (\bibinfo {year} {2020})},\ \Eprint
  {https://arxiv.org/abs/2001.11412} {arXiv:2001.11412 [gr-qc]} \BibitemShut
  {NoStop}%
\bibitem [{\citenamefont {O'Leary}\ \emph {et~al.}(2009)\citenamefont
  {O'Leary}, \citenamefont {Kocsis},\ and\ \citenamefont
  {Loeb}}]{OLeary:2008myb}%
  \BibitemOpen
  \bibfield  {author} {\bibinfo {author} {\bibfnamefont {R.~M.}\ \bibnamefont
  {O'Leary}}, \bibinfo {author} {\bibfnamefont {B.}~\bibnamefont {Kocsis}},\
  and\ \bibinfo {author} {\bibfnamefont {A.}~\bibnamefont {Loeb}},\ }\bibfield
  {title} {\bibinfo {title} {{Gravitational waves from scattering of
  stellar-mass black holes in galactic nuclei}},\ }\href
  {https://doi.org/10.1111/j.1365-2966.2009.14653.x} {\bibfield  {journal}
  {\bibinfo  {journal} {Mon. Not. Roy. Astron. Soc.}\ }\textbf {\bibinfo
  {volume} {395}},\ \bibinfo {pages} {2127} (\bibinfo {year} {2009})},\ \Eprint
  {https://arxiv.org/abs/0807.2638} {arXiv:0807.2638 [astro-ph]} \BibitemShut
  {NoStop}%
\bibitem [{\citenamefont {Lee}\ \emph {et~al.}(2010)\citenamefont {Lee},
  \citenamefont {Ramirez-Ruiz},\ and\ \citenamefont {van~de Ven}}]{Lee:2009ca}%
  \BibitemOpen
  \bibfield  {author} {\bibinfo {author} {\bibfnamefont {W.~H.}\ \bibnamefont
  {Lee}}, \bibinfo {author} {\bibfnamefont {E.}~\bibnamefont {Ramirez-Ruiz}},\
  and\ \bibinfo {author} {\bibfnamefont {G.}~\bibnamefont {van~de Ven}},\
  }\bibfield  {title} {\bibinfo {title} {{Short gamma-ray bursts from
  dynamically-assembled compact binaries in globular clusters: pathways, rates,
  hydrodynamics and cosmological setting}},\ }\href
  {https://doi.org/10.1088/0004-637X/720/1/953} {\bibfield  {journal} {\bibinfo
   {journal} {Astrophys. J.}\ }\textbf {\bibinfo {volume} {720}},\ \bibinfo
  {pages} {953} (\bibinfo {year} {2010})},\ \Eprint
  {https://arxiv.org/abs/0909.2884} {arXiv:0909.2884 [astro-ph.HE]}
  \BibitemShut {NoStop}%
\bibitem [{\citenamefont {Lee}(1995)}]{Lee:1994nq}%
  \BibitemOpen
  \bibfield  {author} {\bibinfo {author} {\bibfnamefont {H.~M.}\ \bibnamefont
  {Lee}},\ }\bibfield  {title} {\bibinfo {title} {{Evolution of galactic nuclei
  with 10 x solar mass black holes}},\ }\href
  {https://doi.org/10.1093/mnras/272.3.605} {\bibfield  {journal} {\bibinfo
  {journal} {Mon. Not. Roy. Astron. Soc.}\ }\textbf {\bibinfo {volume} {272}},\
  \bibinfo {pages} {605} (\bibinfo {year} {1995})},\ \Eprint
  {https://arxiv.org/abs/astro-ph/9409073} {arXiv:astro-ph/9409073}
  \BibitemShut {NoStop}%
\bibitem [{\citenamefont {Hong}\ and\ \citenamefont
  {Lee}(2015)}]{Hong:2015aba}%
  \BibitemOpen
  \bibfield  {author} {\bibinfo {author} {\bibfnamefont {J.}~\bibnamefont
  {Hong}}\ and\ \bibinfo {author} {\bibfnamefont {H.~M.}\ \bibnamefont {Lee}},\
  }\bibfield  {title} {\bibinfo {title} {{Black Hole Binaries in Galactic
  Nuclei and Gravitational Wave Sources}},\ }\href
  {https://doi.org/10.1093/mnras/stv035} {\bibfield  {journal} {\bibinfo
  {journal} {Mon. Not. Roy. Astron. Soc.}\ }\textbf {\bibinfo {volume} {448}},\
  \bibinfo {pages} {754} (\bibinfo {year} {2015})},\ \Eprint
  {https://arxiv.org/abs/1501.02717} {arXiv:1501.02717 [astro-ph.GA]}
  \BibitemShut {NoStop}%
\bibitem [{\citenamefont {Nishizawa}\ \emph {et~al.}(2016)\citenamefont
  {Nishizawa}, \citenamefont {Berti}, \citenamefont {Klein},\ and\
  \citenamefont {Sesana}}]{Nishizawa:2016jji}%
  \BibitemOpen
  \bibfield  {author} {\bibinfo {author} {\bibfnamefont {A.}~\bibnamefont
  {Nishizawa}}, \bibinfo {author} {\bibfnamefont {E.}~\bibnamefont {Berti}},
  \bibinfo {author} {\bibfnamefont {A.}~\bibnamefont {Klein}},\ and\ \bibinfo
  {author} {\bibfnamefont {A.}~\bibnamefont {Sesana}},\ }\bibfield  {title}
  {\bibinfo {title} {{eLISA eccentricity measurements as tracers of binary
  black hole formation}},\ }\href {https://doi.org/10.1103/PhysRevD.94.064020}
  {\bibfield  {journal} {\bibinfo  {journal} {Phys. Rev. D}\ }\textbf {\bibinfo
  {volume} {94}},\ \bibinfo {pages} {064020} (\bibinfo {year} {2016})},\
  \Eprint {https://arxiv.org/abs/1605.01341} {arXiv:1605.01341 [gr-qc]}
  \BibitemShut {NoStop}%
\bibitem [{\citenamefont {Nishizawa}\ \emph {et~al.}(2017)\citenamefont
  {Nishizawa}, \citenamefont {Sesana}, \citenamefont {Berti},\ and\
  \citenamefont {Klein}}]{Nishizawa:2016eza}%
  \BibitemOpen
  \bibfield  {author} {\bibinfo {author} {\bibfnamefont {A.}~\bibnamefont
  {Nishizawa}}, \bibinfo {author} {\bibfnamefont {A.}~\bibnamefont {Sesana}},
  \bibinfo {author} {\bibfnamefont {E.}~\bibnamefont {Berti}},\ and\ \bibinfo
  {author} {\bibfnamefont {A.}~\bibnamefont {Klein}},\ }\bibfield  {title}
  {\bibinfo {title} {{Constraining stellar binary black hole formation
  scenarios with eLISA eccentricity measurements}},\ }\href
  {https://doi.org/10.1093/mnras/stw2993} {\bibfield  {journal} {\bibinfo
  {journal} {Mon. Not. Roy. Astron. Soc.}\ }\textbf {\bibinfo {volume} {465}},\
  \bibinfo {pages} {4375} (\bibinfo {year} {2017})},\ \Eprint
  {https://arxiv.org/abs/1606.09295} {arXiv:1606.09295 [astro-ph.HE]}
  \BibitemShut {NoStop}%
\bibitem [{\citenamefont {Breivik}\ \emph {et~al.}(2016)\citenamefont
  {Breivik}, \citenamefont {Rodriguez}, \citenamefont {Larson}, \citenamefont
  {Kalogera},\ and\ \citenamefont {Rasio}}]{Breivik:2016ddj}%
  \BibitemOpen
  \bibfield  {author} {\bibinfo {author} {\bibfnamefont {K.}~\bibnamefont
  {Breivik}}, \bibinfo {author} {\bibfnamefont {C.~L.}\ \bibnamefont
  {Rodriguez}}, \bibinfo {author} {\bibfnamefont {S.~L.}\ \bibnamefont
  {Larson}}, \bibinfo {author} {\bibfnamefont {V.}~\bibnamefont {Kalogera}},\
  and\ \bibinfo {author} {\bibfnamefont {F.~A.}\ \bibnamefont {Rasio}},\
  }\bibfield  {title} {\bibinfo {title} {{Distinguishing Between Formation
  Channels for Binary Black Holes with LISA}},\ }\href
  {https://doi.org/10.3847/2041-8205/830/1/L18} {\bibfield  {journal} {\bibinfo
   {journal} {Astrophys. J. Lett.}\ }\textbf {\bibinfo {volume} {830}},\
  \bibinfo {pages} {L18} (\bibinfo {year} {2016})},\ \Eprint
  {https://arxiv.org/abs/1606.09558} {arXiv:1606.09558 [astro-ph.GA]}
  \BibitemShut {NoStop}%
\bibitem [{\citenamefont {Zevin}\ \emph {et~al.}(2021)\citenamefont {Zevin},
  \citenamefont {Romero-Shaw}, \citenamefont {Kremer}, \citenamefont {Thrane},\
  and\ \citenamefont {Lasky}}]{Zevin:2021rtf}%
  \BibitemOpen
  \bibfield  {author} {\bibinfo {author} {\bibfnamefont {M.}~\bibnamefont
  {Zevin}}, \bibinfo {author} {\bibfnamefont {I.~M.}\ \bibnamefont
  {Romero-Shaw}}, \bibinfo {author} {\bibfnamefont {K.}~\bibnamefont {Kremer}},
  \bibinfo {author} {\bibfnamefont {E.}~\bibnamefont {Thrane}},\ and\ \bibinfo
  {author} {\bibfnamefont {P.~D.}\ \bibnamefont {Lasky}},\ }\bibfield  {title}
  {\bibinfo {title} {{Implications of Eccentric Observations on Binary Black
  Hole Formation Channels}},\ }\href {https://doi.org/10.3847/2041-8213/ac32dc}
  {\bibfield  {journal} {\bibinfo  {journal} {Astrophys. J. Lett.}\ }\textbf
  {\bibinfo {volume} {921}},\ \bibinfo {pages} {L43} (\bibinfo {year}
  {2021})},\ \Eprint {https://arxiv.org/abs/2106.09042} {arXiv:2106.09042
  [astro-ph.HE]} \BibitemShut {NoStop}%
\bibitem [{\citenamefont {Silsbee}\ and\ \citenamefont
  {Tremaine}(2017)}]{Silsbee:2016djf}%
  \BibitemOpen
  \bibfield  {author} {\bibinfo {author} {\bibfnamefont {K.}~\bibnamefont
  {Silsbee}}\ and\ \bibinfo {author} {\bibfnamefont {S.}~\bibnamefont
  {Tremaine}},\ }\bibfield  {title} {\bibinfo {title} {{Lidov-Kozai Cycles with
  Gravitational Radiation: Merging Black Holes in Isolated Triple Systems}},\
  }\href {https://doi.org/10.3847/1538-4357/aa5729} {\bibfield  {journal}
  {\bibinfo  {journal} {Astrophys. J.}\ }\textbf {\bibinfo {volume} {836}},\
  \bibinfo {pages} {39} (\bibinfo {year} {2017})},\ \Eprint
  {https://arxiv.org/abs/1608.07642} {arXiv:1608.07642 [astro-ph.HE]}
  \BibitemShut {NoStop}%
\bibitem [{\citenamefont {Antonini}\ \emph {et~al.}(2017)\citenamefont
  {Antonini}, \citenamefont {Toonen},\ and\ \citenamefont
  {Hamers}}]{Antonini:2017ash}%
  \BibitemOpen
  \bibfield  {author} {\bibinfo {author} {\bibfnamefont {F.}~\bibnamefont
  {Antonini}}, \bibinfo {author} {\bibfnamefont {S.}~\bibnamefont {Toonen}},\
  and\ \bibinfo {author} {\bibfnamefont {A.~S.}\ \bibnamefont {Hamers}},\
  }\bibfield  {title} {\bibinfo {title} {{Binary black hole mergers from field
  triples: properties, rates and the impact of stellar evolution}},\ }\href
  {https://doi.org/10.3847/1538-4357/aa6f5e} {\bibfield  {journal} {\bibinfo
  {journal} {Astrophys. J.}\ }\textbf {\bibinfo {volume} {841}},\ \bibinfo
  {pages} {77} (\bibinfo {year} {2017})},\ \Eprint
  {https://arxiv.org/abs/1703.06614} {arXiv:1703.06614 [astro-ph.GA]}
  \BibitemShut {NoStop}%
\bibitem [{\citenamefont {Liu}\ \emph {et~al.}(2019)\citenamefont {Liu},
  \citenamefont {Lai},\ and\ \citenamefont {Wang}}]{Liu:2019gdc}%
  \BibitemOpen
  \bibfield  {author} {\bibinfo {author} {\bibfnamefont {B.}~\bibnamefont
  {Liu}}, \bibinfo {author} {\bibfnamefont {D.}~\bibnamefont {Lai}},\ and\
  \bibinfo {author} {\bibfnamefont {Y.-H.}\ \bibnamefont {Wang}},\ }\bibfield
  {title} {\bibinfo {title} {{Black Hole and Neutron Star Binary Mergers in
  Triple Systems: II. Merger Eccentricity and Spin-Orbit Misalignment}}\ }\href
  {https://doi.org/10.3847/1538-4357/ab2dfb} {10.3847/1538-4357/ab2dfb}
  (\bibinfo {year} {2019}),\ \Eprint {https://arxiv.org/abs/1905.00427}
  {arXiv:1905.00427 [astro-ph.HE]} \BibitemShut {NoStop}%
\bibitem [{\citenamefont {Sun}\ \emph {et~al.}(2015)\citenamefont {Sun},
  \citenamefont {Cao}, \citenamefont {Wang},\ and\ \citenamefont
  {Yeh}}]{Sun:2015bva}%
  \BibitemOpen
  \bibfield  {author} {\bibinfo {author} {\bibfnamefont {B.}~\bibnamefont
  {Sun}}, \bibinfo {author} {\bibfnamefont {Z.}~\bibnamefont {Cao}}, \bibinfo
  {author} {\bibfnamefont {Y.}~\bibnamefont {Wang}},\ and\ \bibinfo {author}
  {\bibfnamefont {H.-C.}\ \bibnamefont {Yeh}},\ }\bibfield  {title} {\bibinfo
  {title} {{Parameter estimation of eccentric inspiraling compact binaries
  using an enhanced post circular model for ground-based detectors}},\ }\href
  {https://doi.org/10.1103/PhysRevD.92.044034} {\bibfield  {journal} {\bibinfo
  {journal} {Phys. Rev. D}\ }\textbf {\bibinfo {volume} {92}},\ \bibinfo
  {pages} {044034} (\bibinfo {year} {2015})}\BibitemShut {NoStop}%
\bibitem [{\citenamefont {Ma}\ \emph {et~al.}(2017)\citenamefont {Ma},
  \citenamefont {Cao}, \citenamefont {Lin}, \citenamefont {Pan},\ and\
  \citenamefont {Yo}}]{Ma:2017bux}%
  \BibitemOpen
  \bibfield  {author} {\bibinfo {author} {\bibfnamefont {S.}~\bibnamefont
  {Ma}}, \bibinfo {author} {\bibfnamefont {Z.}~\bibnamefont {Cao}}, \bibinfo
  {author} {\bibfnamefont {C.-Y.}\ \bibnamefont {Lin}}, \bibinfo {author}
  {\bibfnamefont {H.-P.}\ \bibnamefont {Pan}},\ and\ \bibinfo {author}
  {\bibfnamefont {H.-J.}\ \bibnamefont {Yo}},\ }\bibfield  {title} {\bibinfo
  {title} {{Gravitational wave source localization for eccentric binary
  coalesce with a ground-based detector network}},\ }\href
  {https://doi.org/10.1103/PhysRevD.96.084046} {\bibfield  {journal} {\bibinfo
  {journal} {Phys. Rev. D}\ }\textbf {\bibinfo {volume} {96}},\ \bibinfo
  {pages} {084046} (\bibinfo {year} {2017})},\ \Eprint
  {https://arxiv.org/abs/1710.02965} {arXiv:1710.02965 [gr-qc]} \BibitemShut
  {NoStop}%
\bibitem [{\citenamefont {Pan}\ \emph {et~al.}(2019)\citenamefont {Pan},
  \citenamefont {Lin}, \citenamefont {Cao},\ and\ \citenamefont
  {Yo}}]{Pan:2019anf}%
  \BibitemOpen
  \bibfield  {author} {\bibinfo {author} {\bibfnamefont {H.-P.}\ \bibnamefont
  {Pan}}, \bibinfo {author} {\bibfnamefont {C.-Y.}\ \bibnamefont {Lin}},
  \bibinfo {author} {\bibfnamefont {Z.}~\bibnamefont {Cao}},\ and\ \bibinfo
  {author} {\bibfnamefont {H.-J.}\ \bibnamefont {Yo}},\ }\bibfield  {title}
  {\bibinfo {title} {{Accuracy of source localization for eccentric inspiraling
  binary mergers using a ground-based detector network}},\ }\href
  {https://doi.org/10.1103/PhysRevD.100.124003} {\bibfield  {journal} {\bibinfo
   {journal} {Phys. Rev. D}\ }\textbf {\bibinfo {volume} {100}},\ \bibinfo
  {pages} {124003} (\bibinfo {year} {2019})},\ \Eprint
  {https://arxiv.org/abs/1912.04455} {arXiv:1912.04455 [gr-qc]} \BibitemShut
  {NoStop}%
\bibitem [{\citenamefont {Mikoczi}\ \emph {et~al.}(2012)\citenamefont
  {Mikoczi}, \citenamefont {Kocsis}, \citenamefont {Forgacs},\ and\
  \citenamefont {Vasuth}}]{Mikoczi:2012qy}%
  \BibitemOpen
  \bibfield  {author} {\bibinfo {author} {\bibfnamefont {B.}~\bibnamefont
  {Mikoczi}}, \bibinfo {author} {\bibfnamefont {B.}~\bibnamefont {Kocsis}},
  \bibinfo {author} {\bibfnamefont {P.}~\bibnamefont {Forgacs}},\ and\ \bibinfo
  {author} {\bibfnamefont {M.}~\bibnamefont {Vasuth}},\ }\bibfield  {title}
  {\bibinfo {title} {{Parameter estimation for inspiraling eccentric compact
  binaries including pericenter precession}},\ }\href
  {https://doi.org/10.1103/PhysRevD.86.104027} {\bibfield  {journal} {\bibinfo
  {journal} {Phys. Rev. D}\ }\textbf {\bibinfo {volume} {86}},\ \bibinfo
  {pages} {104027} (\bibinfo {year} {2012})},\ \Eprint
  {https://arxiv.org/abs/1206.5786} {arXiv:1206.5786 [gr-qc]} \BibitemShut
  {NoStop}%
\bibitem [{\citenamefont {Abbott}\ \emph {et~al.}(2019)\citenamefont {Abbott}
  \emph {et~al.}}]{LIGOScientific:2019dag}%
  \BibitemOpen
  \bibfield  {author} {\bibinfo {author} {\bibfnamefont {B.~P.}\ \bibnamefont
  {Abbott}} \emph {et~al.} (\bibinfo {collaboration} {LIGO Scientific,
  Virgo}),\ }\bibfield  {title} {\bibinfo {title} {{Search for Eccentric Binary
  Black Hole Mergers with Advanced LIGO and Advanced Virgo during their First
  and Second Observing Runs}},\ }\href
  {https://doi.org/10.3847/1538-4357/ab3c2d} {\bibfield  {journal} {\bibinfo
  {journal} {Astrophys. J.}\ }\textbf {\bibinfo {volume} {883}},\ \bibinfo
  {pages} {149} (\bibinfo {year} {2019})},\ \Eprint
  {https://arxiv.org/abs/1907.09384} {arXiv:1907.09384 [astro-ph.HE]}
  \BibitemShut {NoStop}%
\bibitem [{\citenamefont {Romero-Shaw}\ \emph {et~al.}(2019)\citenamefont
  {Romero-Shaw}, \citenamefont {Lasky},\ and\ \citenamefont
  {Thrane}}]{Romero-Shaw:2019itr}%
  \BibitemOpen
  \bibfield  {author} {\bibinfo {author} {\bibfnamefont {I.~M.}\ \bibnamefont
  {Romero-Shaw}}, \bibinfo {author} {\bibfnamefont {P.~D.}\ \bibnamefont
  {Lasky}},\ and\ \bibinfo {author} {\bibfnamefont {E.}~\bibnamefont
  {Thrane}},\ }\bibfield  {title} {\bibinfo {title} {{Searching for
  Eccentricity: Signatures of Dynamical Formation in the First
  Gravitational-Wave Transient Catalogue of LIGO and Virgo}},\ }\href
  {https://doi.org/10.1093/mnras/stz2996} {\bibfield  {journal} {\bibinfo
  {journal} {Mon. Not. Roy. Astron. Soc.}\ }\textbf {\bibinfo {volume} {490}},\
  \bibinfo {pages} {5210} (\bibinfo {year} {2019})},\ \Eprint
  {https://arxiv.org/abs/1909.05466} {arXiv:1909.05466 [astro-ph.HE]}
  \BibitemShut {NoStop}%
\bibitem [{\citenamefont {Nitz}\ \emph {et~al.}(2019)\citenamefont {Nitz},
  \citenamefont {Lenon},\ and\ \citenamefont {Brown}}]{Nitz:2019spj}%
  \BibitemOpen
  \bibfield  {author} {\bibinfo {author} {\bibfnamefont {A.~H.}\ \bibnamefont
  {Nitz}}, \bibinfo {author} {\bibfnamefont {A.}~\bibnamefont {Lenon}},\ and\
  \bibinfo {author} {\bibfnamefont {D.~A.}\ \bibnamefont {Brown}},\ }\bibfield
  {title} {\bibinfo {title} {{Search for Eccentric Binary Neutron Star Mergers
  in the first and second observing runs of Advanced LIGO}},\ }\href
  {https://doi.org/10.3847/1538-4357/ab6611} {\bibfield  {journal} {\bibinfo
  {journal} {Astrophys. J.}\ }\textbf {\bibinfo {volume} {890}},\ \bibinfo
  {pages} {1} (\bibinfo {year} {2019})},\ \Eprint
  {https://arxiv.org/abs/1912.05464} {arXiv:1912.05464 [astro-ph.HE]}
  \BibitemShut {NoStop}%
\bibitem [{\citenamefont {Wu}\ \emph {et~al.}(2020)\citenamefont {Wu},
  \citenamefont {Cao},\ and\ \citenamefont {Zhu}}]{Wu:2020zwr}%
  \BibitemOpen
  \bibfield  {author} {\bibinfo {author} {\bibfnamefont {S.}~\bibnamefont
  {Wu}}, \bibinfo {author} {\bibfnamefont {Z.}~\bibnamefont {Cao}},\ and\
  \bibinfo {author} {\bibfnamefont {Z.-H.}\ \bibnamefont {Zhu}},\ }\bibfield
  {title} {\bibinfo {title} {{Measuring the eccentricity of binary black holes
  in GWTC-1 by using the inspiral-only waveform}},\ }\href
  {https://doi.org/10.1093/mnras/staa1176} {\bibfield  {journal} {\bibinfo
  {journal} {Mon. Not. Roy. Astron. Soc.}\ }\textbf {\bibinfo {volume} {495}},\
  \bibinfo {pages} {466} (\bibinfo {year} {2020})},\ \Eprint
  {https://arxiv.org/abs/2002.05528} {arXiv:2002.05528 [astro-ph.IM]}
  \BibitemShut {NoStop}%
\bibitem [{\citenamefont {Romero-Shaw}\ \emph {et~al.}(2020)\citenamefont
  {Romero-Shaw}, \citenamefont {Lasky}, \citenamefont {Thrane},\ and\
  \citenamefont {Bustillo}}]{Romero-Shaw:2020thy}%
  \BibitemOpen
  \bibfield  {author} {\bibinfo {author} {\bibfnamefont {I.~M.}\ \bibnamefont
  {Romero-Shaw}}, \bibinfo {author} {\bibfnamefont {P.~D.}\ \bibnamefont
  {Lasky}}, \bibinfo {author} {\bibfnamefont {E.}~\bibnamefont {Thrane}},\ and\
  \bibinfo {author} {\bibfnamefont {J.~C.}\ \bibnamefont {Bustillo}},\
  }\bibfield  {title} {\bibinfo {title} {{GW190521: orbital eccentricity and
  signatures of dynamical formation in a binary black hole merger signal}},\
  }\href {https://doi.org/10.3847/2041-8213/abbe26} {\bibfield  {journal}
  {\bibinfo  {journal} {Astrophys. J. Lett.}\ }\textbf {\bibinfo {volume}
  {903}},\ \bibinfo {pages} {L5} (\bibinfo {year} {2020})},\ \Eprint
  {https://arxiv.org/abs/2009.04771} {arXiv:2009.04771 [astro-ph.HE]}
  \BibitemShut {NoStop}%
\bibitem [{\citenamefont {Gayathri}\ \emph {et~al.}(2022)\citenamefont
  {Gayathri}, \citenamefont {Healy}, \citenamefont {Lange}, \citenamefont
  {O'Brien}, \citenamefont {Szczepanczyk}, \citenamefont {Bartos},
  \citenamefont {Campanelli}, \citenamefont {Klimenko}, \citenamefont
  {Lousto},\ and\ \citenamefont {O'Shaughnessy}}]{Gayathri:2020coq}%
  \BibitemOpen
  \bibfield  {author} {\bibinfo {author} {\bibfnamefont {V.}~\bibnamefont
  {Gayathri}}, \bibinfo {author} {\bibfnamefont {J.}~\bibnamefont {Healy}},
  \bibinfo {author} {\bibfnamefont {J.}~\bibnamefont {Lange}}, \bibinfo
  {author} {\bibfnamefont {B.}~\bibnamefont {O'Brien}}, \bibinfo {author}
  {\bibfnamefont {M.}~\bibnamefont {Szczepanczyk}}, \bibinfo {author}
  {\bibfnamefont {I.}~\bibnamefont {Bartos}}, \bibinfo {author} {\bibfnamefont
  {M.}~\bibnamefont {Campanelli}}, \bibinfo {author} {\bibfnamefont
  {S.}~\bibnamefont {Klimenko}}, \bibinfo {author} {\bibfnamefont {C.~O.}\
  \bibnamefont {Lousto}},\ and\ \bibinfo {author} {\bibfnamefont
  {R.}~\bibnamefont {O'Shaughnessy}},\ }\bibfield  {title} {\bibinfo {title}
  {{Eccentricity estimate for black hole mergers with numerical relativity
  simulations}},\ }\href {https://doi.org/10.1038/s41550-021-01568-w}
  {\bibfield  {journal} {\bibinfo  {journal} {Nature Astron.}\ }\textbf
  {\bibinfo {volume} {6}},\ \bibinfo {pages} {344} (\bibinfo {year} {2022})},\
  \Eprint {https://arxiv.org/abs/2009.05461} {arXiv:2009.05461 [astro-ph.HE]}
  \BibitemShut {NoStop}%
\bibitem [{\citenamefont {Kawamura}\ \emph {et~al.}(2006)\citenamefont
  {Kawamura} \emph {et~al.}}]{Kawamura:2006up}%
  \BibitemOpen
  \bibfield  {author} {\bibinfo {author} {\bibfnamefont {S.}~\bibnamefont
  {Kawamura}} \emph {et~al.},\ }\bibfield  {title} {\bibinfo {title} {{The
  Japanese space gravitational wave antenna DECIGO}},\ }\href
  {https://doi.org/10.1088/0264-9381/23/8/S17} {\bibfield  {journal} {\bibinfo
  {journal} {Class. Quant. Grav.}\ }\textbf {\bibinfo {volume} {23}},\ \bibinfo
  {pages} {S125} (\bibinfo {year} {2006})}\BibitemShut {NoStop}%
\bibitem [{\citenamefont {Kawamura}\ \emph {et~al.}(2021)\citenamefont
  {Kawamura} \emph {et~al.}}]{Kawamura:2020pcg}%
  \BibitemOpen
  \bibfield  {author} {\bibinfo {author} {\bibfnamefont {S.}~\bibnamefont
  {Kawamura}} \emph {et~al.},\ }\bibfield  {title} {\bibinfo {title} {{Current
  status of space gravitational wave antenna DECIGO and B-DECIGO}},\ }\href
  {https://doi.org/10.1093/ptep/ptab019} {\bibfield  {journal} {\bibinfo
  {journal} {PTEP}\ }\textbf {\bibinfo {volume} {2021}},\ \bibinfo {pages}
  {05A105} (\bibinfo {year} {2021})},\ \Eprint
  {https://arxiv.org/abs/2006.13545} {arXiv:2006.13545 [gr-qc]} \BibitemShut
  {NoStop}%
\bibitem [{\citenamefont {Harry}\ \emph {et~al.}(2006)\citenamefont {Harry},
  \citenamefont {Fritschel}, \citenamefont {Shaddock}, \citenamefont
  {Folkner},\ and\ \citenamefont {Phinney}}]{Harry:2006fi}%
  \BibitemOpen
  \bibfield  {author} {\bibinfo {author} {\bibfnamefont {G.~M.}\ \bibnamefont
  {Harry}}, \bibinfo {author} {\bibfnamefont {P.}~\bibnamefont {Fritschel}},
  \bibinfo {author} {\bibfnamefont {D.~A.}\ \bibnamefont {Shaddock}}, \bibinfo
  {author} {\bibfnamefont {W.}~\bibnamefont {Folkner}},\ and\ \bibinfo {author}
  {\bibfnamefont {E.~S.}\ \bibnamefont {Phinney}},\ }\bibfield  {title}
  {\bibinfo {title} {{Laser interferometry for the big bang observer}},\ }\href
  {https://doi.org/10.1088/0264-9381/23/15/008} {\bibfield  {journal} {\bibinfo
   {journal} {Class. Quant. Grav.}\ }\textbf {\bibinfo {volume} {23}},\
  \bibinfo {pages} {4887} (\bibinfo {year} {2006})},\ \bibinfo {note}
  {[Erratum: Class.Quant.Grav. 23, 7361 (2006)]}\BibitemShut {NoStop}%
\bibitem [{\citenamefont {Graham}\ \emph {et~al.}(2017)\citenamefont {Graham},
  \citenamefont {Hogan}, \citenamefont {Kasevich}, \citenamefont {Rajendran},\
  and\ \citenamefont {Romani}}]{Graham:2017pmn}%
  \BibitemOpen
  \bibfield  {author} {\bibinfo {author} {\bibfnamefont {P.~W.}\ \bibnamefont
  {Graham}}, \bibinfo {author} {\bibfnamefont {J.~M.}\ \bibnamefont {Hogan}},
  \bibinfo {author} {\bibfnamefont {M.~A.}\ \bibnamefont {Kasevich}}, \bibinfo
  {author} {\bibfnamefont {S.}~\bibnamefont {Rajendran}},\ and\ \bibinfo
  {author} {\bibfnamefont {R.~W.}\ \bibnamefont {Romani}} (\bibinfo
  {collaboration} {MAGIS}),\ }\bibfield  {title} {\bibinfo {title} {{Mid-band
  gravitational wave detection with precision atomic sensors}},\ }\href@noop {}
  {\  (\bibinfo {year} {2017})},\ \Eprint {https://arxiv.org/abs/1711.02225}
  {arXiv:1711.02225 [astro-ph.IM]} \BibitemShut {NoStop}%
\bibitem [{\citenamefont {El-Neaj}\ \emph {et~al.}(2020)\citenamefont {El-Neaj}
  \emph {et~al.}}]{AEDGE:2019nxb}%
  \BibitemOpen
  \bibfield  {author} {\bibinfo {author} {\bibfnamefont {Y.~A.}\ \bibnamefont
  {El-Neaj}} \emph {et~al.} (\bibinfo {collaboration} {AEDGE}),\ }\bibfield
  {title} {\bibinfo {title} {{AEDGE: Atomic Experiment for Dark Matter and
  Gravity Exploration in Space}},\ }\href
  {https://doi.org/10.1140/epjqt/s40507-020-0080-0} {\bibfield  {journal}
  {\bibinfo  {journal} {EPJ Quant. Technol.}\ }\textbf {\bibinfo {volume}
  {7}},\ \bibinfo {pages} {6} (\bibinfo {year} {2020})},\ \Eprint
  {https://arxiv.org/abs/1908.00802} {arXiv:1908.00802 [gr-qc]} \BibitemShut
  {NoStop}%
\bibitem [{\citenamefont {Graham}\ and\ \citenamefont
  {Jung}(2018)}]{Graham:2017lmg}%
  \BibitemOpen
  \bibfield  {author} {\bibinfo {author} {\bibfnamefont {P.~W.}\ \bibnamefont
  {Graham}}\ and\ \bibinfo {author} {\bibfnamefont {S.}~\bibnamefont {Jung}},\
  }\bibfield  {title} {\bibinfo {title} {{Localizing Gravitational Wave Sources
  with Single-Baseline Atom Interferometers}},\ }\href
  {https://doi.org/10.1103/PhysRevD.97.024052} {\bibfield  {journal} {\bibinfo
  {journal} {Phys. Rev. D}\ }\textbf {\bibinfo {volume} {97}},\ \bibinfo
  {pages} {024052} (\bibinfo {year} {2018})},\ \Eprint
  {https://arxiv.org/abs/1710.03269} {arXiv:1710.03269 [gr-qc]} \BibitemShut
  {NoStop}%
\bibitem [{\citenamefont {Yang}\ \emph
  {et~al.}(2022{\natexlab{b}})\citenamefont {Yang}, \citenamefont {Lee},
  \citenamefont {Cai}, \citenamefont {Choi},\ and\ \citenamefont
  {Jung}}]{Yang:2021xox}%
  \BibitemOpen
  \bibfield  {author} {\bibinfo {author} {\bibfnamefont {T.}~\bibnamefont
  {Yang}}, \bibinfo {author} {\bibfnamefont {H.~M.}\ \bibnamefont {Lee}},
  \bibinfo {author} {\bibfnamefont {R.-G.}\ \bibnamefont {Cai}}, \bibinfo
  {author} {\bibfnamefont {H.~G.}\ \bibnamefont {Choi}},\ and\ \bibinfo
  {author} {\bibfnamefont {S.}~\bibnamefont {Jung}},\ }\bibfield  {title}
  {\bibinfo {title} {{Space-borne atom interferometric gravitational wave
  detections. Part II. Dark sirens and finding the one}},\ }\href
  {https://doi.org/10.1088/1475-7516/2022/01/042} {\bibfield  {journal}
  {\bibinfo  {journal} {JCAP}\ }\textbf {\bibinfo {volume} {01}}\bibfield
  {number} {\bibinfo  {number} { (01)},\ \bibinfo {pages} {042}},\ }\Eprint
  {https://arxiv.org/abs/2110.09967} {arXiv:2110.09967 [gr-qc]} \BibitemShut
  {NoStop}%
\bibitem [{\citenamefont {Liu}\ \emph {et~al.}(2022{\natexlab{a}})\citenamefont
  {Liu}, \citenamefont {Liu}, \citenamefont {Hu}, \citenamefont {Shao},\ and\
  \citenamefont {Kang}}]{Liu:2022rvk}%
  \BibitemOpen
  \bibfield  {author} {\bibinfo {author} {\bibfnamefont {M.}~\bibnamefont
  {Liu}}, \bibinfo {author} {\bibfnamefont {C.}~\bibnamefont {Liu}}, \bibinfo
  {author} {\bibfnamefont {Y.-M.}\ \bibnamefont {Hu}}, \bibinfo {author}
  {\bibfnamefont {L.}~\bibnamefont {Shao}},\ and\ \bibinfo {author}
  {\bibfnamefont {Y.}~\bibnamefont {Kang}},\ }\bibfield  {title} {\bibinfo
  {title} {{Dark-siren Cosmology with Decihertz Gravitational-wave
  Detectors}},\ }\href@noop {} {\  (\bibinfo {year} {2022}{\natexlab{a}})},\
  \Eprint {https://arxiv.org/abs/2205.06991} {arXiv:2205.06991 [astro-ph.CO]}
  \BibitemShut {NoStop}%
\bibitem [{\citenamefont {Yang}\ \emph
  {et~al.}(2022{\natexlab{c}})\citenamefont {Yang}, \citenamefont {Cai},\ and\
  \citenamefont {Lee}}]{Yang:2022iwn}%
  \BibitemOpen
  \bibfield  {author} {\bibinfo {author} {\bibfnamefont {T.}~\bibnamefont
  {Yang}}, \bibinfo {author} {\bibfnamefont {R.-G.}\ \bibnamefont {Cai}},\ and\
  \bibinfo {author} {\bibfnamefont {H.~M.}\ \bibnamefont {Lee}},\ }\bibfield
  {title} {\bibinfo {title} {{Space-borne atom interferometric gravitational
  wave detections. Part III. Eccentricity on dark sirens}},\ }\href@noop {} {\
  (\bibinfo {year} {2022}{\natexlab{c}})},\ \Eprint
  {https://arxiv.org/abs/2208.10998} {arXiv:2208.10998 [gr-qc]} \BibitemShut
  {NoStop}%
\bibitem [{\citenamefont {{LIGO Scientific Collaboration}}(2018)}]{lalsuite}%
  \BibitemOpen
  \bibfield  {author} {\bibinfo {author} {\bibnamefont {{LIGO Scientific
  Collaboration}}},\ }\href {https://doi.org/10.7935/GT1W-FZ16} {\bibinfo
  {title} {{LIGO} {A}lgorithm {L}ibrary - {LALS}uite}},\ \bibinfo
  {howpublished} {free software (GPL)} (\bibinfo {year} {2018})\BibitemShut
  {NoStop}%
\bibitem [{\citenamefont {Nitz}\ \emph {et~al.}(2022)\citenamefont {Nitz},
  \citenamefont {Harry}, \citenamefont {Brown}, \citenamefont {Biwer},
  \citenamefont {Willis}, \citenamefont {Canton}, \citenamefont {Capano},
  \citenamefont {Dent}, \citenamefont {Pekowsky}, \citenamefont {Williamson},
  \citenamefont {De}, \citenamefont {Cabero}, \citenamefont {Machenschalk},
  \citenamefont {Macleod}, \citenamefont {Kumar}, \citenamefont {Reyes},
  \citenamefont {dfinstad}, \citenamefont {Pannarale}, \citenamefont {Kumar},
  \citenamefont {Massinger}, \citenamefont {Tápai}, \citenamefont {Singer},
  \citenamefont {Davies}, \citenamefont {Khan}, \citenamefont {Fairhurst},
  \citenamefont {Nielsen}, \citenamefont {Singh}, \citenamefont {Chandra},
  \citenamefont {shasvath},\ and\ \citenamefont {veronica
  villa}}]{alex_nitz_2022_6324278}%
  \BibitemOpen
  \bibfield  {author} {\bibinfo {author} {\bibfnamefont {A.}~\bibnamefont
  {Nitz}}, \bibinfo {author} {\bibfnamefont {I.}~\bibnamefont {Harry}},
  \bibinfo {author} {\bibfnamefont {D.}~\bibnamefont {Brown}}, \bibinfo
  {author} {\bibfnamefont {C.~M.}\ \bibnamefont {Biwer}}, \bibinfo {author}
  {\bibfnamefont {J.}~\bibnamefont {Willis}}, \bibinfo {author} {\bibfnamefont
  {T.~D.}\ \bibnamefont {Canton}}, \bibinfo {author} {\bibfnamefont
  {C.}~\bibnamefont {Capano}}, \bibinfo {author} {\bibfnamefont
  {T.}~\bibnamefont {Dent}}, \bibinfo {author} {\bibfnamefont {L.}~\bibnamefont
  {Pekowsky}}, \bibinfo {author} {\bibfnamefont {A.~R.}\ \bibnamefont
  {Williamson}}, \bibinfo {author} {\bibfnamefont {S.}~\bibnamefont {De}},
  \bibinfo {author} {\bibfnamefont {M.}~\bibnamefont {Cabero}}, \bibinfo
  {author} {\bibfnamefont {B.}~\bibnamefont {Machenschalk}}, \bibinfo {author}
  {\bibfnamefont {D.}~\bibnamefont {Macleod}}, \bibinfo {author} {\bibfnamefont
  {P.}~\bibnamefont {Kumar}}, \bibinfo {author} {\bibfnamefont
  {S.}~\bibnamefont {Reyes}}, \bibinfo {author} {\bibnamefont {dfinstad}},
  \bibinfo {author} {\bibfnamefont {F.}~\bibnamefont {Pannarale}}, \bibinfo
  {author} {\bibfnamefont {S.}~\bibnamefont {Kumar}}, \bibinfo {author}
  {\bibfnamefont {T.}~\bibnamefont {Massinger}}, \bibinfo {author}
  {\bibfnamefont {M.}~\bibnamefont {Tápai}}, \bibinfo {author} {\bibfnamefont
  {L.}~\bibnamefont {Singer}}, \bibinfo {author} {\bibfnamefont {G.~S.~C.}\
  \bibnamefont {Davies}}, \bibinfo {author} {\bibfnamefont {S.}~\bibnamefont
  {Khan}}, \bibinfo {author} {\bibfnamefont {S.}~\bibnamefont {Fairhurst}},
  \bibinfo {author} {\bibfnamefont {A.}~\bibnamefont {Nielsen}}, \bibinfo
  {author} {\bibfnamefont {S.}~\bibnamefont {Singh}}, \bibinfo {author}
  {\bibfnamefont {K.}~\bibnamefont {Chandra}}, \bibinfo {author} {\bibnamefont
  {shasvath}},\ and\ \bibinfo {author} {\bibnamefont {veronica villa}},\ }\href
  {https://doi.org/10.5281/zenodo.6324278} {\bibinfo {title} {gwastro/pycbc:
  v2.0.2 release of pycbc}} (\bibinfo {year} {2022})\BibitemShut {NoStop}%
\bibitem [{\citenamefont {Huerta}\ \emph {et~al.}(2014)\citenamefont {Huerta},
  \citenamefont {Kumar}, \citenamefont {McWilliams}, \citenamefont
  {O'Shaughnessy},\ and\ \citenamefont {Yunes}}]{Huerta:2014eca}%
  \BibitemOpen
  \bibfield  {author} {\bibinfo {author} {\bibfnamefont {E.~A.}\ \bibnamefont
  {Huerta}}, \bibinfo {author} {\bibfnamefont {P.}~\bibnamefont {Kumar}},
  \bibinfo {author} {\bibfnamefont {S.~T.}\ \bibnamefont {McWilliams}},
  \bibinfo {author} {\bibfnamefont {R.}~\bibnamefont {O'Shaughnessy}},\ and\
  \bibinfo {author} {\bibfnamefont {N.}~\bibnamefont {Yunes}},\ }\bibfield
  {title} {\bibinfo {title} {{Accurate and efficient waveforms for compact
  binaries on eccentric orbits}},\ }\href
  {https://doi.org/10.1103/PhysRevD.90.084016} {\bibfield  {journal} {\bibinfo
  {journal} {Phys. Rev. D}\ }\textbf {\bibinfo {volume} {90}},\ \bibinfo
  {pages} {084016} (\bibinfo {year} {2014})},\ \Eprint
  {https://arxiv.org/abs/1408.3406} {arXiv:1408.3406 [gr-qc]} \BibitemShut
  {NoStop}%
\bibitem [{\citenamefont {Buonanno}\ \emph {et~al.}(2009)\citenamefont
  {Buonanno}, \citenamefont {Iyer}, \citenamefont {Ochsner}, \citenamefont
  {Pan},\ and\ \citenamefont {Sathyaprakash}}]{Buonanno:2009zt}%
  \BibitemOpen
  \bibfield  {author} {\bibinfo {author} {\bibfnamefont {A.}~\bibnamefont
  {Buonanno}}, \bibinfo {author} {\bibfnamefont {B.}~\bibnamefont {Iyer}},
  \bibinfo {author} {\bibfnamefont {E.}~\bibnamefont {Ochsner}}, \bibinfo
  {author} {\bibfnamefont {Y.}~\bibnamefont {Pan}},\ and\ \bibinfo {author}
  {\bibfnamefont {B.~S.}\ \bibnamefont {Sathyaprakash}},\ }\bibfield  {title}
  {\bibinfo {title} {{Comparison of post-Newtonian templates for compact binary
  inspiral signals in gravitational-wave detectors}},\ }\href
  {https://doi.org/10.1103/PhysRevD.80.084043} {\bibfield  {journal} {\bibinfo
  {journal} {Phys. Rev. D}\ }\textbf {\bibinfo {volume} {80}},\ \bibinfo
  {pages} {084043} (\bibinfo {year} {2009})},\ \Eprint
  {https://arxiv.org/abs/0907.0700} {arXiv:0907.0700 [gr-qc]} \BibitemShut
  {NoStop}%
\bibitem [{\citenamefont {Yunes}\ \emph {et~al.}(2009)\citenamefont {Yunes},
  \citenamefont {Arun}, \citenamefont {Berti},\ and\ \citenamefont
  {Will}}]{Yunes:2009yz}%
  \BibitemOpen
  \bibfield  {author} {\bibinfo {author} {\bibfnamefont {N.}~\bibnamefont
  {Yunes}}, \bibinfo {author} {\bibfnamefont {K.~G.}\ \bibnamefont {Arun}},
  \bibinfo {author} {\bibfnamefont {E.}~\bibnamefont {Berti}},\ and\ \bibinfo
  {author} {\bibfnamefont {C.~M.}\ \bibnamefont {Will}},\ }\bibfield  {title}
  {\bibinfo {title} {{Post-Circular Expansion of Eccentric Binary Inspirals:
  Fourier-Domain Waveforms in the Stationary Phase Approximation}},\ }\href
  {https://doi.org/10.1103/PhysRevD.80.084001} {\bibfield  {journal} {\bibinfo
  {journal} {Phys. Rev. D}\ }\textbf {\bibinfo {volume} {80}},\ \bibinfo
  {pages} {084001} (\bibinfo {year} {2009})},\ \bibinfo {note} {[Erratum:
  Phys.Rev.D 89, 109901 (2014)]},\ \Eprint {https://arxiv.org/abs/0906.0313}
  {arXiv:0906.0313 [gr-qc]} \BibitemShut {NoStop}%
\bibitem [{\citenamefont {Rubbo}\ \emph {et~al.}(2004)\citenamefont {Rubbo},
  \citenamefont {Cornish},\ and\ \citenamefont {Poujade}}]{Rubbo:2003ap}%
  \BibitemOpen
  \bibfield  {author} {\bibinfo {author} {\bibfnamefont {L.~J.}\ \bibnamefont
  {Rubbo}}, \bibinfo {author} {\bibfnamefont {N.~J.}\ \bibnamefont {Cornish}},\
  and\ \bibinfo {author} {\bibfnamefont {O.}~\bibnamefont {Poujade}},\
  }\bibfield  {title} {\bibinfo {title} {{Forward modeling of space borne
  gravitational wave detectors}},\ }\href
  {https://doi.org/10.1103/PhysRevD.69.082003} {\bibfield  {journal} {\bibinfo
  {journal} {Phys. Rev. D}\ }\textbf {\bibinfo {volume} {69}},\ \bibinfo
  {pages} {082003} (\bibinfo {year} {2004})},\ \Eprint
  {https://arxiv.org/abs/gr-qc/0311069} {arXiv:gr-qc/0311069} \BibitemShut
  {NoStop}%
\bibitem [{\citenamefont {Cutler}\ and\ \citenamefont
  {Flanagan}(1994)}]{Cutler:1994ys}%
  \BibitemOpen
  \bibfield  {author} {\bibinfo {author} {\bibfnamefont {C.}~\bibnamefont
  {Cutler}}\ and\ \bibinfo {author} {\bibfnamefont {E.~E.}\ \bibnamefont
  {Flanagan}},\ }\bibfield  {title} {\bibinfo {title} {{Gravitational waves
  from merging compact binaries: How accurately can one extract the binary's
  parameters from the inspiral wave form?}},\ }\href
  {https://doi.org/10.1103/PhysRevD.49.2658} {\bibfield  {journal} {\bibinfo
  {journal} {Phys. Rev. D}\ }\textbf {\bibinfo {volume} {49}},\ \bibinfo
  {pages} {2658} (\bibinfo {year} {1994})},\ \Eprint
  {https://arxiv.org/abs/gr-qc/9402014} {arXiv:gr-qc/9402014} \BibitemShut
  {NoStop}%
\bibitem [{\citenamefont {Yagi}\ and\ \citenamefont
  {Seto}(2011)}]{Yagi:2011wg}%
  \BibitemOpen
  \bibfield  {author} {\bibinfo {author} {\bibfnamefont {K.}~\bibnamefont
  {Yagi}}\ and\ \bibinfo {author} {\bibfnamefont {N.}~\bibnamefont {Seto}},\
  }\bibfield  {title} {\bibinfo {title} {{Detector configuration of DECIGO/BBO
  and identification of cosmological neutron-star binaries}},\ }\href
  {https://doi.org/10.1103/PhysRevD.83.044011} {\bibfield  {journal} {\bibinfo
  {journal} {Phys. Rev. D}\ }\textbf {\bibinfo {volume} {83}},\ \bibinfo
  {pages} {044011} (\bibinfo {year} {2011})},\ \bibinfo {note} {[Erratum:
  Phys.Rev.D 95, 109901 (2017)]},\ \Eprint {https://arxiv.org/abs/1101.3940}
  {arXiv:1101.3940 [astro-ph.CO]} \BibitemShut {NoStop}%
\bibitem [{\citenamefont {Cutler}(1998)}]{Cutler:1997ta}%
  \BibitemOpen
  \bibfield  {author} {\bibinfo {author} {\bibfnamefont {C.}~\bibnamefont
  {Cutler}},\ }\bibfield  {title} {\bibinfo {title} {{Angular resolution of the
  LISA gravitational wave detector}},\ }\href
  {https://doi.org/10.1103/PhysRevD.57.7089} {\bibfield  {journal} {\bibinfo
  {journal} {Phys. Rev. D}\ }\textbf {\bibinfo {volume} {57}},\ \bibinfo
  {pages} {7089} (\bibinfo {year} {1998})},\ \Eprint
  {https://arxiv.org/abs/gr-qc/9703068} {arXiv:gr-qc/9703068} \BibitemShut
  {NoStop}%
\bibitem [{\citenamefont {Lower}\ \emph {et~al.}(2018)\citenamefont {Lower},
  \citenamefont {Thrane}, \citenamefont {Lasky},\ and\ \citenamefont
  {Smith}}]{Lower:2018seu}%
  \BibitemOpen
  \bibfield  {author} {\bibinfo {author} {\bibfnamefont {M.~E.}\ \bibnamefont
  {Lower}}, \bibinfo {author} {\bibfnamefont {E.}~\bibnamefont {Thrane}},
  \bibinfo {author} {\bibfnamefont {P.~D.}\ \bibnamefont {Lasky}},\ and\
  \bibinfo {author} {\bibfnamefont {R.}~\bibnamefont {Smith}},\ }\bibfield
  {title} {\bibinfo {title} {{Measuring eccentricity in binary black hole
  inspirals with gravitational waves}},\ }\href
  {https://doi.org/10.1103/PhysRevD.98.083028} {\bibfield  {journal} {\bibinfo
  {journal} {Phys. Rev. D}\ }\textbf {\bibinfo {volume} {98}},\ \bibinfo
  {pages} {083028} (\bibinfo {year} {2018})},\ \Eprint
  {https://arxiv.org/abs/1806.05350} {arXiv:1806.05350 [astro-ph.HE]}
  \BibitemShut {NoStop}%
\bibitem [{\citenamefont {Yu}\ \emph {et~al.}(2020)\citenamefont {Yu},
  \citenamefont {Wang}, \citenamefont {Zhao},\ and\ \citenamefont
  {Lu}}]{Yu:2020vyy}%
  \BibitemOpen
  \bibfield  {author} {\bibinfo {author} {\bibfnamefont {J.}~\bibnamefont
  {Yu}}, \bibinfo {author} {\bibfnamefont {Y.}~\bibnamefont {Wang}}, \bibinfo
  {author} {\bibfnamefont {W.}~\bibnamefont {Zhao}},\ and\ \bibinfo {author}
  {\bibfnamefont {Y.}~\bibnamefont {Lu}},\ }\bibfield  {title} {\bibinfo
  {title} {{Hunting for the host galaxy groups of binary black holes and the
  application in constraining Hubble constant}},\ }\href
  {https://doi.org/10.1093/mnras/staa2465} {\bibfield  {journal} {\bibinfo
  {journal} {Mon. Not. Roy. Astron. Soc.}\ }\textbf {\bibinfo {volume} {498}},\
  \bibinfo {pages} {1786} (\bibinfo {year} {2020})},\ \Eprint
  {https://arxiv.org/abs/2003.06586} {arXiv:2003.06586 [astro-ph.CO]}
  \BibitemShut {NoStop}%
\bibitem [{\citenamefont {Chen}\ and\ \citenamefont
  {Holz}(2016)}]{Chen:2016tys}%
  \BibitemOpen
  \bibfield  {author} {\bibinfo {author} {\bibfnamefont {H.-Y.}\ \bibnamefont
  {Chen}}\ and\ \bibinfo {author} {\bibfnamefont {D.~E.}\ \bibnamefont
  {Holz}},\ }\bibfield  {title} {\bibinfo {title} {{Finding the One:
  Identifying the Host Galaxies of Gravitational-Wave Sources}},\ }\href@noop
  {} {\  (\bibinfo {year} {2016})},\ \Eprint {https://arxiv.org/abs/1612.01471}
  {arXiv:1612.01471 [astro-ph.HE]} \BibitemShut {NoStop}%
\bibitem [{\citenamefont {Usman}\ \emph {et~al.}(2019)\citenamefont {Usman},
  \citenamefont {Mills},\ and\ \citenamefont {Fairhurst}}]{Usman:2018imj}%
  \BibitemOpen
  \bibfield  {author} {\bibinfo {author} {\bibfnamefont {S.~A.}\ \bibnamefont
  {Usman}}, \bibinfo {author} {\bibfnamefont {J.~C.}\ \bibnamefont {Mills}},\
  and\ \bibinfo {author} {\bibfnamefont {S.}~\bibnamefont {Fairhurst}},\
  }\bibfield  {title} {\bibinfo {title} {{Constraining the Inclinations of
  Binary Mergers from Gravitational-wave Observations}},\ }\href
  {https://doi.org/10.3847/1538-4357/ab0b3e} {\bibfield  {journal} {\bibinfo
  {journal} {Astrophys. J.}\ }\textbf {\bibinfo {volume} {877}},\ \bibinfo
  {pages} {82} (\bibinfo {year} {2019})},\ \Eprint
  {https://arxiv.org/abs/1809.10727} {arXiv:1809.10727 [gr-qc]} \BibitemShut
  {NoStop}%
\bibitem [{\citenamefont {Vitale}\ \emph {et~al.}(2019)\citenamefont {Vitale},
  \citenamefont {Farr}, \citenamefont {Ng},\ and\ \citenamefont
  {Rodriguez}}]{Vitale:2018yhm}%
  \BibitemOpen
  \bibfield  {author} {\bibinfo {author} {\bibfnamefont {S.}~\bibnamefont
  {Vitale}}, \bibinfo {author} {\bibfnamefont {W.~M.}\ \bibnamefont {Farr}},
  \bibinfo {author} {\bibfnamefont {K.}~\bibnamefont {Ng}},\ and\ \bibinfo
  {author} {\bibfnamefont {C.~L.}\ \bibnamefont {Rodriguez}},\ }\bibfield
  {title} {\bibinfo {title} {{Measuring the star formation rate with
  gravitational waves from binary black holes}},\ }\href
  {https://doi.org/10.3847/2041-8213/ab50c0} {\bibfield  {journal} {\bibinfo
  {journal} {Astrophys. J. Lett.}\ }\textbf {\bibinfo {volume} {886}},\
  \bibinfo {pages} {L1} (\bibinfo {year} {2019})},\ \Eprint
  {https://arxiv.org/abs/1808.00901} {arXiv:1808.00901 [astro-ph.HE]}
  \BibitemShut {NoStop}%
\bibitem [{\citenamefont {Madau}\ and\ \citenamefont
  {Dickinson}(2014)}]{Madau:2014bja}%
  \BibitemOpen
  \bibfield  {author} {\bibinfo {author} {\bibfnamefont {P.}~\bibnamefont
  {Madau}}\ and\ \bibinfo {author} {\bibfnamefont {M.}~\bibnamefont
  {Dickinson}},\ }\bibfield  {title} {\bibinfo {title} {{Cosmic Star Formation
  History}},\ }\href {https://doi.org/10.1146/annurev-astro-081811-125615}
  {\bibfield  {journal} {\bibinfo  {journal} {Ann. Rev. Astron. Astrophys.}\
  }\textbf {\bibinfo {volume} {52}},\ \bibinfo {pages} {415} (\bibinfo {year}
  {2014})},\ \Eprint {https://arxiv.org/abs/1403.0007} {arXiv:1403.0007
  [astro-ph.CO]} \BibitemShut {NoStop}%
\bibitem [{\citenamefont {Kowalska}\ \emph {et~al.}(2011)\citenamefont
  {Kowalska}, \citenamefont {Bulik}, \citenamefont {Belczynski}, \citenamefont
  {Dominik},\ and\ \citenamefont {Gondek-Rosinska}}]{Kowalska:2010qg}%
  \BibitemOpen
  \bibfield  {author} {\bibinfo {author} {\bibfnamefont {I.}~\bibnamefont
  {Kowalska}}, \bibinfo {author} {\bibfnamefont {T.}~\bibnamefont {Bulik}},
  \bibinfo {author} {\bibfnamefont {K.}~\bibnamefont {Belczynski}}, \bibinfo
  {author} {\bibfnamefont {M.}~\bibnamefont {Dominik}},\ and\ \bibinfo {author}
  {\bibfnamefont {D.}~\bibnamefont {Gondek-Rosinska}},\ }\bibfield  {title}
  {\bibinfo {title} {{The eccentricity distribution of compact binaries}},\
  }\href {https://doi.org/10.1051/0004-6361/201015777} {\bibfield  {journal}
  {\bibinfo  {journal} {Astron. Astrophys.}\ }\textbf {\bibinfo {volume}
  {527}},\ \bibinfo {pages} {A70} (\bibinfo {year} {2011})},\ \Eprint
  {https://arxiv.org/abs/1010.0511} {arXiv:1010.0511 [astro-ph.CO]}
  \BibitemShut {NoStop}%
\bibitem [{\citenamefont {Tak\'atsy}\ \emph {et~al.}(2019)\citenamefont
  {Tak\'atsy}, \citenamefont {B\'ecsy},\ and\ \citenamefont
  {Raffai}}]{Takatsy:2018euo}%
  \BibitemOpen
  \bibfield  {author} {\bibinfo {author} {\bibfnamefont {J.}~\bibnamefont
  {Tak\'atsy}}, \bibinfo {author} {\bibfnamefont {B.}~\bibnamefont {B\'ecsy}},\
  and\ \bibinfo {author} {\bibfnamefont {P.}~\bibnamefont {Raffai}},\
  }\bibfield  {title} {\bibinfo {title} {{Eccentricity distributions of
  eccentric binary black holes in galactic nuclei}},\ }\href
  {https://doi.org/10.1093/mnras/stz820} {\bibfield  {journal} {\bibinfo
  {journal} {Mon. Not. Roy. Astron. Soc.}\ }\textbf {\bibinfo {volume} {486}},\
  \bibinfo {pages} {570} (\bibinfo {year} {2019})},\ \Eprint
  {https://arxiv.org/abs/1812.04012} {arXiv:1812.04012 [astro-ph.HE]}
  \BibitemShut {NoStop}%
\bibitem [{\citenamefont {Samsing}\ and\ \citenamefont
  {Ramirez-Ruiz}(2017)}]{Samsing:2017rat}%
  \BibitemOpen
  \bibfield  {author} {\bibinfo {author} {\bibfnamefont {J.}~\bibnamefont
  {Samsing}}\ and\ \bibinfo {author} {\bibfnamefont {E.}~\bibnamefont
  {Ramirez-Ruiz}},\ }\bibfield  {title} {\bibinfo {title} {{On the Assembly
  Rate of Highly Eccentric Binary Black Hole Mergers}},\ }\href
  {https://doi.org/10.3847/2041-8213/aa6f0b} {\bibfield  {journal} {\bibinfo
  {journal} {Astrophys. J. Lett.}\ }\textbf {\bibinfo {volume} {840}},\
  \bibinfo {pages} {L14} (\bibinfo {year} {2017})},\ \Eprint
  {https://arxiv.org/abs/1703.09703} {arXiv:1703.09703 [astro-ph.HE]}
  \BibitemShut {NoStop}%
\bibitem [{\citenamefont {Wang}\ \emph {et~al.}(2022)\citenamefont {Wang},
  \citenamefont {Ruan}, \citenamefont {Yang}, \citenamefont {Guo},
  \citenamefont {Cai},\ and\ \citenamefont {Hu}}]{Wang:2020dkc}%
  \BibitemOpen
  \bibfield  {author} {\bibinfo {author} {\bibfnamefont {R.}~\bibnamefont
  {Wang}}, \bibinfo {author} {\bibfnamefont {W.-H.}\ \bibnamefont {Ruan}},
  \bibinfo {author} {\bibfnamefont {Q.}~\bibnamefont {Yang}}, \bibinfo {author}
  {\bibfnamefont {Z.-K.}\ \bibnamefont {Guo}}, \bibinfo {author} {\bibfnamefont
  {R.-G.}\ \bibnamefont {Cai}},\ and\ \bibinfo {author} {\bibfnamefont
  {B.}~\bibnamefont {Hu}},\ }\bibfield  {title} {\bibinfo {title} {{Hubble
  parameter estimation via dark sirens with the LISA-Taiji network}},\ }\href
  {https://doi.org/10.1093/nsr/nwab054} {\bibfield  {journal} {\bibinfo
  {journal} {Natl. Sci. Rev.}\ }\textbf {\bibinfo {volume} {9}},\ \bibinfo
  {pages} {nwab054} (\bibinfo {year} {2022})},\ \Eprint
  {https://arxiv.org/abs/2010.14732} {arXiv:2010.14732 [astro-ph.CO]}
  \BibitemShut {NoStop}%
\bibitem [{\citenamefont {Zhu}\ \emph {et~al.}(2022{\natexlab{a}})\citenamefont
  {Zhu}, \citenamefont {Hu}, \citenamefont {Wang}, \citenamefont {Zhang},
  \citenamefont {Li}, \citenamefont {Hendry},\ and\ \citenamefont
  {Mei}}]{Zhu:2021aat}%
  \BibitemOpen
  \bibfield  {author} {\bibinfo {author} {\bibfnamefont {L.-G.}\ \bibnamefont
  {Zhu}}, \bibinfo {author} {\bibfnamefont {Y.-M.}\ \bibnamefont {Hu}},
  \bibinfo {author} {\bibfnamefont {H.-T.}\ \bibnamefont {Wang}}, \bibinfo
  {author} {\bibfnamefont {J.-d.}\ \bibnamefont {Zhang}}, \bibinfo {author}
  {\bibfnamefont {X.-D.}\ \bibnamefont {Li}}, \bibinfo {author} {\bibfnamefont
  {M.}~\bibnamefont {Hendry}},\ and\ \bibinfo {author} {\bibfnamefont
  {J.}~\bibnamefont {Mei}},\ }\bibfield  {title} {\bibinfo {title}
  {{Constraining the cosmological parameters using gravitational wave
  observations of massive black hole binaries and statistical redshift
  information}},\ }\href {https://doi.org/10.1103/PhysRevResearch.4.013247}
  {\bibfield  {journal} {\bibinfo  {journal} {Phys. Rev. Res.}\ }\textbf
  {\bibinfo {volume} {4}},\ \bibinfo {pages} {013247} (\bibinfo {year}
  {2022}{\natexlab{a}})},\ \Eprint {https://arxiv.org/abs/2104.11956}
  {arXiv:2104.11956 [astro-ph.CO]} \BibitemShut {NoStop}%
\bibitem [{\citenamefont {Zhu}\ \emph {et~al.}(2022{\natexlab{b}})\citenamefont
  {Zhu}, \citenamefont {Xie}, \citenamefont {Hu}, \citenamefont {Liu},
  \citenamefont {Li}, \citenamefont {Napolitano}, \citenamefont {Tang},
  \citenamefont {Zhang},\ and\ \citenamefont {Mei}}]{Zhu:2021bpp}%
  \BibitemOpen
  \bibfield  {author} {\bibinfo {author} {\bibfnamefont {L.-G.}\ \bibnamefont
  {Zhu}}, \bibinfo {author} {\bibfnamefont {L.-H.}\ \bibnamefont {Xie}},
  \bibinfo {author} {\bibfnamefont {Y.-M.}\ \bibnamefont {Hu}}, \bibinfo
  {author} {\bibfnamefont {S.}~\bibnamefont {Liu}}, \bibinfo {author}
  {\bibfnamefont {E.-K.}\ \bibnamefont {Li}}, \bibinfo {author} {\bibfnamefont
  {N.~R.}\ \bibnamefont {Napolitano}}, \bibinfo {author} {\bibfnamefont
  {B.-T.}\ \bibnamefont {Tang}}, \bibinfo {author} {\bibfnamefont {J.-d.}\
  \bibnamefont {Zhang}},\ and\ \bibinfo {author} {\bibfnamefont
  {J.}~\bibnamefont {Mei}},\ }\bibfield  {title} {\bibinfo {title}
  {{Constraining the Hubble constant to a precision of about 1\% using
  multi-band dark standard siren detections}},\ }\href
  {https://doi.org/10.1007/s11433-021-1859-9} {\bibfield  {journal} {\bibinfo
  {journal} {Sci. China Phys. Mech. Astron.}\ }\textbf {\bibinfo {volume}
  {65}},\ \bibinfo {pages} {259811} (\bibinfo {year} {2022}{\natexlab{b}})},\
  \Eprint {https://arxiv.org/abs/2110.05224} {arXiv:2110.05224 [astro-ph.CO]}
  \BibitemShut {NoStop}%
\bibitem [{\citenamefont {Hirata}\ \emph {et~al.}(2010)\citenamefont {Hirata},
  \citenamefont {Holz},\ and\ \citenamefont {Cutler}}]{Hirata:2010ba}%
  \BibitemOpen
  \bibfield  {author} {\bibinfo {author} {\bibfnamefont {C.~M.}\ \bibnamefont
  {Hirata}}, \bibinfo {author} {\bibfnamefont {D.~E.}\ \bibnamefont {Holz}},\
  and\ \bibinfo {author} {\bibfnamefont {C.}~\bibnamefont {Cutler}},\
  }\bibfield  {title} {\bibinfo {title} {{Reducing the weak lensing noise for
  the gravitational wave Hubble diagram using the non-Gaussianity of the
  magnification distribution}},\ }\href
  {https://doi.org/10.1103/PhysRevD.81.124046} {\bibfield  {journal} {\bibinfo
  {journal} {Phys. Rev. D}\ }\textbf {\bibinfo {volume} {81}},\ \bibinfo
  {pages} {124046} (\bibinfo {year} {2010})},\ \Eprint
  {https://arxiv.org/abs/1004.3988} {arXiv:1004.3988 [astro-ph.CO]}
  \BibitemShut {NoStop}%
\bibitem [{\citenamefont {Tamanini}\ \emph {et~al.}(2016)\citenamefont
  {Tamanini}, \citenamefont {Caprini}, \citenamefont {Barausse}, \citenamefont
  {Sesana}, \citenamefont {Klein},\ and\ \citenamefont
  {Petiteau}}]{Tamanini:2016zlh}%
  \BibitemOpen
  \bibfield  {author} {\bibinfo {author} {\bibfnamefont {N.}~\bibnamefont
  {Tamanini}}, \bibinfo {author} {\bibfnamefont {C.}~\bibnamefont {Caprini}},
  \bibinfo {author} {\bibfnamefont {E.}~\bibnamefont {Barausse}}, \bibinfo
  {author} {\bibfnamefont {A.}~\bibnamefont {Sesana}}, \bibinfo {author}
  {\bibfnamefont {A.}~\bibnamefont {Klein}},\ and\ \bibinfo {author}
  {\bibfnamefont {A.}~\bibnamefont {Petiteau}},\ }\bibfield  {title} {\bibinfo
  {title} {{Science with the space-based interferometer eLISA. III: Probing the
  expansion of the Universe using gravitational wave standard sirens}},\ }\href
  {https://doi.org/10.1088/1475-7516/2016/04/002} {\bibfield  {journal}
  {\bibinfo  {journal} {JCAP}\ }\textbf {\bibinfo {volume} {04}},\ \bibinfo
  {pages} {002}},\ \Eprint {https://arxiv.org/abs/1601.07112} {arXiv:1601.07112
  [astro-ph.CO]} \BibitemShut {NoStop}%
\bibitem [{\citenamefont {Speri}\ \emph {et~al.}(2021)\citenamefont {Speri},
  \citenamefont {Tamanini}, \citenamefont {Caldwell}, \citenamefont {Gair},\
  and\ \citenamefont {Wang}}]{Speri:2020hwc}%
  \BibitemOpen
  \bibfield  {author} {\bibinfo {author} {\bibfnamefont {L.}~\bibnamefont
  {Speri}}, \bibinfo {author} {\bibfnamefont {N.}~\bibnamefont {Tamanini}},
  \bibinfo {author} {\bibfnamefont {R.~R.}\ \bibnamefont {Caldwell}}, \bibinfo
  {author} {\bibfnamefont {J.~R.}\ \bibnamefont {Gair}},\ and\ \bibinfo
  {author} {\bibfnamefont {B.}~\bibnamefont {Wang}},\ }\bibfield  {title}
  {\bibinfo {title} {{Testing the Quasar Hubble Diagram with LISA Standard
  Sirens}},\ }\href {https://doi.org/10.1103/PhysRevD.103.083526} {\bibfield
  {journal} {\bibinfo  {journal} {Phys. Rev. D}\ }\textbf {\bibinfo {volume}
  {103}},\ \bibinfo {pages} {083526} (\bibinfo {year} {2021})},\ \Eprint
  {https://arxiv.org/abs/2010.09049} {arXiv:2010.09049 [astro-ph.CO]}
  \BibitemShut {NoStop}%
\bibitem [{\citenamefont {Kocsis}\ \emph {et~al.}(2006)\citenamefont {Kocsis},
  \citenamefont {Frei}, \citenamefont {Haiman},\ and\ \citenamefont
  {Menou}}]{Kocsis:2005vv}%
  \BibitemOpen
  \bibfield  {author} {\bibinfo {author} {\bibfnamefont {B.}~\bibnamefont
  {Kocsis}}, \bibinfo {author} {\bibfnamefont {Z.}~\bibnamefont {Frei}},
  \bibinfo {author} {\bibfnamefont {Z.}~\bibnamefont {Haiman}},\ and\ \bibinfo
  {author} {\bibfnamefont {K.}~\bibnamefont {Menou}},\ }\bibfield  {title}
  {\bibinfo {title} {{Finding the electromagnetic counterparts of cosmological
  standard sirens}},\ }\href {https://doi.org/10.1086/498236} {\bibfield
  {journal} {\bibinfo  {journal} {Astrophys. J.}\ }\textbf {\bibinfo {volume}
  {637}},\ \bibinfo {pages} {27} (\bibinfo {year} {2006})},\ \Eprint
  {https://arxiv.org/abs/astro-ph/0505394} {arXiv:astro-ph/0505394}
  \BibitemShut {NoStop}%
\bibitem [{\citenamefont {Torrado}\ and\ \citenamefont
  {Lewis}(2021)}]{Torrado:2020dgo}%
  \BibitemOpen
  \bibfield  {author} {\bibinfo {author} {\bibfnamefont {J.}~\bibnamefont
  {Torrado}}\ and\ \bibinfo {author} {\bibfnamefont {A.}~\bibnamefont
  {Lewis}},\ }\bibfield  {title} {\bibinfo {title} {{Cobaya: Code for Bayesian
  Analysis of hierarchical physical models}},\ }\href
  {https://doi.org/10.1088/1475-7516/2021/05/057} {\bibfield  {journal}
  {\bibinfo  {journal} {JCAP}\ }\textbf {\bibinfo {volume} {05}},\ \bibinfo
  {pages} {057}},\ \Eprint {https://arxiv.org/abs/2005.05290} {arXiv:2005.05290
  [astro-ph.IM]} \BibitemShut {NoStop}%
\bibitem [{\citenamefont {{Torrado}}\ and\ \citenamefont
  {{Lewis}}(2019)}]{2019ascl.soft10019T}%
  \BibitemOpen
  \bibfield  {author} {\bibinfo {author} {\bibfnamefont {J.}~\bibnamefont
  {{Torrado}}}\ and\ \bibinfo {author} {\bibfnamefont {A.}~\bibnamefont
  {{Lewis}}},\ }\href@noop {} {\bibinfo {title} {{Cobaya: Bayesian analysis in
  cosmology}}},\ \bibinfo {howpublished} {Astrophysics Source Code Library,
  record ascl:1910.019} (\bibinfo {year} {2019}),\ \Eprint
  {https://arxiv.org/abs/1910.019} {ascl:1910.019} \BibitemShut {NoStop}%
\bibitem [{\citenamefont {Lewis}(2019)}]{Lewis:2019xzd}%
  \BibitemOpen
  \bibfield  {author} {\bibinfo {author} {\bibfnamefont {A.}~\bibnamefont
  {Lewis}},\ }\bibfield  {title} {\bibinfo {title} {{GetDist: a Python package
  for analysing Monte Carlo samples}},\ }\href@noop {} {\  (\bibinfo {year}
  {2019})},\ \Eprint {https://arxiv.org/abs/1910.13970} {arXiv:1910.13970
  [astro-ph.IM]} \BibitemShut {NoStop}%
\bibitem [{\citenamefont {Belgacem}\ \emph
  {et~al.}(2018{\natexlab{b}})\citenamefont {Belgacem}, \citenamefont {Dirian},
  \citenamefont {Foffa},\ and\ \citenamefont {Maggiore}}]{Belgacem:2018lbp}%
  \BibitemOpen
  \bibfield  {author} {\bibinfo {author} {\bibfnamefont {E.}~\bibnamefont
  {Belgacem}}, \bibinfo {author} {\bibfnamefont {Y.}~\bibnamefont {Dirian}},
  \bibinfo {author} {\bibfnamefont {S.}~\bibnamefont {Foffa}},\ and\ \bibinfo
  {author} {\bibfnamefont {M.}~\bibnamefont {Maggiore}},\ }\bibfield  {title}
  {\bibinfo {title} {{Modified gravitational-wave propagation and standard
  sirens}},\ }\href {https://doi.org/10.1103/PhysRevD.98.023510} {\bibfield
  {journal} {\bibinfo  {journal} {Phys. Rev. D}\ }\textbf {\bibinfo {volume}
  {98}},\ \bibinfo {pages} {023510} (\bibinfo {year} {2018}{\natexlab{b}})},\
  \Eprint {https://arxiv.org/abs/1805.08731} {arXiv:1805.08731 [gr-qc]}
  \BibitemShut {NoStop}%
\bibitem [{\citenamefont {Nishizawa}(2018)}]{Nishizawa:2017nef}%
  \BibitemOpen
  \bibfield  {author} {\bibinfo {author} {\bibfnamefont {A.}~\bibnamefont
  {Nishizawa}},\ }\bibfield  {title} {\bibinfo {title} {{Generalized framework
  for testing gravity with gravitational-wave propagation. I. Formulation}},\
  }\href {https://doi.org/10.1103/PhysRevD.97.104037} {\bibfield  {journal}
  {\bibinfo  {journal} {Phys. Rev. D}\ }\textbf {\bibinfo {volume} {97}},\
  \bibinfo {pages} {104037} (\bibinfo {year} {2018})},\ \Eprint
  {https://arxiv.org/abs/1710.04825} {arXiv:1710.04825 [gr-qc]} \BibitemShut
  {NoStop}%
\bibitem [{\citenamefont {Arai}\ and\ \citenamefont
  {Nishizawa}(2018)}]{Arai:2017hxj}%
  \BibitemOpen
  \bibfield  {author} {\bibinfo {author} {\bibfnamefont {S.}~\bibnamefont
  {Arai}}\ and\ \bibinfo {author} {\bibfnamefont {A.}~\bibnamefont
  {Nishizawa}},\ }\bibfield  {title} {\bibinfo {title} {{Generalized framework
  for testing gravity with gravitational-wave propagation. II. Constraints on
  Horndeski theory}},\ }\href {https://doi.org/10.1103/PhysRevD.97.104038}
  {\bibfield  {journal} {\bibinfo  {journal} {Phys. Rev. D}\ }\textbf {\bibinfo
  {volume} {97}},\ \bibinfo {pages} {104038} (\bibinfo {year} {2018})},\
  \Eprint {https://arxiv.org/abs/1711.03776} {arXiv:1711.03776 [gr-qc]}
  \BibitemShut {NoStop}%
\bibitem [{\citenamefont {Nishizawa}\ and\ \citenamefont
  {Arai}(2019)}]{Nishizawa:2019rra}%
  \BibitemOpen
  \bibfield  {author} {\bibinfo {author} {\bibfnamefont {A.}~\bibnamefont
  {Nishizawa}}\ and\ \bibinfo {author} {\bibfnamefont {S.}~\bibnamefont
  {Arai}},\ }\bibfield  {title} {\bibinfo {title} {{Generalized framework for
  testing gravity with gravitational-wave propagation. III. Future prospect}},\
  }\href {https://doi.org/10.1103/PhysRevD.99.104038} {\bibfield  {journal}
  {\bibinfo  {journal} {Phys. Rev. D}\ }\textbf {\bibinfo {volume} {99}},\
  \bibinfo {pages} {104038} (\bibinfo {year} {2019})},\ \Eprint
  {https://arxiv.org/abs/1901.08249} {arXiv:1901.08249 [gr-qc]} \BibitemShut
  {NoStop}%
\bibitem [{\citenamefont {Belgacem}\ \emph
  {et~al.}(2019{\natexlab{b}})\citenamefont {Belgacem} \emph
  {et~al.}}]{LISACosmologyWorkingGroup:2019mwx}%
  \BibitemOpen
  \bibfield  {author} {\bibinfo {author} {\bibfnamefont {E.}~\bibnamefont
  {Belgacem}} \emph {et~al.} (\bibinfo {collaboration} {LISA Cosmology Working
  Group}),\ }\bibfield  {title} {\bibinfo {title} {{Testing modified gravity at
  cosmological distances with LISA standard sirens}},\ }\href
  {https://doi.org/10.1088/1475-7516/2019/07/024} {\bibfield  {journal}
  {\bibinfo  {journal} {JCAP}\ }\textbf {\bibinfo {volume} {07}},\ \bibinfo
  {pages} {024}},\ \Eprint {https://arxiv.org/abs/1906.01593} {arXiv:1906.01593
  [astro-ph.CO]} \BibitemShut {NoStop}%
\bibitem [{\citenamefont {D'Agostino}\ and\ \citenamefont
  {Nunes}(2019)}]{DAgostino:2019hvh}%
  \BibitemOpen
  \bibfield  {author} {\bibinfo {author} {\bibfnamefont {R.}~\bibnamefont
  {D'Agostino}}\ and\ \bibinfo {author} {\bibfnamefont {R.~C.}\ \bibnamefont
  {Nunes}},\ }\bibfield  {title} {\bibinfo {title} {{Probing observational
  bounds on scalar-tensor theories from standard sirens}},\ }\href
  {https://doi.org/10.1103/PhysRevD.100.044041} {\bibfield  {journal} {\bibinfo
   {journal} {Phys. Rev. D}\ }\textbf {\bibinfo {volume} {100}},\ \bibinfo
  {pages} {044041} (\bibinfo {year} {2019})},\ \Eprint
  {https://arxiv.org/abs/1907.05516} {arXiv:1907.05516 [gr-qc]} \BibitemShut
  {NoStop}%
\bibitem [{\citenamefont {Bonilla}\ \emph {et~al.}(2020)\citenamefont
  {Bonilla}, \citenamefont {D'Agostino}, \citenamefont {Nunes},\ and\
  \citenamefont {de~Araujo}}]{Bonilla:2019mbm}%
  \BibitemOpen
  \bibfield  {author} {\bibinfo {author} {\bibfnamefont {A.}~\bibnamefont
  {Bonilla}}, \bibinfo {author} {\bibfnamefont {R.}~\bibnamefont {D'Agostino}},
  \bibinfo {author} {\bibfnamefont {R.~C.}\ \bibnamefont {Nunes}},\ and\
  \bibinfo {author} {\bibfnamefont {J.~C.~N.}\ \bibnamefont {de~Araujo}},\
  }\bibfield  {title} {\bibinfo {title} {{Forecasts on the speed of
  gravitational waves at high $z$}},\ }\href
  {https://doi.org/10.1088/1475-7516/2020/03/015} {\bibfield  {journal}
  {\bibinfo  {journal} {JCAP}\ }\textbf {\bibinfo {volume} {03}},\ \bibinfo
  {pages} {015}},\ \Eprint {https://arxiv.org/abs/1910.05631} {arXiv:1910.05631
  [gr-qc]} \BibitemShut {NoStop}%
\bibitem [{\citenamefont {Mukherjee}\ \emph
  {et~al.}(2021{\natexlab{b}})\citenamefont {Mukherjee}, \citenamefont
  {Wandelt},\ and\ \citenamefont {Silk}}]{Mukherjee:2020mha}%
  \BibitemOpen
  \bibfield  {author} {\bibinfo {author} {\bibfnamefont {S.}~\bibnamefont
  {Mukherjee}}, \bibinfo {author} {\bibfnamefont {B.~D.}\ \bibnamefont
  {Wandelt}},\ and\ \bibinfo {author} {\bibfnamefont {J.}~\bibnamefont
  {Silk}},\ }\bibfield  {title} {\bibinfo {title} {{Testing the general theory
  of relativity using gravitational wave propagation from dark standard
  sirens}},\ }\href {https://doi.org/10.1093/mnras/stab001} {\bibfield
  {journal} {\bibinfo  {journal} {Mon. Not. Roy. Astron. Soc.}\ }\textbf
  {\bibinfo {volume} {502}},\ \bibinfo {pages} {1136} (\bibinfo {year}
  {2021}{\natexlab{b}})},\ \Eprint {https://arxiv.org/abs/2012.15316}
  {arXiv:2012.15316 [astro-ph.CO]} \BibitemShut {NoStop}%
\bibitem [{\citenamefont {Kalomenopoulos}\ \emph {et~al.}(2021)\citenamefont
  {Kalomenopoulos}, \citenamefont {Khochfar}, \citenamefont {Gair},\ and\
  \citenamefont {Arai}}]{Kalomenopoulos:2020klp}%
  \BibitemOpen
  \bibfield  {author} {\bibinfo {author} {\bibfnamefont {M.}~\bibnamefont
  {Kalomenopoulos}}, \bibinfo {author} {\bibfnamefont {S.}~\bibnamefont
  {Khochfar}}, \bibinfo {author} {\bibfnamefont {J.}~\bibnamefont {Gair}},\
  and\ \bibinfo {author} {\bibfnamefont {S.}~\bibnamefont {Arai}},\ }\bibfield
  {title} {\bibinfo {title} {{Mapping the inhomogeneous Universe with standard
  sirens: degeneracy between inhomogeneity and modified gravity theories}},\
  }\href {https://doi.org/10.1093/mnras/stab557} {\bibfield  {journal}
  {\bibinfo  {journal} {Mon. Not. Roy. Astron. Soc.}\ }\textbf {\bibinfo
  {volume} {503}},\ \bibinfo {pages} {3179} (\bibinfo {year} {2021})},\ \Eprint
  {https://arxiv.org/abs/2007.15020} {arXiv:2007.15020 [astro-ph.CO]}
  \BibitemShut {NoStop}%
\bibitem [{\citenamefont {Mastrogiovanni}\ \emph
  {et~al.}(2021{\natexlab{b}})\citenamefont {Mastrogiovanni}, \citenamefont
  {Haegel}, \citenamefont {Karathanasis}, \citenamefont {Hernandez},\ and\
  \citenamefont {Steer}}]{Mastrogiovanni:2020mvm}%
  \BibitemOpen
  \bibfield  {author} {\bibinfo {author} {\bibfnamefont {S.}~\bibnamefont
  {Mastrogiovanni}}, \bibinfo {author} {\bibfnamefont {L.}~\bibnamefont
  {Haegel}}, \bibinfo {author} {\bibfnamefont {C.}~\bibnamefont
  {Karathanasis}}, \bibinfo {author} {\bibfnamefont {I.~M.~n.}\ \bibnamefont
  {Hernandez}},\ and\ \bibinfo {author} {\bibfnamefont {D.~A.}\ \bibnamefont
  {Steer}},\ }\bibfield  {title} {\bibinfo {title} {{Gravitational wave
  friction in light of GW170817 and GW190521}},\ }\href
  {https://doi.org/10.1088/1475-7516/2021/02/043} {\bibfield  {journal}
  {\bibinfo  {journal} {JCAP}\ }\textbf {\bibinfo {volume} {02}},\ \bibinfo
  {pages} {043}},\ \Eprint {https://arxiv.org/abs/2010.04047} {arXiv:2010.04047
  [gr-qc]} \BibitemShut {NoStop}%
\bibitem [{\citenamefont {Mastrogiovanni}\ \emph {et~al.}(2020)\citenamefont
  {Mastrogiovanni}, \citenamefont {Steer},\ and\ \citenamefont
  {Barsuglia}}]{Mastrogiovanni:2020gua}%
  \BibitemOpen
  \bibfield  {author} {\bibinfo {author} {\bibfnamefont {S.}~\bibnamefont
  {Mastrogiovanni}}, \bibinfo {author} {\bibfnamefont {D.}~\bibnamefont
  {Steer}},\ and\ \bibinfo {author} {\bibfnamefont {M.}~\bibnamefont
  {Barsuglia}},\ }\bibfield  {title} {\bibinfo {title} {{Probing modified
  gravity theories and cosmology using gravitational-waves and associated
  electromagnetic counterparts}},\ }\href
  {https://doi.org/10.1103/PhysRevD.102.044009} {\bibfield  {journal} {\bibinfo
   {journal} {Phys. Rev. D}\ }\textbf {\bibinfo {volume} {102}},\ \bibinfo
  {pages} {044009} (\bibinfo {year} {2020})},\ \Eprint
  {https://arxiv.org/abs/2004.01632} {arXiv:2004.01632 [gr-qc]} \BibitemShut
  {NoStop}%
\bibitem [{\citenamefont {Beutler}\ \emph {et~al.}(2011)\citenamefont
  {Beutler}, \citenamefont {Blake}, \citenamefont {Colless}, \citenamefont
  {Jones}, \citenamefont {Staveley-Smith}, \citenamefont {Campbell},
  \citenamefont {Parker}, \citenamefont {Saunders},\ and\ \citenamefont
  {Watson}}]{Beutler:2011hx}%
  \BibitemOpen
  \bibfield  {author} {\bibinfo {author} {\bibfnamefont {F.}~\bibnamefont
  {Beutler}}, \bibinfo {author} {\bibfnamefont {C.}~\bibnamefont {Blake}},
  \bibinfo {author} {\bibfnamefont {M.}~\bibnamefont {Colless}}, \bibinfo
  {author} {\bibfnamefont {D.~H.}\ \bibnamefont {Jones}}, \bibinfo {author}
  {\bibfnamefont {L.}~\bibnamefont {Staveley-Smith}}, \bibinfo {author}
  {\bibfnamefont {L.}~\bibnamefont {Campbell}}, \bibinfo {author}
  {\bibfnamefont {Q.}~\bibnamefont {Parker}}, \bibinfo {author} {\bibfnamefont
  {W.}~\bibnamefont {Saunders}},\ and\ \bibinfo {author} {\bibfnamefont
  {F.}~\bibnamefont {Watson}},\ }\bibfield  {title} {\bibinfo {title} {{The 6dF
  Galaxy Survey: Baryon Acoustic Oscillations and the Local Hubble Constant}},\
  }\href {https://doi.org/10.1111/j.1365-2966.2011.19250.x} {\bibfield
  {journal} {\bibinfo  {journal} {Mon. Not. Roy. Astron. Soc.}\ }\textbf
  {\bibinfo {volume} {416}},\ \bibinfo {pages} {3017} (\bibinfo {year}
  {2011})},\ \Eprint {https://arxiv.org/abs/1106.3366} {arXiv:1106.3366
  [astro-ph.CO]} \BibitemShut {NoStop}%
\bibitem [{\citenamefont {Ross}\ \emph {et~al.}(2015)\citenamefont {Ross},
  \citenamefont {Samushia}, \citenamefont {Howlett}, \citenamefont {Percival},
  \citenamefont {Burden},\ and\ \citenamefont {Manera}}]{Ross:2014qpa}%
  \BibitemOpen
  \bibfield  {author} {\bibinfo {author} {\bibfnamefont {A.~J.}\ \bibnamefont
  {Ross}}, \bibinfo {author} {\bibfnamefont {L.}~\bibnamefont {Samushia}},
  \bibinfo {author} {\bibfnamefont {C.}~\bibnamefont {Howlett}}, \bibinfo
  {author} {\bibfnamefont {W.~J.}\ \bibnamefont {Percival}}, \bibinfo {author}
  {\bibfnamefont {A.}~\bibnamefont {Burden}},\ and\ \bibinfo {author}
  {\bibfnamefont {M.}~\bibnamefont {Manera}},\ }\bibfield  {title} {\bibinfo
  {title} {{The clustering of the SDSS DR7 main Galaxy sample \textendash{} I.
  A 4 per cent distance measure at $z = 0.15$}},\ }\href
  {https://doi.org/10.1093/mnras/stv154} {\bibfield  {journal} {\bibinfo
  {journal} {Mon. Not. Roy. Astron. Soc.}\ }\textbf {\bibinfo {volume} {449}},\
  \bibinfo {pages} {835} (\bibinfo {year} {2015})},\ \Eprint
  {https://arxiv.org/abs/1409.3242} {arXiv:1409.3242 [astro-ph.CO]}
  \BibitemShut {NoStop}%
\bibitem [{\citenamefont {Ross}\ \emph {et~al.}(2017)\citenamefont {Ross} \emph
  {et~al.}}]{BOSS:2016apd}%
  \BibitemOpen
  \bibfield  {author} {\bibinfo {author} {\bibfnamefont {A.~J.}\ \bibnamefont
  {Ross}} \emph {et~al.} (\bibinfo {collaboration} {BOSS}),\ }\bibfield
  {title} {\bibinfo {title} {{The clustering of galaxies in the completed
  SDSS-III Baryon Oscillation Spectroscopic Survey: Observational systematics
  and baryon acoustic oscillations in the correlation function}},\ }\href
  {https://doi.org/10.1093/mnras/stw2372} {\bibfield  {journal} {\bibinfo
  {journal} {Mon. Not. Roy. Astron. Soc.}\ }\textbf {\bibinfo {volume} {464}},\
  \bibinfo {pages} {1168} (\bibinfo {year} {2017})},\ \Eprint
  {https://arxiv.org/abs/1607.03145} {arXiv:1607.03145 [astro-ph.CO]}
  \BibitemShut {NoStop}%
\bibitem [{\citenamefont {Vargas-Maga\~na}\ \emph {et~al.}(2018)\citenamefont
  {Vargas-Maga\~na} \emph {et~al.}}]{Vargas-Magana:2016imr}%
  \BibitemOpen
  \bibfield  {author} {\bibinfo {author} {\bibfnamefont {M.}~\bibnamefont
  {Vargas-Maga\~na}} \emph {et~al.},\ }\bibfield  {title} {\bibinfo {title}
  {{The clustering of galaxies in the completed SDSS-III Baryon Oscillation
  Spectroscopic Survey: theoretical systematics and Baryon Acoustic
  Oscillations in the galaxy correlation function}},\ }\href
  {https://doi.org/10.1093/mnras/sty571} {\bibfield  {journal} {\bibinfo
  {journal} {Mon. Not. Roy. Astron. Soc.}\ }\textbf {\bibinfo {volume} {477}},\
  \bibinfo {pages} {1153} (\bibinfo {year} {2018})},\ \Eprint
  {https://arxiv.org/abs/1610.03506} {arXiv:1610.03506 [astro-ph.CO]}
  \BibitemShut {NoStop}%
\bibitem [{\citenamefont {Beutler}\ \emph {et~al.}(2017)\citenamefont {Beutler}
  \emph {et~al.}}]{BOSS:2016hvq}%
  \BibitemOpen
  \bibfield  {author} {\bibinfo {author} {\bibfnamefont {F.}~\bibnamefont
  {Beutler}} \emph {et~al.} (\bibinfo {collaboration} {BOSS}),\ }\bibfield
  {title} {\bibinfo {title} {{The clustering of galaxies in the completed
  SDSS-III Baryon Oscillation Spectroscopic Survey: baryon acoustic
  oscillations in the Fourier space}},\ }\href
  {https://doi.org/10.1093/mnras/stw2373} {\bibfield  {journal} {\bibinfo
  {journal} {Mon. Not. Roy. Astron. Soc.}\ }\textbf {\bibinfo {volume} {464}},\
  \bibinfo {pages} {3409} (\bibinfo {year} {2017})},\ \Eprint
  {https://arxiv.org/abs/1607.03149} {arXiv:1607.03149 [astro-ph.CO]}
  \BibitemShut {NoStop}%
\bibitem [{\citenamefont {Scolnic}\ \emph {et~al.}(2018)\citenamefont {Scolnic}
  \emph {et~al.}}]{Pan-STARRS1:2017jku}%
  \BibitemOpen
  \bibfield  {author} {\bibinfo {author} {\bibfnamefont {D.~M.}\ \bibnamefont
  {Scolnic}} \emph {et~al.} (\bibinfo {collaboration} {Pan-STARRS1}),\
  }\bibfield  {title} {\bibinfo {title} {{The Complete Light-curve Sample of
  Spectroscopically Confirmed SNe Ia from Pan-STARRS1 and Cosmological
  Constraints from the Combined Pantheon Sample}},\ }\href
  {https://doi.org/10.3847/1538-4357/aab9bb} {\bibfield  {journal} {\bibinfo
  {journal} {Astrophys. J.}\ }\textbf {\bibinfo {volume} {859}},\ \bibinfo
  {pages} {101} (\bibinfo {year} {2018})},\ \Eprint
  {https://arxiv.org/abs/1710.00845} {arXiv:1710.00845 [astro-ph.CO]}
  \BibitemShut {NoStop}%
\bibitem [{\citenamefont {Cao}\ and\ \citenamefont {Han}(2017)}]{Cao:2017ndf}%
  \BibitemOpen
  \bibfield  {author} {\bibinfo {author} {\bibfnamefont {Z.}~\bibnamefont
  {Cao}}\ and\ \bibinfo {author} {\bibfnamefont {W.-B.}\ \bibnamefont {Han}},\
  }\bibfield  {title} {\bibinfo {title} {{Waveform model for an eccentric
  binary black hole based on the effective-one-body-numerical-relativity
  formalism}},\ }\href {https://doi.org/10.1103/PhysRevD.96.044028} {\bibfield
  {journal} {\bibinfo  {journal} {Phys. Rev. D}\ }\textbf {\bibinfo {volume}
  {96}},\ \bibinfo {pages} {044028} (\bibinfo {year} {2017})},\ \Eprint
  {https://arxiv.org/abs/1708.00166} {arXiv:1708.00166 [gr-qc]} \BibitemShut
  {NoStop}%
\bibitem [{\citenamefont {Liu}\ \emph {et~al.}(2020)\citenamefont {Liu},
  \citenamefont {Cao},\ and\ \citenamefont {Shao}}]{Liu:2019jpg}%
  \BibitemOpen
  \bibfield  {author} {\bibinfo {author} {\bibfnamefont {X.}~\bibnamefont
  {Liu}}, \bibinfo {author} {\bibfnamefont {Z.}~\bibnamefont {Cao}},\ and\
  \bibinfo {author} {\bibfnamefont {L.}~\bibnamefont {Shao}},\ }\bibfield
  {title} {\bibinfo {title} {{Validating the Effective-One-Body
  Numerical-Relativity Waveform Models for Spin-aligned Binary Black Holes
  along Eccentric Orbits}},\ }\href
  {https://doi.org/10.1103/PhysRevD.101.044049} {\bibfield  {journal} {\bibinfo
   {journal} {Phys. Rev. D}\ }\textbf {\bibinfo {volume} {101}},\ \bibinfo
  {pages} {044049} (\bibinfo {year} {2020})},\ \Eprint
  {https://arxiv.org/abs/1910.00784} {arXiv:1910.00784 [gr-qc]} \BibitemShut
  {NoStop}%
\bibitem [{\citenamefont {Liu}\ \emph {et~al.}(2022{\natexlab{b}})\citenamefont
  {Liu}, \citenamefont {Cao},\ and\ \citenamefont {Zhu}}]{Liu:2021pkr}%
  \BibitemOpen
  \bibfield  {author} {\bibinfo {author} {\bibfnamefont {X.}~\bibnamefont
  {Liu}}, \bibinfo {author} {\bibfnamefont {Z.}~\bibnamefont {Cao}},\ and\
  \bibinfo {author} {\bibfnamefont {Z.-H.}\ \bibnamefont {Zhu}},\ }\bibfield
  {title} {\bibinfo {title} {{A higher-multipole gravitational waveform model
  for an eccentric binary black holes based on the
  effective-one-body-numerical-relativity formalism}},\ }\href
  {https://doi.org/10.1088/1361-6382/ac4119} {\bibfield  {journal} {\bibinfo
  {journal} {Class. Quant. Grav.}\ }\textbf {\bibinfo {volume} {39}},\ \bibinfo
  {pages} {035009} (\bibinfo {year} {2022}{\natexlab{b}})},\ \Eprint
  {https://arxiv.org/abs/2102.08614} {arXiv:2102.08614 [gr-qc]} \BibitemShut
  {NoStop}%
\bibitem [{\citenamefont {Ramos-Buades}\ \emph {et~al.}(2022)\citenamefont
  {Ramos-Buades}, \citenamefont {Buonanno}, \citenamefont {Khalil},\ and\
  \citenamefont {Ossokine}}]{Ramos-Buades:2021adz}%
  \BibitemOpen
  \bibfield  {author} {\bibinfo {author} {\bibfnamefont {A.}~\bibnamefont
  {Ramos-Buades}}, \bibinfo {author} {\bibfnamefont {A.}~\bibnamefont
  {Buonanno}}, \bibinfo {author} {\bibfnamefont {M.}~\bibnamefont {Khalil}},\
  and\ \bibinfo {author} {\bibfnamefont {S.}~\bibnamefont {Ossokine}},\
  }\bibfield  {title} {\bibinfo {title} {{Effective-one-body multipolar
  waveforms for eccentric binary black holes with nonprecessing spins}},\
  }\href {https://doi.org/10.1103/PhysRevD.105.044035} {\bibfield  {journal}
  {\bibinfo  {journal} {Phys. Rev. D}\ }\textbf {\bibinfo {volume} {105}},\
  \bibinfo {pages} {044035} (\bibinfo {year} {2022})},\ \Eprint
  {https://arxiv.org/abs/2112.06952} {arXiv:2112.06952 [gr-qc]} \BibitemShut
  {NoStop}%
\bibitem [{\citenamefont {Hinder}\ \emph {et~al.}(2018)\citenamefont {Hinder},
  \citenamefont {Kidder},\ and\ \citenamefont {Pfeiffer}}]{Hinder:2017sxy}%
  \BibitemOpen
  \bibfield  {author} {\bibinfo {author} {\bibfnamefont {I.}~\bibnamefont
  {Hinder}}, \bibinfo {author} {\bibfnamefont {L.~E.}\ \bibnamefont {Kidder}},\
  and\ \bibinfo {author} {\bibfnamefont {H.~P.}\ \bibnamefont {Pfeiffer}},\
  }\bibfield  {title} {\bibinfo {title} {{Eccentric binary black hole
  inspiral-merger-ringdown gravitational waveform model from numerical
  relativity and post-Newtonian theory}},\ }\href
  {https://doi.org/10.1103/PhysRevD.98.044015} {\bibfield  {journal} {\bibinfo
  {journal} {Phys. Rev. D}\ }\textbf {\bibinfo {volume} {98}},\ \bibinfo
  {pages} {044015} (\bibinfo {year} {2018})},\ \Eprint
  {https://arxiv.org/abs/1709.02007} {arXiv:1709.02007 [gr-qc]} \BibitemShut
  {NoStop}%
\bibitem [{\citenamefont {Moore}\ and\ \citenamefont
  {Yunes}(2019)}]{Moore:2019xkm}%
  \BibitemOpen
  \bibfield  {author} {\bibinfo {author} {\bibfnamefont {B.}~\bibnamefont
  {Moore}}\ and\ \bibinfo {author} {\bibfnamefont {N.}~\bibnamefont {Yunes}},\
  }\bibfield  {title} {\bibinfo {title} {{A 3PN Fourier Domain Waveform for
  Non-Spinning Binaries with Moderate Eccentricity}},\ }\href
  {https://doi.org/10.1088/1361-6382/ab3778} {\bibfield  {journal} {\bibinfo
  {journal} {Class. Quant. Grav.}\ }\textbf {\bibinfo {volume} {36}},\ \bibinfo
  {pages} {185003} (\bibinfo {year} {2019})},\ \Eprint
  {https://arxiv.org/abs/1903.05203} {arXiv:1903.05203 [gr-qc]} \BibitemShut
  {NoStop}%
\bibitem [{\citenamefont {Joshi}\ \emph {et~al.}(2022)\citenamefont {Joshi},
  \citenamefont {Rosofsky}, \citenamefont {Haas},\ and\ \citenamefont
  {Huerta}}]{Joshi:2022ocr}%
  \BibitemOpen
  \bibfield  {author} {\bibinfo {author} {\bibfnamefont {A.~V.}\ \bibnamefont
  {Joshi}}, \bibinfo {author} {\bibfnamefont {S.~G.}\ \bibnamefont {Rosofsky}},
  \bibinfo {author} {\bibfnamefont {R.}~\bibnamefont {Haas}},\ and\ \bibinfo
  {author} {\bibfnamefont {E.~A.}\ \bibnamefont {Huerta}},\ }\bibfield  {title}
  {\bibinfo {title} {{Numerical relativity higher order gravitational waveforms
  of eccentric, spinning, non-precessing binary black hole mergers}},\
  }\href@noop {} {\  (\bibinfo {year} {2022})},\ \Eprint
  {https://arxiv.org/abs/2210.01852} {arXiv:2210.01852 [gr-qc]} \BibitemShut
  {NoStop}%
\end{thebibliography}%

\end{document}